\documentclass[review]{elsarticle}
\usepackage[letterpaper, margin=1.in, top=1.1in]{geometry}
\usepackage{lineno,hyperref}

\usepackage{bm}
\usepackage{graphicx}
\usepackage{subfig}
\usepackage[mathscr]{euscript}
\usepackage[overload]{empheq}
\usepackage{pifont}
\newcommand{\cmark}{\ding{51}}%
\newcommand{\xmark}{\ding{55}}%

\makeatletter
\def\Cline#1#2{\@Cline#1#2\@nil}
\def\@Cline#1-#2#3\@nil{%
  \omit
  \@multicnt#1%
  \advance\@multispan\m@ne
  \ifnum\@multicnt=\@ne\@firstofone{&\omit}\fi
  \@multicnt#2%
  \advance\@multicnt-#1%
  \advance\@multispan\@ne
  \leaders\hrule\@height#3\hfill
  \cr}

\usepackage{ etoolbox, xparse}   
\DeclarePairedDelimiterX{\abs}[1]\lvert\rvert{\ifblank{#1}{\,\cdot\,}{#1}}
\let\oldabs\abs
\def\abs{\futurelet\testchar\MaybeOptArgAbs}
\def\MaybeOptArgAbs{\ifx[\testchar\let\next\OptArgAbs
\else \let\next\NoOptArgAbs\fi \next}
\def\OptArgAbs[#1]#2{\oldabs[#1]{#2}}
\def\NoOptArgAbs#1{\ifblank{#1}{\oldabs{}}{\oldabs[\big]{#1}}}

\DeclarePairedDelimiterX{\set}[1]\{\}{\setargs{#1}}
\NewDocumentCommand{\setargs}{>{\SplitArgument{1}{;}}m}
{\setargsaux#1}
\NewDocumentCommand{\setargsaux}{mm}
{\IfNoValueTF{#2}{#1}{\nonscript\,#1\nonscript\;\delimsize\vert\nonscript\:\allowbreak #2\nonscript\,}}
\let\oldset\set
\def\set{\futurelet\testchar\MaybeOptArgSet}
\def\MaybeOptArgSet{\ifx[\testchar \let\next\OptArgSet
\else \let\next\NoOptArgSet \fi \next}
\def\OptArgSet[#1]#2{\oldset[#1]{#2}}
\def\NoOptArgSet#1{\OptArgSet[\big]{#1}}

\begin{document}
%

\begin{frontmatter}

\title{A comprehensive high-order solver verification methodology \\for free fluid flows}

\author[]{Farshad Navah\corref{mycorrespondingauthor}}
\ead{farshad.navah@mail.mcgill.ca}

\author[]{Siva Nadarajah}
\address{Mechanical Engineering Department, \\McGill University,
Montr\'{e}al, Qu\'{e}bec, Canada, H3A 0C3}

\cortext[mycorrespondingauthor]{Corresponding author}




\begin{abstract}
{The aim of this article is to present a comprehensive methodology for the verification of computational fluid dynamics (CFD) solvers with a special attention to aspects pertinent to discretizations with orders of accuracy (OOAs) higher than two. The method of manufactured solutions (MMS) is adopted and a series of manufactured solutions (MSs) is introduced that examines various components of CFD solvers for free flows (not bounded by walls), including inviscid, laminar and  turbulent problems when the latter are modelled by the Reynolds-averaged Navier-Stokes (RANS) equations. The treatment of curved elements is also examined. These MSs are furthermore conceived with demonstrated suitability for the verification of OOAs up to the sixth. Each MS is as well utilized to discuss salient aspects useful to the code verification methodology such as the relative qualities of the most useful norms in measuring the discretization error, the sensitivity analysis of the verification process to forcing function terms, the relation between residual minimization and discretization error convergence in iterative solutions and finally the sensitivity of high-order discretizations to grid stretching and self-similarity. Furthermore, scripts and code are provided as accompanying material to assist the interested reader in reproducing the verification results of each manufactured solution (MS).}
\end{abstract}


\end{frontmatter}


\section{Introduction}
\label{sec:intro}
{Code verification is a crucial step prior to the application of a scientific simulation software to the solution of practical problems as it aims at examining the soundness of the implementation of the governing equations in the numerical framework. With the increasing interest of the research community in the design and application of high-order-of-accuracy discretization methods for CFD problems, there is an imperative need to extend the verification methodology to this class of methods. Code verification is in fact even more critical for higher-order methods since it is the only means to provide assurance that the effort invested in their design and development is justified by the delivery of the expected higher performance in terms of accuracy per computational effort. We hence  present in this paper the fundamental aspects towards a comprehensive code verification methodology for CFD solvers with all orders of accuracy. 

To carry out the demonstration of the methodology and without loss of generality, we choose the numerical framework composed of the compressible RANS equations closed by the original and modified versions of the Spalart-Allmaras (SA) model of turbulence and discretized by the correction procedure via flux reconstruction (CPR) scheme.}

The article is structured as this: in Section \ref{sec:intro}, the context and contributions of this work are introduced, followed by a comprehensive presentation of the theoretical background in verification and validation (V\&V) in Section \ref{sec:theo}. The governing equations as well as the compact high-order numerical method are respectively exhibited in Sections \ref{sec:goveq} and \ref{sec:num_meth}, including a precise description of all the employed boundary conditions. The application results of the verification and the discussion of the salient aspects of the methodology appear in Section  \ref{sec:cases} and the article ends with conclusions in the last Section.

\subsection{Contributions}
A series of trigonometric manufactured solutions for the sequential verification of high-order RANS solvers is devised such that it demonstrably achieves all OOAs up to the sixth order on moderately fine isotropic grids, without being trivially reached on the coarsest ones. Attention is invested in ensuring that the MSs produce a fair balance between different terms of the governing equations. The sequence of MSs targets constitutive components of solvers in an isolated fashion and with incremental complexity such that systematic debugging is enabled and gathering cumulative evidence on the soundness of high-order CFD solver implementation is made possible. The MSs serve thus to examine the implementation of Euler, Navier-Stokes (NS) and RANS equations along with the original and also with the modified SA model, for free flows, i.e., flows that are not bounded by walls. The set of MSs is as well employed to explore the following concepts:
\begin{itemize}
\item The comparative description of different norms and a demonstration of the importance of $L_\infty$ norm in code verification;
\item The need for the inclusion of a relatively high order of accuracy in code verification;
\item The significance of the balancing of forcing function terms of the MMS and the sensitivity analysis of the verification process to terms with the lowest magnitude in the forcing functions;
\item The verification of both the original and the modified SA models of turbulence;
\item The relation of residual convergence level with regards to discretization error magnitude and insight on the necessary level of residual convergence;
\item The examination of the treatment of non-affine mapping of curved elements;
\item The effect of grid self-similarity and stretching on grid convergence of solutions with smooth gradients.
\end{itemize}

Accompanying (see \cite{navah017_github}) IPython \cite{PER-GRA:2007} notebook and C routine facilitate the application of the described verification methodology through the reproduction of the manufactured fields and forcing functions of the presented MSs.

\section{Theoretical background}
\label{sec:theo}
In this section, first the terminology involved in V\&V is completed and defined further, the MMS is formalized then and finally, a short review of the previous works with a focus on verification via the MMS in CFD is presented.

\subsection{Terminology in V\&V}
\begin{figure}[!hbt]
\centering
\includegraphics[trim = 8mm 83mm 0mm 0mm, clip,width=0.5\linewidth]
{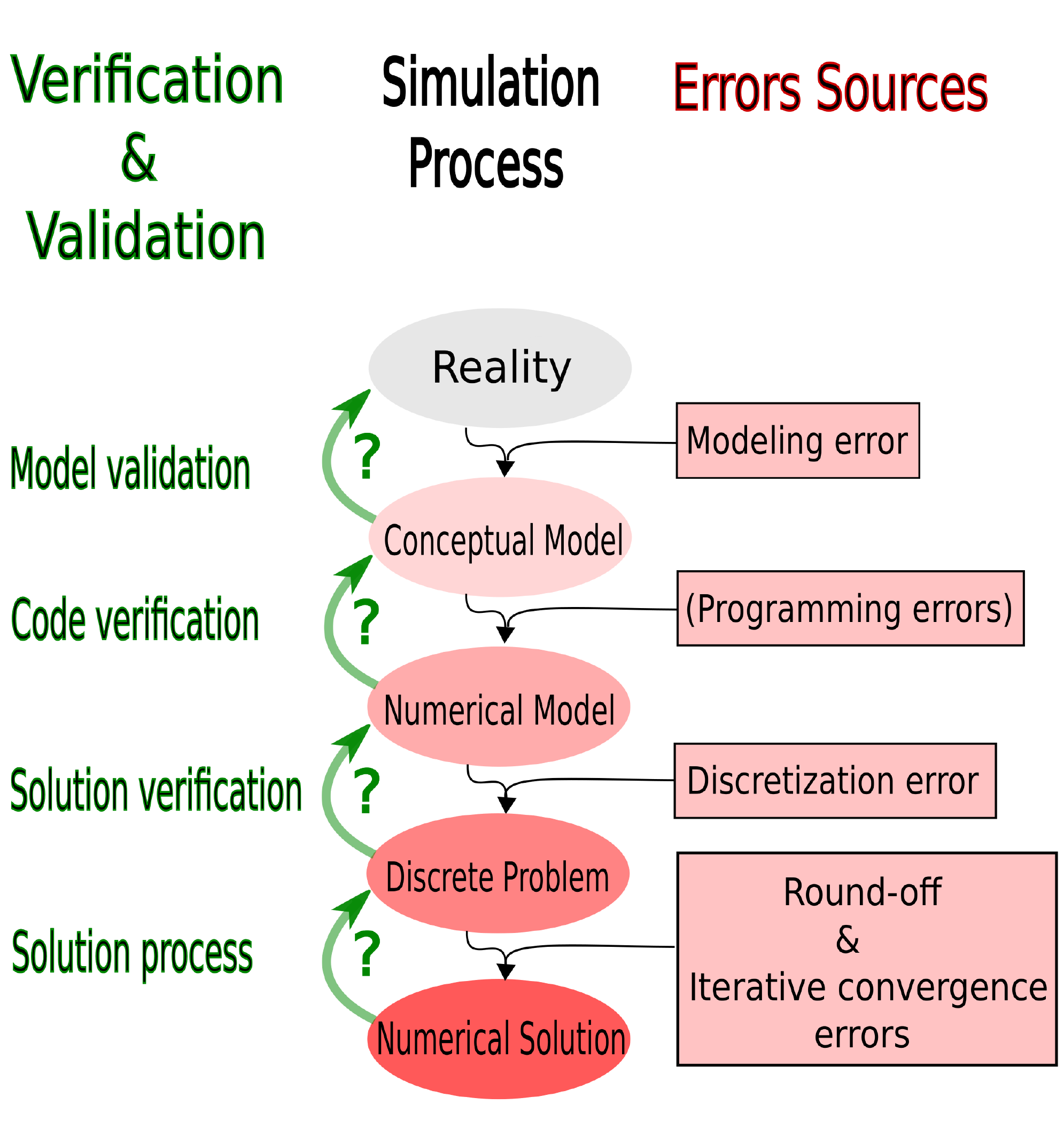}
\caption{Verification and validation in relation to sources of error in a scientific simulation}
\label{fig:V&V}
\end{figure}

Figure \ref{fig:V&V} illustrates the relation between major concepts of interest under the three themes of \textit{simulation process}, \textit{error sources}, as well as \textit{verification and validation}. 

As a scientific simulation process takes place, errors from various sources slip into its different steps, contaminating incrementally the outcome of the process. Abstractly, as the reality that we aim to capture cascades through a simulation, it diminishes at each step of the process. The role of V\&V is hence to ensure that the amount of original reality captured by the simulation is sufficient for the purpose that the simulation is meant to serve, by ideally providing a dependable measurement of the discrepancies. In what follows, we describe more precisely these ideas with reference to Fig. \ref{fig:V&V}.

Any scientific \textit{simulation process} starts from a \textit{reality}, a physical phenomenon that it aims at reproducing. Based on experimental data and previous theoretical knowledge, the relation between various quantities playing a role in the physical process is described by a \textit{conceptual model}, i.e., a series of mathematical equations such as the partial differentials equations (PDEs) of Euler, NS or those of RANS-based turbulence models. Almost in all cases, these models are a mere, yet hopefully reliable, approximation of the inherent complexity of physics and as such they contain a \textit{modelling error}. In this regard, \textit{model validation} questions how well the phenomenon of interest is approached by the conceptual model. For complex problems such as those encountered in CFD, this is however only considerable once the simulation process has ended and a numerical solution is available. In order to solve the conceptual model, a numerical algorithm, a scheme, is often needed. The application of this numerical recipe to the conceptual model yields a \textit{numerical model} such as the discretization of the RANS-SA system of PDEs by the CPR scheme. As the size of the discrete problem is increased for well-posed and smooth solutions, a numerical model is expected to tend towards the conceptual model with a rate known as the \textit{formal order of accuracy}. A numerical model with suitable properties such as stability and efficiency is translated to a computer code. Considering the complexity of the numerical model, \textit{programming errors} often occur at this step.  Indeed, according to an exhaustive analysis of the quality of scientific computing codes: "There were about 8 serious static faults per 1000 lines of executable lines in C, and about 12 serious faults per 1000 lines in Fortran"\cite{Oberkampf2002}.  \textit{Code verification} has for purpose to identify and eliminate the mistakes affecting the correspondence between the scientific software and the conceptual model via the formal order of accuracy. Similarly to model validation, code verification relies on the numerical solution of specific problems. To solve a given problem, the spatial and temporal domains are discretized by a set of points, called \textit{degrees of freedom} (DOFs), to which the discrete solution is associated. The \textit{discretization error} is the difference between the continuous and discrete solutions and the \textit{solution verification} is the estimation of this error for a given solution. The \textit{solution process} refers to the minimization of discrete residual equations by iterative methods, mandatory for tackling non-linear systems, and by algorithms handling linear algebraic systems. The truncation of real values for representation on computer architectures, by double precision types for example, introduces a \textit{round-off error} that affects the numerical solution by propagating through the discrete equations. On the other hand, a lack of sufficient minimization of the discrete residual equations results in an \textit{iterative convergence error} that imposes a gap between the achieved numerical solution and the actual solution of the discrete problem. Both the round-off and iterative convergence errors need to be controlled during the solution process to ensure that these sources of error are minimized such that the discretization error is isolated as the major source of numerical error. This condition enables code and solution verifications to be carried out since they operate only on the discretization error and its rate of reduction for increasing DOFs.

\subsubsection{Code verification methods}
The evaluation of OOAs in code verification requires the knowledge of the exact solution which could be provided by devising a problem for which such a solution exists. This approach is called the method of analytic solutions (MAS) which has the advantage of avoiding code modification and hence being non-intrusive. However, MAS is not always applicable as in the case of RANS equations for which analytic solutions of interest do not exist. Whenever applicable, MAS often involves simplified solutions that might not properly exercise all the governing equation terms. An example of an analytic solution for the NS equations is the Couette problem \cite{vermeire-2014}. 

An alternative and more powerful approach is via the MMS which enables the verification of any arbitrary set of governing equations but which requires point-wise code modifications and hence is considered to be an intrusive method. Other advantages of the MMS are:
\begin{itemize}
\item The possibility of debugging by systematic simplification and elimination of the governing equation terms;
\item The examination of arbitrary boundary conditions;
\item The verification of specific scheme properties by devising specific MSs.
\end{itemize} 

For these reasons, we adopt the MMS as the verification method of choice.

\subsection{Verification via the MMS}
Consider a linear differential operator, $\mathcal{A}$, producing a homogeneous output, uniquely when applied to $Q^{\mathrm{ex}}$, a time-invariant and smooth scalar field on a bounded spatial domain, viz.,
\begin{equation}
\mathcal{A} (Q^{\mathrm{ex}})= 0.
\label{eq:sample_pde}
\end{equation}
For a tessellation of the domain augmented by a polynomial expansion of degree P resulting in a representative resolution size, $h$, the equivalent discrete equation reads
\begin{equation}
\mathcal{A}^{h}(Q^{h}) = \epsilon\approx0,
\label{eq:sample_pde2}
\end{equation}
where $\mathcal{A}^{h}$ and $Q^{h}$  are respectively the discrete operator and solution and
$\epsilon$ is a finite precision representation of zero such as the \textit{machine precision}. The truncation error is defined as
\begin{equation}
\Theta_\mathcal{A}^{{h}} = \mathcal{A}^{h} (Q^{\mathrm{ex}}) - \mathcal{A} (Q^{\mathrm{ex}}) = \mathcal{A}^{h} (Q^{\mathrm{ex}}).
\label{eq:sample_pde3}
\end{equation}

According to the Lax equivalence theorem, if $\mathcal{A}^{h}$ is stable, i.e., its variations with regards to finite variations of $Q^{h}$ are bounded, and if $\mathcal{A}^{h}$ is furthermore consistent, i.e., \[\mathrm{lim}_{{h}\rightarrow 0} \Theta_\mathcal{A}^{{h}} = \mathcal{A} (Q^{\mathrm{ex}})=0,\] sufficient conditions are fulfilled for the discrete solution to be convergent, i.e., 
\[\mathrm{lim}_{{h}\rightarrow 0} Q^{{h}} = Q^{\mathrm{ex}},\]
and hence the discretization error, $\mathcal{E}_Q= (Q^{{h}} - Q^{\mathrm{ex}} )$,   tends to zero with $\|\mathcal{E}\| \sim \mathcal{O}(h^o)$ where $\|\cdot\|$ is a measurement in an appropriate norm and $o$ is the formal rate of convergence (order of accuracy).

Now, let's apply the operator $\mathcal{A}$ to an arbitrary smooth scalar field, $Q^{\mathrm{MS}} \neq Q^{\mathrm{ex}}$. Obviously,
\begin{equation*}
\mathcal{A} (Q^{\mathrm{MS}}) \neq 0.
\label{eq:sample_pde4}
\end{equation*}
But, Eq. \eqref{eq:sample_pde} can be modified by the addition of a suitable \textit{forcing function}, $\mathcal{F}(Q^{\mathrm{MS}})$, such that
\begin{equation}
\mathcal{R}(Q^{\mathrm{MS}})  = \mathcal{A} (Q^{\mathrm{MS}})  - \mathcal{F}(Q^{\mathrm{MS}})= 0.
\label{eq:sample_pde1b}
\end{equation}
where $\mathcal{R}$ designates the modified operator. An immediate choice of forcing function is $\mathcal{F}(Q^{\mathrm{MS}})= \mathcal{A} (Q^{\mathrm{MS}}) $. The modified counterparts of the discrete equation \eqref{eq:sample_pde2} and the truncation error  \eqref{eq:sample_pde3} respectively are 
\begin{equation}
\mathcal{R}^{h}(Q^{h}) = \mathcal{A}^{h}(Q^{h}) - \mathcal{F}(Q^{\mathrm{MS}})  = \epsilon\approx0,
\label{eq:sample_pde2b}
\end{equation}
and
\begin{equation}
\Theta_\mathcal{R}^{{h}} = \mathcal{R}^{h} (Q^{\mathrm{\mathrm{MS}}}) - \mathcal{R} (Q^{\mathrm{\mathrm{MS}}}) = \mathcal{R}^{h} (Q^{\mathrm{\mathrm{MS}}}),
\label{eq:sample_pde3b}
\end{equation}
via Eq. \eqref{eq:sample_pde1b}.

The stability and consistency of $\mathcal{R}^{h}$ follow from those of $\mathcal{A}^{h}$ and hence the convergence to the manufactured solution is established as
\[\mathrm{lim}_{h\rightarrow 0} Q^{{h}} = Q^{\mathrm{MS}},\]
and $\|\mathcal{E}_Q\| = \|Q^{{h}} - Q^{\mathrm{MS}}\| \sim \mathcal{O}(h^o)$ is expected. The reader is referred to the literature for further information on the value of the formal OOAs in finite difference and finite volume frameworks. As for continuous finite element and discontinuous compact variational methods, we refer to the comprehensive analyses in \cite{Arnold-et-al_2002,RalfHartmann2008}. A generic result for such methods applied to linear (and by extension to non-linear) advection-diffusion problems is that the expected order in $L_1$, $L_2$ and $L_\infty$ norms is $o=\mathrm{P}+1$, i.e., $\mathcal{E} \sim \mathcal{O}(h^{\mathrm{P}+1})$, whereas it rather is $o=\mathrm{P}$ in $H_1$ norm and semi-norm. The definition of each norm is provided in  \ref{sec:norms}.

Let's note that the focus of this work is on the verification of the steady-state RANS equations since a semi-discrete scheme is considered. In other words, the integration in time is segregated from the spatial discretization and could hence be verified in an independent fashion.
\subsection{Review of the literature}
The V\&V in CFD originates from pioneering and comprehensive contributions of Roache \cite{roache-1998}, Oberkampf et al. \cite{Oberkampf2002,oberkampf2010verification}, Salari and Knupp \cite{Salari} as well as Roy et al. \cite{Roy2002,Roy2005}. The applications of the MMS via trigonometric manufactured solutions are reported for Euler and NS solvers \cite{Roy2002,Roy2004,Salari,Bond2007,Shunn2007,Vedovoto2011,Ulerich2012}, for RANS solvers \cite{Roy2007,Bond2007,Veluri2012,Oliver-et-al_2012},  for fluid-structure interaction \cite{tremblay2006} and more recently for multiphase flows \cite{Choudhary2016}. Physically realistic manufactured solutions for the verification of RANS-based turbulence models, that mimic a turbulent boundary layer in the vicinity of no-slip wall, are introduced by E\c{c}a et al. \cite{Eca2007a,Eca2007,Eca2012,Eca2014,Eca2016} as well as by Oliver and colleagues \cite{Oliver-et-al_2012}. Finally, a library for verification via MMS is presented in \cite{Malaya2013}. The mentioned applications involve low-order finite volume and finite element codes. Applications of the MMS to the verification of high-order spatial discretizations are found in \cite{Fidkowski2013,Manuel2014,Galbraith2015,Navah_eccomas_016} for DG/FR/CPR methods and in \cite{silva2010} for mixed finite differences methods. However, these applications are of limited scope as each targets a few specific solver features at a time. 


\section{Governing equations}\label{sec:goveq}

The governing equations are the steady-state, compressible RANS equations assuming a Newtonian and calorically perfect gas along with the conservative form of the original and revised SA models of turbulence \cite{Allmaras-et-al_2012}. The SA model is chosen due to its simplicity and proven effectiveness for the simulation of aerodynamic flows. The revised SA model supplements the original model by conditional variations of some of its terms in order to provide enhanced numerical stability despite the occurrence of negative values of the model's working variable, especially in presence of coarse spatial discretizations of the boundary layer in high-order solutions.

We cast the governing PDEs into the following compact form for general unsteady advection-diffusion models:
\begin{equation}
\partial_t(Q_k) +\partial_j(F^{inv}_{kj}) - \partial_j(F^{vis}_{kj}) = S_k,
\label{eq:NS}
\end{equation}
where $Q_k$ is a state variable and the solution of the $k^{th}$ partial differential equation (PDE) with $k \in [1\,..\,{N_\mathrm{eq}}]$ where  $N_\mathrm{eq}=N_\mathrm{d}+3$ is the number of equations based on  $N_\mathrm{d}$  space dimensions; $F^{inv}_{kj}$ and $F^{vis}_{kj}$ respectively express the $k^{th}$ inviscid (advective) and viscous (diffusive) fluxes for $j\in [1\,..\,{N_\mathrm{d}}]$,  $S_k$  refers to the source term for equation $k$ and we use the Einstein's summation convention for repeated indices. The expanded form of the conservation laws is retrieved by substituting the following expressions and values in Eq. \eqref{eq:NS}:

\begin{itemize}
\item[-] Conservation of mass ($k=1$)
\begin{equation}
Q_k=\rho,\qquad\qquad F^{inv}_{kj}=\rho u_j,\qquad F^{vis}_{kj}=0,\qquad S_k=0;\label{eq:cont}
\end{equation}
\item[-] Conservation of momentum ($k \in [2\,..\,{N_\mathrm{d}+1}]$ and $i=k-1$)
\begin{equation}
Q_k=\rho u_i,\qquad F^{inv}_{kj}=\rho u_ju_i +p \delta_{ij},\qquad F^{vis}_{kj}=\tau_{ij},\qquad S_k=0;\label{eq:mom}
\end{equation}
\item[-] Conservation of energy ($k=N_\mathrm{d}+2$)
\begin{equation}
Q_k=\rho E, \qquad  F^{inv}_{kj}=\rho u_j H, \qquad  F^{vis}_{kj}=u_i \tau_{ij} +\omega_j, \qquad  S_k=0;\label{eq:ener}
\end{equation}
\item[-] Transport of the turbulent working variable ($k=N_\mathrm{d}+3$)
\begin{equation}
\begin{split}
Q_k=\rho \tilde{\nu}, \qquad F^{inv}_{kj}=\rho u_j \tilde{\nu},\qquad F^{vis}_{kj}=\frac{1}{\sigma}(\mu+\rho \tilde{\nu}f_n)\,\partial_j \tilde{\nu}, \qquad
S_k=  \rho \,\mathcal{P} - \rho \,\mathcal{D} + \rho \,\mathcal{T} \\+\frac{c_{b2}}{\sigma} \rho \,\partial_j \tilde{\nu} \, \partial_j \tilde{\nu} 
 - \frac{1}{\sigma}(\nu+\tilde \nu f_n)\, \partial_j (\rho \,\partial_j \tilde \nu) \label{eq:sa}.\
\end{split}
\end{equation}
%
\end{itemize}
The quantities appearing in equations \eqref{eq:cont} through \eqref{eq:sa} are defined as follows: $\rho$ is the density, ${\bm{u}} = {\bf{e}}_i u_i$ is the velocity vector with ${\bf{e}}_i$ being the $i^{th}$ orthonormal basis vector of the Euclidean spatial system, $E$ is the total energy per mass defined as $E=e+\frac{1}{2} (u_iu_i)$ where $e$ is the internal energy which for a calorically perfect gas is defined as $e=\frac{R}{\gamma-1}T$ where $R$ is the gas constant and $T$ is the temperature. The total enthalpy is defined as $H=E + \frac{p}{\rho}$ with $p$ denoting the pressure related to the energy via the ideal gas law as
\begin{equation}
{\footnotesize p=(\gamma-1)\rho\left(E-\frac{1}{2}(u_iu_i)\right)},
\label{eq:state}
\end{equation}
where $\gamma$ is the specific heat ratio ($\gamma=1.4$ for air).

In Eq. \eqref{eq:mom}, $\tau_{ij}$ are the components of the viscous stress tensor, $\underline{\underline{\tau}}$, which for compressible Newtonian fluids read
\begin{equation*}
\tau_{ij}=2 \, \mu_{\mathrm{eff}} \, S_{ij},
\quad \mathrm{with} \;\;
S_{ij} = \frac{1}{2}(\partial_i u_j + \partial_j u_i)-\frac{1}{3}\partial_k u_k \delta_{ij},
\end{equation*}
where $\mu_{\mathrm{eff}}$ is the effective viscosity, defined as the sum of the dynamic viscosity, $\mu$, and the eddy viscosity, $\mu_t$, viz., $\mu_{\mathrm{eff}}= \mu+\mu_t$; and $\delta_{ij}$ represents the Kronecker delta. Note that the dynamic viscosity is assumed to be spatially constant throughout this work.

In Eq. \eqref{eq:ener}, $\omega_j =\lambda_{\mathrm{eff}}\, \partial_j T$ is the $j^{th}$ component of the heat flux vector where $\lambda_{\mathrm{eff}}$ is the effective thermal conductivity defined as $\lambda_{\mathrm{eff}} = \lambda + \lambda_t$, with $\lambda=\frac{\gamma R}{(\gamma-1)}\frac{\mu}{\mathrm{Pr}}$, the molecular conductivity,  and $\lambda_t=\frac{\gamma R}{(\gamma-1)} \frac{\mu_t}{\mathrm{Pr}_t}$, the eddy conductivity. The laminar and turbulent Prandtl numbers are respectively set to $\mathrm{Pr}=0.7$ and $\mathrm{Pr}_t=0.9$ unless specified.


In the SA model, $\mu_t$, the turbulent (eddy) viscosity, is expressed by

\begin{subequations}
\begin{empheq}
[left={\mu_t= \rho \nu_t =\empheqlbrace}]{alignat=2}
&\rho \tilde{\nu} f_{v1}\qquad & \tilde \nu \geq 0, \label{eq:mu_t+}\\
&0                      \qquad & \tilde \nu < 0,    \label{eq:mu_t-}
\end{empheq}
\end{subequations}
where
\begin{equation*}
f_{v1}=\frac{\chi^3}{\chi^3+c_{v1}^3},
\quad
\chi=\tilde{\nu}/\nu,
\quad
c_{v1}=7.1,
\end{equation*}
and $\tilde{\nu}$ is the working variable of the SA model that represents a turbulent kinematic viscosity.

In Eq. \eqref{eq:sa}, the production term, $\mathcal{P}$, is defined as
\begin{subequations}
\begin{empheq}
[left={\mathcal{P}=\empheqlbrace}]{alignat=2}
& c_{b1}(1-f_{t2})\tilde s \tilde \nu\qquad &  \tilde \nu \geq 0, \label{eq:prod+}\\
& c_{b1}(1-c_{t3})       s \tilde \nu\qquad &  \tilde \nu < 0,    \label{eq:prod-}
\end{empheq}
\end{subequations}
where $c_{b1} = 0.1355$, $f_{t2}=c_{t3}\,\mathrm{exp}(-c_{t4}\,\chi^2)$ with $c_{t3}=1.2$ and $c_{t4}=0.5$ is the laminar suppression term, $s=\lvert\varepsilon_{ijk}\partial_j u_k \rvert$ is the vorticity magnitude with $\varepsilon_{ijk}$ standing for the Levi-Civita symbol of permutation, and $\tilde s$ is the modified vorticity defined as
\begin{subequations}
\begin{empheq}
[left={\tilde s=\empheqlbrace}]{alignat=2}
& s+\bar{s}      \qquad &  \bar{s} \geq -c_{v2} s, \label{eq:S+}\\
& s+\frac{s(c_{v2}^2s+c_{v3}\bar{s})}{(c_{v3}-2c_{v2})s-\bar{s}}\qquad &   \bar{s} < -c_{v2}s, \label{eq:S-}
\end{empheq}
\end{subequations}
where
\begin{gather*}
\begin{aligned}
&\bar{s}=\frac{\tilde{\nu} f_{\nu 2}}{\kappa^2 d_w^2},\quad f_{v2} = 1 - \frac{\chi}{1+\chi f_{v1}},
&c_{v2}=0.7, \quad c_{v3}=0.9, \quad \kappa = 0.41,
       \end{aligned}
\end{gather*}
and $d_w$ being the distance to the closest wall. In Eq. \eqref{eq:sa}, the destruction term, $\mathcal{D}$, is defined as
\begin{subequations}
\begin{empheq}
[left={\mathcal{D=}\empheqlbrace}]{alignat=2}
& \left(c_{w1} f_w -\frac{c_{b1}}{\kappa^2}f_{t2}\right)\frac{ {\tilde\nu}^2 }{d_w^2}\qquad &  \tilde \nu \geq 0, \label{eq:D+}\\
& - c_{w1} \frac{ {\tilde\nu}^2 }{d_w^2}  \qquad &   \tilde \nu < 0, \label{eq:D-}
\end{empheq}
\end{subequations}
where
\begin{equation*}
\begin{aligned}
c_{w1} = \frac{c_{b1}}{\kappa^2}+\frac{1+c_{b2}}{\sigma},\phantom{\frac{\frac{1}{1}}{\frac{1^2}{1^\frac{1}{2}}}}
c_{b2} = 0.622, \quad \sigma =2/3, \;\,  \mathrm{and} \;\, f_w = g\left(\frac{1+c_{w3}^6}{g^6+c_{w3}^6}\right)^{1/6}.
\end{aligned}
\label{eq:dist}
\end{equation*}

We refer the reader to \cite{Allmaras-et-al_2012} for the full definition of the trip term, $\mathcal{T}$, in Eq. \eqref{eq:sa} which serves to mimic the effect of a forced transition. A value of $\mathcal{T}=0$ is considered throughout this work. The remaining closure functions and constants of the SA model are
\begin{subequations}
\begin{empheq}
[left={f_n=\empheqlbrace}]{alignat=2}
& 1                                  \qquad &  \tilde \nu \geq 0, \label{eq:fn+}\\
& \frac{c_{n1}+\chi^3}{c_{n1}-\chi^3}\qquad &   \tilde \nu < 0, \label{eq:fn-}
\end{empheq}
\end{subequations}
\begin{equation*}
\begin{aligned}
g = r + c_{w2}(r^6-r),
\quad
r = \mathrm{min} \left(\frac{\tilde \nu}{\tilde s \kappa^2 d_w^2}, r_\mathrm{lim} \right),
\quad
r_\mathrm{lim}=10,
\quad
c_{n1}=16,
\quad
c_{w2} = 0.3,
\quad \mathrm{and}\quad
c_{w3} = 2.
\end{aligned}
\end{equation*}

The original SA model is represented by equations \ref{eq:sa}, \ref{eq:mu_t+}, \ref{eq:prod+}, \ref{eq:S+}, \ref{eq:D+} and \ref{eq:fn+} whereas the modified SA model corresponds to equations \ref{eq:sa}, \ref{eq:mu_t-}, \ref{eq:prod-}, \ref{eq:S-}, \ref{eq:D-} and \ref{eq:fn-}.

%

\section{Compact high-order numerical method}
\label{sec:num_meth}
The high-order discretization scheme adopted in this work is the CPR \cite{Wang-Gao_2009}, that is an extension of the flux reconstruction (FR) \cite{Huynh_2009a} scheme to simplices by a lifting collocation penalty technique. Variations of the correction function at the core of this scheme allow to recover a number of prominent compact high-order methods such as the spectral difference, the spectral volume and the discontinuous Galerkin (DG). Hence, the FR/CPR method is regarded as a unifying compact high-order scheme and a representative member of this class.

\subsection{CPR Scheme}

\label{sec:CPR}
To discuss the formulation of the CPR scheme, let's consider a scalar hyperbolic conservation law,
\begin{equation}
\partial_t Q + \partial_j F_{j}=0,
\label{eq:con_law}
\end{equation}
where $Q$ is the state variable and $\bm{F}={\bf{e}}_q F_q(Q)\in{\rm I\!R}^{N_\mathrm{d}}$ is a generic flux vector which could be or not function of state derivatives.
Multiplying by an arbitrary test function, $\phi$, integrating and applying the divergence theorem, the variational formulation of Eq. \eqref{eq:con_law} in Green's form is obtained
\begin{equation*}
\int_\Omega\partial_t(Q) \, \phi \, d\Omega - \int_\Omega F_{j}\,\partial_j(\phi) \,d\Omega + \int_\Gamma F_{j}  n_j \phi\, d\Gamma=0,
\end{equation*}
where $\Omega$ is a bounded spatial domain in ${\rm I\!R}^{N_\mathrm{d}}$, $\Gamma$ is its frontier and $\bm{n}= {\bf{e}}_q n_q$ is the local unit outward normal. Considering a tessellation, $\Xi$, of $\Omega$, into a set of continuous and non-overlapping elements, ${e_i}\in \Xi$, the variational formulation reads
\begin{align*}
\sum_{e_i}\left(\int_{\Omega_{e_i}}\partial_t(Q)  \,\phi \, d\Omega 
- \int_{\Omega_{e_i}} F_{j}\,\partial_j(\phi) \,d\Omega  + \int_{\Gamma_{e_i}} F_{j} n_j \phi\, d\Gamma \right)=0.
\end{align*}
Taking into account the inner, $Q^-$, and the outer, $Q^+$, states defined with regards to the direction of $\bm{n}$ at a given element boundary, the inter-element coupling of discontinuous solutions can be achieved via a common Riemann flux, $\hat{{F}}^n(Q^-,Q^+,\bm{n}^-,\bm{n}^+)$,  such that
\begin{equation}
F_jn_j \equiv F^n \approx \hat{F}^{n}(Q^-,Q^+,\bm{n}),
\label{eq:num_flux}
\end{equation}
since for a well-constructed mesh, $\bm{n}\equiv\bm{n}^-$=$-\bm{n}^+$ where $-$ and $+$ exponents respectively refer to the unit outward normal from the element $e_i$ and from its neighbour at the same point on the interface.

Integrating by parts and applying the divergence theorem once again yields the variational formulation in the divergence form,
\begin{align*}
\sum_{e_i}\left(\int_{\Omega_{e_i}}\partial_t(Q)  \,\phi  \,d\Omega + \int_{\Omega_{e_i}} \partial_j(F_{j})\,\phi \,d\Omega  + \int_{\Gamma_{e_i}} (\hat{F}^{n}-{F}^{n})  \phi\, d\Gamma \right)=0.
\end{align*}
The boundary term can be projected into the element. This is achieved by the following \textit{lifting operator} that provides a correction field, $\psi$, as output:
\begin{equation}
\int_{\Omega_{e_i}} \,\psi \, \phi\, d\Omega=\int_{\Gamma_{e_i}} (\hat{F}^{n}-{F}^{n}) \phi\, d\Gamma.
\label{eq:lift_op}
\end{equation}
We are interested in a correction that belongs to the space of polynomials of degree $N_\mathrm{P}$ or less, viz., $\psi \in {\rm I\!P}^{N_\mathrm{P}}$ since
the solution and the correction on element $e_i$ are discretized by a set of interpolation polynomials $\varphi_l$ with the property that $\varphi_l \in {\rm I\!P}^{N_\mathrm{P}}$ on element $e_i$ and identically 0 on the others, such that
\begin{equation}
Q_{e_i} =  Q_l  \,\varphi_l,
\label{Q_interp}
\end{equation}
and
\begin{equation}
\psi_{e_i} = \psi_l \, \varphi_l,
\label{delta_interp}
\end{equation}
where $Q_l$ and $\psi_l$ are respectively the state and correction values at the solution point $l \in [1\,..\,{N_\mathrm{node}}]$ where ${N_\mathrm{node}}$ is the number of solution points per element (${N_\mathrm{node}}=({N_\mathrm{P}}+1)^{N_\mathrm{d}}$ for a tensor-product element). Throughout this article, the notation $\mathrm{P}$ is used equivalently to ${N_\mathrm{P}}$. The short notation, P3, P4, etc. will also signify P=3, P=4, etc. Remark that integral operators in Eq. \eqref{eq:lift_op} could be computed via quadratures on a reference element once and stored for use on physical elements, assuming that a valid geometrical mapping exists.

In the case where the flux is a non-linear function of the state, the flux divergence will not necessarily belong to the space ${\rm I\!P}^{N_\mathrm{P}}$. We therefore consider the following projection of the flux divergence onto the space ${\rm I\!P}^{N_\mathrm{P}}$:
\begin{equation}
\int_{\Omega_{e_i}} \prod_{e_i} (\partial_j F_j) \, \phi \, d \Omega = \int_{\Omega_{e_i}} \partial_j F_j \phi \, d \Omega,
\label{eq:fdiv_proj}
\end{equation}
where $\prod_{e_i} (\partial_j F_j)$ is the projected flux divergence.
The resulting discrete formulation is
\begin{equation*}
\sum_{e_i}\int_{\Omega_{e_i}}\left(\partial_t(Q_{e_i}) + \prod_{e_i} (\partial_j F_j)\,+  \psi_{e_i}  \right) \phi \, d \Omega=0,
\end{equation*}
where all terms are of degree ${N_\mathrm{P}}$ or less. Choosing a proper test space that guarantees solution uniqueness yields
\begin{equation}
\sum_{e_i}\left(\partial_t(Q_{e_i}) + \prod_{e_i} (\partial_j F_j)\,+  \psi_{e_i}  \right)=0,
\label{eq:CPR_scheme}
\end{equation}
which is the variational formulation in the divergence form of Eq. \eqref{eq:con_law} and which can be cast into a purely differential scheme, thus avoiding costly explicit quadratures. In fact, the direct projection of the divergence term via Eq. \eqref{eq:fdiv_proj} still requires the use of quadratures. Two more efficient alternatives could be considered \cite{Wang-Gao_2009}: the Lagrange polynomial (LP) and the chain rule (CR) approaches. 

The LP method consists of interpolating the flux before the application of the divergence operator, as
\begin{equation*}
(F_j)_{e_i} = {F_j}_l \, \varphi_l,
\end{equation*}
where ${F_j}_l$ is the $j^{th}$ spatial component of the flux evaluated at the solution point $l$. The projection of the divergence can then be computed as
\begin{equation*}
\prod_{e_i} (\partial_j F_j) = \partial_j (F_j)_{e_i} =  \partial_j(\varphi_l) \, {F_j}_l. 
\end{equation*}

The CR approach employs the flux Jacobian, $A_{j} = \frac{\partial{F_{j}}}{\partial Q}$, and the spatial derivatives of the state variable evaluated via
\begin{equation*}
\partial_j Q_{e_i}=\partial_j(\varphi_l{Q_l}) = \partial_j(\varphi_l)\,{Q_l},
\end{equation*}
for a projection onto the space ${\rm I\!P}^{N_\mathrm{P}}$ that reads
\begin{equation*}
\prod_{e_i} (\partial_j F_j) = A_{j} \, \partial_j Q_{e_i}.
\end{equation*}

Numerical experiments have shown \cite{Wang-Gao_2009} that while the LP is fully conservative, the CR is more accurate at the expense of a slight loss in strict conservation. A fix to retrieve the full conservation of the CR has been proposed in \cite{Gao-Wang_2013}. In this work, the CR and LP methods are used for the evaluation of the divergence of inviscid and viscous fluxes respectively. Furthermore, we choose the Lagrange polynomials as both test functions ($\phi$ in Eq. \eqref{eq:lift_op}) and interpolation functions ($\varphi$ in Eqs. \eqref{Q_interp} and  \eqref{delta_interp}), thus retrieving a pre-integrated version of the discontinuous Galerkin method. Finally, we utilize the Gauss-Legendre-Lobatto (GLL) set, also known as Gauss-Lobatto, as solution nodes on quadrangular elements.

\subsection{Numerical fluxes}
For finite number of DOFs, the scheme described in the precedent section yields discontinuous solutions at element interfaces. The common interface flux of Eq. \eqref{eq:num_flux} couples these solutions by taking the following form for the considered governing equations:
\begin{equation}
\hat{F}^{n}=\mathcal{H}^{inv}(Q^-,Q^+,\bm{n}) + \mathcal{H}^{vis}(Q^-,Q^+,\bm{n},{\bf{e}}_q(\partial_q Q)^-,{\bf{e}}_q(\partial_q Q)^+),
\label{eq:num_fluxes}
\end{equation}
where $\mathcal{H}^{inv}$ and $\mathcal{H}^{vis}$ are proper numerical fluxes for respectively the inviscid and viscous components of the equations and $\cdot^-$ and $\cdot^+$ refer to the traces of a given quantity from the inner and outer sides of the interface respectively. The choices of numerical fluxes are further described in what follows.
\subsubsection{Inviscid flux}
We opt for the Roe's approximate Riemann solver \cite{Roe_1981} as $\mathcal{H}^{inv}$ in Eq. \eqref{eq:num_fluxes}, expressed as
\begin{equation}
\mathcal{H}^{inv} = \frac{1}{2}\left(F^{inv}_{j}(Q^-)\,n_j+F^{inv}_{j}(Q^+)\,n_j\,+D\right),
\label{eq:inv_flux}
\end{equation}
where $D$ is the dissipation added to the central flux, stabilizing it through upwinding.

As we solve the RANS and the SA equations in a fully coupled fashion, the Roe numerical flux is re-derived following Appendix D of \cite{Burgess2011} to account for this coupling. The resulting upwinding contrasts with that of decoupled approaches in which only the normal velocity at the surface is considered in the dissipation term of the SA common flux. In the coupled upwinding, the dissipation term accounts as well for sound speed and pressure. The coupling is reported \cite{Burgess2011} to enhance the smoothness and robustness of the RANS-SA solutions.

\subsubsection{Viscous flux}
\label{sec:BR2}
An intuitive approach for deriving a viscous numerical flux would be to consider a central flux based on the element-wise approximation of  $\partial_q Q$, computed as $\partial_q(\varphi_l)Q_{l}$, such as
\begin{equation}
\begin{split}
\mathcal{H}^{vis}  = 
\frac{1}{2}\left( F^{vis}_{j}\left(Q^-,{\bf{e}}_q(\partial_q Q)^-\right) +  F^{vis}_{j}\left(Q^+,{\bf{e}}_q(\partial_q Q)^+\right)   \right) \, {n}_j.
\end{split}
\label{eq:flux_visc_central}
\end{equation}

However, this choice suffers from instabilities for under-resolved elliptic problems  due to the singularity of the resulting discrete viscous operator \cite{Hesthaven-Warburton_2008}. Different approaches have been devised to alleviate this problem that mostly rely on the basic idea of introducing a correction on the solution derivatives that takes into account the data from neighbouring elements and penalizes the derivatives for solution jumps at the interface. We consider a representative method of this class that is the second flux of Bassi and Rebay \cite{Bassi-Rebay_2000a}, referred to as BR2, which employs an auxiliary vector function, $\bm{\theta} = {\bf{e}_q}\theta_q$,  such that
\begin{equation}
\theta_{q} \approx \partial_q Q.
\label{eq:aux}
\end{equation}

Applying the approach described in section \ref{sec:CPR} to Eq. \eqref{eq:aux} yields its variational formulation in the divergence form,
\begin{equation}
\int_{\Omega_{e_i}} \theta_{q}\,\phi\, d\Omega - \int_{\Omega_{e_i}} \partial_q (Q)\,\phi\, d\Omega=\int_{\Gamma_{e_i}} (\hat{Q}-Q) \, \phi\, n_q \,d\Gamma,
\label{eq:aux2}
\end{equation}
where $\hat{Q} =\frac{1}{2}({Q}^-+{Q}^+)$ is the common interface value of the state variable. 

By considering a lifting operation such that
\begin{equation}
\int_{\Omega_{e_i}} R_{q}\,\phi\, d\Omega =\int_{\Gamma_{e_i}} (\hat{Q}-Q) \, \phi\, n_q \,d\Gamma,
\label{eq:correc_full}
\end{equation}
where $R_{q}$ is a correction field accounting for the data from the neighbouring element, the term $\overline{\overline{\partial_q Q}}\equiv\theta_{q}=\partial_q Q+R_{q}$ can be interpreted as a corrected divergence that serves in the computation of the viscous flux, $  {\bm{F}}^{vis} = {\bf{e}}_jF^{vis}_{j}(Q,{\bf{e}}_q\overline{\overline{\partial_q Q}})$ for the discontinuous interface flux value in Eq. \eqref{eq:lift_op} and for flux divergence value in Eq. \eqref{eq:CPR_scheme}. Note that $\mathrm{lim}_{h\rightarrow0} R_q = 0$ is expected and $\theta_{q} = \partial_q Q$ is recovered in the continuum limit. As for the viscous numerical flux in Eq. \eqref{eq:num_fluxes}, such a correction for the divergence results in an extended stencil since it couples the elemental solution to the data from neighbours of its immediate neighbours. Nevertheless,  considering a segmentation of $\Gamma_{e_i}$ into $N_f$ non-overlapping and continuous faces, the viscous numerical flux on face $(\Gamma_{e_i})_f$, with $f \in [1 \,..\, N_f]$, can be evaluated as 

\begin{equation}
\begin{split}
\mathcal{H}^{vis} =
\frac{1}{2}\left( F^{vis}_{j}\left(Q^-,{\bf{e}}_q(\overline{\partial_q Q})^-\right) +  F^{vis}_{j}\left(Q^+,{\bf{e}}_q(\overline{\partial_q Q})^+\right)   \right) \, {n}_j,
\end{split}
\label{eq:flux_visc}
\end{equation}
where $\overline{\partial_q Q}=\partial_q Q+C_{r_{q}}\,r_{q}$ designates a partially corrected derivative. The constant $C_{r_{q}}$ ensures the stability of the scheme for highly diffusive problems and is set to the value of 2 in this work. Finally, the partial correction, $r_q$, is the result of the following lifting operation:
\begin{equation}
\int_{\Omega_{e_i}} r_{q}\,\phi\, d\Omega =\int_{(\Gamma_{e_i})_f} (\hat{Q}-Q) \, \phi\, n_q \,d\Gamma.
\label{eq:correc_part}
\end{equation}
To conserve the efficiency of tensor product operators, Eq. \eqref{eq:correc_part} is applied in a 1D manner in the context of FR/CPR schemes on quadrangular and hexahedral elements \cite{Wang-et-al_2011a}.

\subsection{Boundary conditions}
\label{sec:BC_conds}
The soundness of a CFD implementation can not be established without the verification of its treatment of boundary conditions  (BCs). Therefore, in this section, we precisely define the expression of the BCs that will be examined in this work.

\subsubsection{Riemann BC}
The inviscid boundary conditions for free flows are imposed in the sense of \textit{weak-Riemann} of \cite{Mengaldo2014} via the numerical flux of Eq. \eqref{eq:inv_flux} that takes into account the incoming and outgoing characteristics arising from the inner, $Q_k^-$ and the outer, $Q_k^+=Q_k^\mathrm{BC}$, states where $Q_k^\mathrm{BC}$ is the $k^{th}$ boundary state defined at ghost nodes. 

In the case of a manufactured solution we set
\begin{equation}
Q_k^\mathrm{BC} = Q_k^{\mathrm{MS}}\vert_{\Gamma_\mathrm{BC}},
\label{eq:bc_Riem_MS}
\end{equation}
where $Q_k^{\mathrm{MS}}\vert_{\Gamma_\mathrm{BC}}$ designates the value of the manufactured solution evaluated at the point of the boundary where the BC is imposed.

It is important to note that for verification purposes, the treatment of the boundary conditions should be the same in the manufactured case as in the general flow problem. Hence, for the general flow problem, one would simply provide the desired boundary state at the ghost node instead of $Q_k^{\mathrm{MS}}\vert_{\Gamma_\mathrm{BC}}$. For a farfield BC for example, the free-stream values, $Q_k^{\infty}$, are imposed by setting $Q_k^\mathrm{BC}=Q_k^{\infty}$.

\subsubsection{Viscous BC}
The boundary condition for the viscous terms is enforced through the numerical flux of Eq. \eqref{eq:flux_visc}, by setting the desired values of the state variables and their derivatives at the ghost nodes via $Q_k^+=Q_k^\mathrm{BC}$ and $(\overline{\partial_q Q_k})^+=({\partial_q Q_k})^\mathrm{BC}$, and by taking the common state in Eqs. \eqref{eq:correc_full} and \eqref{eq:correc_part} to be $\hat{Q} = \frac{1}{2}(Q^-+Q^\mathrm{BC})$. 

In the case of manufactured solutions, the boundary value of the state is set via Eq. \eqref{eq:bc_Riem_MS} and its derivatives read
\begin{equation}
({\partial_q Q_k})^\mathrm{BC} = (\partial_qQ_k)^{\mathrm{MS}}\vert_{\Gamma_\mathrm{BC}}.
\label{eq:bc_vis_MS1}
\end{equation}
In the case of a farfield boundary employed in an application problem, the boundary values are set to $Q_k^\mathrm{BC}=Q_k^{-}$ and $({\partial_q Q_k})^\mathrm{BC}=(\overline{\partial_qQ_k})^{-}$, such that $\mathcal{H}^{vis} =
 F^{vis}_{j}\left(Q^-,{\bf{e}}_q(\overline{\partial_q Q})^-\right)  \, {n}_j$ at the farfield boundary.

\subsection{Treatment of curved elements}
We refer the reader to Eq. (37) of Section 2.3 of \cite{Wang-et-al_2011a} for a detailed description of the approach adopted in this work to treat curved elements. Furthermore, all mappings from computational element to physical element are isoparametric, i.e., they are based on polynomial expansions of same degree as the solution.


\subsection{Solution process}
Starting from an initialization by exact values, an implicit Euler scheme, i.e., the relaxed Newton's method with pseudo-time integration, is employed to smoothly converge the non-linear residuals to the vicinity of the final solution by gradually decreasing the relaxation (increasing the pseudo-time step) down to the final stage where the full Newton's method yields a quadratic convergence to the discrete solution. The linearization of the residual equation with regards to the solution (the Jacobian matrix) is required by the Newton's method to determine the direction of descent. In order to ensure optimal residual minimization (to machine precision), an analytic Jacobian is implemented for the fully coupled system of RANS-SA including the original and modified portions of the SA equation. This linearization is then verified to provide significant digits of double precision when compared to a linearization by \textit{complex step} via \textit{operator overloading} \cite{Martins-2000}. The operator overloading automatically ensures that the Jacobian matrix is consistent with the residual equation such that any modification in the latter is inherently accounted for. The effort involved in the achievement of an exact Jacobian  is justified since it enables the full operation of the Newton's method in the minimization of residuals. In fact, as iterative and round-off errors scale with residuals \cite{Roy2005}, their minimization permits to safely direct the focus towards the discretization error. The linear system is solved by a generalized minimal residual (GMRES) method along with a global block-Jacobi and local ILU(0) preconditionners via the PETSc package \cite{petsc-efficient}.


\section{Verification cases and methodology}
\label{sec:cases}

\begin{table}
\centering
\begin{tabular}{ l|l|c|c|c|c|c|c  }
\textbf{Property} &  \textbf{Feature}       & MS-1      & MS-2    & MS-3    & MS-4    & MS-5   &  \textbf{Cum.}   \\ \hline\hline
Re                &  Inviscid       &   \cmark  &  \cmark & \cmark  & \cmark & \cmark  & \cmark \\ \Cline{2-8}{0.5pt}
                  & Viscous        &   \xmark  &  \xmark & \cmark  & \cmark & \cmark  & \cmark \\ \Cline{2-8}{0.5pt}
                  & Turbulent      &   \xmark  &  \xmark & \xmark  & \cmark & \cmark  & \cmark \\ \Cline{1-8}{0.5pt}
Ma                & Supersonic     &   \xmark  &  \cmark & \xmark  & \xmark & \xmark  & \cmark \\ \Cline{2-8}{0.5pt}
                  & Transonic      &   \xmark  &  \xmark & \xmark  & \xmark & \xmark  & \xmark \\ \Cline{2-8}{0.5pt}  
                  & Subsonic       &   \cmark  &  \xmark & \cmark  & \cmark & \cmark  & \cmark \\ \Cline{1-8}{0.5pt}
Boundary          & Riemann        &   \cmark  &  \cmark & \cmark  & \cmark & \cmark  & \cmark \\ \Cline{2-8}{0.5pt}
Conditions        & Viscous        &   \xmark  &  \xmark & \cmark  & \cmark & \cmark  & \cmark \\ \Cline{2-8}{0.5pt}
                  & Slip Wall      &   \xmark  &  \xmark & \xmark  & \xmark & \xmark  & \xmark \\ \Cline{2-8}{0.5pt}
                  & No-slip Wall    &   \xmark  &  \xmark & \xmark  & \xmark & \xmark  & \xmark \\ \Cline{1-8}{0.5pt}                  
Mapping           & Curved Elements &   \xmark  &  \cmark & \xmark  & \xmark & \xmark  & \cmark
\end{tabular}
\caption{List of solver capabilities verified by manufactured solutions for free flows}
\label{tb:MS_prop}
\end{table}

The five manufactured solutions considered in this work, i.e., MS-1 through MS-5 (see Table \ref{tb:MS_prop}) are defined by a combination of trigonometric functions with terms depending exclusively on $x$ or on $y$ (mono terms) along with cross terms depending on both independent variables simultaneously. The cross terms are necessary for the verification of the viscous terms of the governing equations which feature mixed second order partial derivatives  whereas the mono terms are included to maintain the spatial dependency in the case where the cross terms are turned off for debugging purposes by setting their coefficients to zero. The generic form of the MS is
\begin{equation}
\begin{aligned}
\rho^{\mathrm{MS}} &= \rho_0   + \rho_x \mathrm{sin}(a_{\rho_{x}}    \pi   x / L)  + \rho_y \mathrm{cos}(a_{\rho_{y}}   \pi   y / L)  + \rho_{xy} \mathrm{cos}(a_{\rho_{x y}}   \pi   x / L) \,\mathrm{cos}(a_{\rho_{x y}}   \pi   y / L),&\\
u^{\mathrm{MS}}  &= u_0     + u_x   \mathrm{sin}(a_{u_{x}}      \pi   x / L)  + u_y    \mathrm{cos}(a_{u_{y}}      \pi   y / L)   + u_{xy}  \mathrm{cos}(a_{u_{xy}}     \pi   x / L) \,\mathrm{cos}(a_{u_{xy}}     \pi   y / L),&\\
v^{\mathrm{MS}}  &= v_0     + v_x   \mathrm{cos}(a_{v_{x}}      \pi   x / L)  + v_y  \mathrm{sin}(a_{v_{y}}      \pi   y / L)   + v_{xy} \mathrm{cos}(a_{v_{xy}}     \pi   x / L)\, \mathrm{cos}(a_{v_{xy}}     \pi   y / L),&\\
p^{\mathrm{MS}}  &= p_0     + p_x  \mathrm{cos}(a_{p_{x}}      \pi   x / L)  + p_y       \mathrm{sin}(a_{p_{y}}      \pi   y / L)   + p_{xy}  \mathrm{cos}(a_{p_{xy}}     \pi   x / L) \,\mathrm{cos}(a_{p_{xy}}     \pi   y / L),&\\
\tilde{\nu}^{\mathrm{MS}}  &= \tilde{\nu}_0     + \tilde{\nu}_x   \mathrm{cos}(a_{\tilde{\nu}_{x}}      \pi   x / L)  + \tilde{\nu}_y       \mathrm{cos}(a_{\tilde{\nu}_{y}}      \pi   y / L)   + \tilde{\nu}_{xy}       \mathrm{cos}(a_{\tilde{\nu}_{xy}}     \pi   x / L) \,\mathrm{cos}(a_{\tilde{\nu}_{xy}}     \pi   y / L),&\\
\end{aligned}
\label{eq:trigo_MS}
\end{equation}
where $u$ and $v$ respectively refer to the first and second velocity components, $L=1.0$ is a reference length and the manufactured total energy, $E^{\mathrm{MS}}$, is obtained via Eq. \eqref{eq:state} with $\gamma=1.4$.

We choose to define the MS in terms of the primitive variables since this procures a few advantages: firstly, since the Mach (Ma) number features the primitive variables directly, tuning the primitive variables of the MS allows for an easier control over the flow regime (subsonic vs. supersonic, compressible vs. incompressible). Secondly, the derivatives appearing in the forcing functions are those of the primitive variables and the MS definition in terms of the latter spares futile chain-rule operations. Finally, the expression of MS in primitive variables is forthwith compatible with the non-conservative formulation of incompressible solvers.

The constants appearing in Eqs. \eqref{eq:trigo_MS} serve to tune the magnitudes and frequencies of the trigonometric functions of the manufactured primitive variables such that the manufactured solution verifies all the terms of the PDE in the desired flow regime and furthermore features a numerically desirable behaviour. For example, a suitable manufactured solution would yield an asymptotic (monotonic) range that is attainable on reasonably fine grids and that is nevertheless not trivially obtained on the coarsest grids. Finally, let's recall that the manufactured solution is not bounded by strict fidelity to physics since the verification is rather a mathematical exercise; it is however desirable to produce quantities that remain within the range of physically valid variations such that density and absolute temperature are defined positive for example.

The spatial domain for this set of manufactured solutions is $\Omega = [0,1]^2$ and unless specified, all the grids are formed by isotropic quadrangular elements and refined by halving each element in each direction. The typical element size, $h$, is defined as 
\begin{equation}
h=\sqrt[{-N_d}]{N_{\mathrm{DOF}}},
\label{eq:elem_size}
\end{equation}
where $N_{\mathrm{DOF}}=\sum_{e_i}(N_\mathrm{node})_{e_i}$ is the total number of DOFs per equation.

In the next sections, we present manufactured solutions for free (unbounded) flows that target a specific set of solver capabilities. Each manufactured solution serves as well to exemplify and discuss important aspects of the verification process in general and in particular with attention to what is pertinent for high-order solvers. Extensive results are presented in \ref{sec:MS-resl} as reference to facilitate comparison with other high-order codes.

\subsection{Inviscid flows in subsonic regime - MS-1}

MS-1 is devised to verify both the incompressible ($0.13\leq\mathrm{Ma} \leq 0.3$) and  compressible ($0.3 \leq \mathrm{Ma} \leq 0.49$) portions of the subsonic regime. The parameters of Eqs. \ref{eq:trigo_MS} that define this MS are given in Table \ref{tb:MS-1_cons} and MS-1 fields are illustrated in Fig. \ref{fig:MS-1}. The results of the grid refinement study for polynomial degrees P1 to P5 are presented in Figs. \ref{fig:Err_allE_allP_MS-1} in terms of discretization error in $L_1$, $L_2$ and $L_\infty$ norms versus mesh size for all conservative variables. For each norm and P combination, the slope of the observed curve is compared to that of the expected OOAs, i.e., $\mathcal{O}(h^{\mathrm{P}+1}$).  It is as well notable that by increasing the degree of the polynomial discretization, P, the error level decreases for the same number of DOFs as in the previous case. This, in fact, is a demonstration of the higher accuracy/effort performance of high-order methods that constitutes their core interest.

\begin{table}
\centering
\begin{tabular}{ c||c|c|c|c|c|c|c }
$(\cdot)$& $(\cdot)_0$ & $(\cdot)_x$ & $(\cdot)_y$ & $(\cdot)_{xy}$ & $a_{(\cdot)_x}$ & $a_{(\cdot)_y}$ & $a_{(\cdot)_{xy}}$\\ \hline\hline
  $\rho$          &   $1.0$     & $0.3$       & $-0.2$      & $0.3$          &      $1.0$      &       $1.0$     &         $1.0 $     \\ \hline
  $u$             &   $1.0$     & $0.3$       & $0.3$      & $0.3$           &      $3.0$      &       $1.0$     &         $1.0 $     \\ \hline  
  $v$             &   $1.0$     & $0.3$       & $0.3$      & $0.3$           &      $1.0$      &       $1.0$     &         $1.0 $     \\ \hline  
  $p$             &   $18.0$     & $5.0$       & $5.0$     & $0.5$           &      $2.0$      &       $1.0$     &         $1.0 $     \\ \hline  
  $\tilde{\nu}$    &   $0.0$     & $0.0$       & $0.0$     & $0.0$           &      $0.0$      &       $0.0$     &         $0.0 $      
\end{tabular}
\caption{Parameters of MS-1}
\label{tb:MS-1_cons}
\end{table}

For the same spatial distribution of the discretization error, each norm in Fig. \ref{fig:Err_allE_allP_MS-1} outputs a different value. A general trend can be identified, with the $L_\infty$ yielding the highest value, followed by the $L_2$ and $L_1$ norms that are closer together. This is expected since the $L_\infty$ reports the maximal value of the error throughout the domain. We will exhaustively dwell on the importance of this norm  in the following sections. The $L_1$ norm measures the magnitude of the error, averaged over the domain, whereas the $L_2$ norm can be regarded as the magnitude of the error, weighed by itself first, and  averaged then. Hence the $L_2$ norm is more sensitive to the error deviations from the average value. Consequently, the disparity between the $L_1$ and $L_2$ norms reflects the irregularity of the spatial distribution of the error.

In terms of verification, the crucial verdict on grid convergence is declared as soon as the observed slope steadily recovers the expected order on at least three consecutive grids \cite{Eca2016}. However, judging the satisfaction of this criterion merely from the error plots, such as the ones in Fig. \ref{fig:Err_allE_allP_MS-1}, is an arduous task. It is therefore recommended to plot the evolution of the OOAs (the observed slope of the error) versus mesh refinement such as in Fig. \ref{fig:Orders_MS-1}, from which (see plot (c)), one can easily detect for example that in the case of P4, the $\rho v$ variable has still not fully grid-converged on the fourth finest grid ($h \simeq 6 \times 10^{-3}$). One can nevertheless acknowledge that this assessment is much harder to achieve by visually comparing the slopes in the plot (c) of Fig. \ref{fig:Err_allE_allP_MS-1}.

\subsubsection{Solution initialization and residual convergence}

\begin{figure}[!hbt]
\centering
\subfloat[Inexact initial solution]{
\includegraphics[trim = 5mm 4mm 4mm 10mm, clip,width=0.36\linewidth]{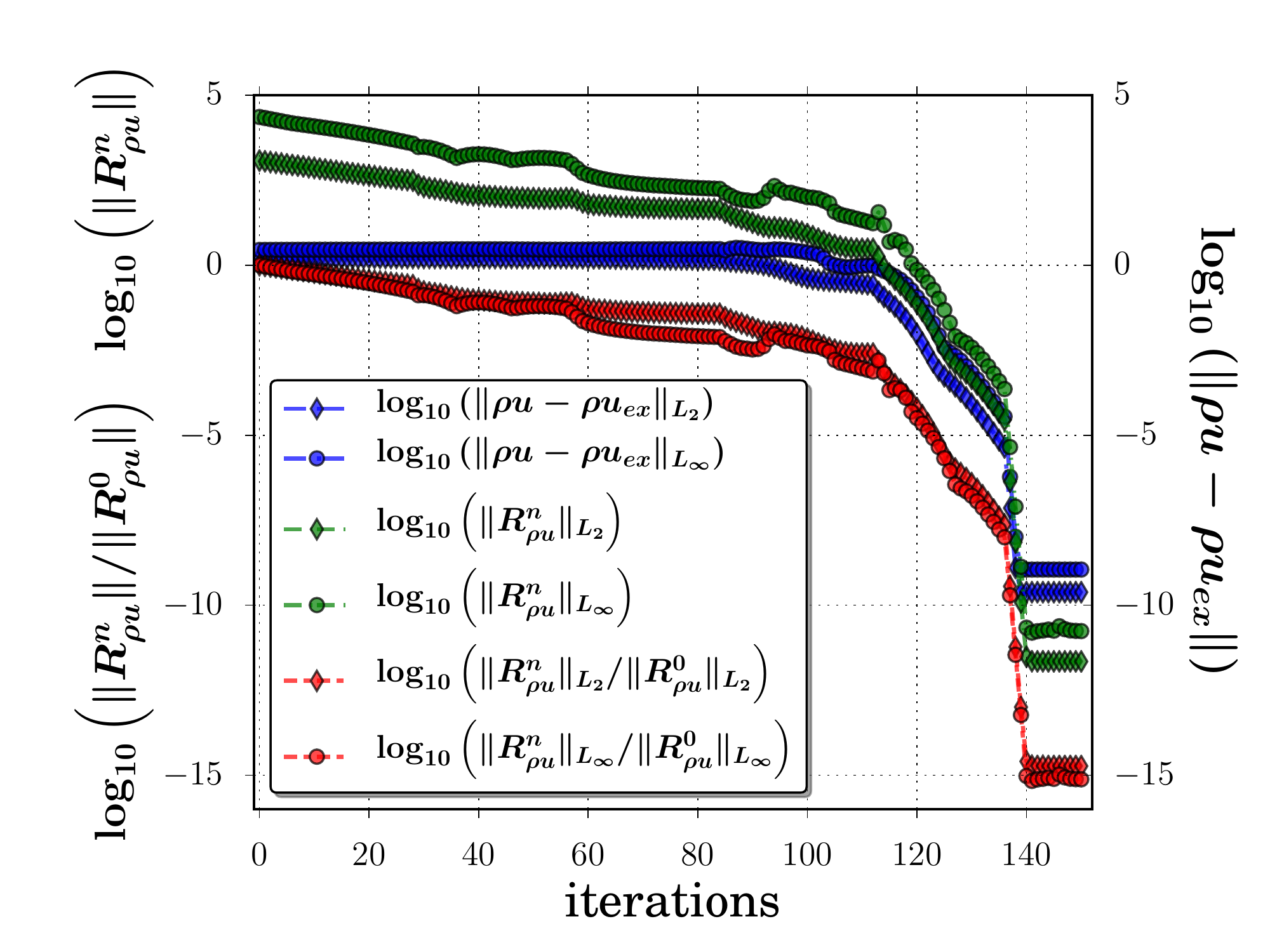}}~~
~
\subfloat[Exact initial solution]{ 
\includegraphics[trim = 5mm 4mm 4mm 10mm, clip,width=0.36\linewidth]
{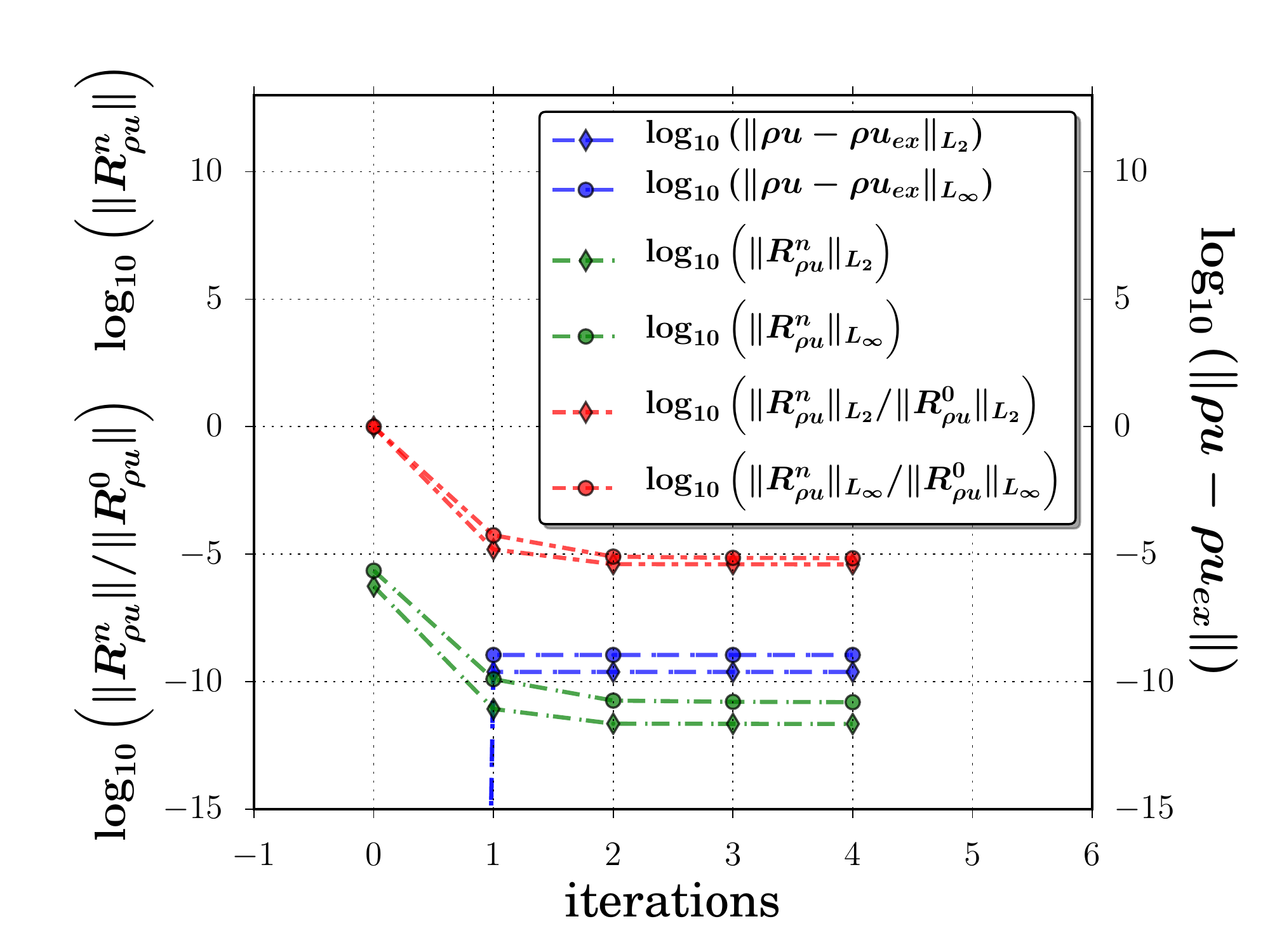}}
\caption{Absolute and relative residuals and discretization error versus number of iterations for MS-1, polynomial degree $\mathrm{P}5$ and grid of $32\times32$ elements}
\label{fig:Res_conv_MS-1}
\end{figure}

In order to study the evolution of the discretization error with mesh refinement in an isolated fashion, other sources of numerical error, such as iterative and floating point errors should be reduced as much as possible, by driving the norm of the residual equations as close to machine precision as possible. This is however a challenging goal in practical situations such as those encountered in external aerodynamics where solution non-linearity, mesh anisotropy and inadequate solution initialization contribute to hinder the convergence of iterative methods such as the Newton's. In the context of verification via the MMS, one could track the evolution of the exact discretization error during the convergence process. To provide insight on the adequate level of iterative convergence, we consider two scenarios: initialization with the exact\footnote{Let's note that for finite number of DOFs, the exact solution of the continuous equation is not the exact solution of the discrete equation.} (manufactured) solution, ($Q^0\,$=$\,Q^{\mathrm{MS}}$), and initialization with an inexact solution, ($Q^0\,$=$\,\alpha \, Q^{\mathrm{MS}}$  with $ \alpha\neq 1$ ).  Figure \ref{fig:Res_conv_MS-1} compares, in either of these cases, the evolution of the discretization error of $\rho u$ as well as that of the absolute and relative residuals of the $x$-momentum equation in $L_2$ and $L_\infty$ norms versus the number of iterations, $n$. For this particular combination of MS and spatial discretization ($\mathrm{P}5$ and $32 \times 32$ elements), an absolute residual norm drop to values below $\mathcal{R}_{\rho u } \simeq 10^{-11}$ is sufficient for the discretization error to converge to its final value ($\mathcal{E}_{\rho u}\simeq10^{-10}$). In terms of relative residuals, the convergence of the discretization error is achieved as soon as the relative residuals are dropped below $\mathcal{R}_{\rho u }^n/\mathcal{R}_{\rho u }^0 \simeq 10^{-10}$ (by 10 orders of magnitude in other words) in the case of inexact initialization and by $\mathcal{R}_{\rho u }^n/\mathcal{R}_{\rho u }^0 \simeq 10^{-5}$ in the case of exact initialization. This indicates that a relative reduction in terms of orders of magnitude of the residual can not serve as a universal convergence criterion in verification exercises due to the inherent dependency of such criterion on the initial residual magnitude. We will dwell further on the necessary level of residual convergence for code verification while discussing laminar cases. 

The residual drop is understandably much faster achieved for the case of exact initialization that constitutes a sufficiently fine guess to ensure the stability and the quadratic convergence of the full Newton's method. Indeed, ensuring the iterative convergence in the case of inexact initialization required the relaxation of the Newton's method by the introduction of pseudo-time integration via the implicit Euler scheme. 
We hence choose to employ exact initialization throughout the rest of this work and consistently reduce discrete residuals to reach a stable minimum.

\subsection{Inviscid flows in supersonic regime - MS-2}

For MS-2, which verifies the implementation of the Euler equations and the farfield boundary conditions in the supersonic regime and on curved domains, we found that the values of parameters in Table \ref{tb:MS-2_cons} result in grid convergence on reasonably fine grids.

\begin{table}
\centering
\begin{tabular}{ c||c|c|c|c|c|c|c }
$(\cdot)$& $(\cdot)_0$ & $(\cdot)_x$ & $(\cdot)_y$ & $(\cdot)_{xy}$ & $a_{(\cdot)_x}$ & $a_{(\cdot)_y}$ & $a_{(\cdot)_{xy}}$\\ \hline\hline
  $\rho$ &   $2.7$     & $0.9$       & $-0.9$      & $1.0$          &      $1.5$      &       $1.5$     &         $1.5 $     \\ \hline
  $u$    &   $2.0$     & $0.7$       & $0.7$       & $0.4$          &      $1.0$      &       $1.0$     &         $1.0 $     \\ \hline  
  $v$    &   $2.0$     & $0.4$       & $0.4$       & $0.4$          &      $1.0$      &       $1.0$     &         $1.0 $     \\ \hline  
  $p$    &   $2.0$     & $1.0$       & $1.0$       & $0.5$          &      $1.0$      &       $1.0$     &         $1.5 $     \\ \hline  
    $\tilde{\nu}$    &   $0.0$     & $0.0$       & $0.0$     & $0.0$           &      $0.0$      &       $0.0$     &         $0.0 $    
\end{tabular}
\caption{Parameters of MS-2}
\label{tb:MS-2_cons}
\end{table}

\begin{figure}[!hbt]
\centering
\includegraphics[trim = 1.5mm 1mm 1mm 1mm, clip,width=0.35\linewidth]{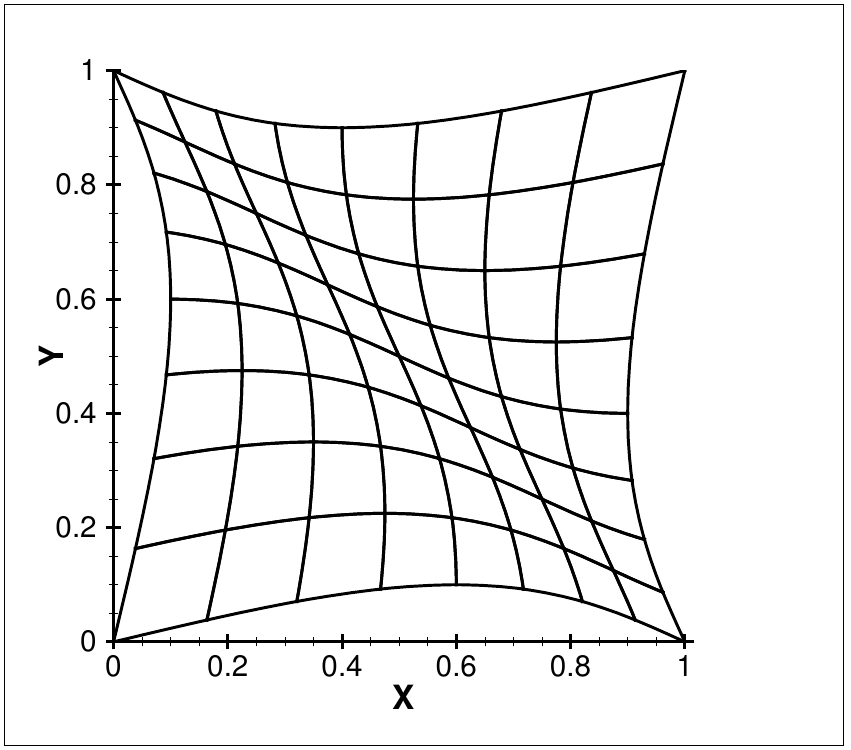}
\caption{Domain and typical grid of MS-2}
\label{fig:grid-MS_2_curved}
\end{figure}

The curved domain is defined via a deformation of the initial domain of $(\mathcal{X}, \mathcal{Y}) \in \Omega_0 = [0,1]^2$ (where isotropic Cartesian grids are created) to the target domain of $(x,y) \in \Omega$ via
\begin{equation}
\begin{split}
x &= \mathcal{X} + 0.1\,\mathrm{sin}(\pi\mathcal{X}+\pi\mathcal{Y}),\\
y &= \mathcal{X} + 0.1\,\mathrm{sin}(\pi\mathcal{X}+\pi\mathcal{Y}),
\end{split}
\label{eq:MS-2_map}
\end{equation}
along with an isoparametric elemental mapping. Figure \ref{fig:grid-MS_2_curved} illustrates the domain and a typical grid of MS-2.

MS-2 fields are plotted in Fig. \ref{fig:MS-2}. The evolution of the errors and the orders versus mesh refinement are given in Figs. \ref{fig:Err_allE_allP_MS-2} and \ref{fig:Orders_MS-2} respectively. The observed OOAs recover the theoretical values for all P and all norms considered except for $L_\infty$ norm of P5 that is affected by the vicinity of the machine precision to the discretization error on the finest grid.

\begin{figure}[!hbt]
\centering
\subfloat[Original domain]{
\includegraphics[trim = 9mm 3mm 18mm 12mm, clip,width=0.33\linewidth]
{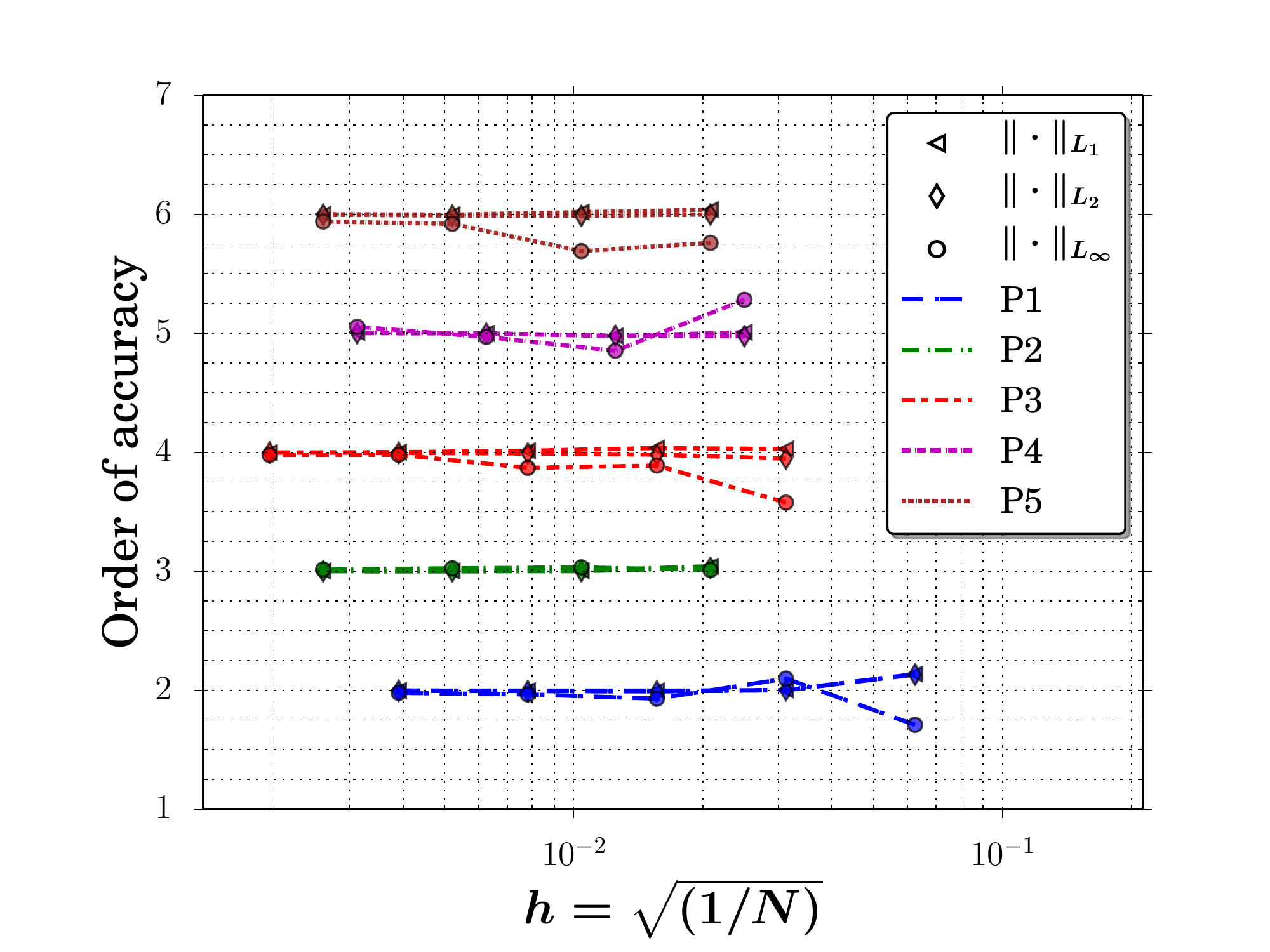}}~~~
\subfloat[Curved domain]{
\includegraphics[trim = 9mm 3mm 18mm 12mm, clip,width=0.33\linewidth]
{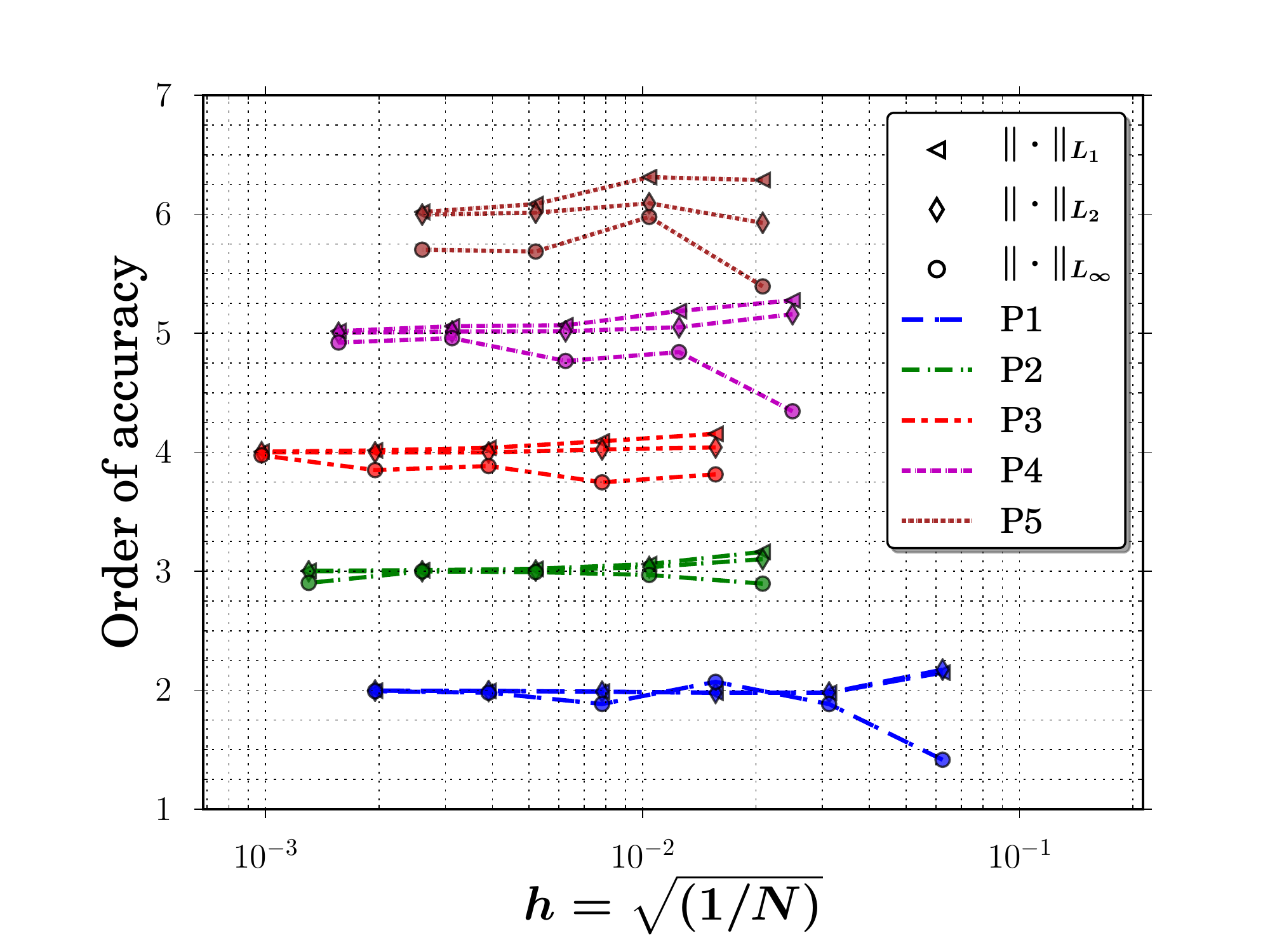}}
\caption{Evolution of the OOAs in $L_1$, $L_2$ and $L_\infty$ norms versus mesh refinement for $\rho u$, polynomial degrees $\mathrm{P}1$--$\mathrm{P}5$ and MS-2 on un-deformed versus curved domains} 
\label{fig:Orders_MS-2_cuv-nocurv}
\end{figure}

The OOAs of $\rho u$ in $L$ norms for the curved domain are compared to the ones for the original domain in Fig. \ref{fig:Orders_MS-2_cuv-nocurv}, showing the occurrence of the asymptotic convergence on coarser grids in the latter case. Two factors are at the source of this difference: first, the domain of definition of the manufactured fields are different since the two domains do not recover each other; second, the curved grid does not exhibit the same isotropy with regards to the solution  as the un-deformed grid. The effect of grid anisotropy is dwelt on at length in Section \ref{sec:grid_eff}. 

In order to avoid delays in the onset of asymptotic convergence with the goal of keeping the required grid sizes reasonable and since the treatment of the curved elements is verified by MS-2, the next manufactured cases will be considered on un-deformed domains along with Cartesian grids.

\subsubsection{Importance of $L_\infty$ norm}
\begin{figure}[!hbt]
\centering
\subfloat[$\mathrm{P}3-16\times16$ elements]{
\includegraphics[trim = 0mm 0mm 0mm 0mm, clip,width=0.27\linewidth]{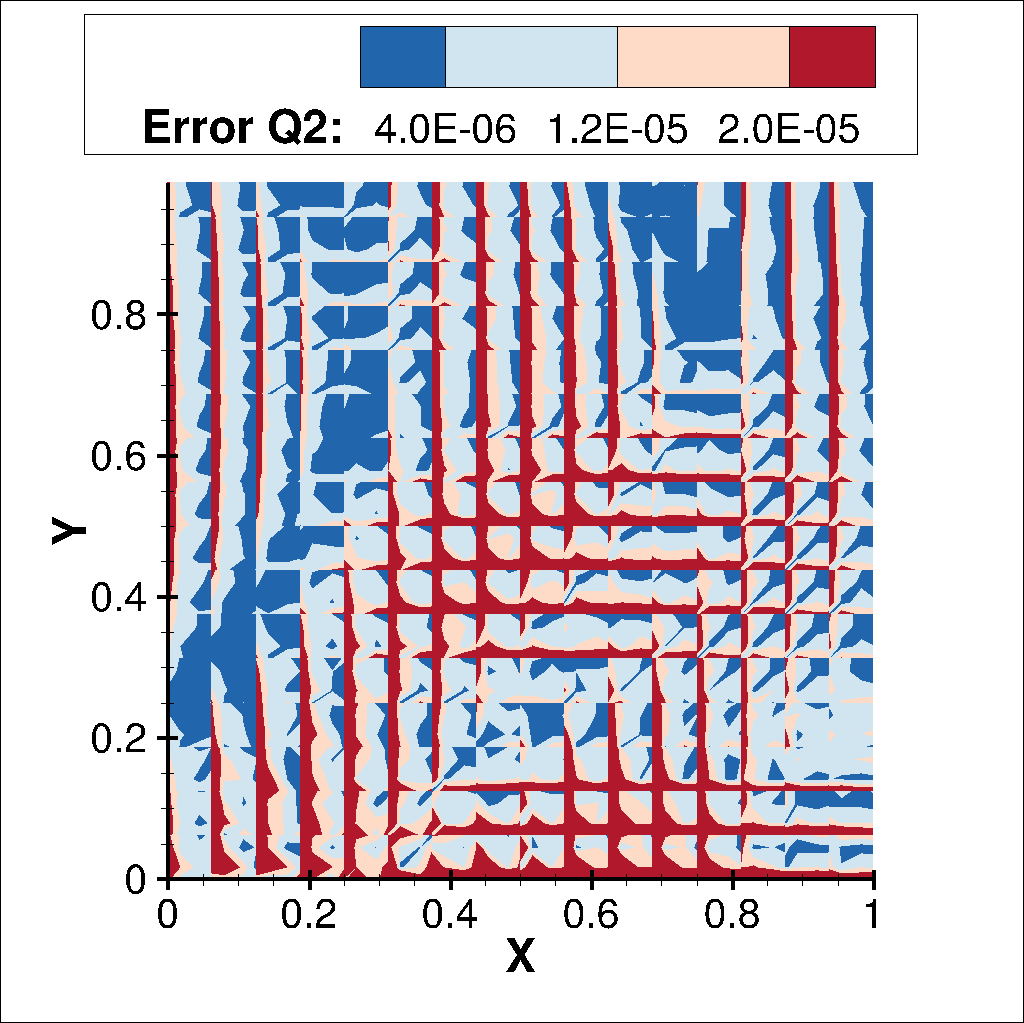}}~~~
\subfloat[$\mathrm{P}3-64\times64$ elements]{
\includegraphics[trim = 0mm 0mm 0mm 0mm, clip,width=0.27\linewidth]
{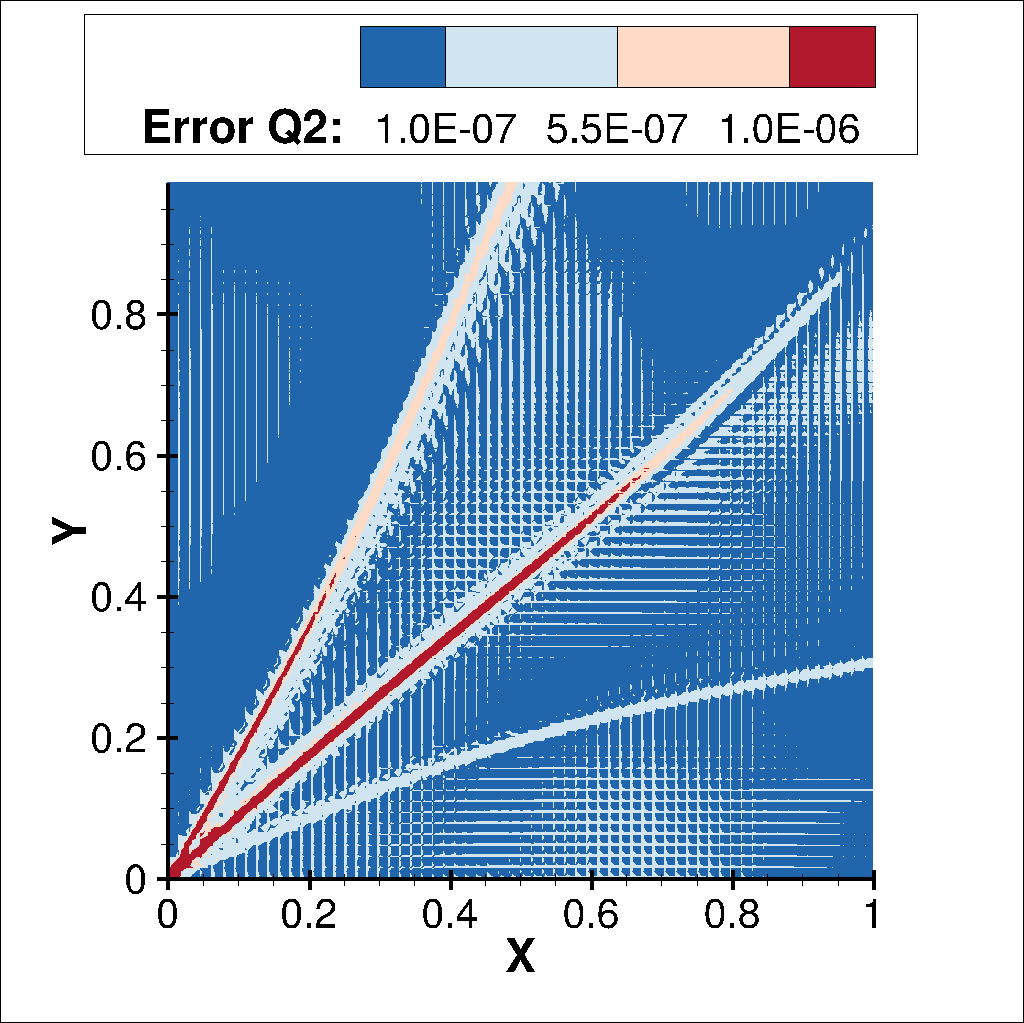}}
\caption{Effect of mesh refinement on the distribution of the $\rho u$ discretization error for the polynomial degree $\mathrm{P}3$ and MS-2  with the spurious boundary condition of $(\rho u )^{BC}=1.000001 \times(\rho^{\mathrm{MS}} u^{\mathrm{MS}}) \vert_{BC}$ at $(x,y)=(0,0)$}
\label{fig:Errordis_MS-2-screw}
\end{figure}

The importance of the $L_\infty$ norm in code verification has previously been discussed in the literature (see \cite{Eca2016} and \cite{Bond2007}) as to be a valuable metric for the detection of localized inconsistencies such as coding errors in the implementation of the boundary conditions. But the discussion has mostly stemmed from theoretical and practical appreciations of the qualities of this norm.
Here, we demonstrate the importance of the $L_\infty$ norm by rather providing a concrete example: a bug affecting the implementation of the boundary condition at a single point is introduced by changing the boundary value of the $\rho u$ variable of MS-2 at $(x,y)=(0,0)$ such that $(\rho u )^{BC}=1.000001 \times(\rho^{\mathrm{MS}} u^{\mathrm{MS}}) \vert_{BC}$, while keeping all other boundary conditions at all other points intact in Eq. \eqref{eq:bc_Riem_MS}. Fig. \ref{fig:Orders_MS-2_screw} shows the evolution of the resulting discretization errors and OOAs versus mesh refinement for the polynomial degrees $\mathrm{P}1$ to $\mathrm{P}3$ on the un-deformed domain of MS-2. For $\mathrm{P}1$, none of the error norms has yet been affected by the punctual boundary error on the finest mesh. Indeed, the discretization error of the $\mathrm{P}1$ is rather large and drops slowly with mesh refinement ($\mathcal{E}\sim\mathcal{O}(h^2)$) when compared to higher orders.  Hence, for the boundary inconsistency to be manifested in a $\mathrm{P}1$ polynomial discretization, one needs to drive the mesh refinement process up to very fine grids for the level of discretization error to be lower than that caused locally by the boundary bug that is around $\mathcal{E}_{\rho u}(0,0)=|\rho u(0,0)-\rho u_{ex}(0,0)|\simeq 2\times 10^{-5}$, see Fig. \ref{fig:Orders_MS-2_screw} (a). For $\mathrm{P}2$ however, the abnormality is detected on the finest mesh but only by the the $L_\infty$ norm. As the degree of the polynomial discretization is increased, the punctual source of stagnating error  dominates the norms for lower mesh sizes since the discretization error diminishes more rapidly on coarser grids for higher $\mathrm{P}$. As such, the $L_\infty$ norm of $\mathrm{P3}$ is affected by the spurious boundary condition as off the third grid with a mesh size of $h\simeq 1.5\times 10^{-2}$ which is the coarsest grid to manifest the bug amongst all the polynomial degrees considered. But still, it takes another level of mesh refinement in $\mathrm{P3}$ for the $L_1$ and $L_2$ norms to pick the error up. The effect of the mesh size on the discretization error is better illustrated in Fig. \ref{fig:Errordis_MS-2-screw} that compares for $\mathrm{P3}$, the higher levels of discretization error, distributed regularly on a $16\times16$ grid, to the lower levels of discretization error on a $32\times32$ grid where  the manufactured bug at $(x,y)=(0,0)$ is exhibited as the global maximum error on the domain. These observations support the idea that the $L_\infty$ norm is a crucial metric to ensure the early detection of localized sources of error on coarse grids. Finally, it is interesting to  note in Fig. \ref{fig:Errordis_MS-2-screw} (b), that the error is being convected in the stream-wise direction from the punctual source.

\subsubsection{Importance of the inclusion of a high-order discretization}
\begin{figure}[!hbt]
\centering
\subfloat[Discretization error ]{
\includegraphics[trim = 9mm 3mm 18mm 12mm, clip,width=0.33\linewidth]
{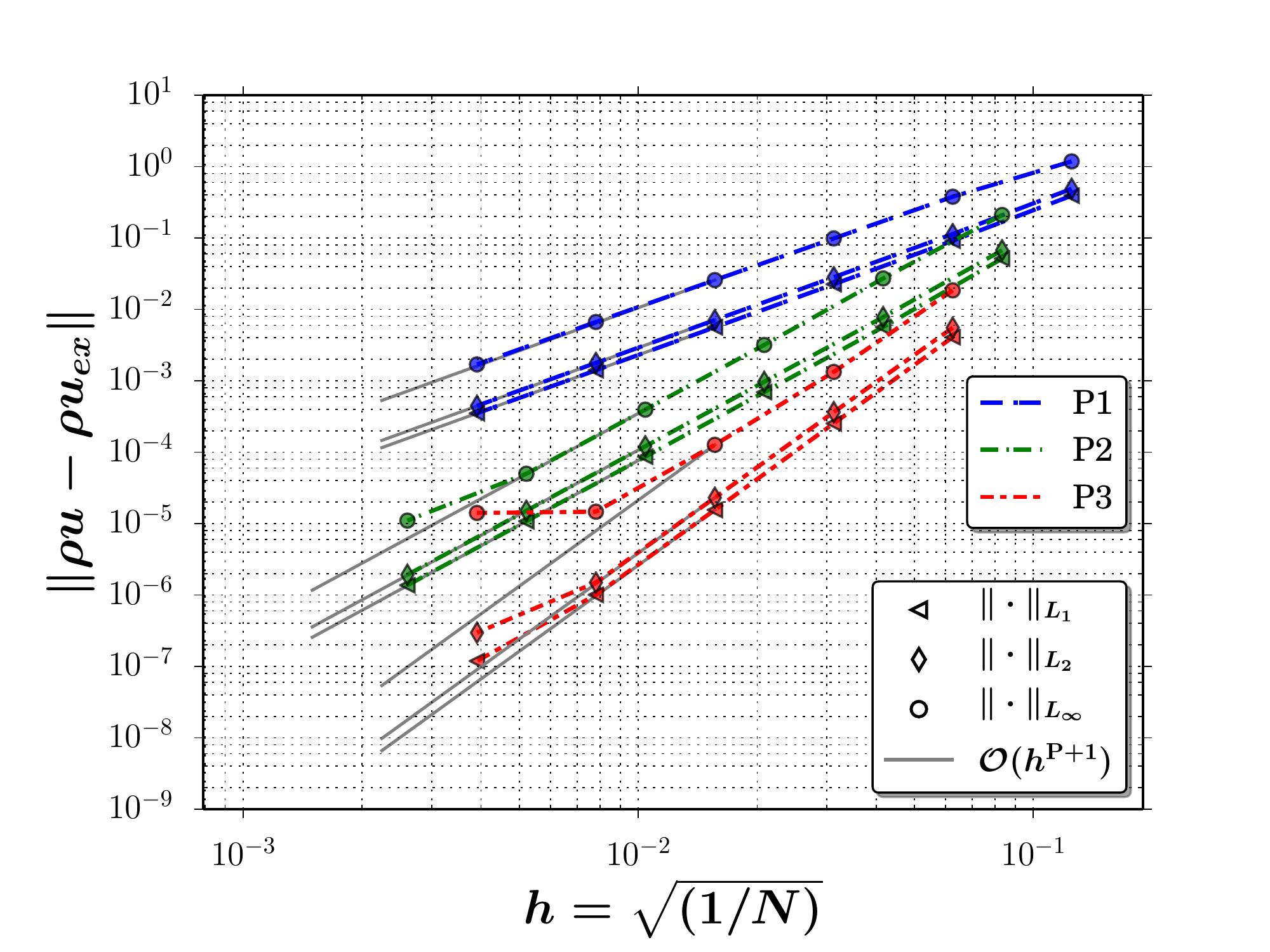}}~~~
\subfloat[Order of accuracy]{
\includegraphics[trim = 9mm 3mm 18mm 12mm, clip,width=0.33\linewidth]{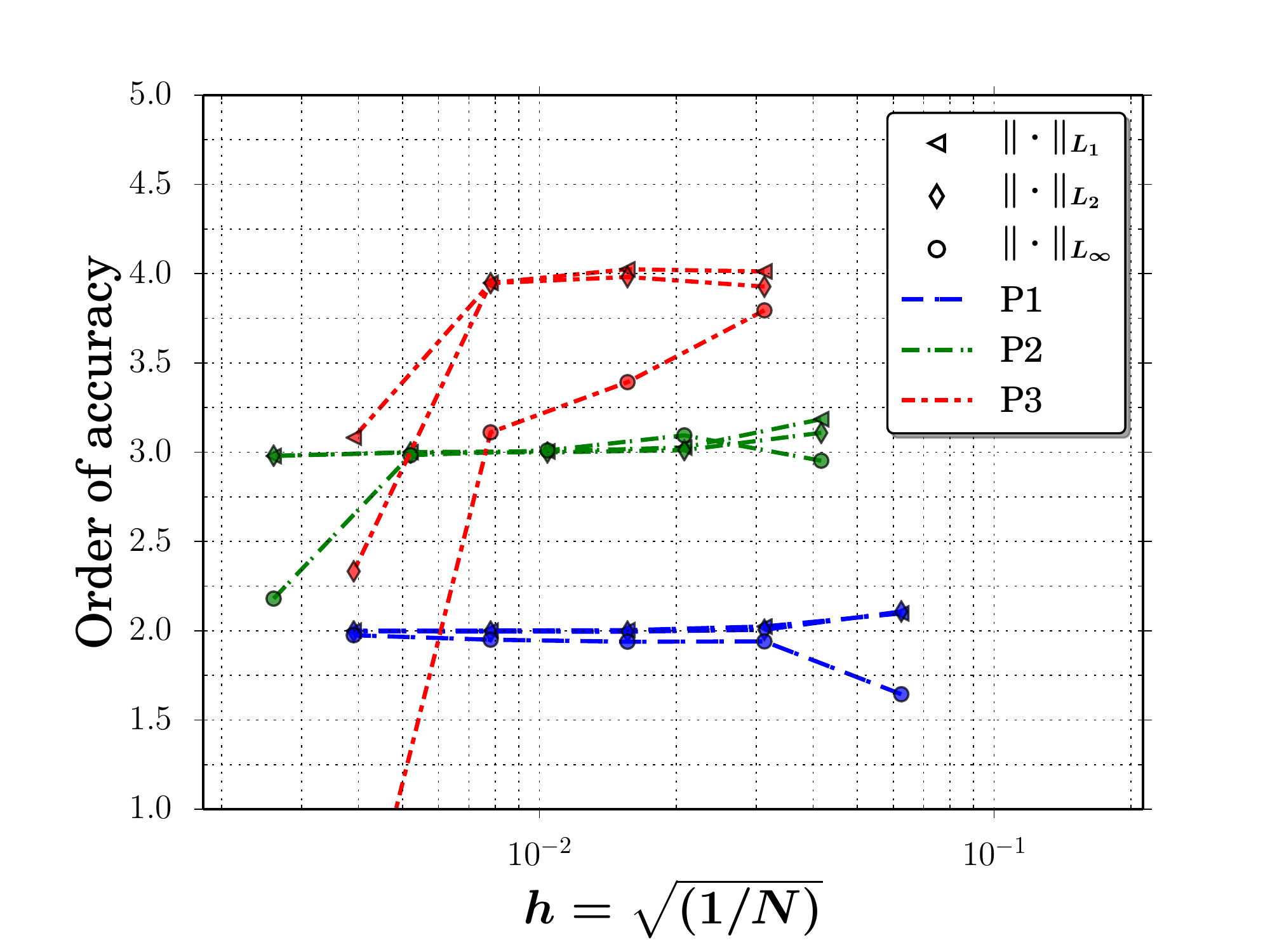}}
\caption{Evolution of the discretization errors and the OOAs  in $L_1$, $L_2$ and $L_\infty$ norms versus mesh refinement for $\rho u$, polynomial degrees $\mathrm{P}1$--$\mathrm{P}3$ and MS-2 with the spurious boundary condition of $(\rho u )^{BC}=1.000001 \times(\rho^{\mathrm{MS}} u^{\mathrm{MS}}) \vert_{BC}$ at $(x,y)=(0,0)$} 
\label{fig:Orders_MS-2_screw}
\end{figure}

One other essential conclusion can be drawn from Fig. \ref{fig:Orders_MS-2_screw} pertaining to the verification of high-order codes: it is desirable to include  a fairly high polynomial degree in the verification process to ensure the early detection of bugs causing errors with small magnitudes. Indeed, since the higher polynomials reduce the discretization error at a higher rate, the presence of the slightest bug can be revealed on coarser grids than if lower orders were used, thus providing a greater confidence in the absence of minor bugs in the code.

Since the Euler equations are verified for the supersonic regime as well, we can focus on the verification of the NS and RANS-SA equations in the subsonic regime. This relies on the fact that flow regime in terms of Ma directly affects the treatment of the inviscid portion of the equations and its corresponding boundary conditions whereas the treatment of the viscous terms is independent of the Ma number.

\subsection{Laminar flows - MS-3}
\label{sec:lam_flows}


\begin{table}
\centering
\begin{tabular}{ c||c|c|c|c|c|c|c }
$(\cdot)$& $(\cdot)_0$ & $(\cdot)_x$ & $(\cdot)_y$ & $(\cdot)_{xy}$ & $a_{(\cdot)_x}$ & $a_{(\cdot)_y}$ & $a_{(\cdot)_{xy}}$\\ \hline\hline
  $\rho$ &   $1.0$     & $0.1$       & $-0.2$     & $0.1$           &      $1.0$      &       $1.0$     &         $1.0 $     \\ \hline
  $u$    &   $2.0$     & $0.3$       & $0.3$      & $0.3$           &      $3.0$      &       $1.0$     &         $1.0 $     \\ \hline  
  $v$    &   $2.0$     & $0.3$       & $0.3$      & $0.3$           &      $1.0$      &       $1.0$     &         $1.0 $     \\ \hline  
  $p$    &   $10.0$     & $1.0$      & $1.0$      & $0.5$           &      $2.0$      &       $1.0$     &         $1.0 $    \\ \hline  
    $\tilde{\nu}$    &   $0.0$     & $0.0$       & $0.0$     & $0.0$           &      $0.0$      &       $0.0$     &         $0.0 $ 
\end{tabular}
\caption{Parameters of MS-3}
\label{tb:MS-3_cons}
\end{table}

\begin{figure}[!hbt]
\centering
\subfloat[MS-1]{
\includegraphics[trim = 16mm 3mm 18mm 13mm, clip,width=0.33\linewidth]
{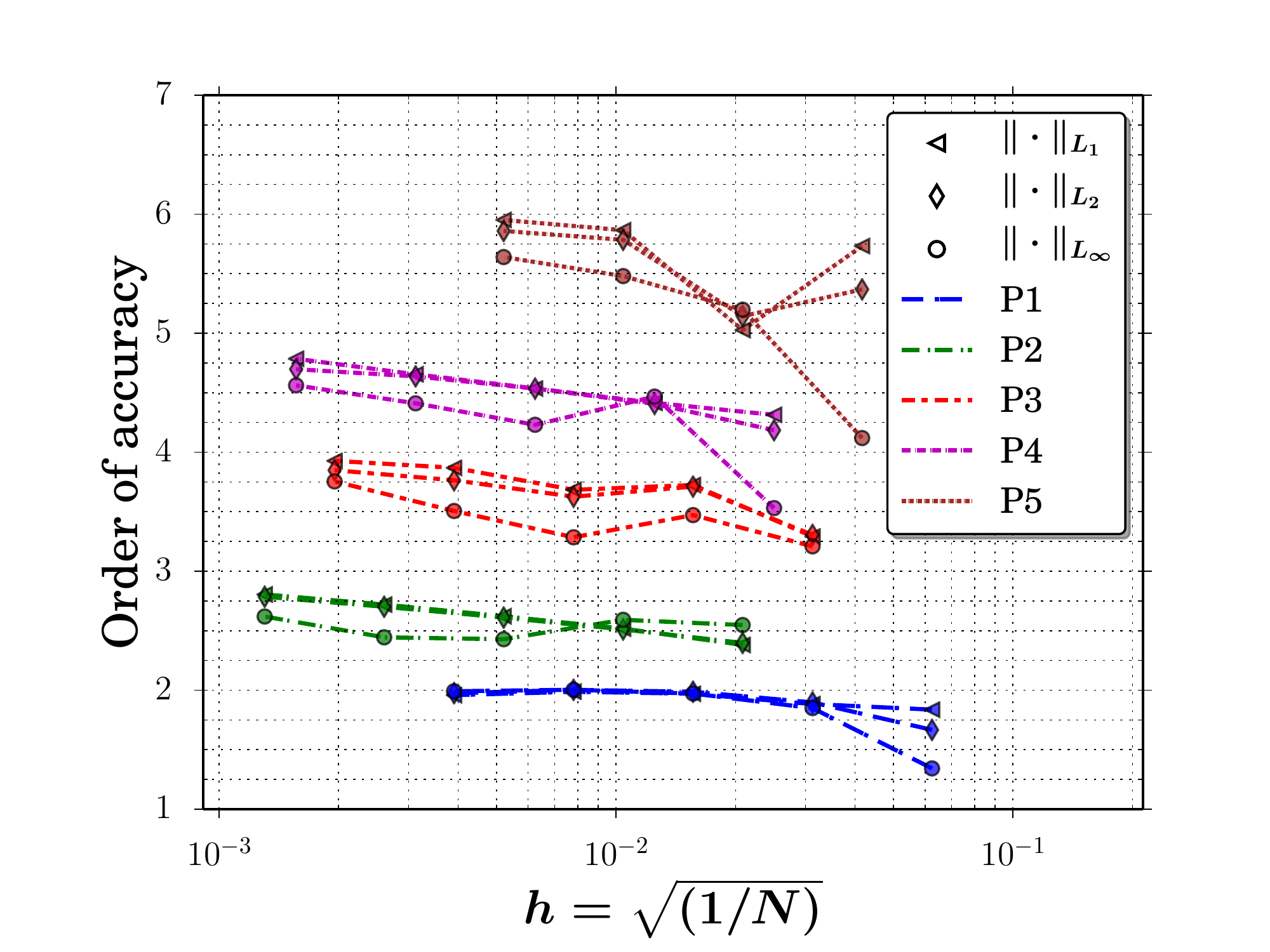}}~~~
\subfloat[MS-3]{
\includegraphics[trim = 16mm 3mm 18mm 13mm, clip,width=0.33\linewidth]
{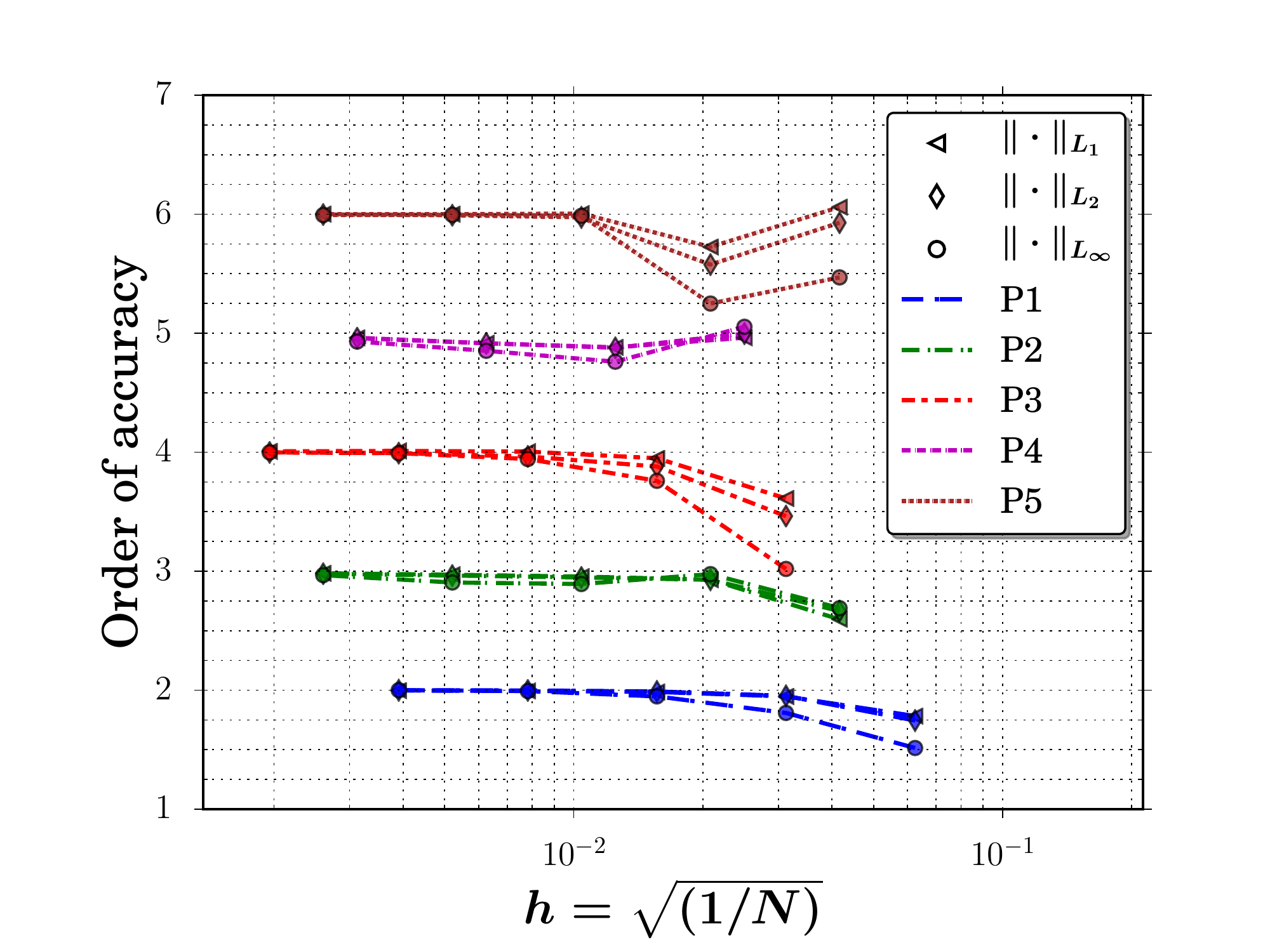}}
\caption{Evolution of the OOAs of $\rho$  in $L_1$, $L_2$ and $L_\infty$ norms versus mesh refinement for MS-1 in the viscous mode and MS-3, for $\mu = 1\times10^{-1}$  and polynomial degrees $\mathrm{P}1$--$\mathrm{P}5$}
\label{fig:Orders_MS-1_vs_MS-3}
\end{figure}

In a first attempt, MS-1, that served to verify the Euler equations in subsonic and simultaneously compressible and incompressible regimes, was extended to viscous mode by adding the diffusion operator of the NS equations to the forcing functions. Although attaining the asymptotic range on reasonably fine grids for both lightly ($\mu = 1 \times 10^{-4}$) and highly ($\mu = 1 \times 10^{+2}$) viscous flows, MS-1 resulted in very delayed asymptotic ranges for intermediary values of dynamic viscosity ($\mu = 1 \times 10^{-1}$) as shown in Fig. \ref{fig:Orders_MS-1_vs_MS-3} (a) for errors in $\rho$. In this moderately viscous mode, the inviscid and viscous terms of the forcing functions of the momentum and energy equations take similar magnitudes in some regions of the domain whereas for more extreme values of $\mu$, the forcing functions are dominated by either. The superposition of first and second order derivatives of the primitive variables, emanating respectively from the  inviscid and viscous terms, translates to the combination of a large number of sinusoidal functions that could explain the delay in the manifestation of the monotonic convergence range. In fact, to capture the multi-modal forcing function, finer grids are necessary, especially for high-order polynomial discretizations that exhibit a higher sensitivity to sub-cell spatial variations, even on coarse grids. Nevertheless, by tuning the trigonometric MS parameters, one could devise a numerically benign MS that yields a reasonably fast asymptotic range for largely distinct values of $\mu$. The search for such properties resulted in the elaboration of MS-3, that verifies the NS equations in the compressible portion of the subsonic regime. MS-3 is defined by Eqs. \ref{eq:trigo_MS} along with the parameters of Table \ref{tb:MS-3_cons}. The spatial distributions of its primitive variables and Ma number are illustrated in Fig. \ref{fig:MS-3}. For this MS, the results of the grid convergence study in terms of the error and order curves are respectively presented in Figs. \ref{fig:Err_allE_allP_MS-3} and \ref{fig:Orders_MS-3}. Figure \ref{fig:Orders_MS-1_vs_MS-3} compares the evolution of OOAs for $\rho$ versus mesh size for MS-1 and MS-3. One can appreciate the faster convergence (in $h$) of MS-3, that is the result of the proper tuning of its parameters. It is interesting to note that although the convergence of the higher orders is delayed for MS-1, its observed orders for P1 converge to the value of $2.0$ on rather coarse grids, demonstrating the particular difficulty in deriving numerically benign manufactured solutions for the verification of high OOAs. 


\subsubsection{Balancing of the forcing function terms}
\begin{figure}[!hbt]
\centering
\subfloat[$\mu = 1 \times 10^{-4}$]{
\includegraphics[trim = 0mm 0mm 0mm 0mm, clip,width=0.27\linewidth]{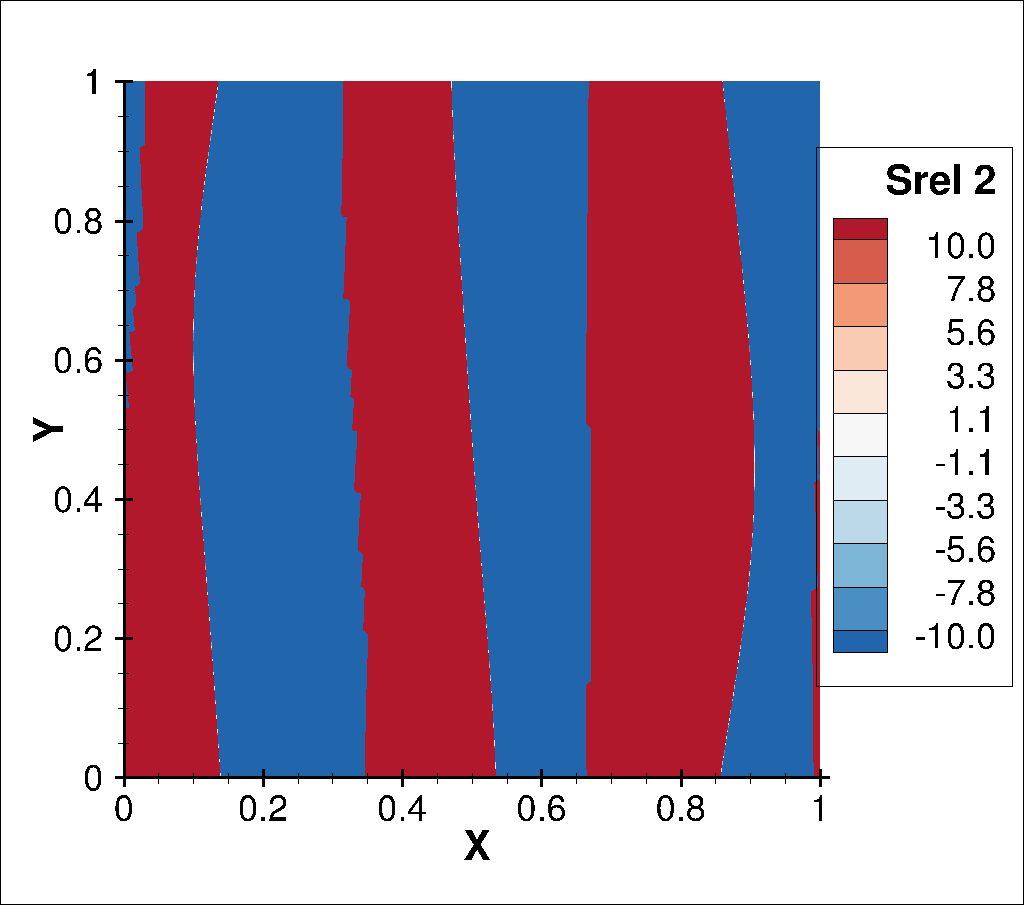}}~~
\subfloat[$\mu = 1 \times 10^{-1}$]{
\includegraphics[trim = 0mm 0mm 0mm 0mm, clip,width=0.27\linewidth]
{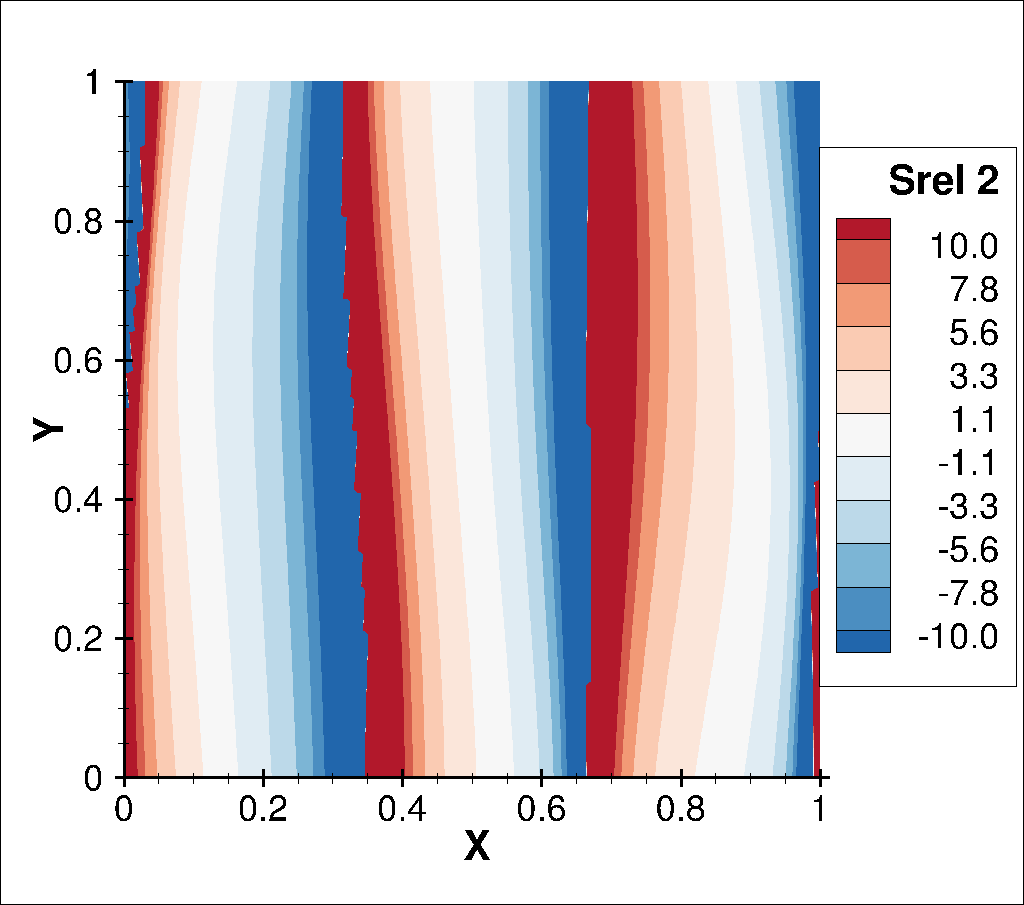}}~~
\subfloat[$\mu = 1 \times 10^{+2}$]{
\includegraphics[trim = 0mm 0mm 0mm 0mm, clip,width=0.27\linewidth]
{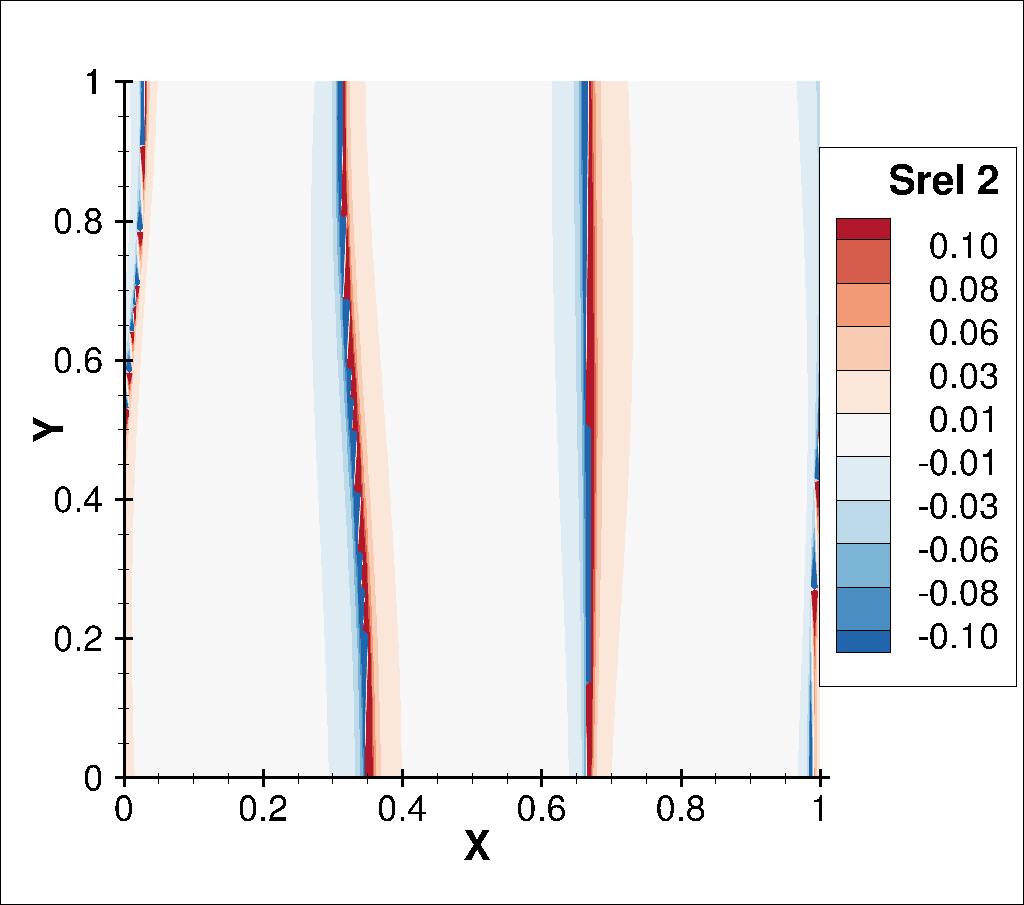}}
\caption{Relative magnitude of the inviscid over viscous terms of the forcing function of the $x$-momentum equation for MS-3 with different $\mu$ values}
\label{fig:SR2_mueffect}
\end{figure}

In realistic flow problems, the Reynolds number (Re) provides a measure of the relative magnitude of advective over the diffusive forces, that is also related to the ratio of the inviscid over the viscous terms of the NS equations. The magnitude of the viscous terms depends on both the value of the viscosity and the strain rate which involves the velocity gradients. The latter are intrinsically adjusted by flow features such as the boundary layer thickness, the free-stream  velocity, the characteristic length, etc. that are themselves correlated with Re (and consequently with $\mu$). However, since a manufactured solution is not necessarily realistic, it might be absent of these features and the velocity gradients could take arbitrary values independently of the value of viscosity. Hence, to generalize the concept of Reynolds number to manufactured cases, we directly look into $\mathrm{S^\mathrm{rel}}(\mathbf{x}) = \mathrm{S^{inv}}(\mathbf{x})/\mathrm{S^{vis}}(\mathbf{x}) $, the ratio of the inviscid over the viscous terms of the manufactured forcing functions in the domain as in \cite{Roy2002}. Figure \ref{fig:SR2_mueffect} compares the $\mathrm{S^\mathrm{rel}}$ distributions for the $x$-momentum equation of MS-3 and three different non-dimensional values of dynamic viscosity, i.e.,  $\mu = 1 \times 10^{-4}$, $\mu = 1 \times 10^{-1}$ and $\mu = 1 \times 10^{+2}$. Using the same MS fixes the velocity gradients while  $\mu$ is changed. To ensure the verification of both the inviscid and viscous terms by the MS, the $\mathrm{S^\mathrm{rel}}(\mathbf{x})$ distribution should simultaneously include regions of viscous preponderance ($\vert \mathrm{S^\mathrm{rel}}(\mathbf{x})\vert \ll 1$) and inviscid prevalence ($1 \ll \vert \mathrm{S^\mathrm{rel}}(\mathbf{x})\vert$) along with non-negligible regions of equivalence ($\vert \mathrm{S^\mathrm{rel}}(\mathbf{x})\vert \approx1$). This is an important criterion, since when one mode outweighs the other by much, the verification process becomes biased and the detection of minor bugs by the MMS in the outweighed mode is jeopardized. Figure  \ref{fig:SR2_mueffect} suggests that for MS-3, a value of $\mu = 1 \times 10^{-1}$ guarantees a balanced leverage since for this value, non-negligible regions of inviscid and viscous dominance exist in different regions of the domain simultaneously.

\begin{figure}[!hbt]
\centering
\subfloat[$\mu=1\times 10^{-4}$]{
\includegraphics[trim = 11mm 3mm 18mm 12mm, clip,width=0.33\linewidth]
{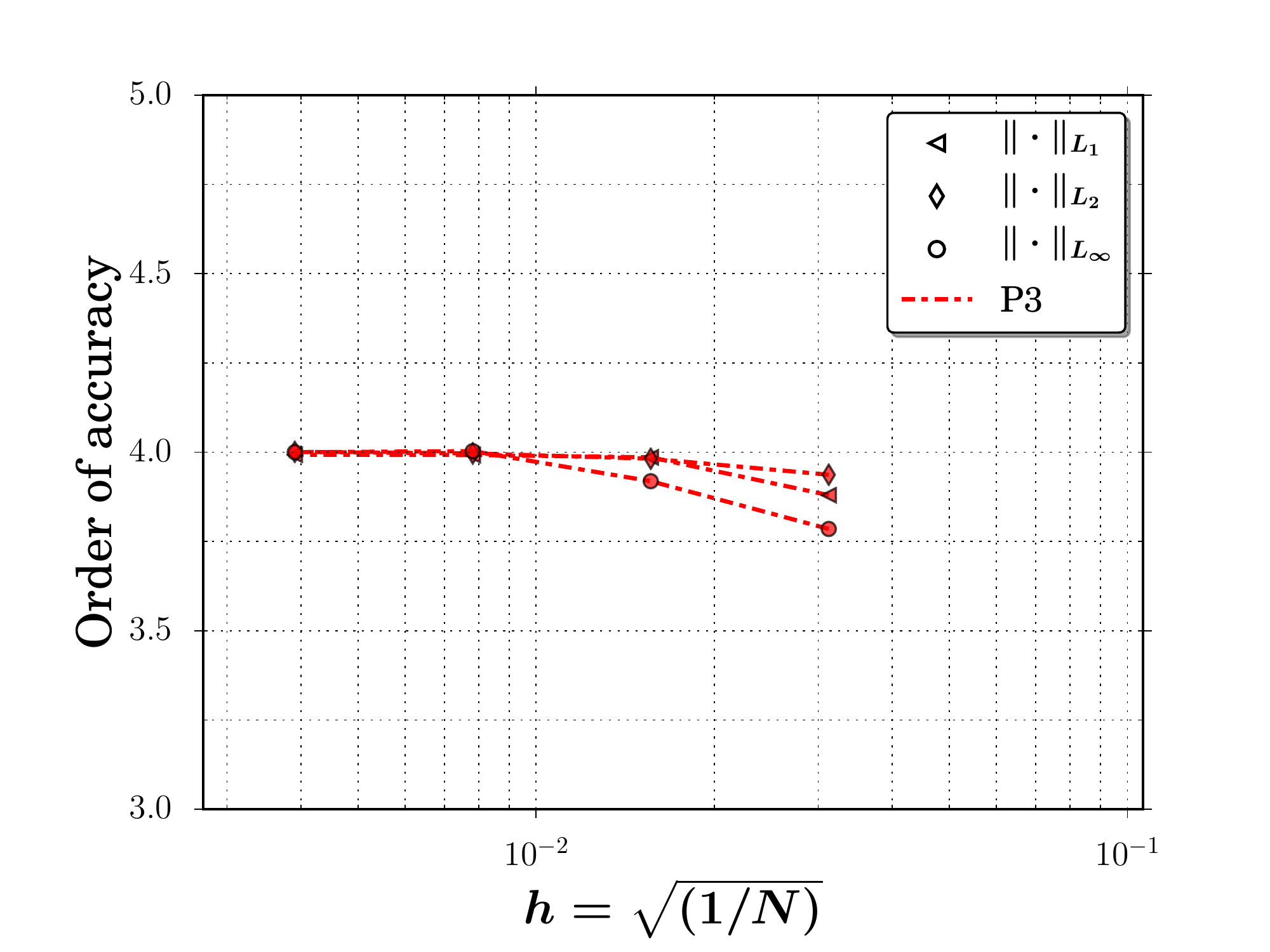}}~~~
\subfloat[$\mu=1\times 10^{-1}$]{
\includegraphics[trim = 11mm 3mm 18mm 12mm, clip,width=0.33\linewidth]
{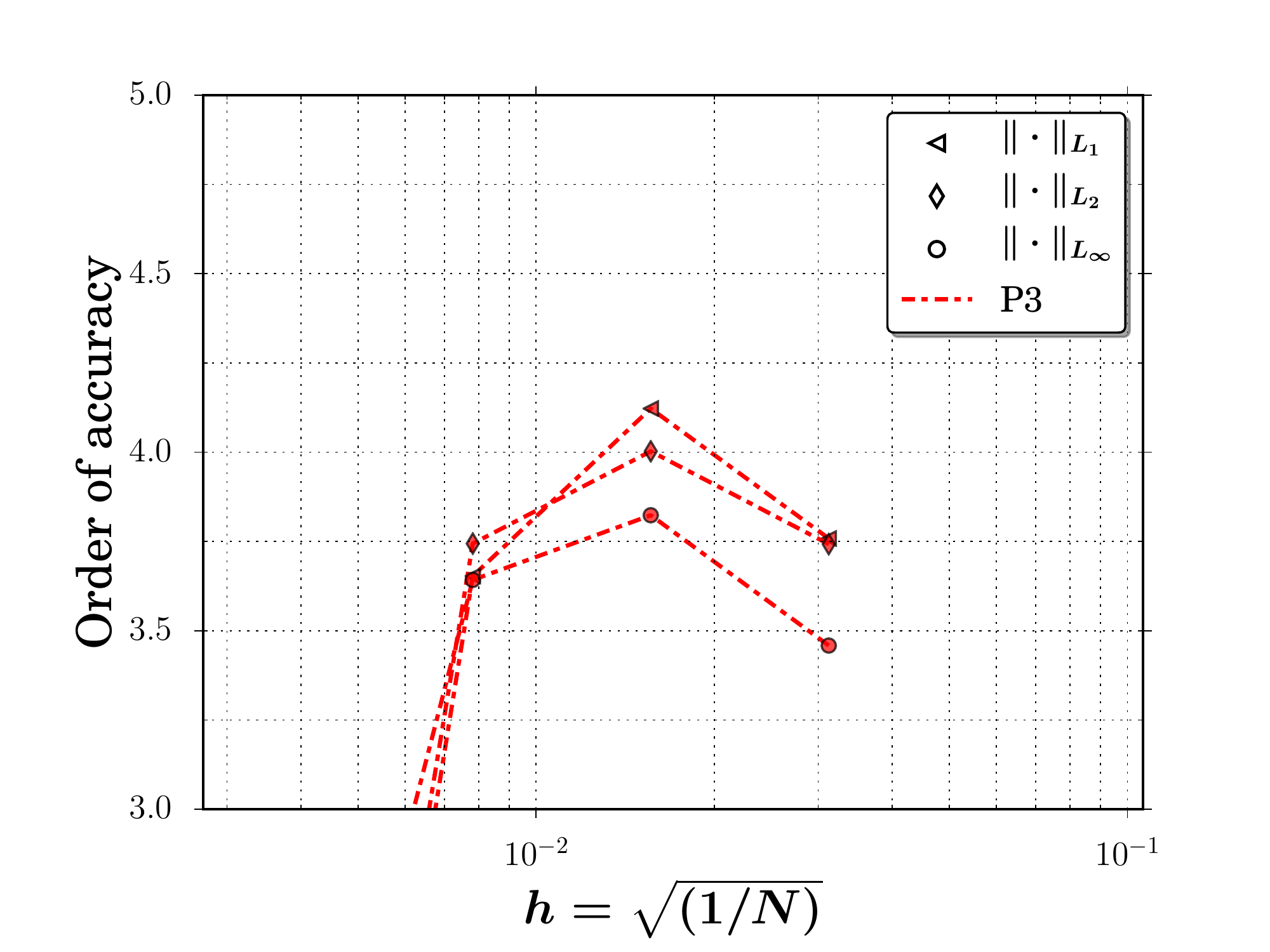}}
\caption{Evolution of the OOAs of $\rho E$ in $L_1$, $L_2$ and $L_\infty$ norms versus mesh refinement for polynomial degree $\mathrm{P}3$ and MS-3 with $\mu=1 \times 10^{-4}$ and $\mu=1\times 10^{-1}$ under the effect of a spurious heat flux term}
\label{fig:Orders_heat_bug}
\end{figure}

To further demonstrate the importance of balancing different terms of the manufactured forcing functions in the verification process, we have introduced a bug by modifying the heat flux in Eq. \eqref{eq:ener}, such that $\omega_j =-\alpha\lambda_{\mathrm{eff}}\, \partial_j T$, where the value of the factor $\alpha$ is changed from its original value of $\alpha=1.0$ to $\alpha=1.0001$. This bug can go undetected for $\mu=1\times 10^{-4}$ and only revealed when the viscosity is increased to $\mu=1\times 10^{-1}$ as shown in Fig \ref{fig:Orders_heat_bug} for the error in $\rho E$ of MS-3, discretized by P3.

\subsubsection{Residual convergence}
\begin{figure}[!hbt]
\centering
\subfloat[$\mu = 1 \times 10^{-4}$]{
\includegraphics[trim = 5mm 4mm 4mm 10mm, clip,width=0.36\linewidth]
{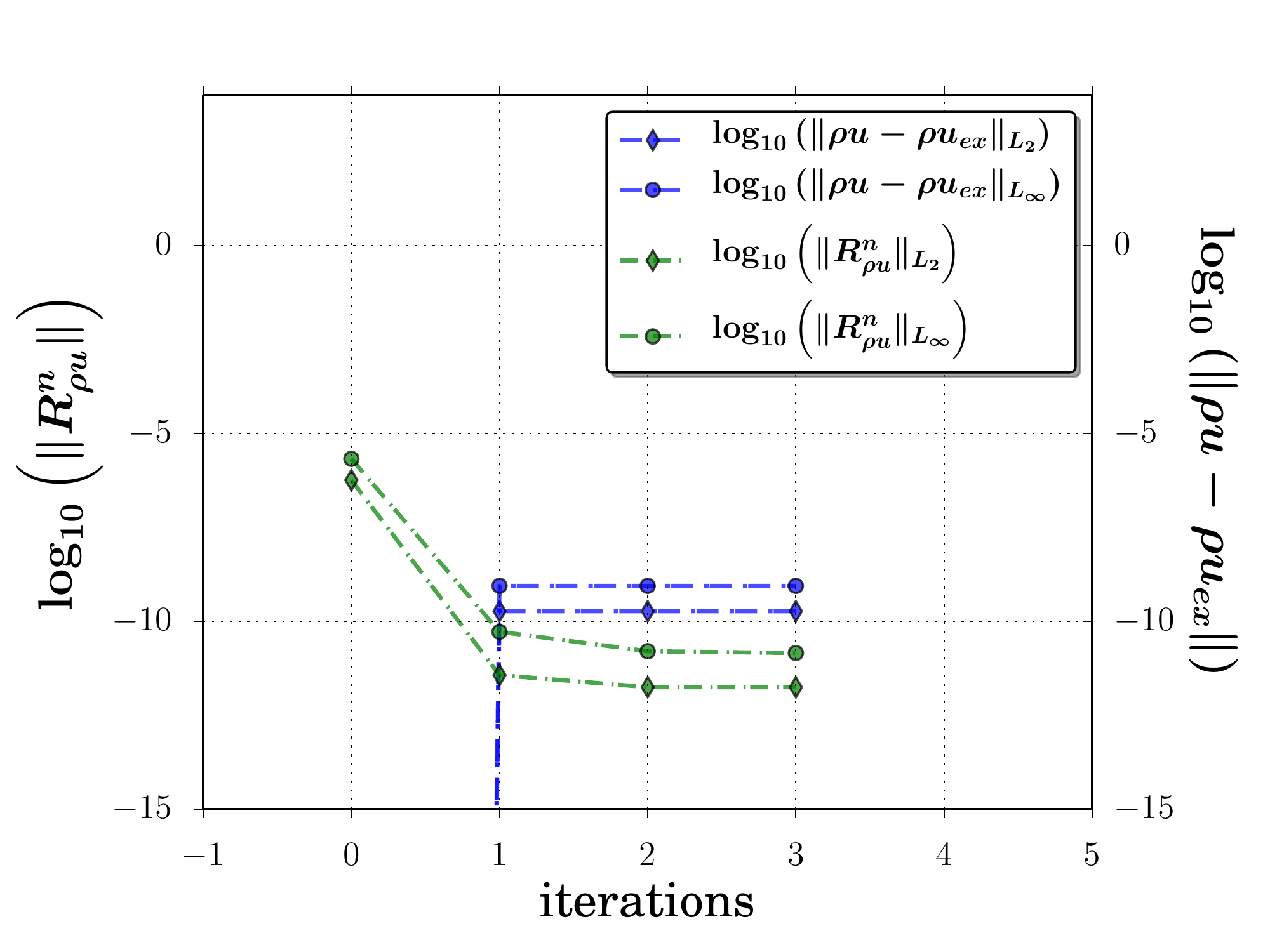}}
~~~
\subfloat[$\mu = 1 \times 10^{-1}$]{
\includegraphics[trim = 5mm 4mm 4mm 10mm, clip,width=0.36\linewidth]{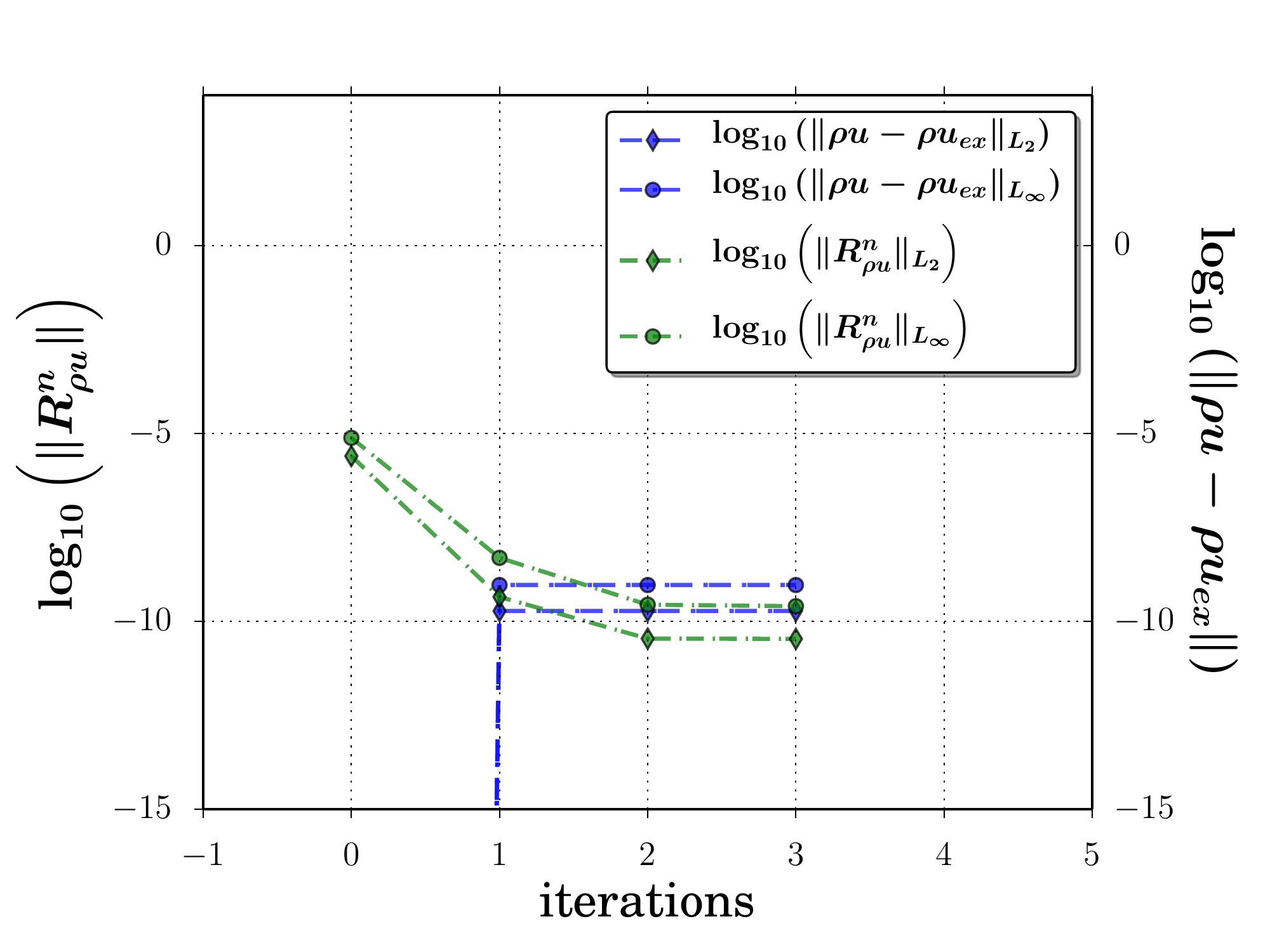}}
\vfill
\subfloat[$\mu = 1 \times 10^{+2}$]{
\includegraphics[trim = 5mm 4mm 4mm 10mm, clip,width=0.36\linewidth]
{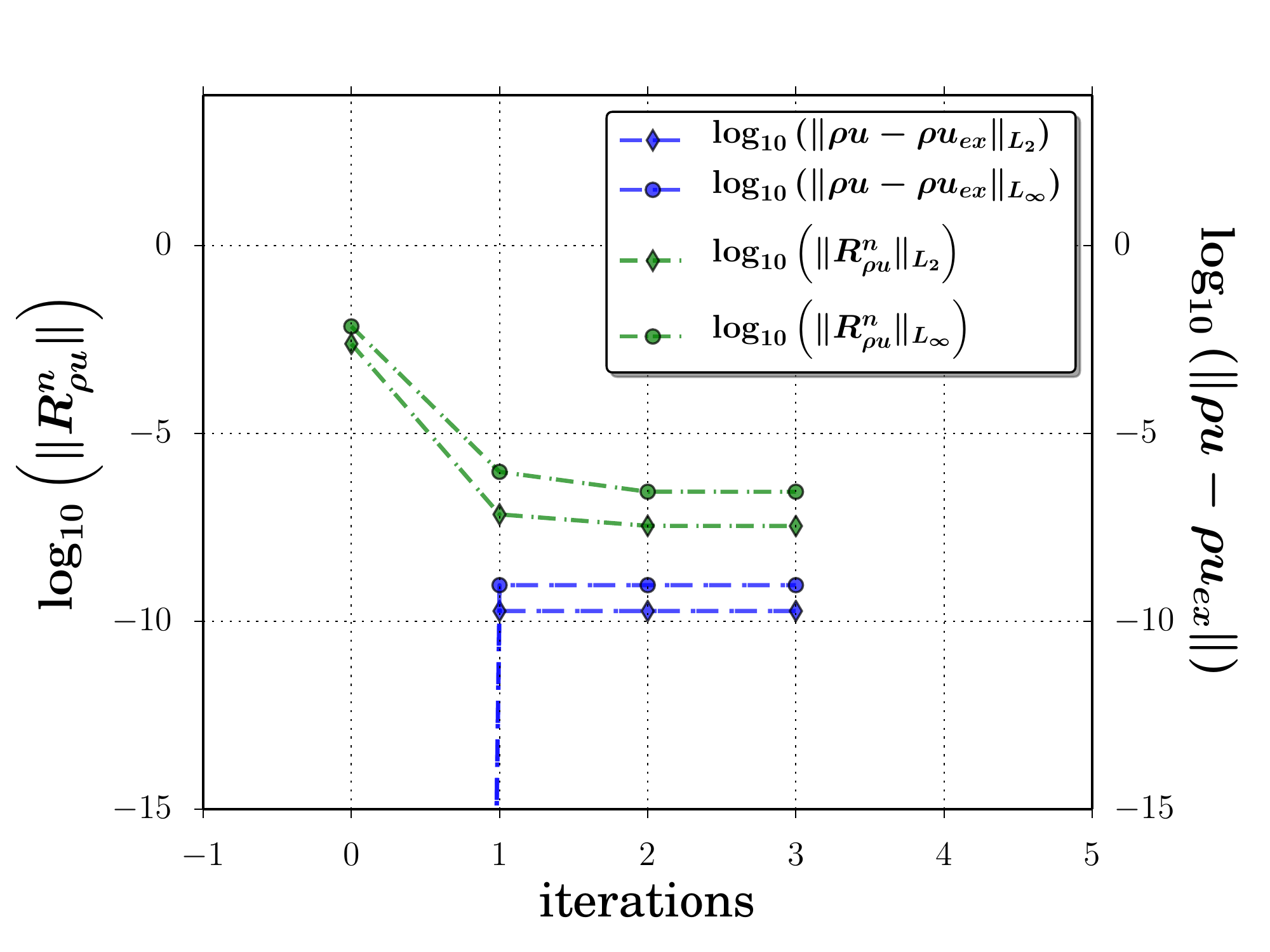}}
~~~
\subfloat[$\mu = 1 \times 10^{+4}$]{
\includegraphics[trim = 5mm 4mm 4mm 10mm, clip,width=0.36\linewidth]
{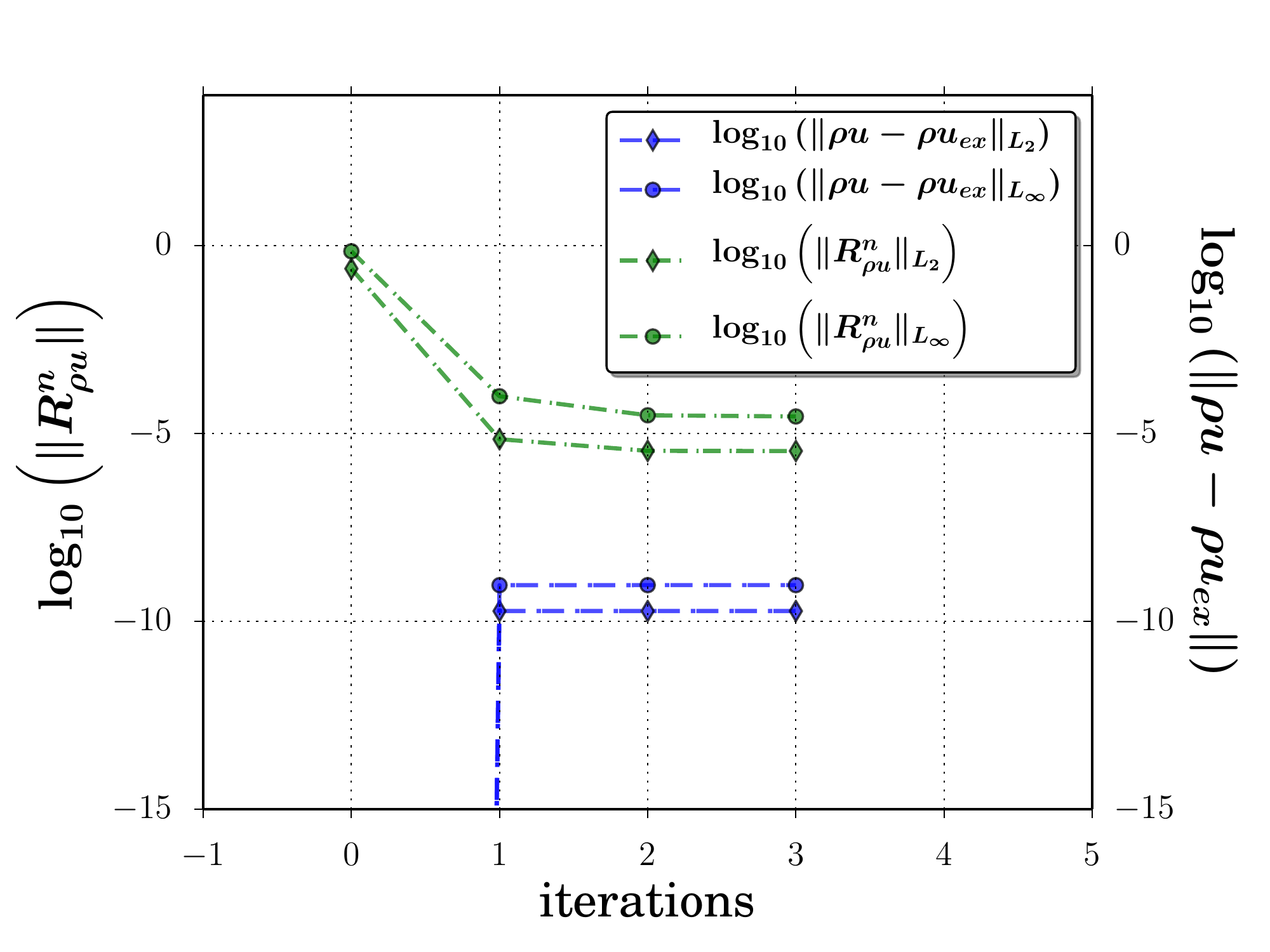}}
\caption{Residual convergence and discretization error of $x$-momentum versus number of iterations for MS-3 with different $\mu$ values, polynomial degree $\mathrm{P}5$ and grid of $32\times32$ elements}
\label{fig:res_conv_mueffect}
\end{figure}

The residual convergence is affected by the inviscid/viscous ratio in two fashions: firstly, as $\mu$ is increased for the same MS, the convergence of iterative linear system solvers such as GMRES is hindered since the prevalence of the viscous terms reduces the diagonal dominance of the system and consequently the system conditioning degrades. This translates to an increase in the number of GMRES iterations needed to satisfy its internal convergence criterion. Secondly, for an exact initialization, the initial residual takes larger values as $\mu$ is increased. This effect is shown in Fig. \ref{fig:res_conv_mueffect} that compares the residual and discretization error convergences of MS-3 for viscosity values of $\mu = 1 \times 10^{-4}$, $\mu = 1 \times 10^{-1}$, $\mu = 1 \times 10^{+2}$ and $\mu = 1 \times 10^{+4}$ discretized by P5 and a grid of $32 \times 32$ elements. The increase in the initial residual with $\mu$ reflects the magnifying effect of the viscosity on the truncation error of the viscous terms since any norm of the discrete residual computed using the exact solution is in fact a measure of the truncation error (see Eq. \ref{eq:sample_pde3b}). 

Once the viscous terms dominate the residual equation, increasing the viscosity any further just scales the level of the residual norm as it ensues from the comparison of residuals for $\mu = 1 \times 10^{+2}$ and $\mu = 1 \times 10^{+4}$ in Fig. \ref{fig:res_conv_mueffect}. It is noteworthy, in the same figure, that although the residual level is sensitive, the norms of the discretization error are quasi invariable with regards to viscosity. This suggests that the level of the discretization error has a higher sensitivity to the construction of the manufactured variables and their derivatives than to the terms of the governing equations considered in this study.

Another important observation can be made form Fig. \ref{fig:res_conv_mueffect} with regards to the required level of iterative convergence. It indeed is argued in \cite{Roy2005} that the iterative errors scale closely with the level of the discrete residual in a wide range of flow problems and in \cite{Roy2004} that the iterative errors should be driven down to two to three orders of magnitude lower than the discretization error with the goal of ensuring that the iterative and round-off errors are negligible and do not affect the verification process. Hence, from Fig. \ref{fig:res_conv_mueffect}, it is obvious that the final value of the residual at convergence is a function of $\mu$ and could be higher than the discretization error. Consequently, it seems that a convergence criterion requiring the residual norm to be lower than the discretization error is overly conservative and in fact unrealizable in some cases. Based on our experience, for numerically benign problems such as the free flows considered in this work, reaching the minimum residual possible, by a fully implicit method via exact initialization, is sufficient for the realization of the expected OOAs. For stiff problems and in the case where the OOAs can not be achieved, one could rather verify their sensitivity to the residual convergence by under-converging a few cases by one or two orders of residual magnitude and recomputing the OOAs. If the latter change significantly, it could be an indication of a need for further residual convergence.

\subsubsection{$H_1$ semi-norms}
In addition to the previously mentioned metrics, the verification of the viscous terms can be complemented by tracking the order of convergence in the $H_1$ semi-norm that monitors the discretization errors in state derivatives. For MS-3 with $\mu = 1 \times 10^{-1}$, the evolution of the errors in this metric versus mesh refinement  is provided in Fig. \ref{fig:Err_allE_allP_H_MS-3} for both the uncorrected ($\partial_q Q_k$) and fully corrected ($\overline{\overline{\partial_q Q_k}}$)  derivatives, for which the semi-norms are respectively labelled $H_1$ and $\overline{\overline{H_1}}$. One can firstly appreciate that whenever there is a significant difference between them, the fully corrected derivatives are more accurate than the uncorrected derivatives. Secondly, the error in this semi-norm recovers convergence slope of $\mathcal{O}(h^\mathrm{P})$ as expected. Note that the regularity of its convergence to the theoretical OOAs can be better assessed in Fig. \ref{fig:Orders_H_MS-3}. 

In order to evaluate the relative importance of $H_1$ semi-norms versus the $L$ norms, we have investigated the ability of each in detecting various types of spurious alterations to the implementation. In all cases, the performances of these types of norms were comparable and as such $H_1$ semi-norms showed no particular prevalence in exhibiting the presence of the bugs.

\begin{figure}[!hbt]
\centering
\subfloat[$\mu = 1 \times 10^{-4}$]{
\includegraphics[trim = 16mm 3mm 18mm 12mm, clip,width=0.33\linewidth]
{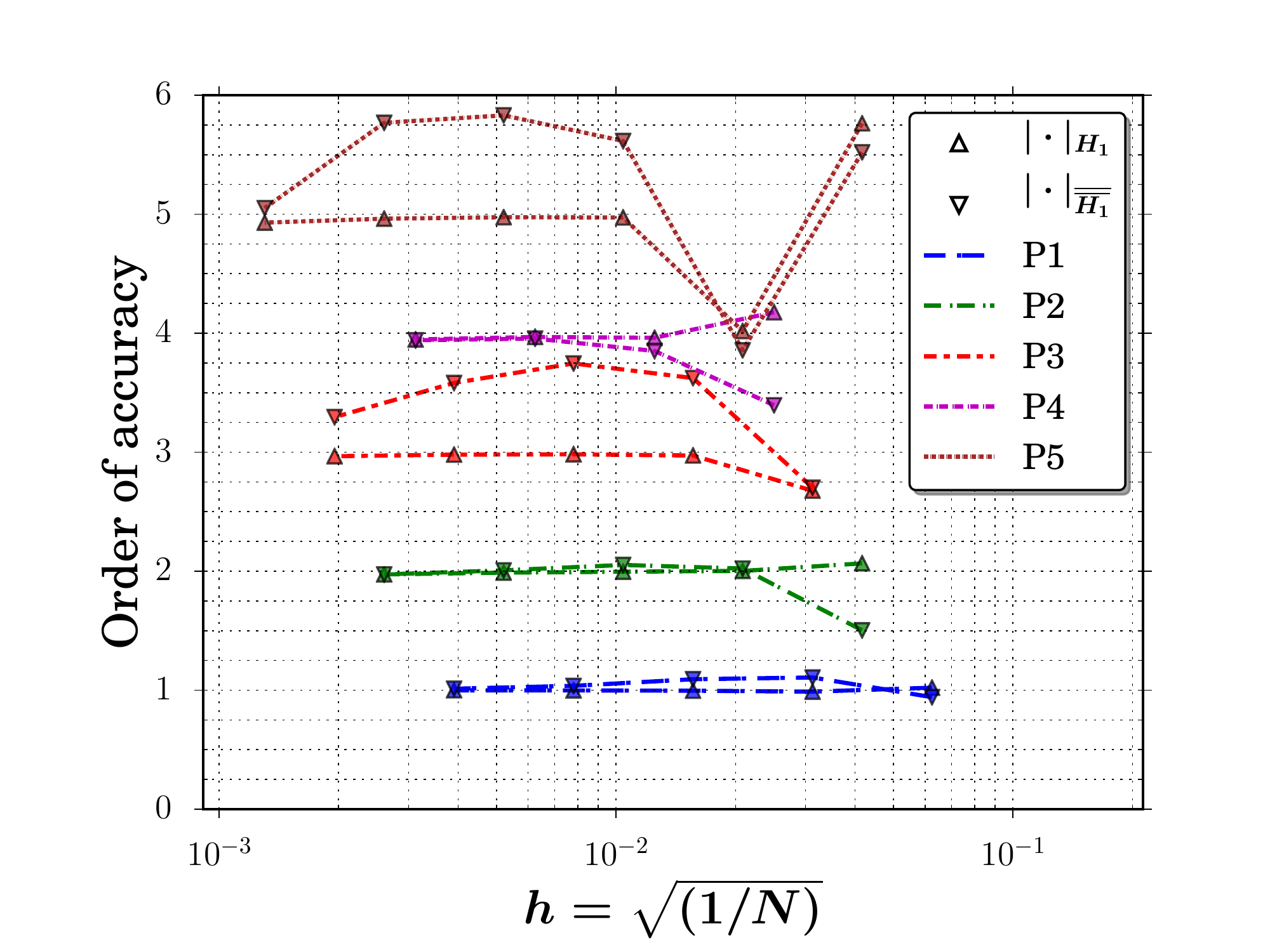}}~~~
\subfloat[$\mu = 1 \times 10^{+2}$]{
\includegraphics[trim = 16mm 3mm 18mm 12mm, clip,width=0.33\linewidth]
{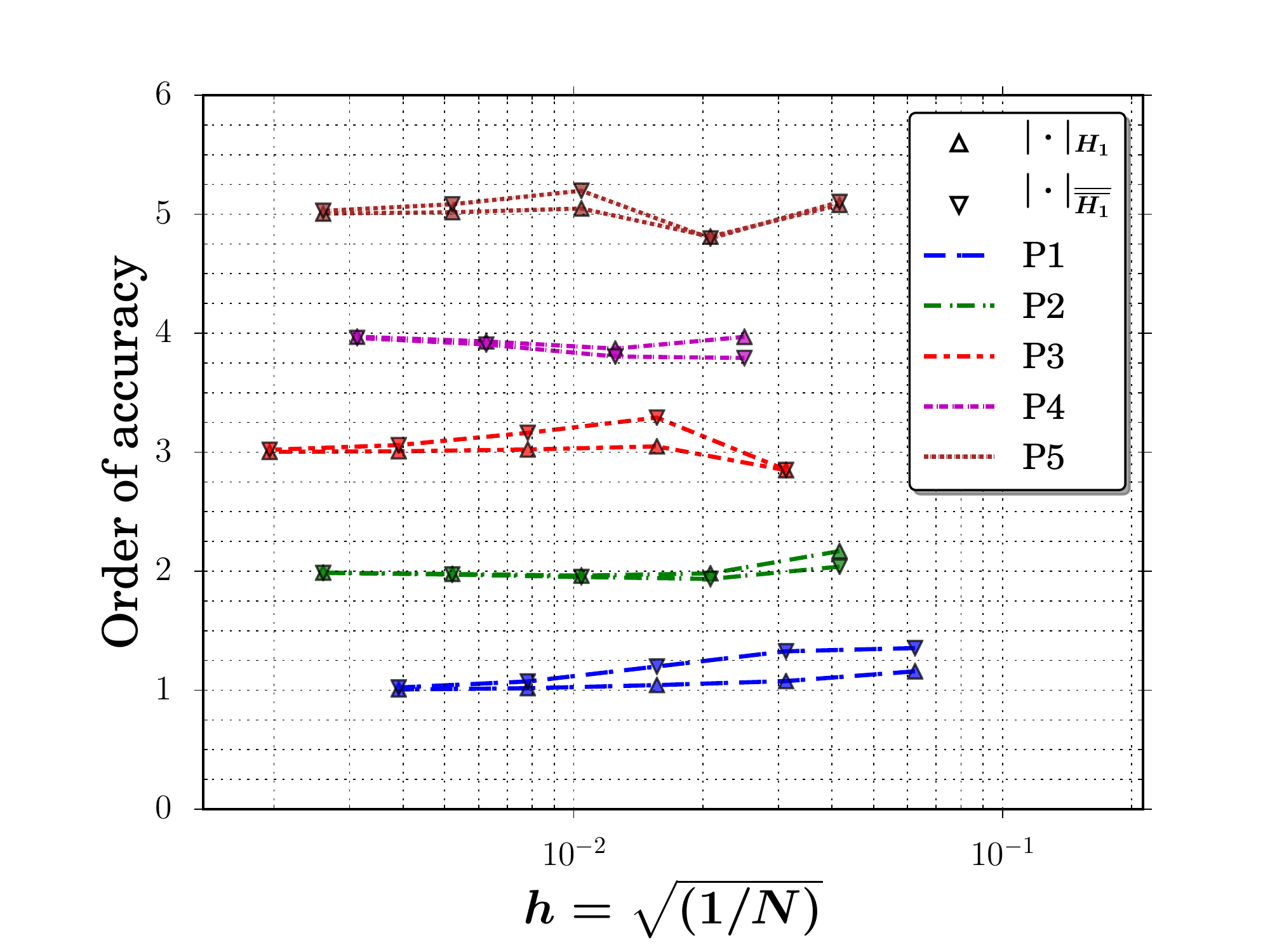}}
\caption{Evolution of the OOAs of $\rho v$ in $H_1$ semi-norm (for uncorrected and fully corrected derivatives) versus mesh refinement, for MS-3 with different $\mu$ values and polynomial degrees $\mathrm{P}1$--$\mathrm{P}5$}
\label{fig:Orders_mueffect-MS3-H}
\end{figure}

\begin{figure}[!hbt]
\centering
\subfloat[$\mu = 1 \times 10^{-4}$]{
\includegraphics[trim = 16mm 3mm 18mm 13mm, clip,width=0.33\linewidth]
{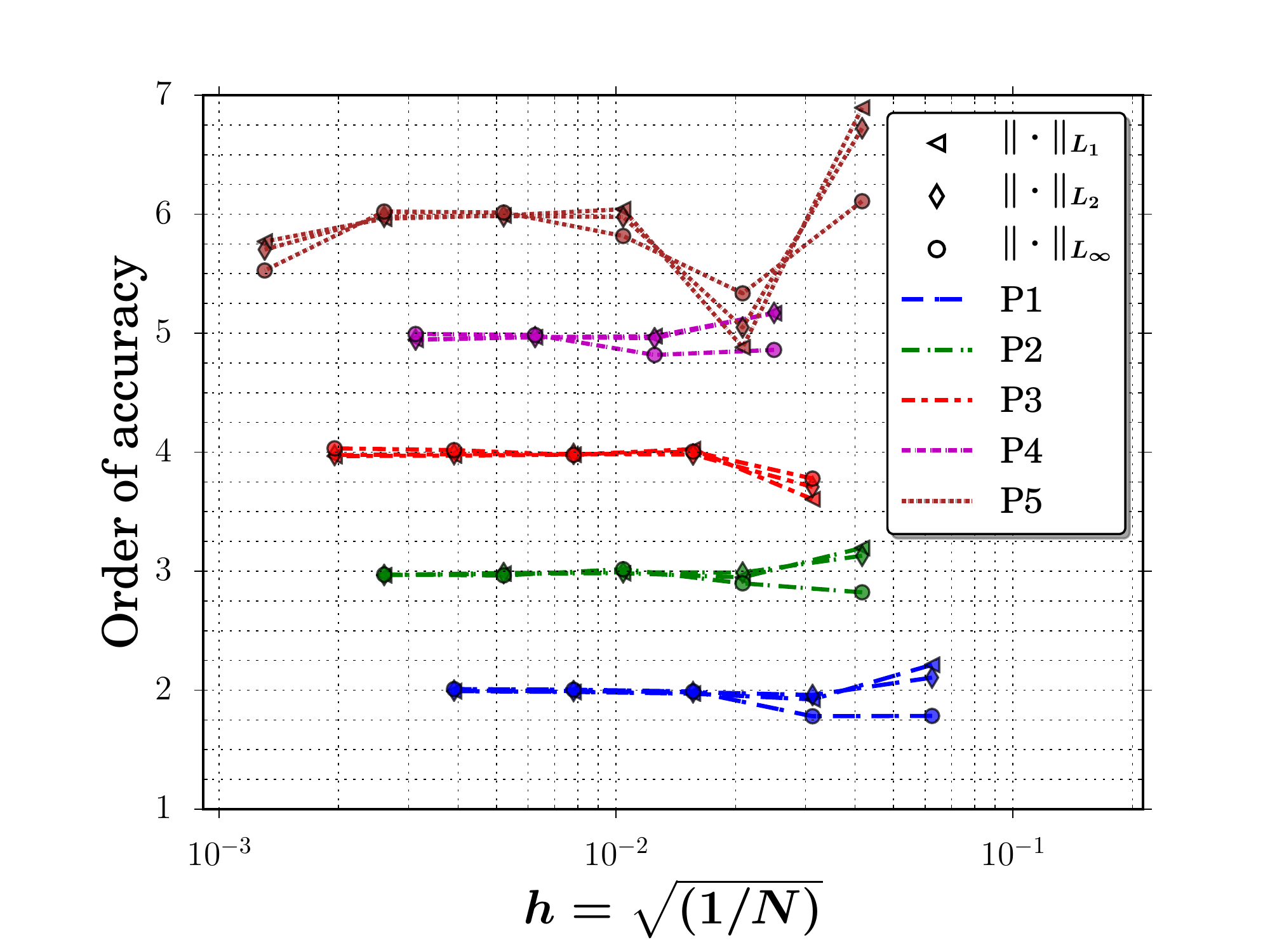}}~~~
\subfloat[$\mu = 1 \times 10^{+2}$]{
\includegraphics[trim = 16mm 3mm 18mm 13mm, clip,width=0.33\linewidth]
{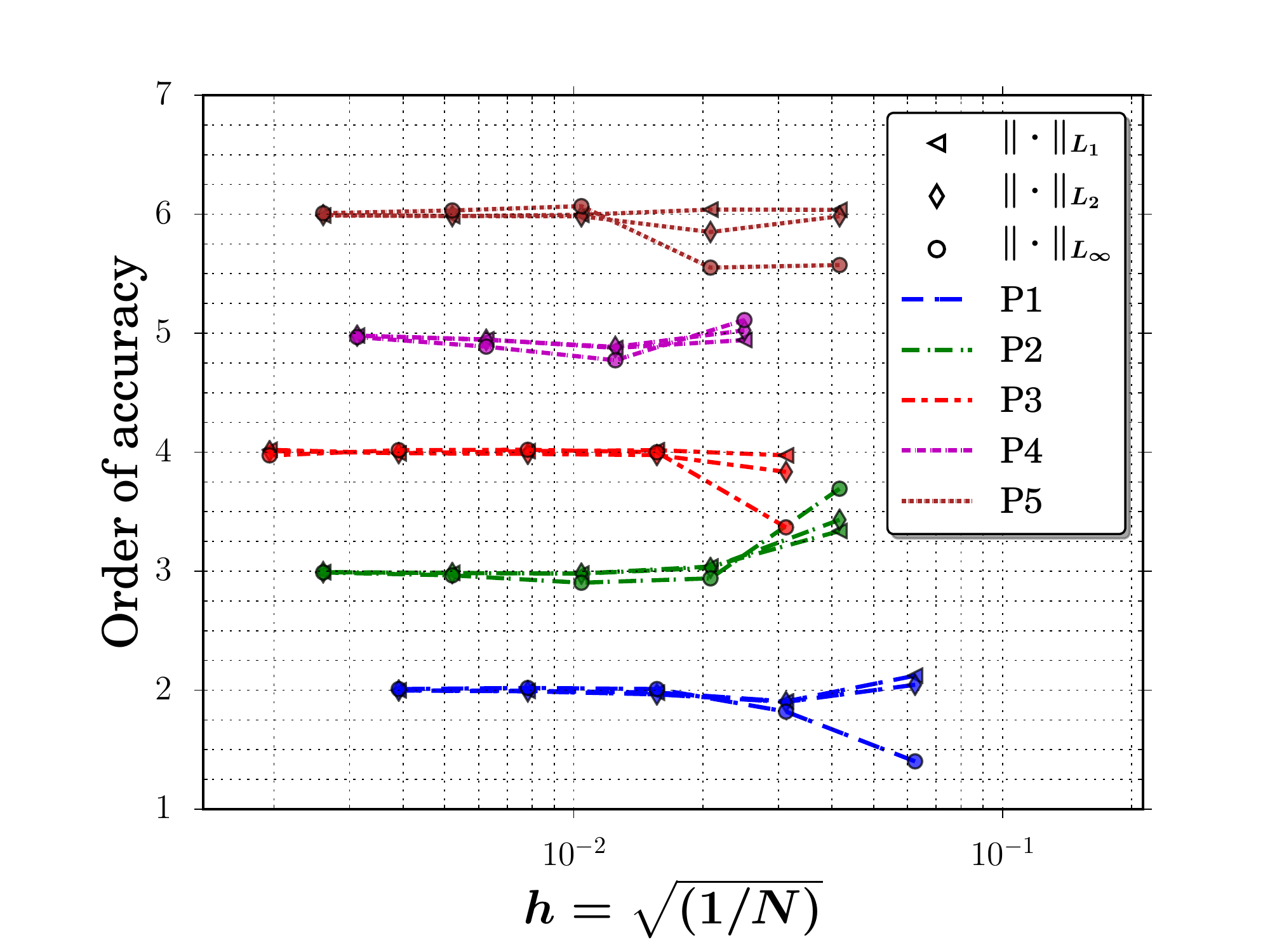}}
\caption{Evolution of the OOAs of $\rho v$ in $L_1$, $L_2$ and $L_\infty$ norms versus mesh refinement, for MS-3 with different $\mu$ values and polynomial degrees $\mathrm{P}1$--$\mathrm{P}5$}
\label{fig:Orders_mueffect-MS3-L}
\end{figure}

Figure \ref{fig:Orders_mueffect-MS3-H} shows that as the viscosity is lowered from $\mu = 1\times10^{+2}$ to $\mu = 1 \times 10^{-4}$, and consequently the diffusive terms loose their relative magnitude compared to the advective terms, the corrected derivatives undergo a significant delay in the attainment of the asymptotic range for P3 and P5 discretizations, whereas the OOAs in $L$ norms, illustrated in Fig. \ref{fig:Orders_mueffect-MS3-L}, are not noticeably affected. Note in the same Fig. that there is a drop in the $L$ norms of the finest P5 solutions which is due to the proximity of the error level to machine precision. The presence of large discrepancies between the OOAs of corrected and uncorrected derivatives in $H_1$ semi-norm seems correlated to the insignificance of the relative magnitude of viscous terms compared to inviscid terms and hence suggests that an examination of the balancing of the forcing function terms is required.


\subsection{Turbulent flows - MS-4 (original SA) and MS-5 (modified SA)}
\label{sec:turb_flow}

\paragraph{MS-4}

\begin{table}
\centering
\begin{tabular}{ c||c|c|c|c|c|c|c }
$(\cdot)$& $(\cdot)_0$ & $(\cdot)_x$ & $(\cdot)_y$ & $(\cdot)_{xy}$ & $a_{(\cdot)_x}$ & $a_{(\cdot)_y}$ & $a_{(\cdot)_{xy}}$\\ \hline\hline
  $\rho$ &   $1.0$     & $0.1$       & $-0.2$     & $0.1$           &      $1.0$      &       $1.0$     &         $1.0 $     \\ \hline
  $u$    &   $2.0$     & $0.3$       & $0.3$      & $0.3$           &      $3.0$      &       $1.0$     &         $1.0 $     \\ \hline  
  $v$    &   $2.0$     & $0.3$       & $0.3$      & $0.3$           &      $1.0$      &       $1.0$     &         $1.0 $     \\ \hline  
  $p$    &   $10.0$     & $1.0$      & $1.0$      & $0.5$           &      $2.0$      &       $1.0$     &         $1.0 $    \\ \hline  
    $\tilde{\nu}$    &   $0.6$     & $-0.03$       & $-0.02$     & $0.02$           &      $2.0$      &       $1.0$     &         $3.0 $ 
\end{tabular}
\caption{Parameters of MS-4}
\label{tb:MS-4_cons}
\end{table}

We consider an extension of MS-3 to ensemble-averaged turbulent flows by activating the $\tilde{\nu}$ field, first such that the original portion of the SA model (Eqs. \ref{eq:sa}, \ref{eq:mu_t+}, \ref{eq:prod+}, \ref{eq:S+}, \ref{eq:D+} and \ref{eq:fn+}) is verified for positive $\tilde{\nu}$ values. This manufactured solution is labelled MS-4 and is determined by parameters presented in Table \ref{tb:MS-4_cons} along with Eqs. \eqref{eq:trigo_MS} and a value of dynamic viscosity of $\mu = 1 \times 10^{-3}$ chosen such that the eddy viscosity be preponderant in comparison. The distance to the wall is set to $d_w=y+1$, for $y \in [0,1]$, to avoid large values in SA source terms with dependencies on $\frac{1}{d_w^2}$, that could otherwise saturate the entire forcing function. 

The balancing of different SA terms is discussed at length further in this section. The manufactured $\tilde{\nu}$ field of MS-4 is presented in Fig. \ref{fig:MS-4}. The other fields are the same as those of MS-3 (Fig. \ref{fig:MS-3}). The errors and OOAs are respectively presented in Figs. \ref{fig:Err_allE_allP_MS-4} and \ref{fig:Orders_MS-4} for $L$ norms and in Figs. \ref{fig:Err_allE_allP_H_MS-4} and \ref{fig:Orders_H_MS-4} for $H_1$ semi-norm.

\subsubsection{Balancing of the forcing functions}
As discussed previously for laminar flows, a balance between different terms of the forcing function is desirable with the goal of enabling the detection of slightest bugs on coarse grids, thus reducing the risk of a false \textit{pass} verdict. We hence extend the budget analysis that involves the relative magnitudes of inviscid versus viscous terms in laminar flows, to include all the terms of the SA source function. In the laminar case, achieving a balance between the advective and diffusive terms is a relatively easy task since the value of $\mu$ directly scales their ratio. For turbulent flows however, the presence of source terms within the RANS model, involving  a range of quantities, makes this balancing a more challenging task. For the assessment of the sought balance, we propose considering a relative absolute magnitude, $\mathrm{S^\mathrm{rel}_{{term}}} = \lvert \mathrm{S_{term}} \rvert \,/\,\mathrm{S^\mathrm{sum}_{SA}}$, as metric of the sensitivity of the verification process to each SA term where $\lvert \mathrm{S_{term}} \rvert$ is the absolute magnitude of the SA term and
\begin{equation}
\begin{aligned}
\mathrm{S^\mathrm{sum}_{SA}} = \sum_{term} \lvert \mathrm{S_{term}} \rvert = \,
&\abs[\Big]{\underbrace{\partial_j (\rho u_j \tilde{\nu})}_\text{Advection}  }  +
\abs[\bigg]{ \underbrace{-\,\partial_j \left(\frac{1}{\sigma}(\mu+\rho \tilde{\nu})\partial_j \tilde{\nu}\right)}_\text{Diffusion}  }+
\abs[\Big]{ \underbrace{\rho \, c_{b1}(1-f_{t2}) \, \tilde s \, \tilde \nu}_\text{Production} } +  \\
&\abs[\bigg]{ \underbrace{-\rho \left(c_{w1} f_w -\frac{c_{b1}}{\kappa^2}f_{t2}\right)\frac{ {\tilde\nu}^2 }{d_w^2}}_\text{Destruction} } + 
\abs[\Big]{ \underbrace{\frac{c_{b2}}{\sigma} \rho \,\partial_j \tilde{\nu} \, \partial_j \tilde{\nu}}_\text{Distribution} } + 
\abs[\Big]{ \underbrace{\frac{1}{\sigma}(\nu+\tilde \nu f_n) \, \partial_j (\rho \, \partial_j \tilde \nu)}_\text{Conservation} }.
\end{aligned}
\label{eq:S_posSAsum}
\end{equation}

\begin{figure}[!hbt]
\centering
\subfloat[Advection]{
\includegraphics[trim = 0mm 0mm 0mm 0mm, clip,width=0.25\linewidth]
{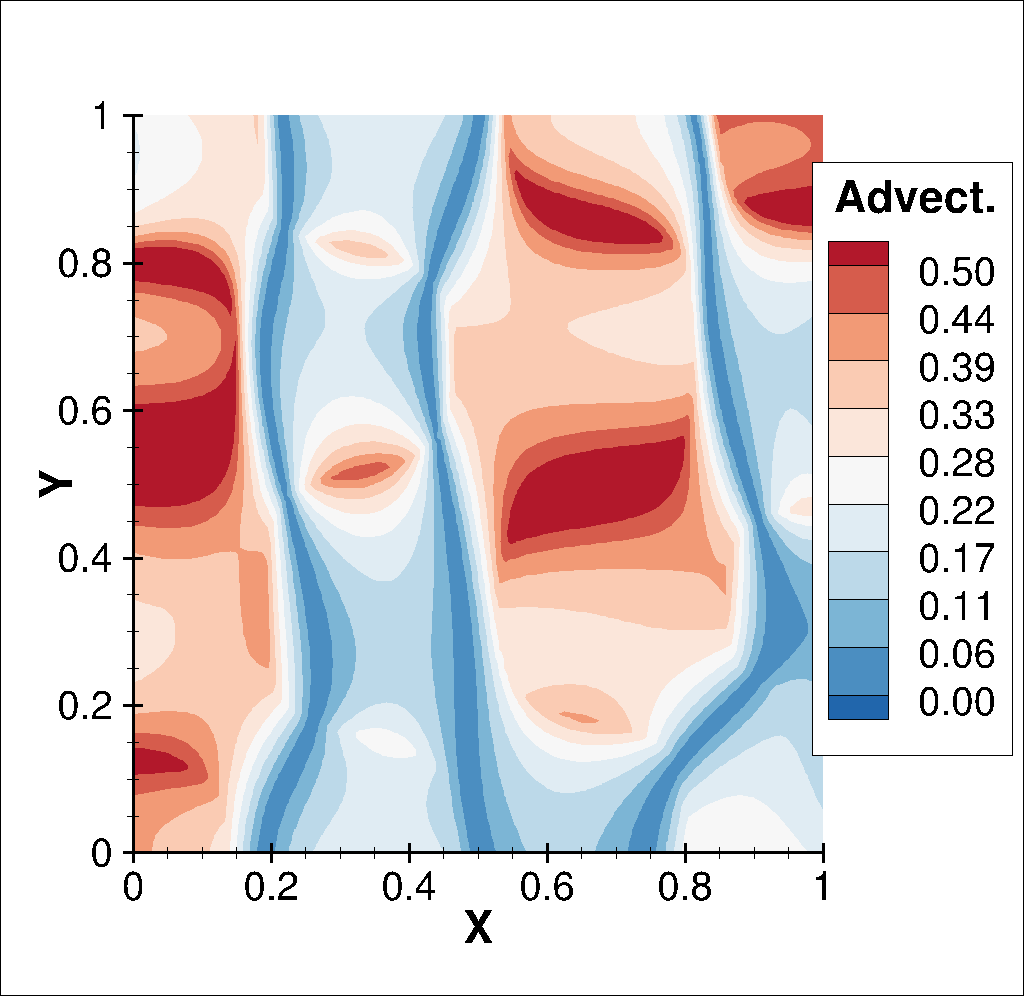}}
~~
\subfloat[Diffusion]{
\includegraphics[trim = 0mm 0mm 0mm 0mm, clip,width=0.25\linewidth]
{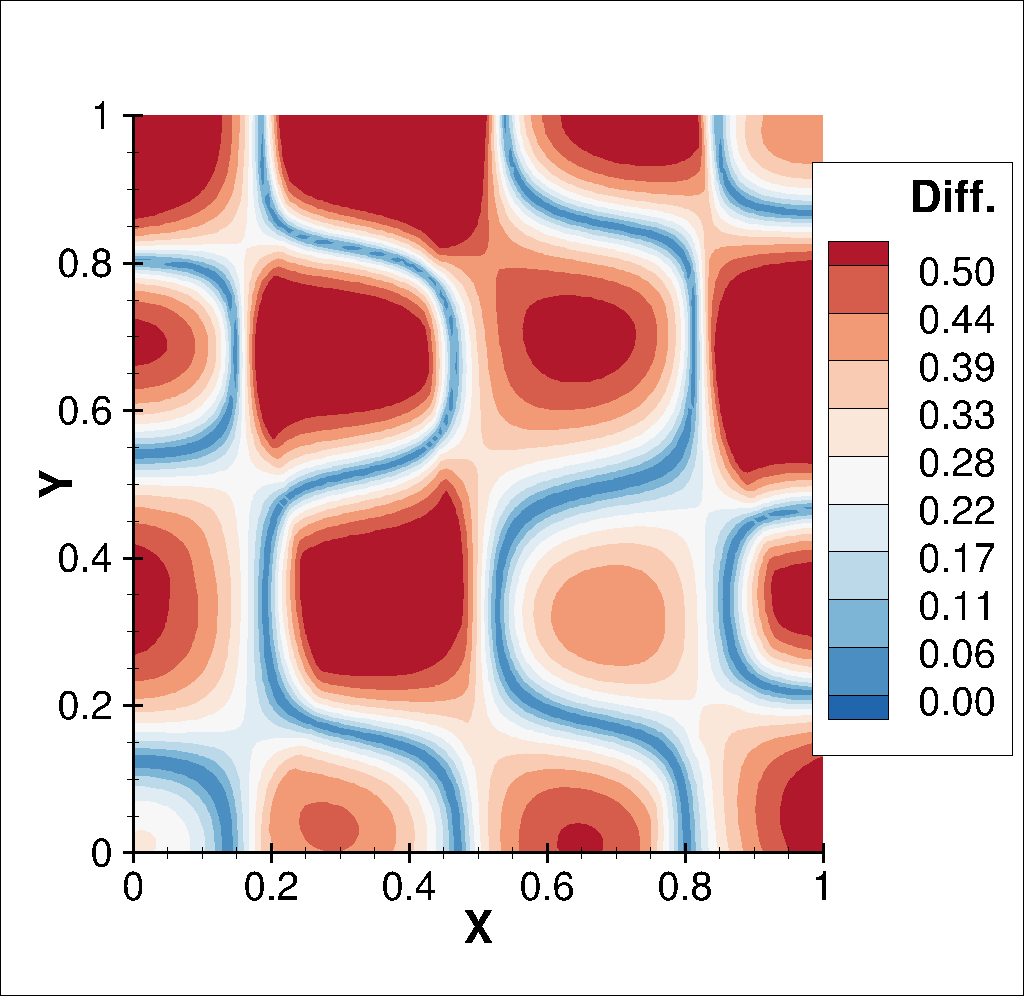}}
~~
\subfloat[Production]{
\includegraphics[trim = 0mm 0mm 0mm 0mm, clip,width=0.25\linewidth]{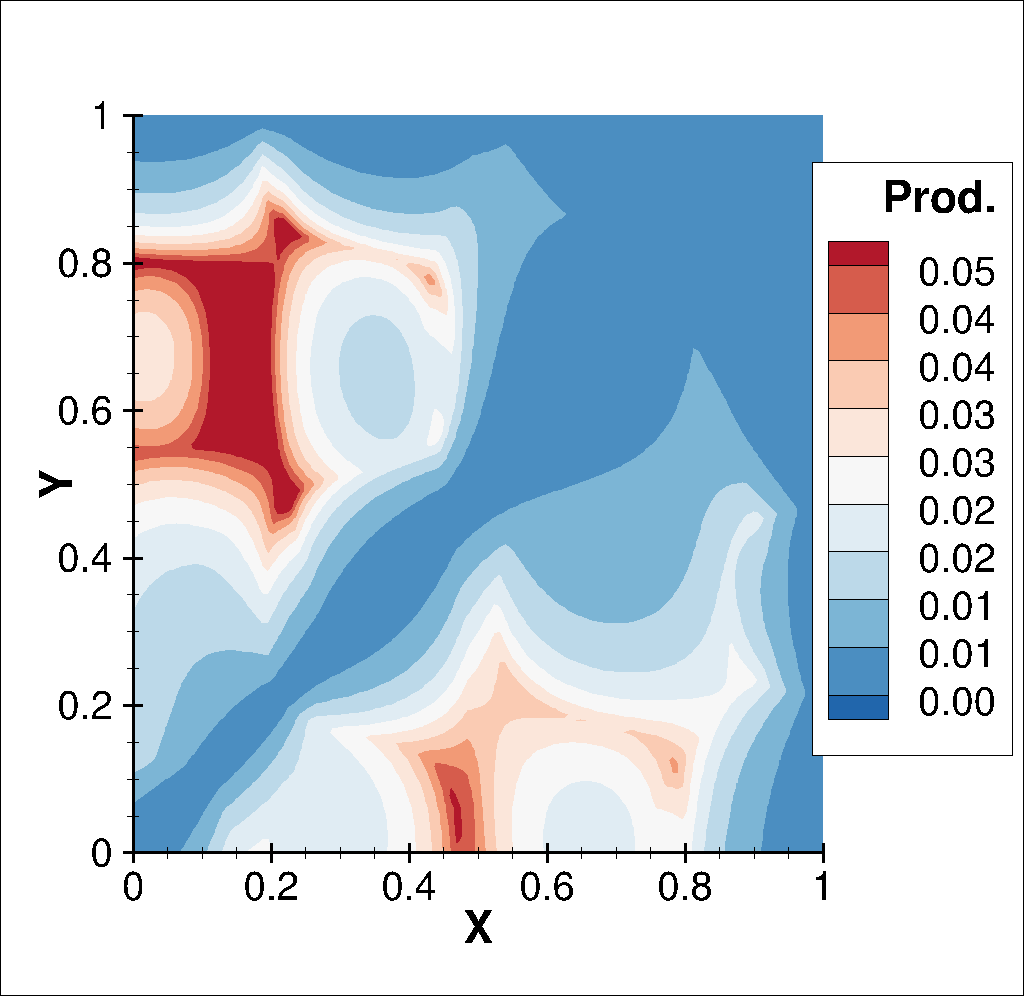}}
\\
\subfloat[Destruction]{
\includegraphics[trim = 0mm 0mm 0mm 0mm, clip,width=0.25\linewidth]{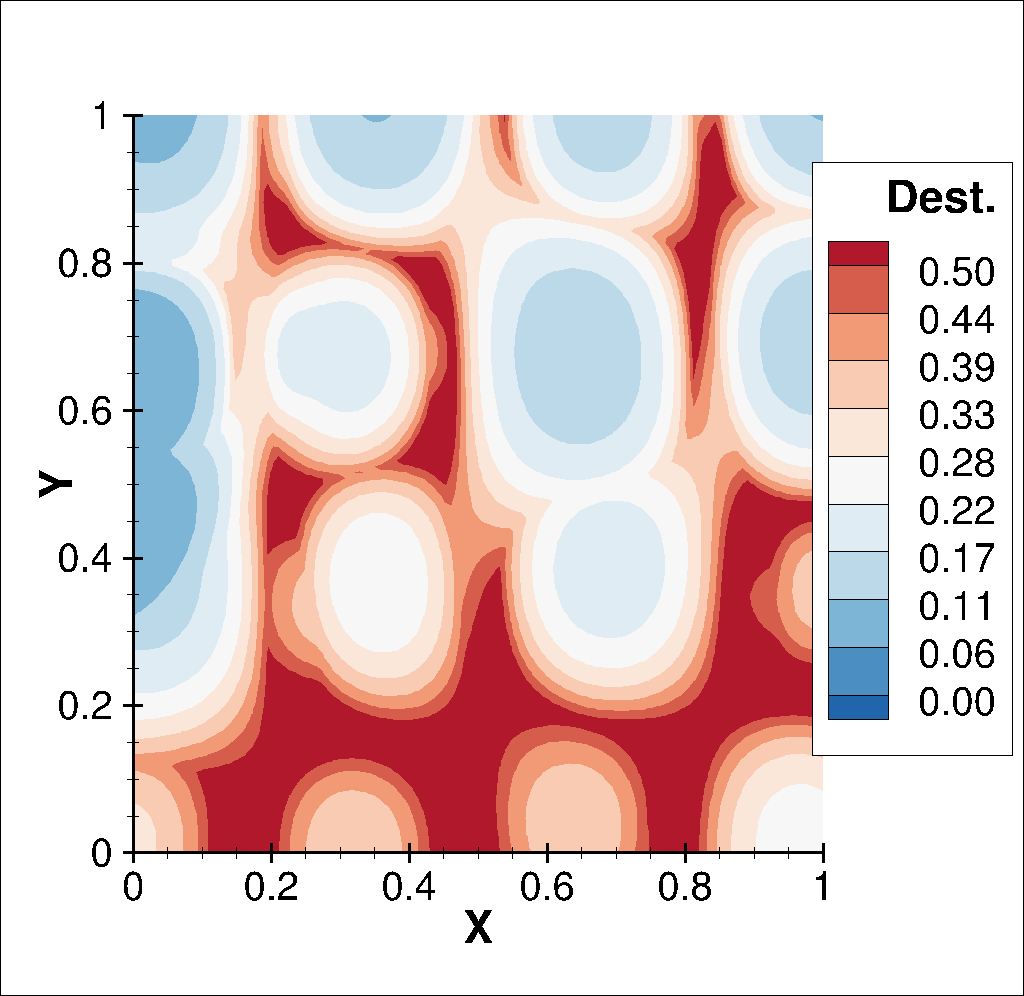}}
~~
\subfloat[Distribution]{
\includegraphics[trim = 0mm 0mm 0mm 0mm, clip,width=0.25\linewidth]{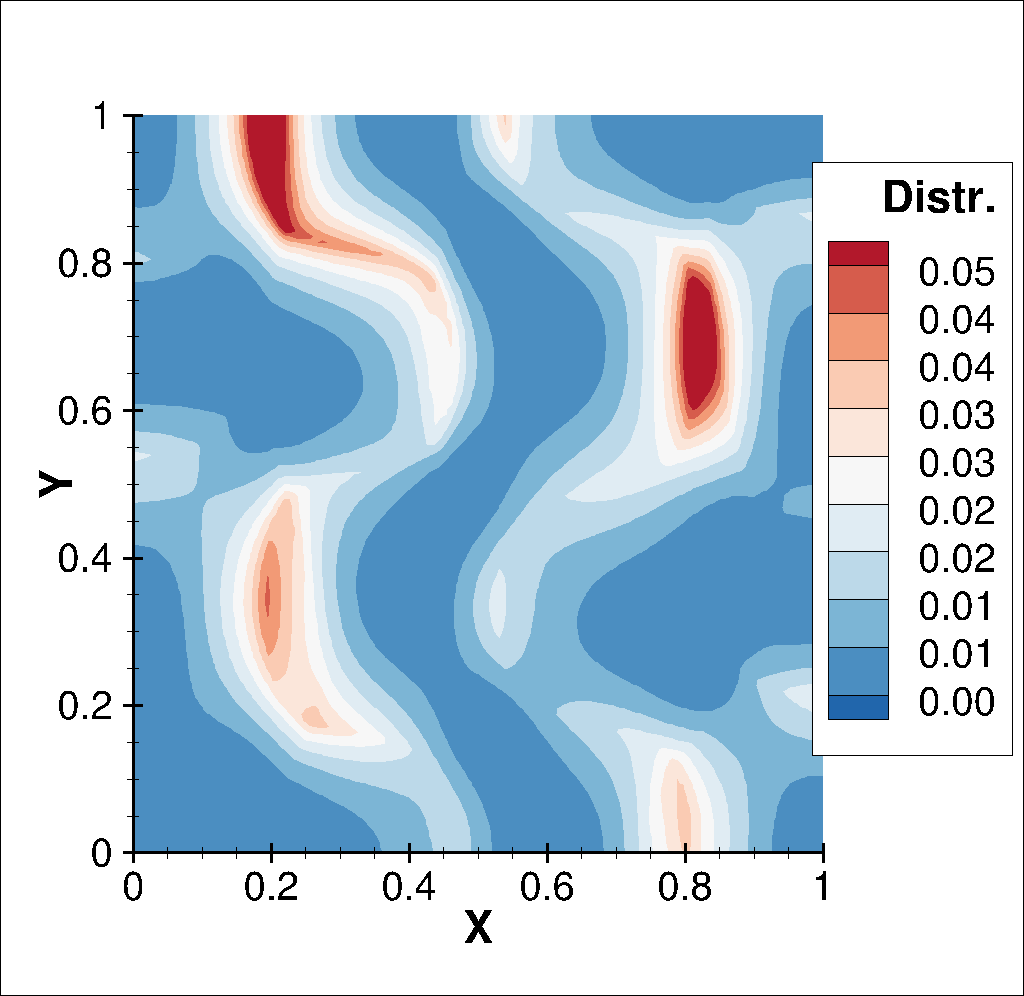}}
~
\subfloat[Conservation]{
\includegraphics[trim = 0mm 0mm 0mm 0mm, clip,width=0.25\linewidth]{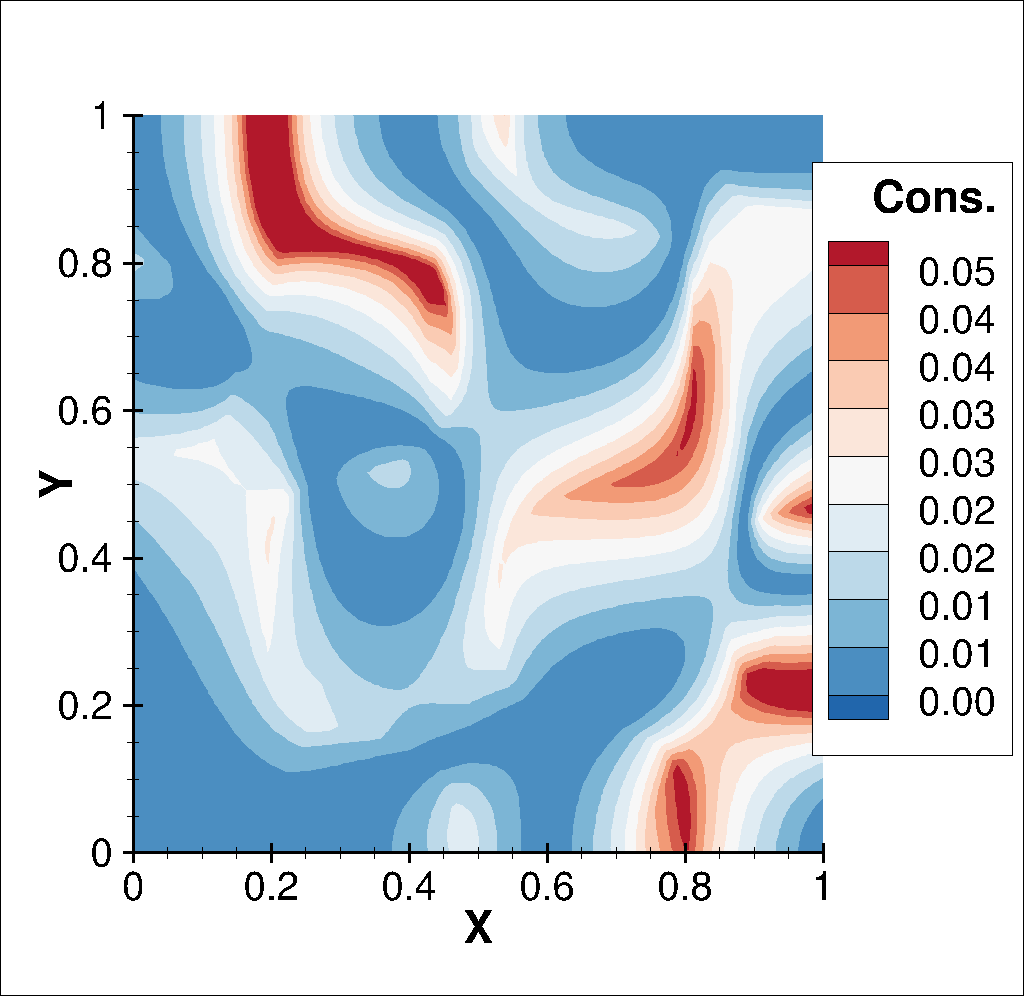}}
\caption{Sensitivity of the verification process to SA forcing function terms of MS-4, measured by $\mathrm{S^\mathrm{rel}_{{term}}}$}
\label{fig:MS-4-Sa-budget}
\end{figure}

\begin{figure}[!hbt]
\centering
\subfloat[Discretization error]{
\includegraphics[trim = 5mm 3mm 18mm 13mm, clip,width=0.33\linewidth]
{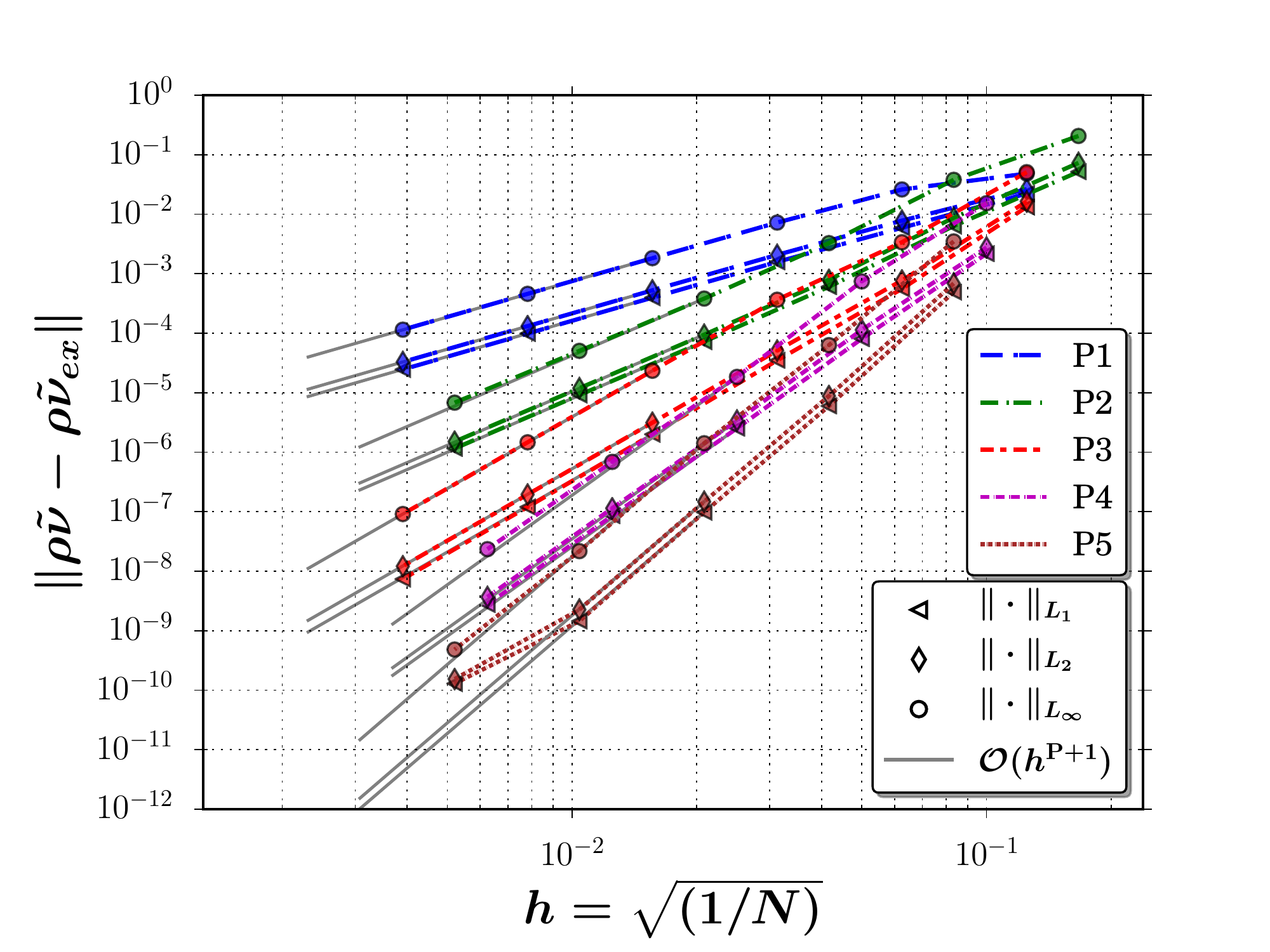}}
~~~
\subfloat[Order of accuracy]{
\includegraphics[trim = 5mm 3mm 18mm 13mm, clip,width=0.33\linewidth]
{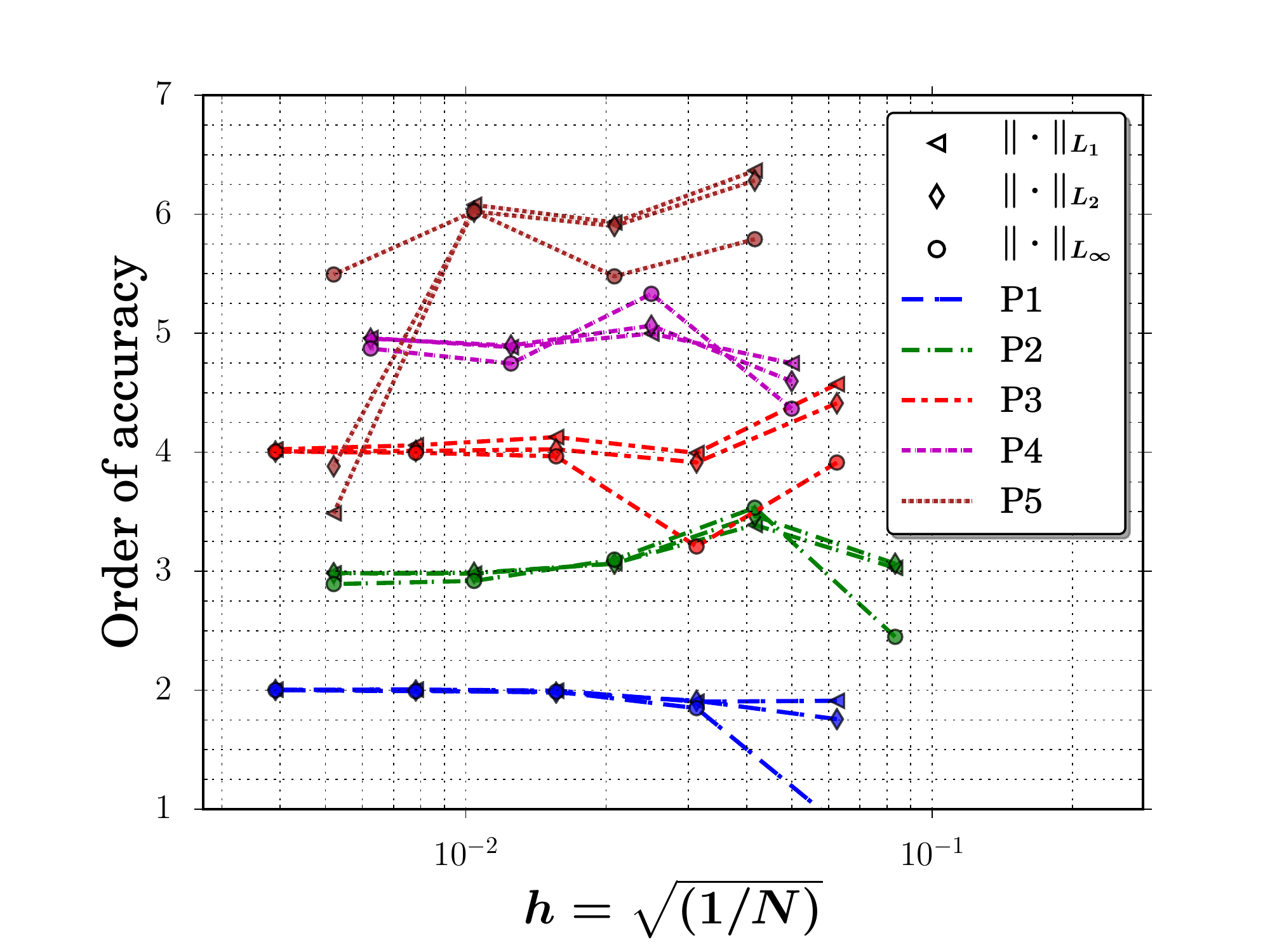}}
\caption{Evolution of the discretization errors and the OOAs  in $L_1$, $L_2$ and $L_\infty$ norms versus mesh refinement for $\rho \tilde{\nu}$, polynomial degrees $\mathrm{P}1$--$\mathrm{P}5$ and MS-4 with the spurious SA distribution term, $(1+d\alpha) \frac{c_{b2}}{\sigma} \rho \,\partial_j \tilde{\nu} \, \partial_j \tilde{\nu}$} 
\label{fig:Orders_MS-4_bug_inSA}
\end{figure}

The six terms of the SA forcing function appearing in Eq. \eqref{eq:S_posSAsum} are those of the original portion of the SA model, Eqs. \ref{eq:sa}, \ref{eq:mu_t+}, \ref{eq:prod+}, \ref{eq:S+}, \ref{eq:D+} and \ref{eq:fn+}.
For each of these terms, the spatial distribution of the $\mathrm{S^\mathrm{rel}_{{term}}}$ is provided in Fig. \ref{fig:MS-4-Sa-budget} showing that the advection, diffusion and destruction terms exhibit  spatial maxima of $\mathrm{max}(\mathrm{S^\mathrm{rel}_{{term}}}) \approx50\%$ on the domain whereas the sensitivity to production, distribution, and conservation terms only accounts to  $\mathrm{max}(\mathrm{S^\mathrm{rel}_{{term}}})\approx 5\%$. To establish whether such a low sensitivity is acceptable and how it relates to actual sensitivity to a given bug, we introduce an error in the distribution term of the SA model, $(1+d\alpha) \frac{c_{b2}}{\sigma} \rho \,\partial_j \tilde{\nu} \, \partial_j \tilde{\nu}$, such that the original term ($d\alpha=0$) is modified sequentially by increasing $d\alpha$ from $d\alpha=1 \times 10^{-15}$ by an order of magnitude ($\times 10$) in each step and the OOAs are recomputed until the bug is detected. Consequently, a significant degradation in the orders of the P5 discretization was detected for a value of $d\alpha=1 \times 10^{-7}$ (see Fig. \ref{fig:Orders_MS-4_bug_inSA}) demonstrating that considerably small inconsistencies in the implementation of the distribution term can be detected and hence $\mathrm{max}(\mathrm{S^\mathrm{rel}_{{term}}})\approx5\%$  is a sufficient criterion to ensure that all terms are correctly verified. From Fig. \ref{fig:Orders_MS-4_bug_inSA}, it can again be  emphasized that including a reasonably high (such as P5) polynomial degree in the verification process is crucial for the detection of minor inconsistencies in the implementation. 

Based on the sensitivity analysis presented and the results obtained for OOAs in Figs. \ref{fig:Orders_MS-4} ($L$ norms) and \ref{fig:Orders_H_MS-4} ($H_1$ semi-norm), it is possible to conclude that the considered implementation of the original portion of the SA model is verified via  MS-4 and hence the focus can be moved on the modified portion of the SA model. 

\paragraph{MS-5}
We propose MS-5 for the verification of the modified portion of the SA model; recalling that the revised model is intended to tackle the instabilities generated when negative values of the SA working variable occur in the numerical solution. Like MS-4, MS-5 is an extension of MS-3 by that it inherits the latter's manufactured fields which it complements by a manufactured solution for $\tilde{\nu}$.

\begin{table}
\centering
\begin{tabular}{ c||c|c|c|c|c|c|c }
$(\cdot)$& $(\cdot)_0$ & $(\cdot)_x$ & $(\cdot)_y$ & $(\cdot)_{xy}$ & $a_{(\cdot)_x}$ & $a_{(\cdot)_y}$ & $a_{(\cdot)_{xy}}$\\ \hline\hline
  $\rho$ &   $1.0$     & $0.1$       & $-0.2$     & $0.1$           &      $1.0$      &       $1.0$     &         $1.0 $     \\ \hline
  $u$    &   $2.0$     & $0.3$       & $0.3$      & $0.3$           &      $3.0$      &       $1.0$     &         $1.0 $     \\ \hline  
  $v$    &   $2.0$     & $0.3$       & $0.3$      & $0.3$           &      $1.0$      &       $1.0$     &         $1.0 $     \\ \hline  
  $p$    &   $10.0$     & $1.0$      & $1.0$      & $0.5$           &      $2.0$      &       $1.0$     &         $1.0 $    \\ \hline  
    $\tilde{\nu}$       &   $-6.0$   & $-0.3$     & $-0.2$          &      $0.2$      &      $2.0$      &       $1.0$     &         $3.0 $ 
\end{tabular}
\caption{Parameters of MS-5}
\label{tb:MS-5_cons}
\end{table}

Table \ref{tb:MS-5_cons} presents the parameters that define MS-5. The dynamic viscosity is set to $\mu=1 \times 10^{-1}$. For negative $\tilde{\nu}$ values, the RANS equations are decoupled from the SA equation by eliminating the effect of the latter on the former via $\mu_t = 0$. However, the SA equation remains unilaterally coupled to the RANS equations. Therefore, MS-5 only verifies the implementation of the SA model and regarding the state variables of all other equations, it should result in the same exact error and orders as observed for MS-3 in Figs. \ref{fig:Err_allE_allP_MS-3}, \ref{fig:Orders_MS-3} for $L$ norms and in Figs. \ref{fig:Err_allE_allP_H_MS-3} and \ref{fig:Orders_H_MS-3} for $H_1$ semi-norm.  As for the SA equation, its manufactured field as well as errors and orders in $L$ and $H_1$ norms are respectively presented in Figs. \ref{fig:MS-5}, \ref{fig:L_MS-5} and \ref{fig:H_MS-5}.

A similar budget analysis of the SA forcing function to the one for MS-4 is conducted here by rather considering the modified portion of the SA model (Eqs. \ref{eq:sa}, \ref{eq:mu_t-}, \ref{eq:prod-}, \ref{eq:S-}, \ref{eq:D-} and \ref{eq:fn-}) such that:
\begin{equation}
\begin{aligned}
\mathrm{S^\mathrm{sum}_{SA}} = \sum_{term} \lvert \mathrm{S_{term}} \rvert = \,
&\abs[\Big]{\underbrace{\partial_j (\rho u_j \tilde{\nu})}_\text{Advection}  }  +
\abs[\bigg]{ \underbrace{-\,\partial_j \left(\frac{1}{\sigma}(\mu+\rho \tilde{\nu}\frac{c_{n1}+\chi^3}{c_{n1}-\chi^3})\partial_j \tilde{\nu}\right)}_\text{Diffusion}  }+
\abs[\Big]{ \underbrace{\rho \, c_{b1}(1-c_{t3}) \, s \, \tilde \nu}_\text{Production} } +  \\
&\abs[\bigg]{ \underbrace{ c_{w1} \frac{ {\tilde\nu}^2 }{d_w^2}}_\text{Destruction} } + 
\abs[\Big]{ \underbrace{\frac{c_{b2}}{\sigma} \rho \,\partial_j \tilde{\nu} \, \partial_j \tilde{\nu}}_\text{Distribution} } + 
\abs[\Big]{ \underbrace{\frac{1}{\sigma}(\nu+\tilde \nu \frac{c_{n1}+\chi^3}{c_{n1}-\chi^3}) \, \partial_j (\rho \, \partial_j \tilde \nu)}_\text{Conservation} }.
\end{aligned}
\label{eq:S_negSAsum}
\end{equation}

\begin{figure}[!hbt]
\centering
\subfloat[Advection]{
\includegraphics[trim = 0mm 0mm 0mm 0mm, clip,width=0.25\linewidth]
{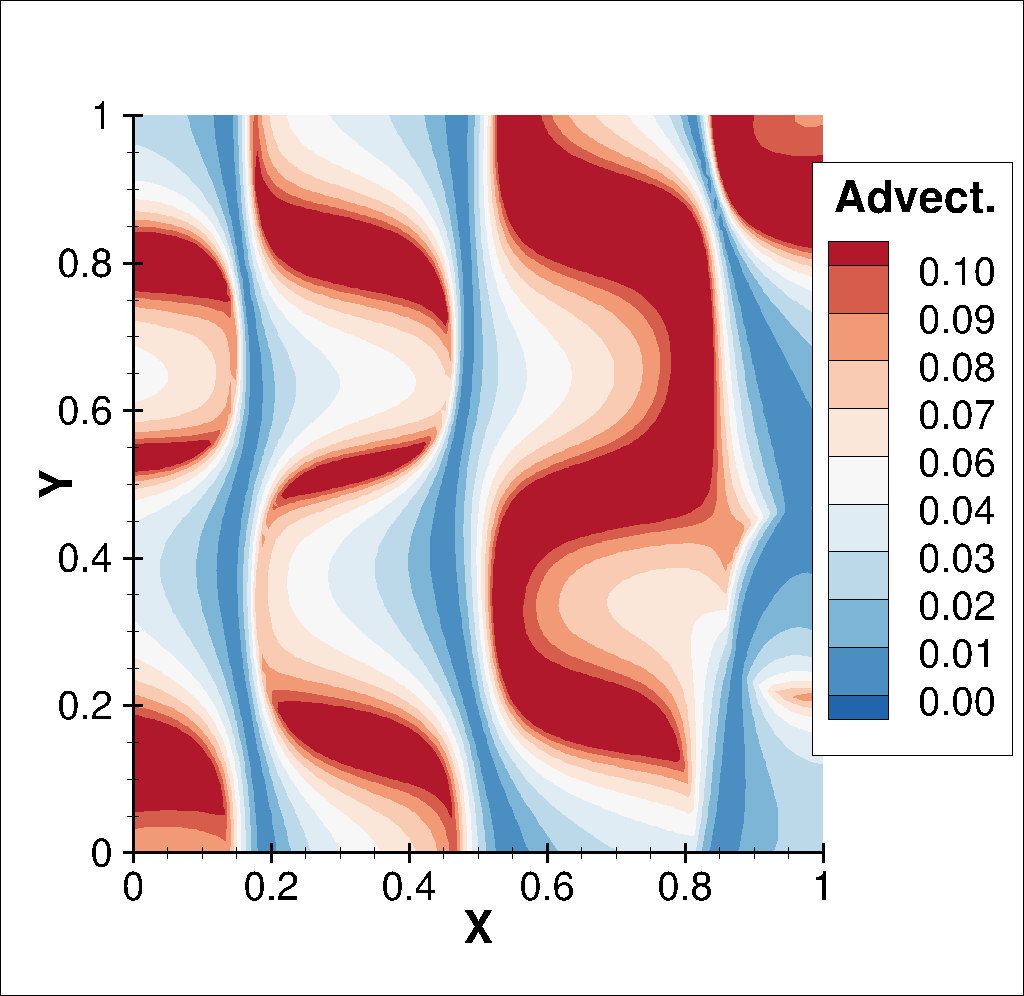}}
~~
\subfloat[Diffusion]{
\includegraphics[trim = 0mm 0mm 0mm 0mm, clip,width=0.25\linewidth]
{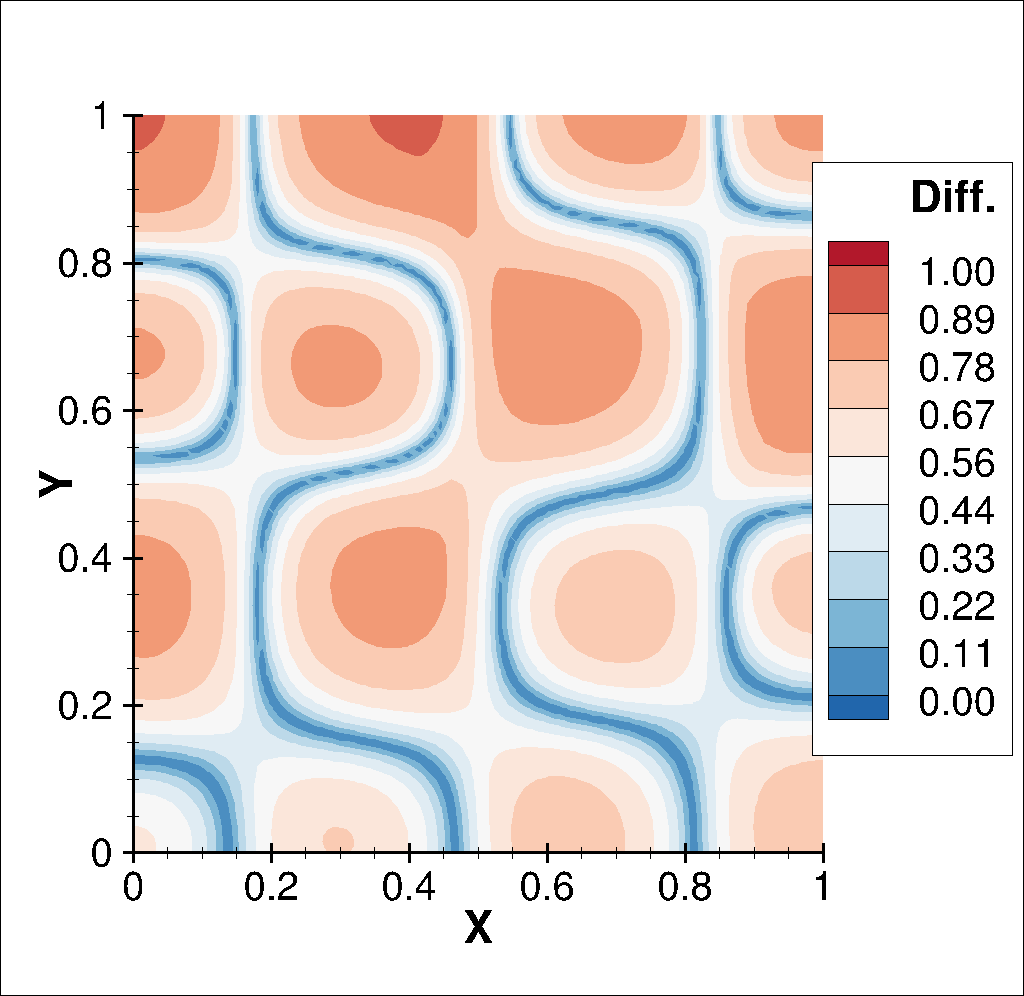}}
~~
\subfloat[Production]{
\includegraphics[trim = 0mm 0mm 0mm 0mm, clip,width=0.25\linewidth]{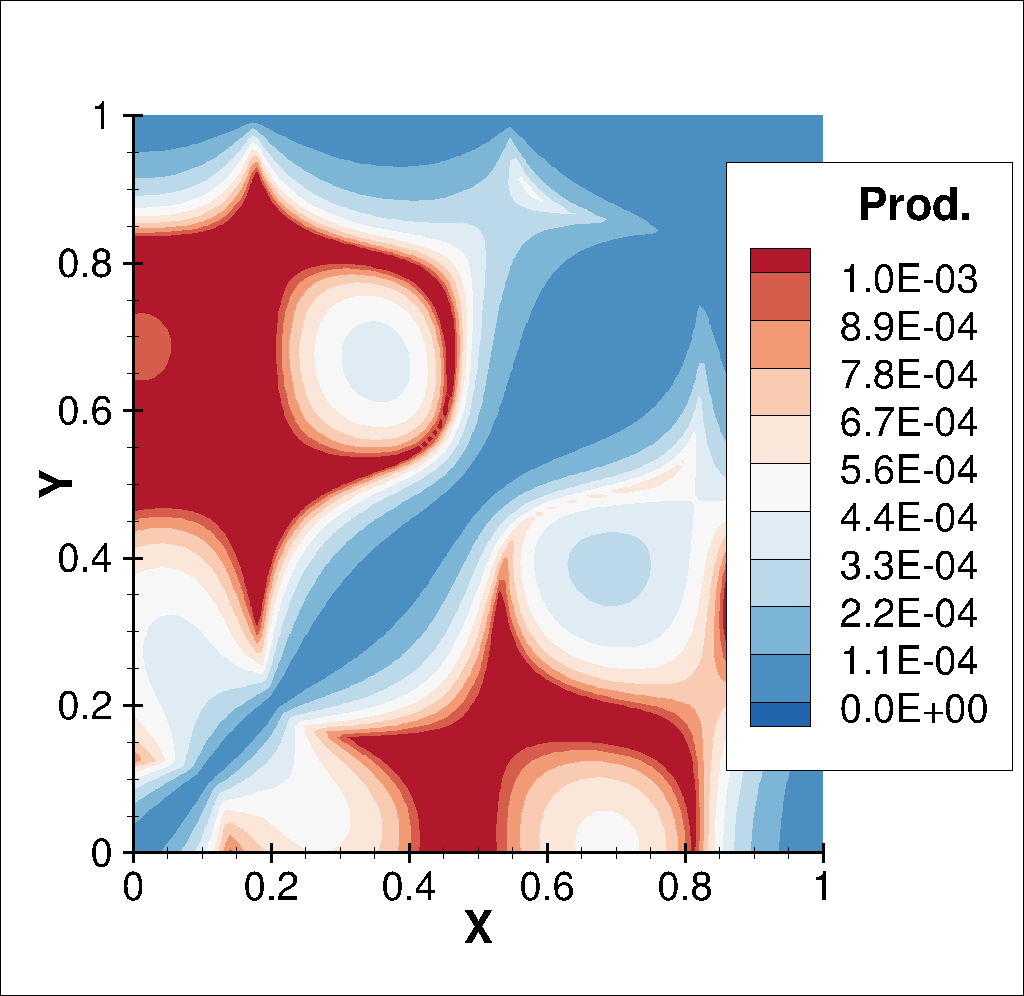}}
\\
\subfloat[Destruction]{
\includegraphics[trim = 0mm 0mm 0mm 0mm, clip,width=0.25\linewidth]{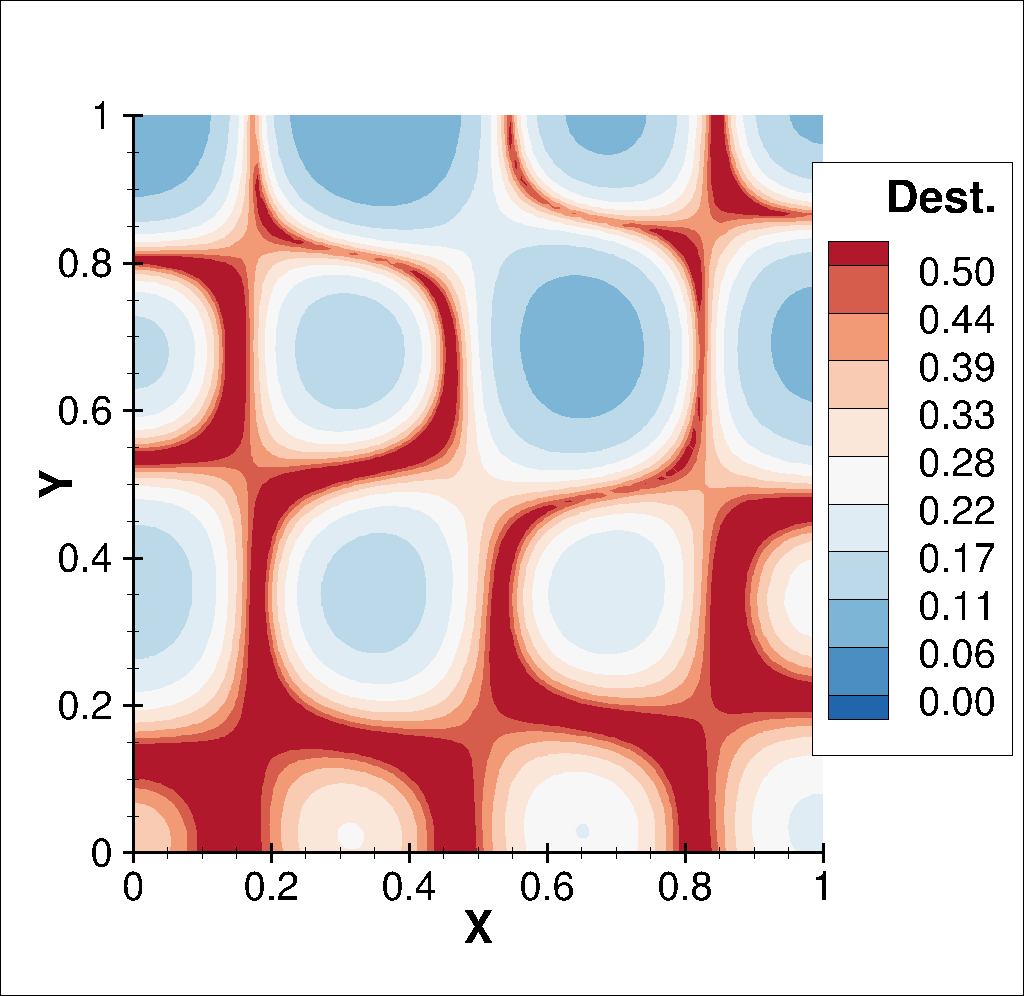}}
~~
\subfloat[Distribution]{
\includegraphics[trim = 0mm 0mm 0mm 0mm, clip,width=0.25\linewidth]{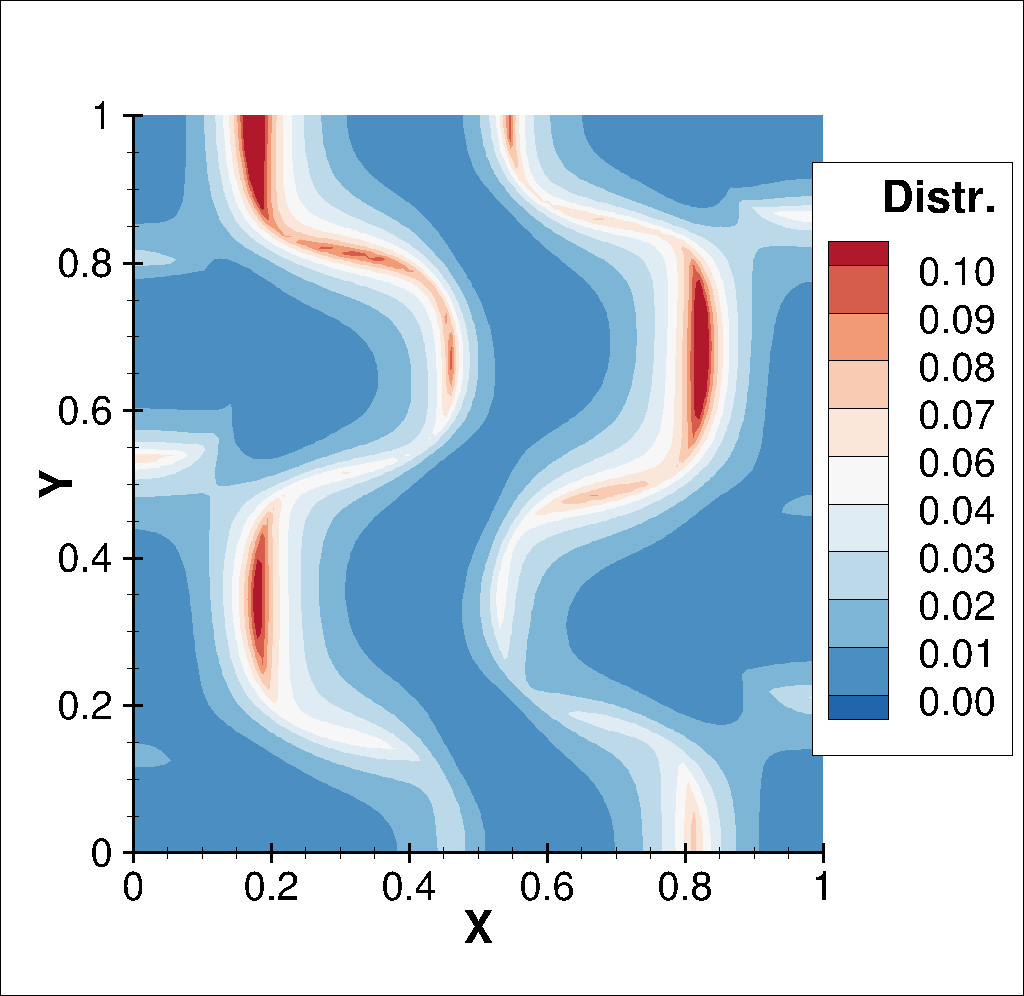}}
~~
\subfloat[Conservation]{
\includegraphics[trim = 0mm 0mm 0mm 0mm, clip,width=0.25\linewidth]{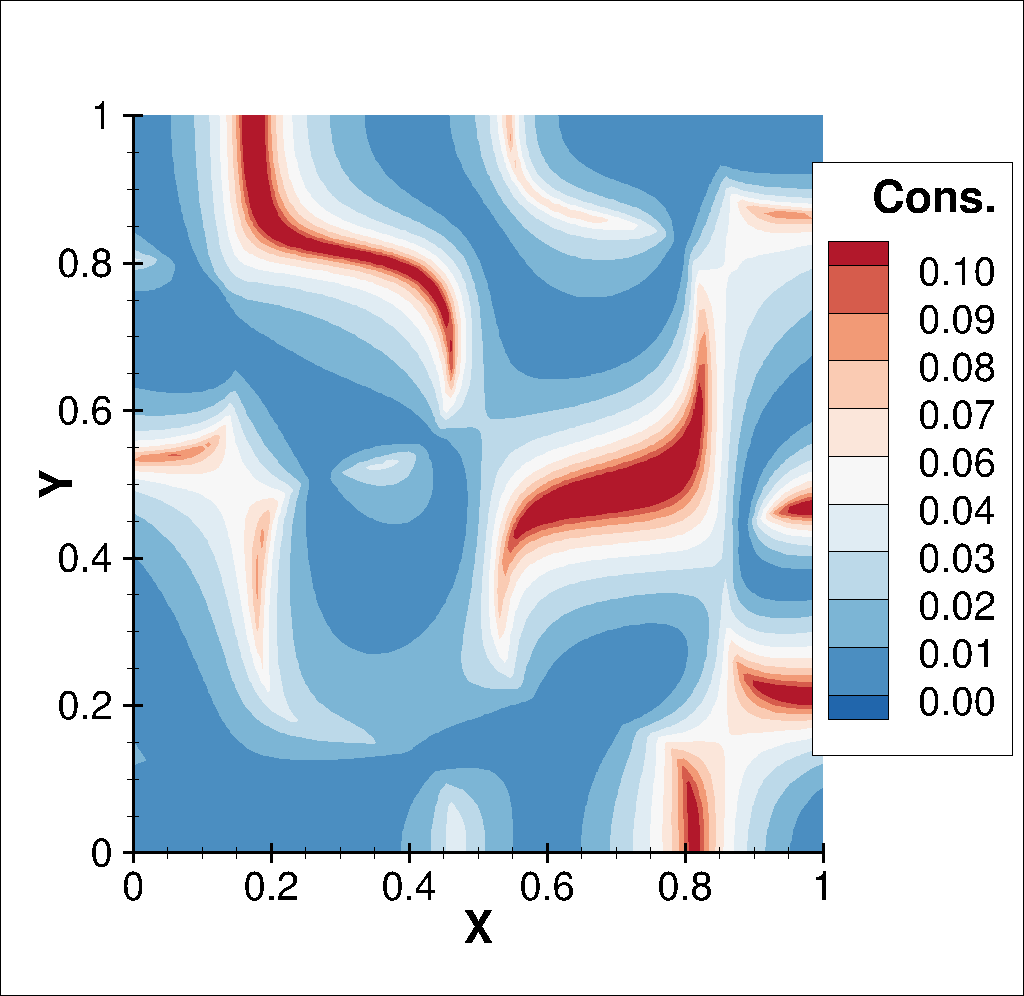}}
\caption{Sensitivity of the verification process to SA forcing function terms of MS-5, measured by $\mathrm{S^\mathrm{rel}_{{term}}}$}
\label{fig:MS-5-Sa-budget}
\end{figure}

This analysis resulted in the spatial sensitivity distributions of Fig. \ref{fig:MS-5-Sa-budget} showing that in descending order of magnitude, the SA forcing function depends on the diffusion, destruction,  advection, conservation, distribution and production terms. The sensitivities to the latter term are very low, $\mathrm{max}(\mathrm{S^\mathrm{rel}_{{term}}})\approx0.1\%$,  and are particularly under the criterion of $\mathrm{max}(\mathrm{S^\mathrm{rel}_{{term}}})\approx5\%$, established previously. It is hence prudent to conduct an actual sensitivity analysis by introducing an inconsistency in the implementation of the production term of the SA model as $(1+d\alpha) \left(\rho \, c_{b1}(1-c_{t3}) \, s \, \tilde \nu \right)$ and computing the OOAs for an initially small and increasingly larger values of $d\alpha$ until a difference in the orders is observed. This exercise resulted in the detection of the bug as soon as $d\alpha = 1 \times 10^{-6}$. Therefore, the production term is verified by MS-5 even for minor bugs and MS-5 can be considered for the verification of the modified SA equation. 

The only term of the governing equations that remains unverified throughout the manufactured cases presented so far is the modified vorticity magnitude, $\tilde s$ of Eq. \ref{eq:S-}. In fact, this is the only term of the modified SA that is not activated by a condition on the sign of $\tilde \nu$ values but rather by a criterion on the vorticity magnitude, $s$, that is often satisfied in the vicinity of solid walls. The verification of this term is hence better suited for wall-bounded manufactured cases.

\subsubsection{Effect of grid stretching and self-similarity}
\label{sec:grid_eff}
Manufactured solutions based on trigonometric functions do not feature sharp gradients and thereby constitute an ideal candidate for code verification on isotropic grids. Nevertheless, mesh stretching (element anisotropy and non-uniformity of element size distribution) is often mandatory in the case of realistic flows such as  boundary layers, jets or shocks in order to tackle sharp variations while ensuring residual convergence and solution accuracy. Determining the necessary and sufficient level of stretching  in every region of the domain is a complex subject. For RANS, one could adopt a heuristic approach by following general guidelines on the wall element size such as $y^+ \leq5$ where $y^+$ is a dimensionless quantity reflecting the flow intensity in the vicinity of solid wall. Although sufficient to ensure the validity of the numerical solution with regards to the turbulence model, this criterion does not guarantee that the grid is appropriate to realize the optimal OOAs for integrated quantities at the wall or for the smoother solution in outer boundary layer region. Furthermore, generating strictly self-similar grid sets for grid convergence study and uncertainty analysis is not always possible for complex geometries. Self-similarity of a grid set refers to the property of featuring a constant refinement ratio of the grid resolution (typical element size), $h$, throughout the domain, for any two consecutive grids of the set. Note that the effect of various common measures of $h$ on the appearance of the asymptotic convergence rate and the achievement of optimal OOAs is discussed in \cite{salas2007,eca-et-al_2013} for low-order discretizations and self-similar and dissimilar grid sets. We choose the definition \eqref{eq:elem_size} that is shown in \cite{eca-et-al_2013} to produce the optimal rates for self-similar grid sets. In this section, we investigate the effect of grid stretching  as well as the impact of self-similarity of a grid set on the OOAs of trigonometric solutions. 

The effect of stretching and self-similarity of a sequence of grids  on the OOAs produced by a grid convergence study is analyzed in \cite{Eca2016} for a low-order solver and $L$ norms. We extend the analysis to higher OOAs and $H_1$ semi-norm by considering:
\begin{enumerate}
\item the effect of self-similarity on uniform grid distributions by applying two random perturbations of element sizes, with increasing amplitudes, around a uniform element size distribution;
\item the effect of grid anisotropy and non-uniformity on trigonometric manufactured cases by generating an expanded element size distribution via a geometric series with fixed size ratio of $r=1.015$;
\item the effect of self-similarity on stretched grids by applying two random perturbations of element sizes, with increasing amplitudes, around the expanded element size distribution via a geometric series with fixed size ratio of $r=1.015$.
\end{enumerate}

In all these cases, the $y$ coordinates are produced by replicating the $x$ coordinates such that the resulting grid is symmetrical. Coarser grids are generated from the finest grid of the sequence by removing every other grid line repeatedly until the desired level of coarsening is reached. As a result, the size distribution pattern of the coarser grid is  similar to the finer grid, strictly if no random perturbation is applied. The random perturbation is applied to the  element $e_i$ with size $\Delta x_i$ as $\Delta x'_i = \alpha_i \,\Delta x_i$
where $\Delta x'_i$ is the perturbed size and $\alpha_i$ is a random number bounded by $1-d\alpha_m\leq\alpha_i\leq1+d\alpha_m$. We consider three values of perturbation intensity, $d\alpha_m$:
\begin{itemize}
\item $d\alpha_0 = 0$, the unperturbed grid set;
\item $d\alpha_1 = 0.25$, moderately perturbed grid set;
\item $d\alpha_2 = 0.5$, most intensely perturbed grid set.
\end{itemize}

\begin{figure}[!hbt]
\centering
\subfloat[Uniform distribution]{
\includegraphics[trim = 5mm 3mm 18mm 13mm, clip,width=0.34\linewidth]{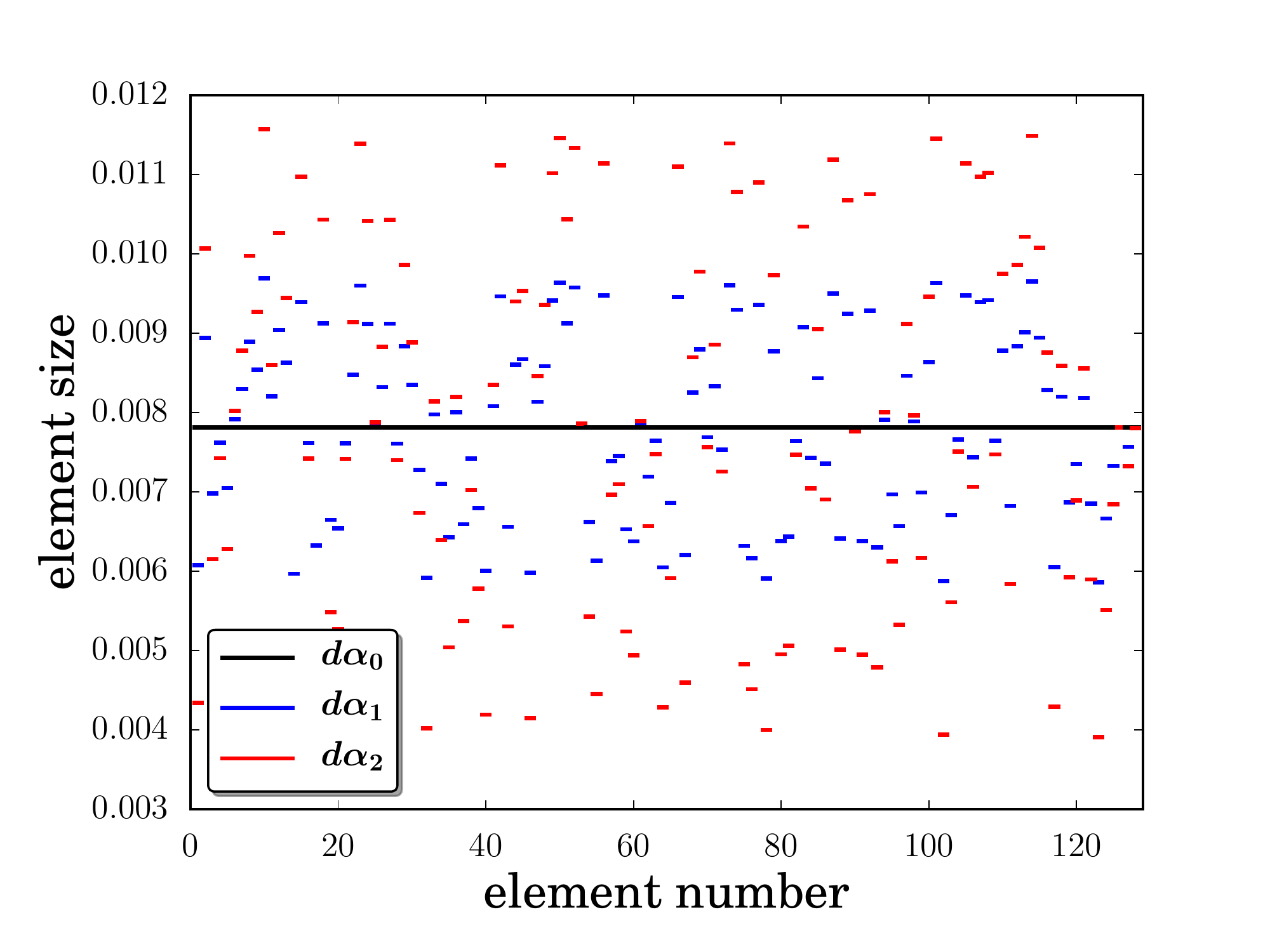}}
~~~
\subfloat[Expanded distribution]{
\includegraphics[trim = 5mm 3mm 18mm 13mm, clip,width=0.34\linewidth]{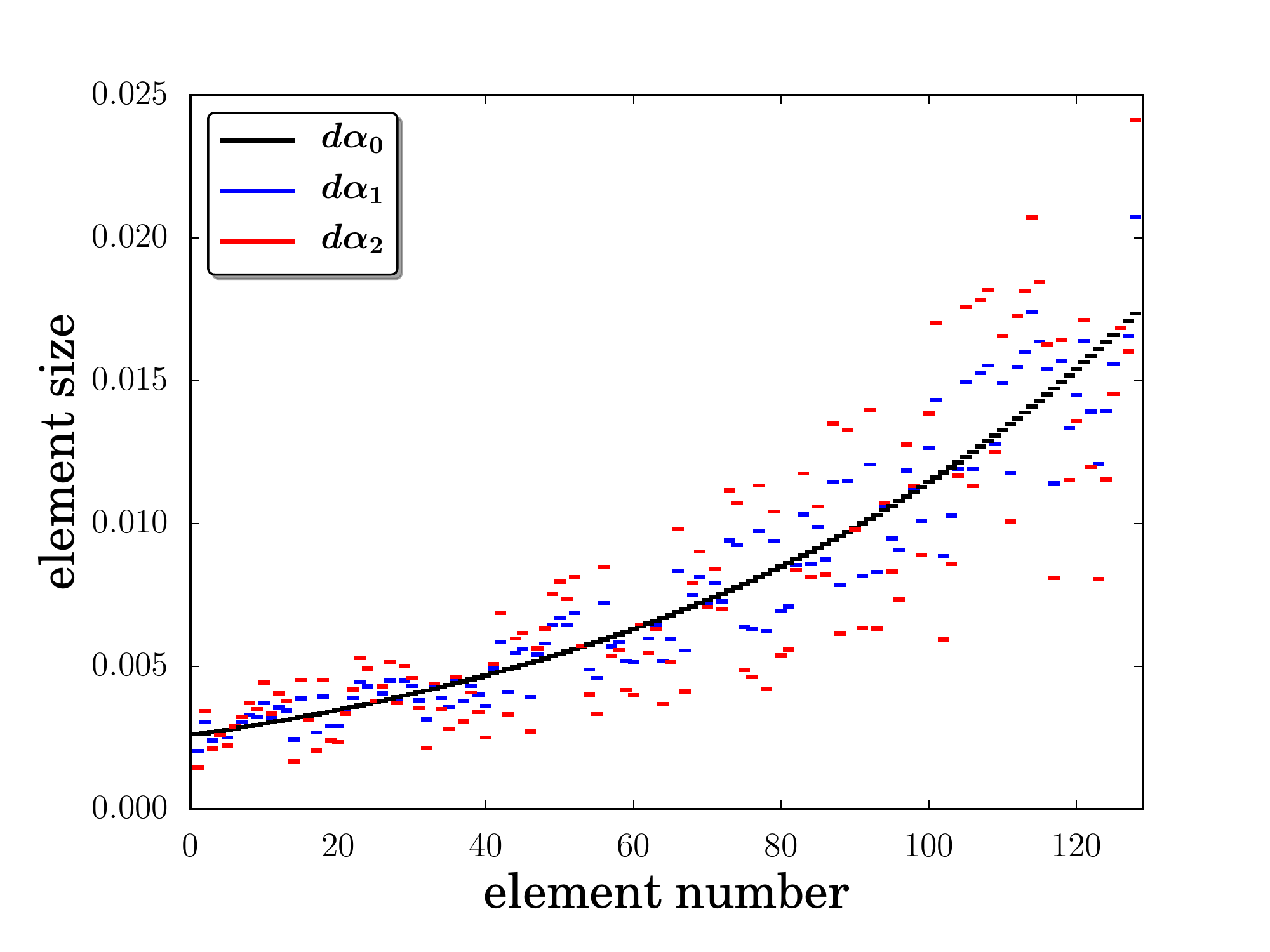}}
\caption{One-dimensional element size distributions for the finest meshes of the uniform and expanded grid sets with $128 \times 128$ elements}
\label{fig:grid_effect}
\end{figure}

Figure \ref{fig:grid_effect} shows  the resulting six element size distributions corresponding to the finest grid of each set (uniform and expanded with three perturbation intensities each). MS-5 is solved on each grid set and the normalized discrepancy between observed, $\mathcal{O}(h^o)$, and  theoretical, $\mathcal{O}(h^{\mathrm{P}+1})$, OOAs for P1--P4 and $\rho \tilde \nu$ is measured by $|(\mathrm{P}+1)-o|/(\mathrm{P}+1)$ and tracked as the mesh is refined. The results are presented in Figs. \ref{fig:grid_sens_unif_L}--\ref{fig:grid_sens_expnd_H} and leading to the following remarks: 
\begin{itemize}
\item The observed order discrepancy in $L_2$ norm for the unperturbed uniform set ($d\alpha_0$ in Fig. \ref{fig:grid_sens_unif_L}) does not reduce as regularly in the present two-dimensional case as in a one-dimensional case \cite{Eca2016}. This suggests that the discretization error in multiple spatial dimensions does not scale as smoothly with mesh refinement as in a single dimension.
\item As the uniform grid is perturbed, the order discrepancy in $L$ (Fig. \ref{fig:grid_sens_unif_L}) and $H_1$ (Fig. \ref{fig:grid_sens_unif_H}) norms often increases rather than reducing with mesh refinement. This trend is more pronounced as the perturbation intensity, $d\alpha$, is increased. In other words, the order departs from the asymptotic range with mesh refinement under the effect of grid dissimilarity.
\item The perturbations of the uniform grid set often affect the order discrepancy in $L_\infty$, $L_2$ and $L_1$ norms most adversely in the descending order (Fig. \ref{fig:grid_sens_unif_L}). This demonstrates the higher sensitivity of $L_\infty$ norm followed by $L_2$ norm with regards to grid self-dissimilarity that is manifested here through the irregularity of discretization error reduction rate.
\item The $H_1$ norm of corrected derivatives ($\overline{\overline{\partial_q Q_k}}$) undergoes larger order discrepancies than that of uncorrected derivatives ($\partial_q Q_k$) for both unperturbed uniform (Fig. \ref{fig:grid_sens_unif_H}) and unperturbed expanded (Fig. \ref{fig:grid_sens_expnd_H}) sets. Hence, the corrected derivatives reach the asymptotic range with a delay with regards to the uncorrected derivatives.
\item Although larger in magnitude compared to the unperturbed uniform grid set ($d\alpha_0$ in Figs. \ref{fig:grid_sens_unif_L} and \ref{fig:grid_sens_unif_H}), the order discrepancy of the unperturbed expanded grid set ($d\alpha_0$ in Figs. \ref{fig:grid_sens_expnd_L} and  \ref{fig:grid_sens_expnd_H}) diminishes with mesh refinement for all $L$ and $H_1$ norms (except for the $L_\infty$ norm of the P2 discretization). This illustrates that the unnecessary application of grid anisotropy delays the asymptotic range but in contrast with grid self-dissimilarity, it does not disrupt it.
\item{As in the case of the uniform set, the random perturbations of the expanded set disrupt the asymptotic convergence for both the $L$ (Fig. \ref{fig:grid_sens_expnd_L}) and $H_1$ norms (Fig. \ref{fig:grid_sens_expnd_H})} and more so as the perturbation increases from $d\alpha_1$ to $d\alpha_2$.
\end{itemize}

In the light of these observations, it can be concluded that for higher OOAs, consistently with low-order discretizations, the inadequate application of grid stretching (anisotropy and non-uniformity), to regions where solution gradients are rather smooth, only delays the occurrence of the asymptotic range, whereas if the grid set is not self-similar, the monotonic convergence of the OOAs is compromised and hence grid self-similarity needs to be enforced in a grid convergence study. 

\section{Conclusions}\label{concs}
The aim of this article is to extend the code verification methodology to high-order accurate  CFD solvers. To this end, a series of trigonometric manufactured solutions is introduced which enables the incremental examination of CFD solver components for the simulation of free flows in inviscid, laminar and turbulent (via the original and modified SA models) regimes. The capability of each of these MSs for the verification of OOAs from the first up to the sixth is demonstrated on a pre-integrated DG implementation via the FR/CPR scheme. These MSs served as well to reach the following conclusions:
\begin{itemize}
\item The $L_2$ norm is shown to be more sensitive to deviations from the mean than the $L_1$ norm whereas the $L_\infty$ norm is demonstrated to be the most useful in detecting localized inconsistencies and hence its inclusion in code verification is recommended;
\item The $H_1$ semi-norm of uncorrected and fully corrected (via BR2 scheme) derivatives is considered and the latter are found to be often more accurate than the former. Note however that among many attempts via introduction of various types of bugs,  no case showed a significant advantage  of $H_1$ semi-norms over $L$ norms in their detection;
\item The inclusion of a relatively high OOA (such as the sixth) in code verification is recommended since its utility is demonstrated for the detection of minor bugs  which affect higher orders first on coarser grids;
\item A supersonic inviscid MS on deformed domain is introduced to verify the treatment of curved elements;
\item  Manufactured cases are presented to verify the original and modified SA models coupled to compressible RANS equations. The importance of the balancing of forcing function terms in assuring the reliable examination of all terms of the governing equations is illustrated via an example. A sensitivity analysis method is introduced to enforce the conclusions of the verification exercise in cases where some terms are outweighed by others, thus providing assurance in the outcomes of the verification campaign. Let's note that the only terms of the governing equations left unverified through this study are those of the modified vorticity (Eq. \eqref{eq:S-}) of the modified SA model;
\item The necessary level of residual minimization for the convergence of the discretization error is investigated and it is shown that for high $\mu$ values in a laminar manufactured case, the residual magnitude is larger than that of the discretization error, indicating that it is not always necessary to drive the residual level lower than the discretization error for the iterative and round-off errors to be negligible;
\item The effect of stretching and self-similarity of a grid set on grid convergence of solutions with smooth gradients is studied numerically for a turbulent MS and it is concluded that unnecessary grid stretching delays the occurrence of the asymptotic convergence whereas the loss of self-similarity disrupts it.
\end{itemize}

The discussion on the extension of verification methodology of high-order solvers on wall-bounded curved domains and realistic turbulent flows is intended to be pursued in a future work.


\section*{Acknowledgements}
The authors gratefully acknowledge the generous support from the Fonds de recherche du Qu\'ebec -- Nature et technologies (FRQNT), the Natural Sciences and Engineering Research Council (NSERC), and the Department of Mechanical Engineering of McGill University.
%
%
%
%
%


\appendix

\clearpage

\section{Definition of the grid convergence metrics}
\label{sec:norms}
Here are the definitions of the norms employed throughout this work to measure the discretization error. The integrals are computed by GLL quadratures.
\vspace{0.1cm}
\begin{itemize}
\item $L_\infty$ norm:
\begin{equation}
\| \mathcal{E}_Q \|_\infty =
\mathrm{max} \lvert Q_i - Q_i^{\mathrm{ex}} \lvert \;\; \mathrm{for} \;\; {i\,\in \,\left[ 1\,..\,{N_\mathrm{DOF}} \right] }
\end{equation}

\item $L_1$ norm:
\begin{equation}
\| \mathcal{E}_Q \|_{L_1} =
\frac{\int_{\Omega} \, \lvert Q - Q^{\mathrm{ex}}\lvert \, d\Omega}{{\int_{\Omega} d\Omega}}
\end{equation}

\item $L_2$ norm:
\begin{equation}
\| \mathcal{E}_Q \|_{L_2} = \left( \frac{ \int_{\Omega} \, \left( Q - Q^{\mathrm{ex}}\right)^2d\Omega}{{\int_{\Omega} d\Omega}} \right)^{\frac{1}{2}}
\end{equation}
\item $H_1$ norm:
\begin{equation}
\| \mathcal{E}_Q \|_{H_1} = \left( \frac{\int_{\Omega} \, \left( Q - Q^{\mathrm{ex}}\right)^2  d\Omega\,+ \,\int_{\Omega}\,\sum_{q=1}^{N_\mathrm{d}}\left(\partial_q Q - (\partial_q Q)^{\mathrm{ex}}\right)^2 \, d\Omega}{{\int_{\Omega} d\Omega}} \right)^{\frac{1}{2}}
\end{equation}
\item $H_1$ semi-norm of uncorrected gradients:
\begin{equation}
\abs[\big]{ \mathcal{E}_Q }_{H_1} =\left(\frac{\int_{\Omega}\,\sum_{q=1}^{N_\mathrm{d}}\left(\partial_q Q - (\partial_q Q)^{\mathrm{ex}} \right)^2 \, d\Omega}{{\int_{\Omega} d\Omega}}\right)^{\frac{1}{2}}
\end{equation}
\item $H_1$ semi-norm of fully corrected gradients:
\begin{equation}
\abs[\big]{ \mathcal{E}_Q }_{\overline{\overline{H_1}}} =\left(\frac{\int_{\Omega}\,\sum_{q=1}^{N_\mathrm{d}}\left({\overline{\overline{\partial_q Q}}} - (\partial_q Q)^{\mathrm{ex}}\right)^2 \, d\Omega}{{\int_{\Omega} d\Omega}} \right)^{\frac{1}{2}}
\end{equation}
\end{itemize}

\clearpage

\section{Manufactured solution results}
\label{sec:MS-resl}

\subsection{MS-1}

\begin{figure}[!hbt]
\centering
\subfloat[$\rho^{\mathrm{MS}}$]{
\includegraphics[trim = 0mm 0mm 0mm 0mm, clip,width=0.27\linewidth]
{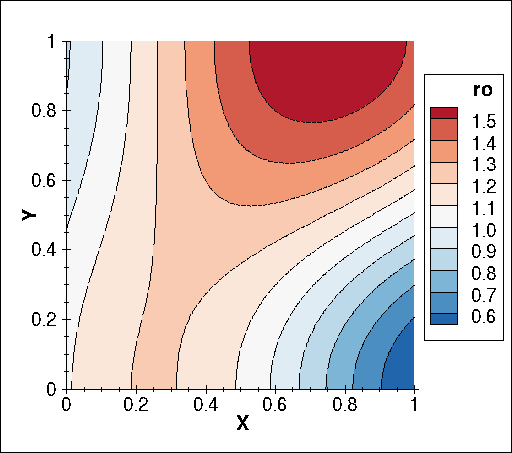}}
~~~
\subfloat[$u^{\mathrm{MS}}$]{
\includegraphics[trim = 0mm 0mm 0mm 0mm, clip,width=0.27\linewidth]
{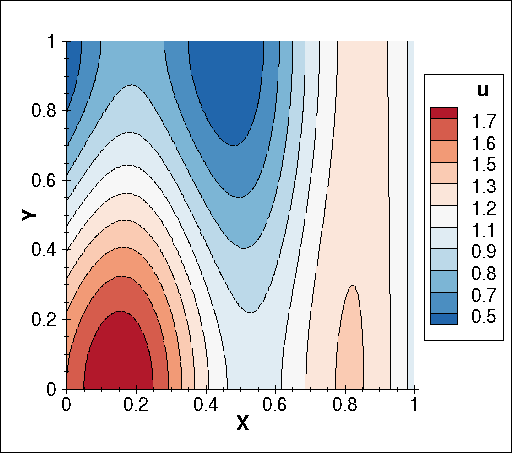}}
\vfill
\subfloat[$v^{\mathrm{MS}}$]{
\includegraphics[trim = 0mm 0mm 0mm 0mm, clip,width=0.27\linewidth]{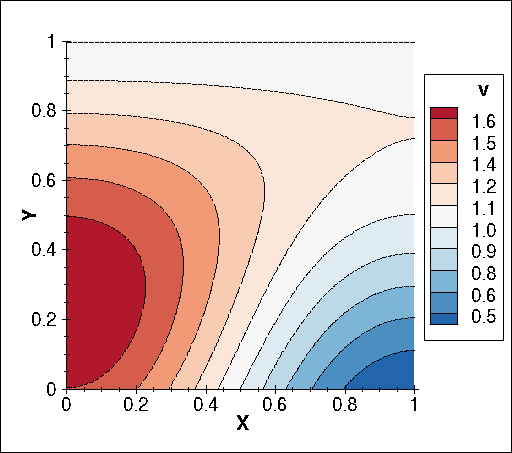}}
~~~
\subfloat[$p^{\mathrm{MS}}$]{
\includegraphics[trim = 0mm 0mm 0mm 0mm, clip,width=0.27\linewidth]{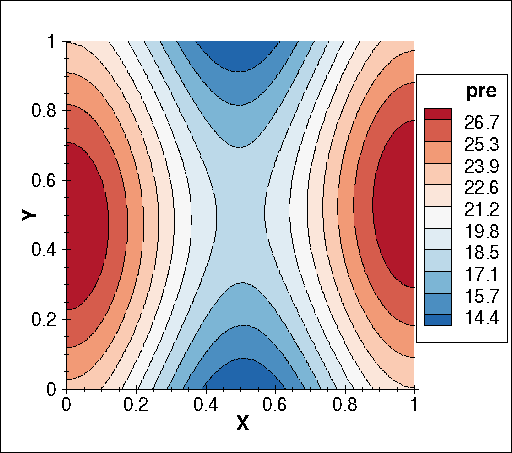}}
\vfill
\subfloat[$\mbox{Ma}^{\mathrm{MS}}$ and ${\bm{u}}^{\mathrm{MS}}$]{
\includegraphics[trim = 0mm 0mm 0mm 0mm, clip,width=0.27\linewidth]{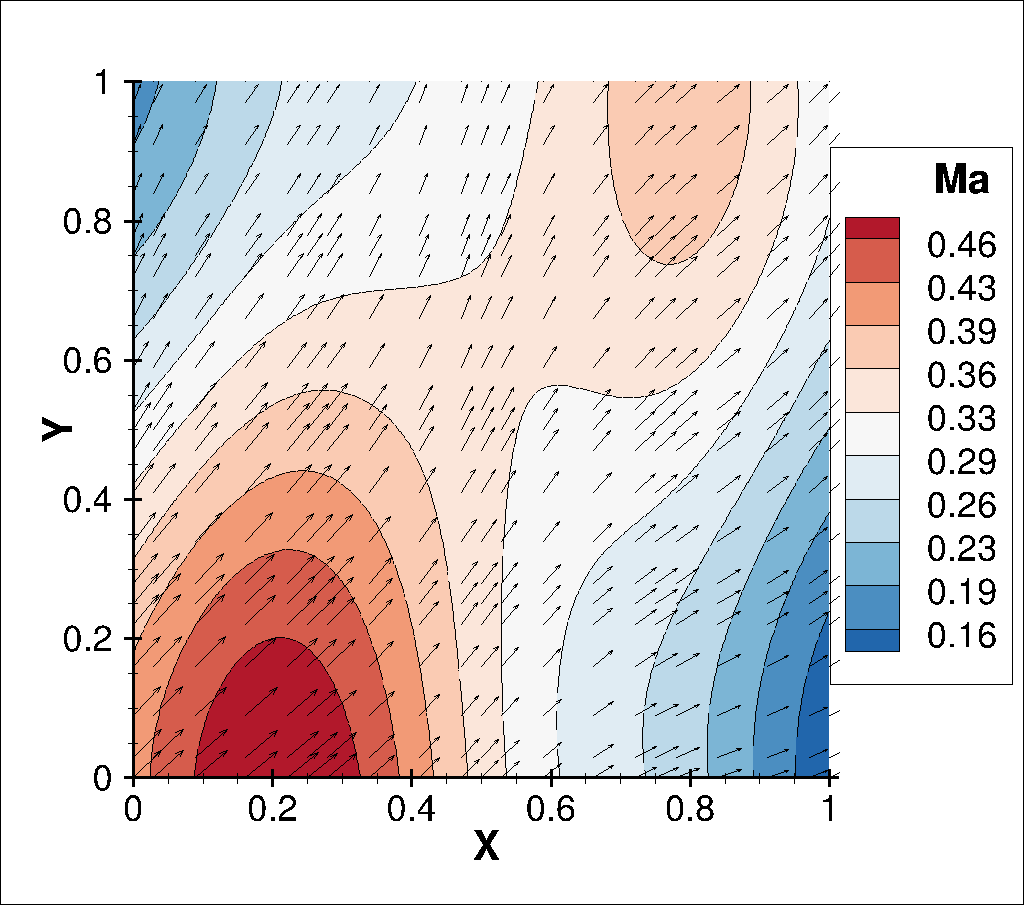}}
\caption{Manufactured solution MS-1}
\label{fig:MS-1}
\end{figure}

\begin{figure}[!hbt]
\centering
\subfloat[$\rho$]{
\includegraphics[trim = 5mm 2mm 18mm 13mm, clip,width=0.32\linewidth]
{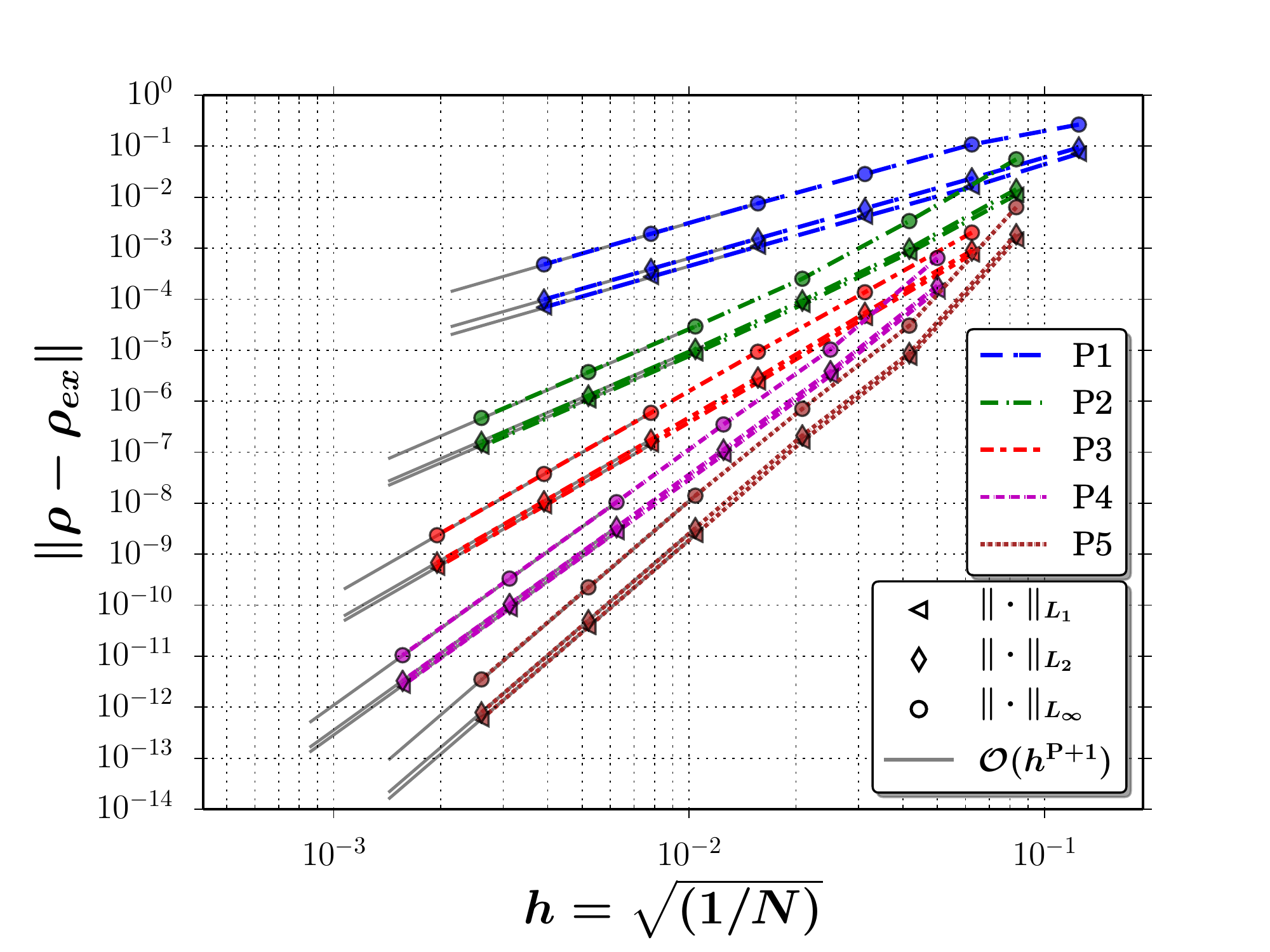}}~~~
\subfloat[$\rho u$]{
\includegraphics[trim = 5mm 2mm 18mm 13mm, clip,width=0.32\linewidth]{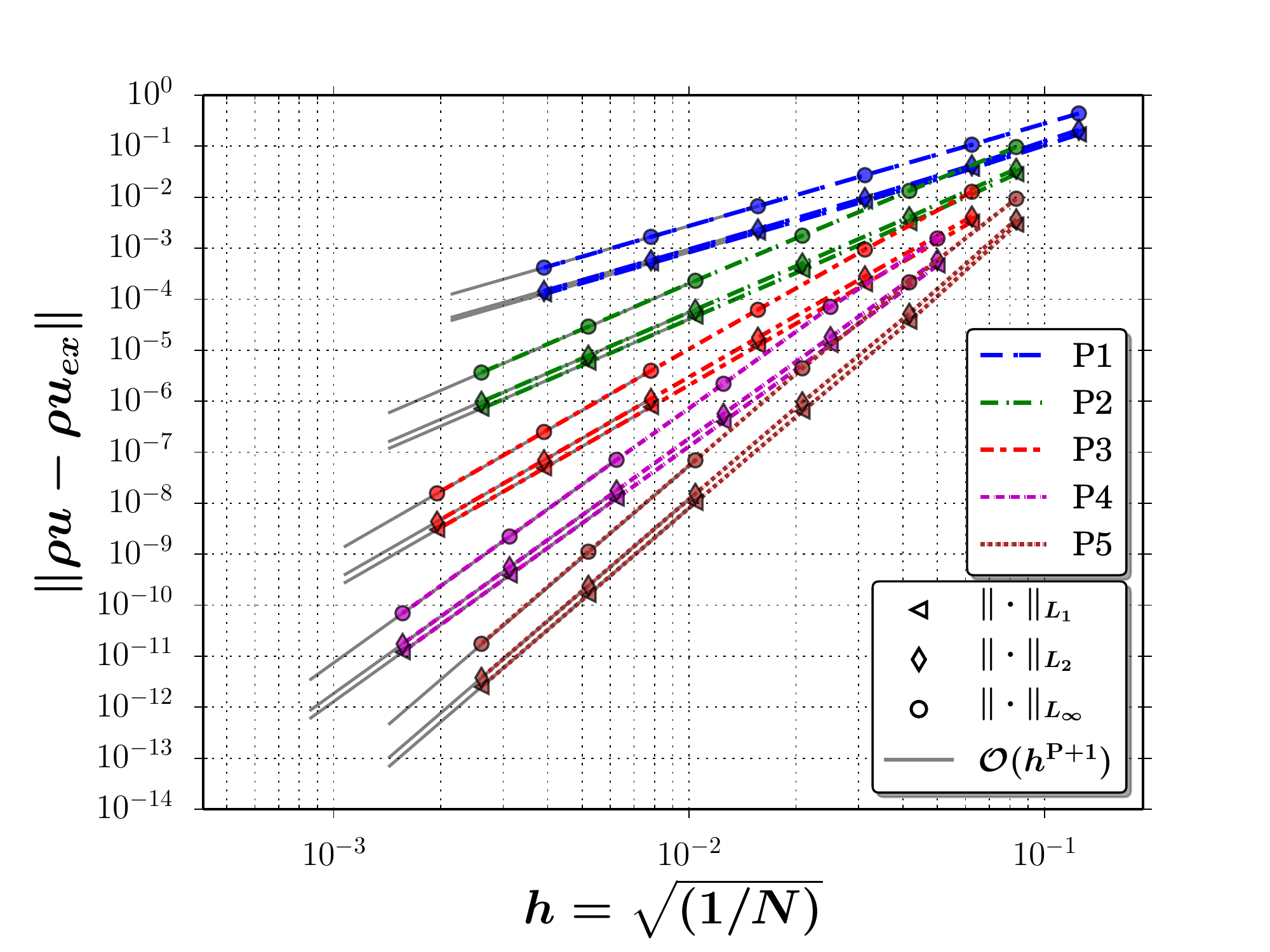}}
\vfill
\subfloat[$\rho v$]{
\includegraphics[trim = 5mm 2mm 18mm 13mm, clip,width=0.32\linewidth]
{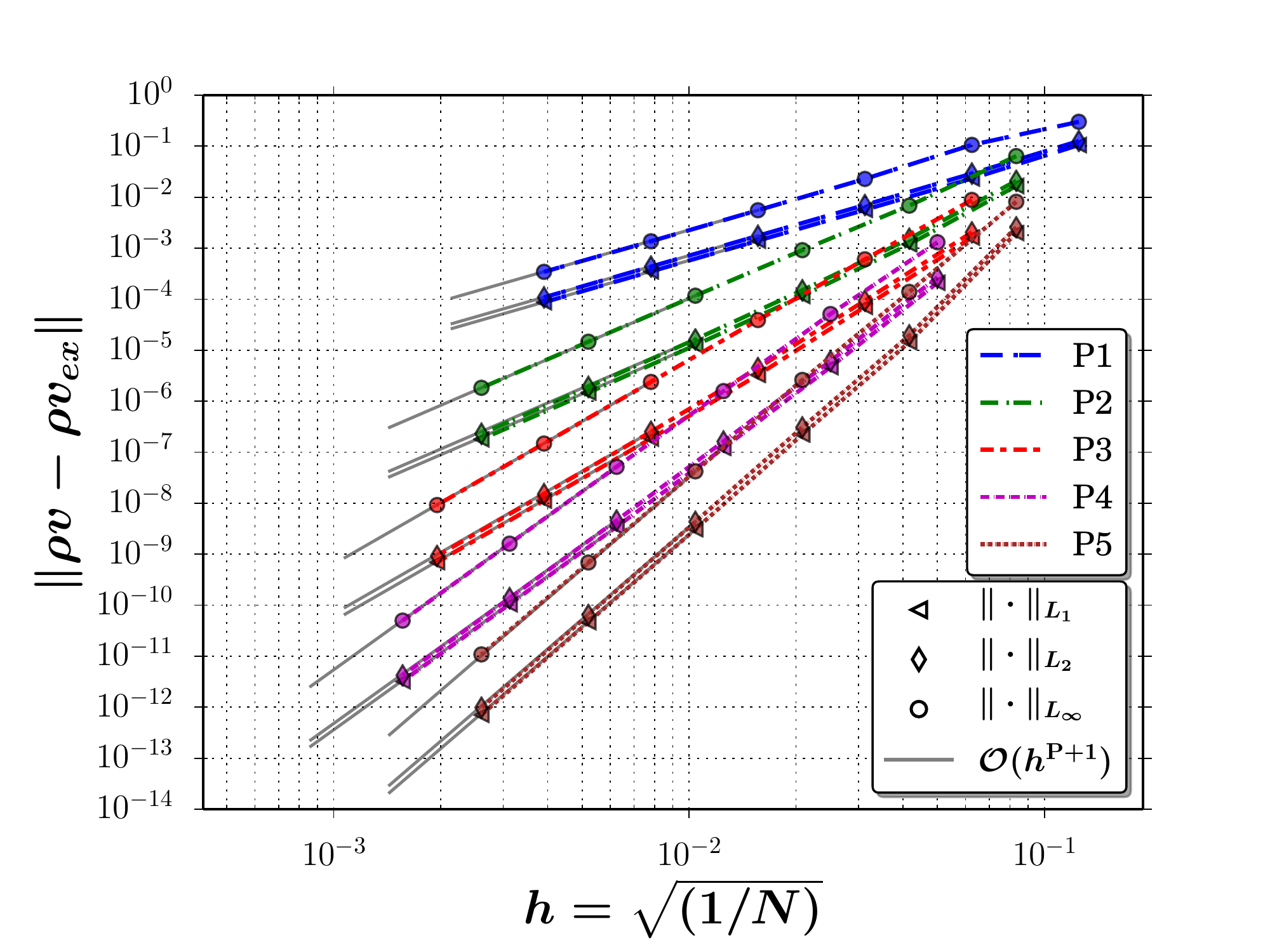}}~~~
\subfloat[$\rho E$]{
\includegraphics[trim = 5mm 2mm 18mm 13mm, clip,width=0.32\linewidth]{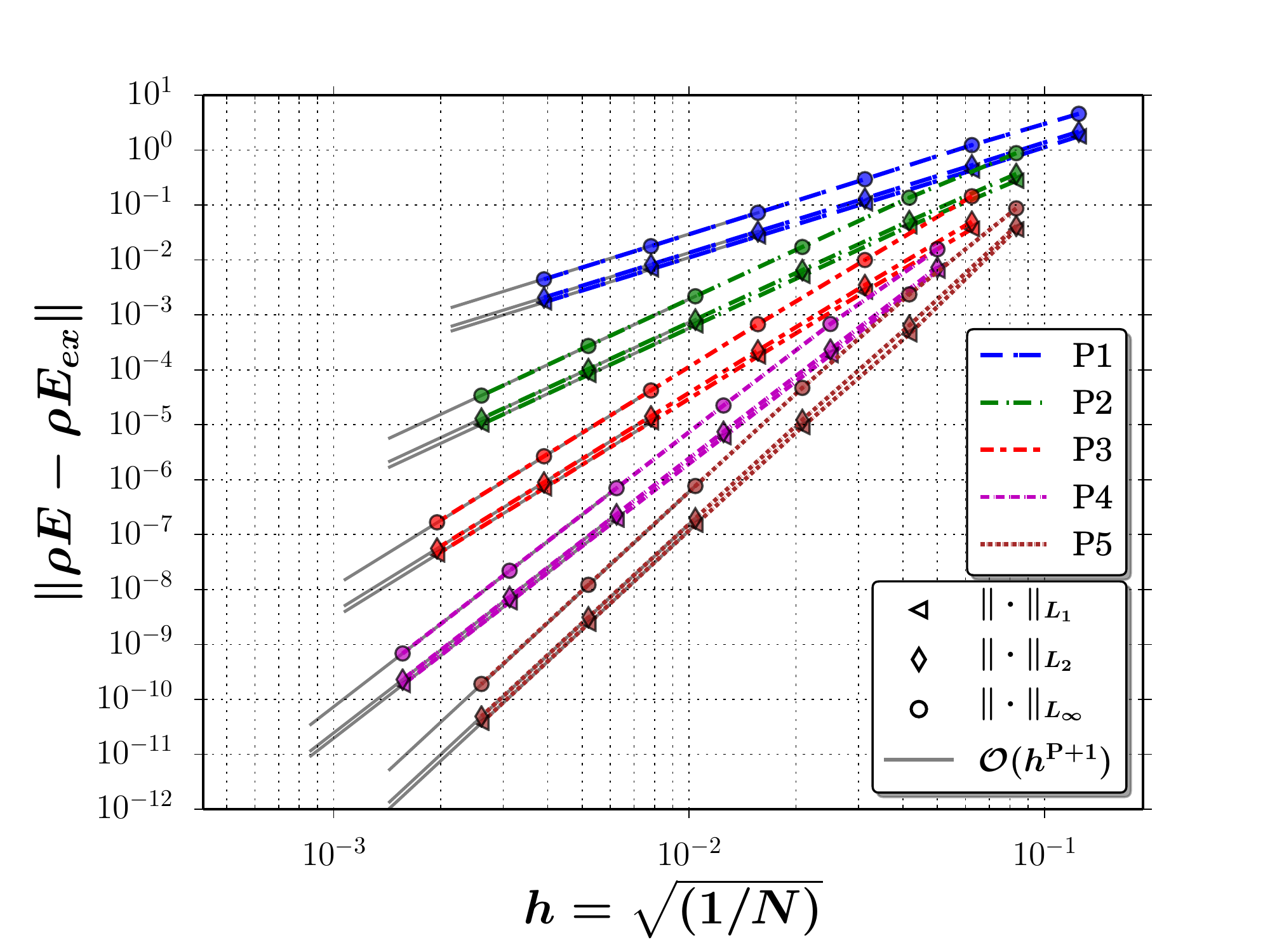}}
\caption{Evolution of the discretization error in $L_1$, $L_2$ and $L_\infty$ norms versus mesh refinement for MS-1 and $\mathrm{P}1$--$\mathrm{P}5$}
\label{fig:Err_allE_allP_MS-1}
\end{figure}

\begin{figure}[!hbt]
\centering
\subfloat[$\rho$]{ 
\includegraphics[trim = 16mm 3mm 18mm 13mm, clip,width=0.3\linewidth]
{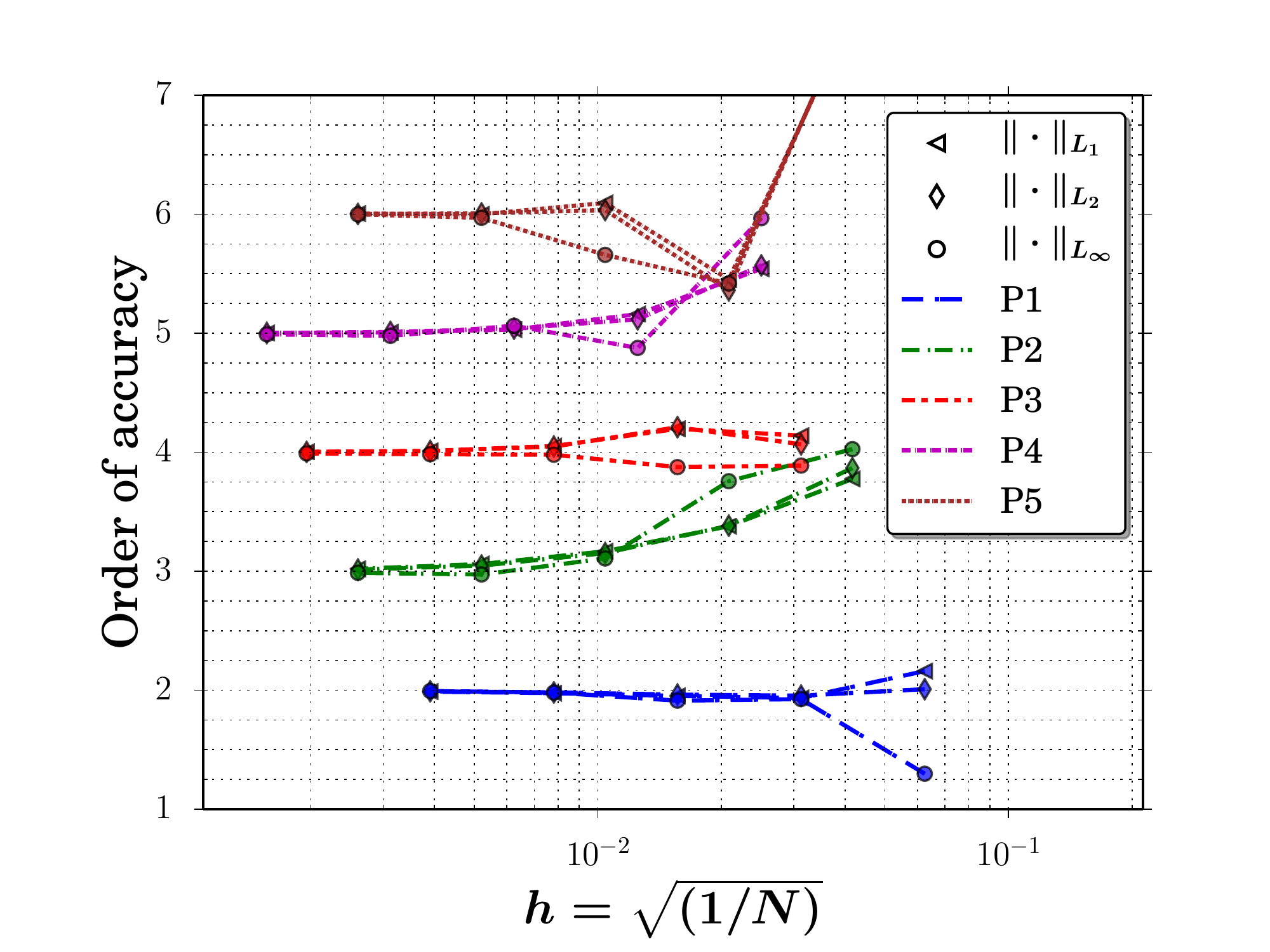}}~~~
\subfloat[$\rho u$]{
\includegraphics[trim = 16mm 3mm 18mm 13mm, clip,width=0.3\linewidth]{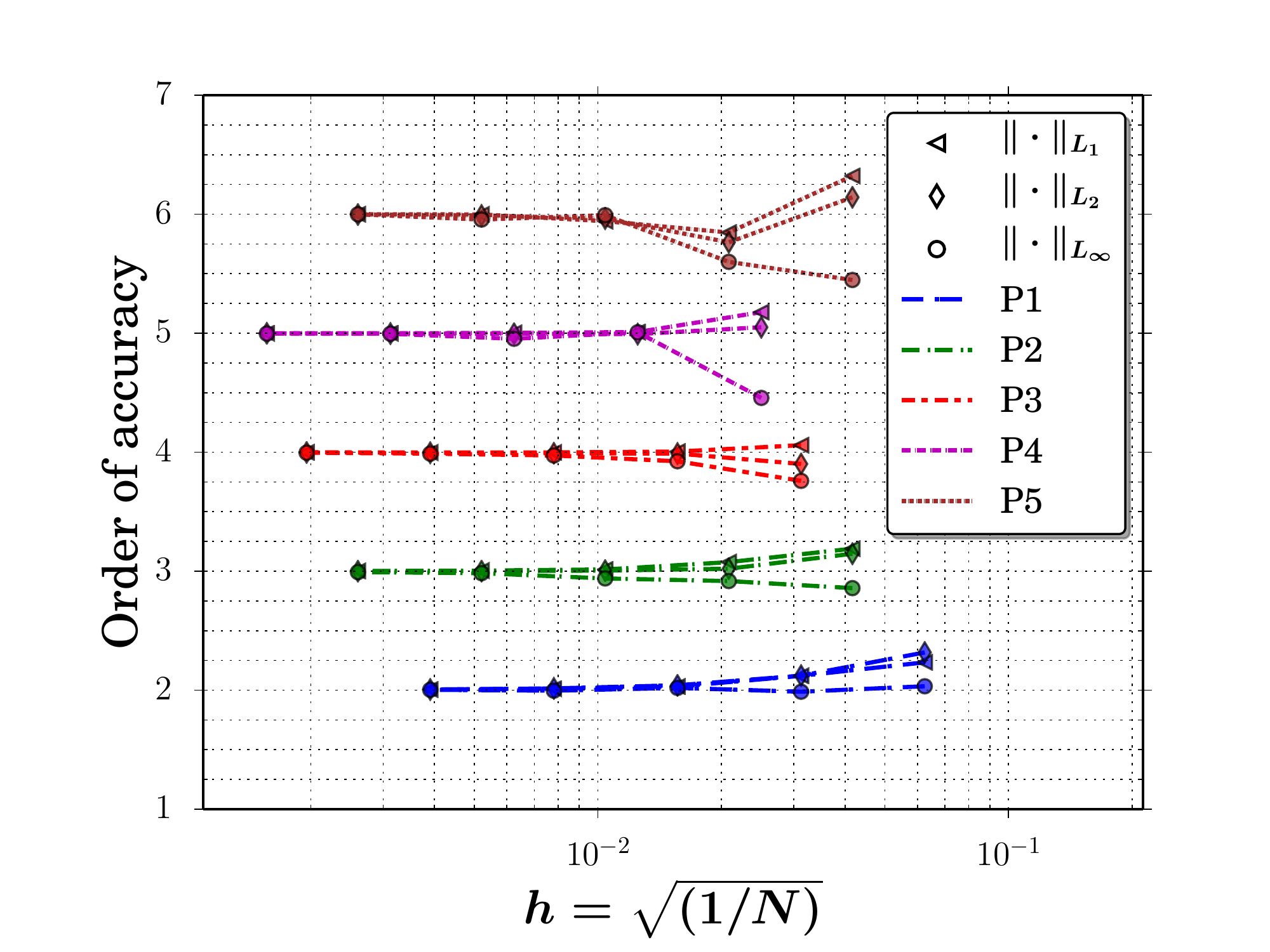}}
\vfill
\subfloat[$\rho v$]{
\includegraphics[trim = 16mm 3mm 18mm 13mm, clip,width=0.3\linewidth]
{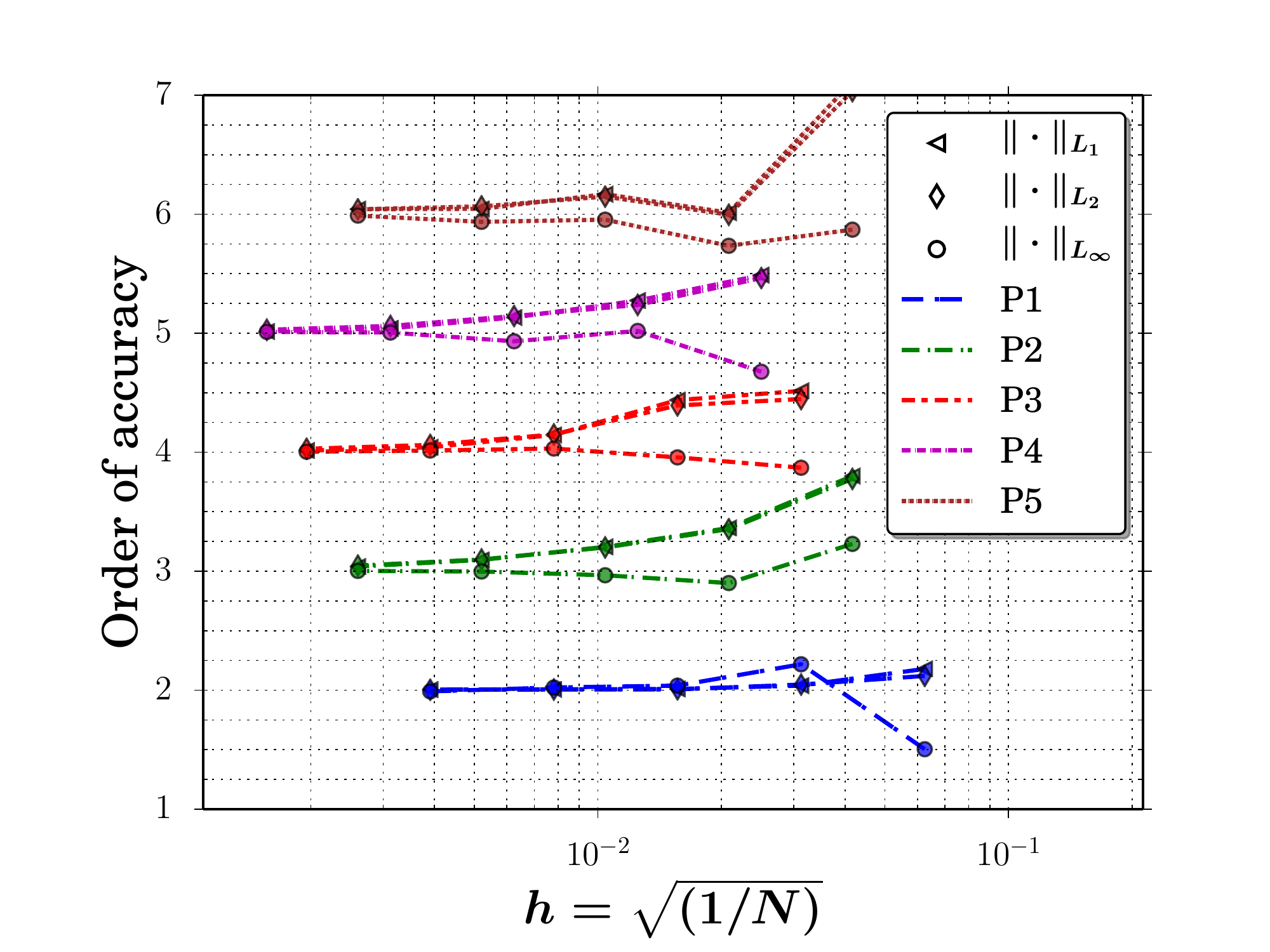}}~~~
\subfloat[$\rho E$]{
\includegraphics[trim = 16mm 3mm 18mm 13mm, clip,width=0.3\linewidth]{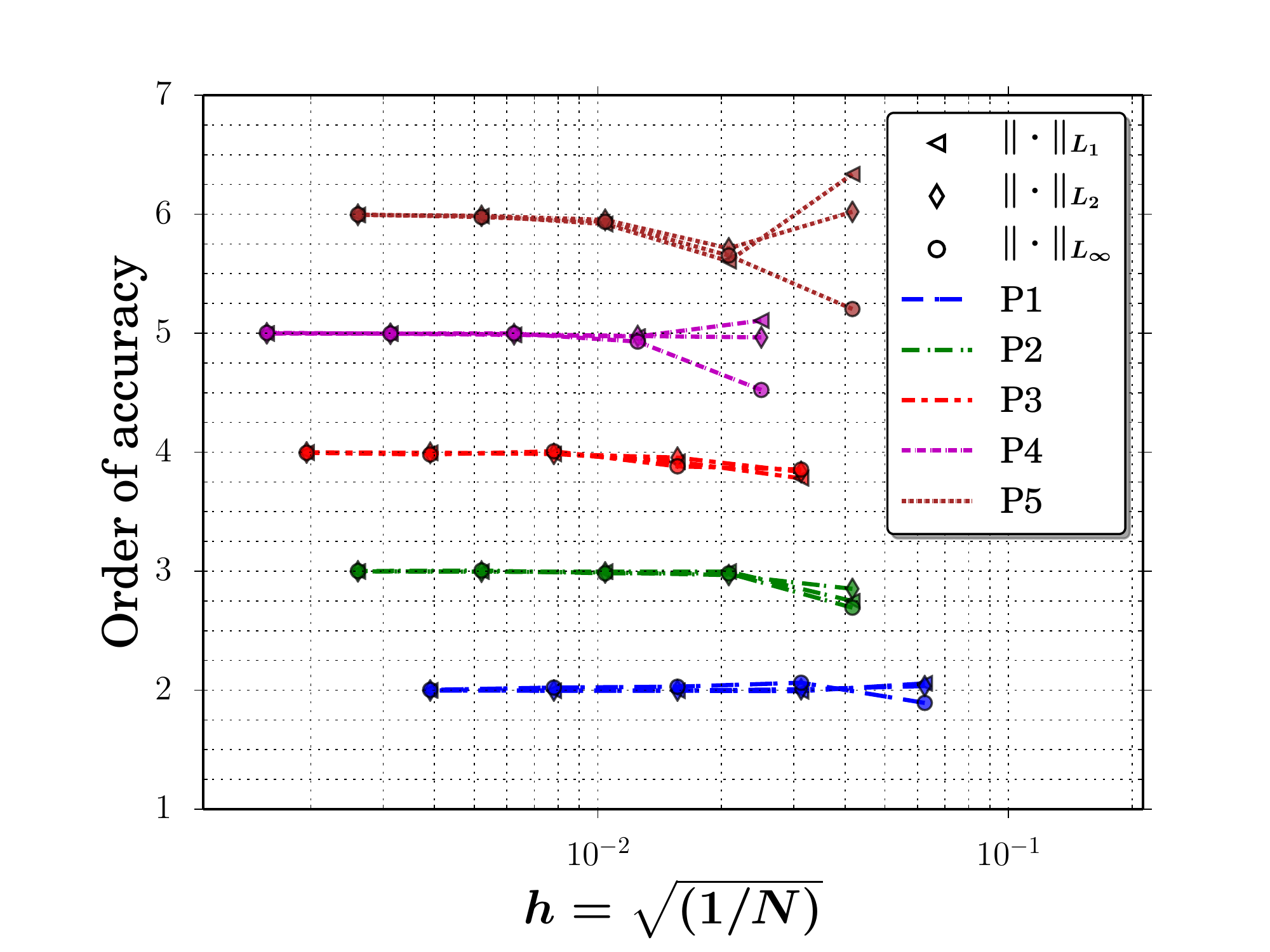}}
\caption{Evolution of the OOAs in $L_1$, $L_2$ and $L_\infty$ norms versus mesh refinement  for MS-1 and $\mathrm{P}1$--$\mathrm{P}5$}
\label{fig:Orders_MS-1}
\end{figure}

\clearpage
\subsection{MS-2}

\begin{figure}[!hbt]
\centering
\subfloat[$\rho^{\mathrm{MS}}$]{
\includegraphics[trim = 0mm 0mm 0mm 0mm, clip,width=0.27\linewidth]
{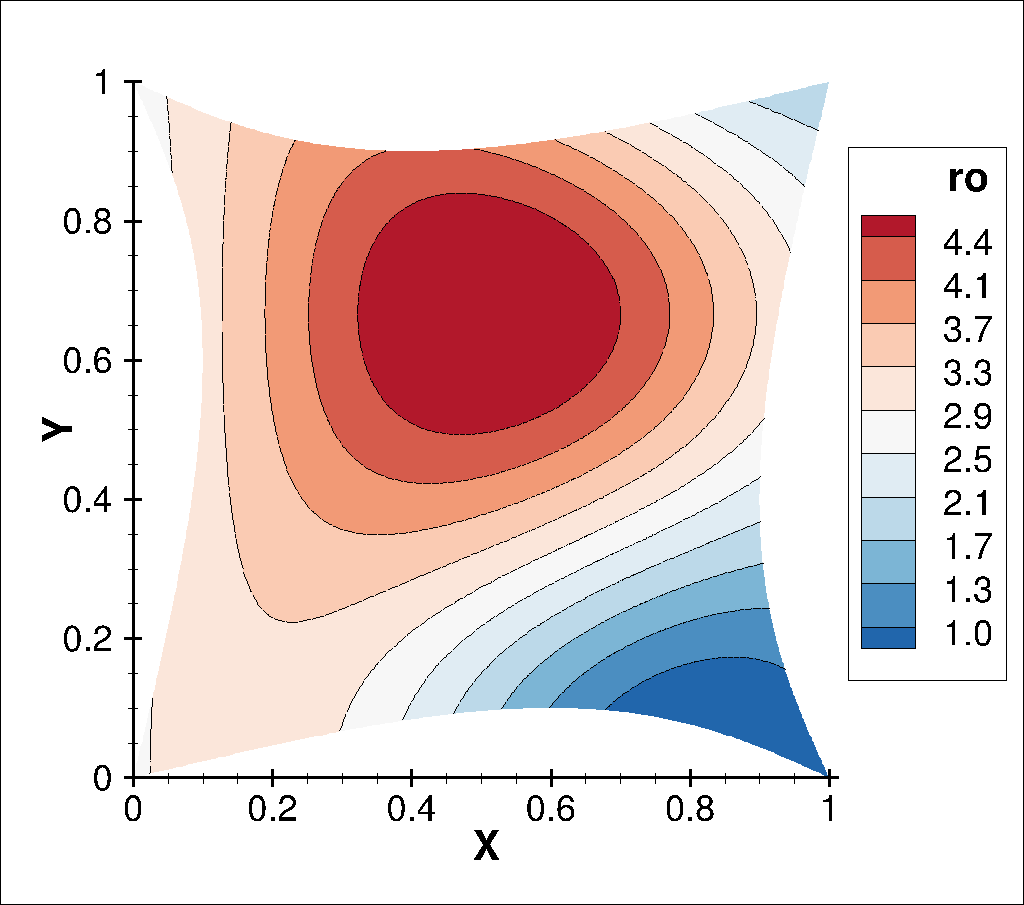}}
~~~
\subfloat[$u^{\mathrm{MS}}$]{
\includegraphics[trim = 0mm 0mm 0mm 0mm, clip,width=0.27\linewidth]
{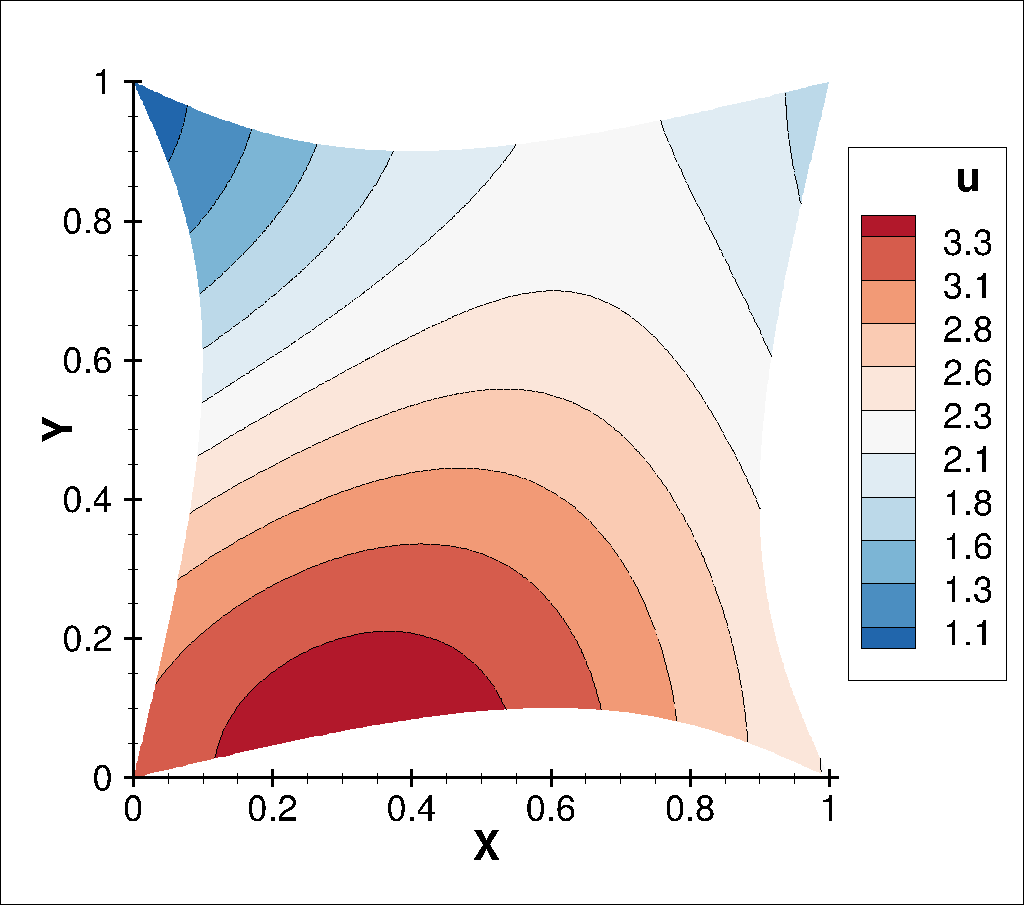}}
\vfill
\subfloat[$v^{\mathrm{MS}}$]{
\includegraphics[trim = 0mm 0mm 0mm 0mm, clip,width=0.27\linewidth]{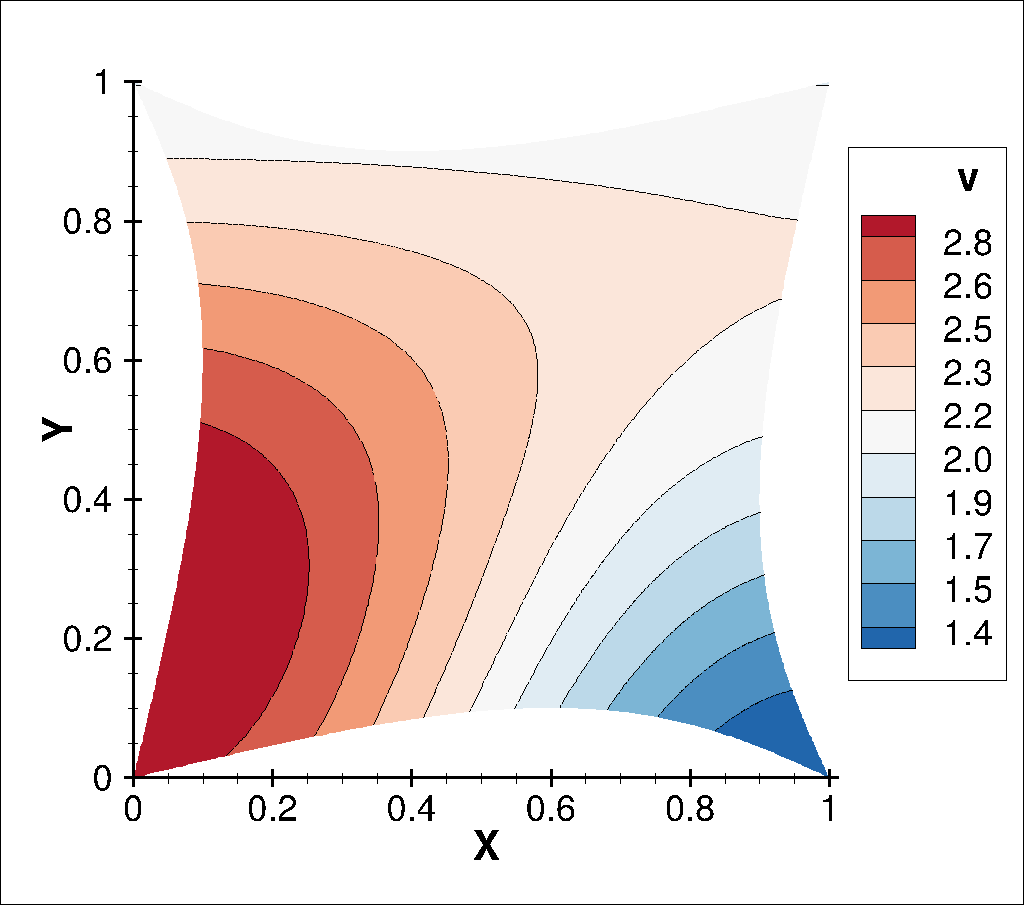}}
~~~
\subfloat[$p^{\mathrm{MS}}$]{
\includegraphics[trim = 0mm 0mm 0mm 0mm, clip,width=0.27\linewidth]{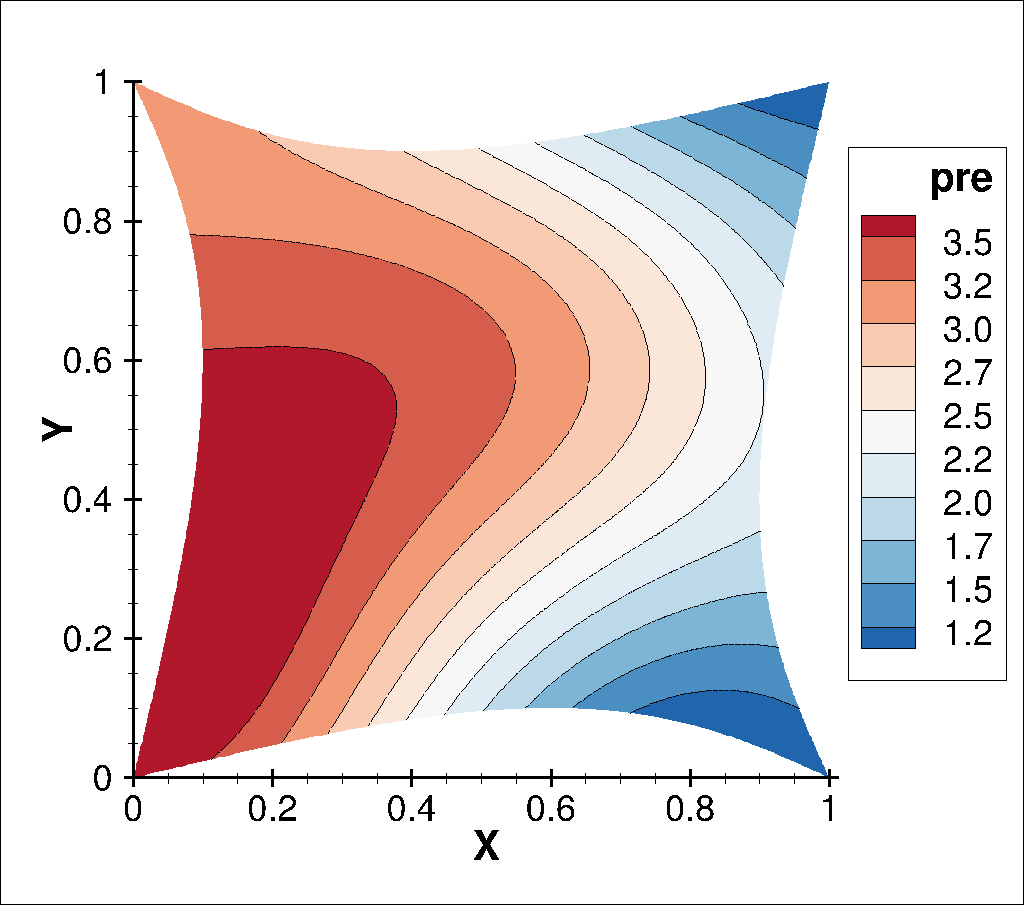}}
\vfill
\subfloat[$\mbox{Ma}^{\mathrm{MS}}$ and ${\bm{u}}^{\mathrm{MS}}$]{
\includegraphics[trim = 0mm 0mm 0mm 0mm, clip,width=0.27\linewidth]{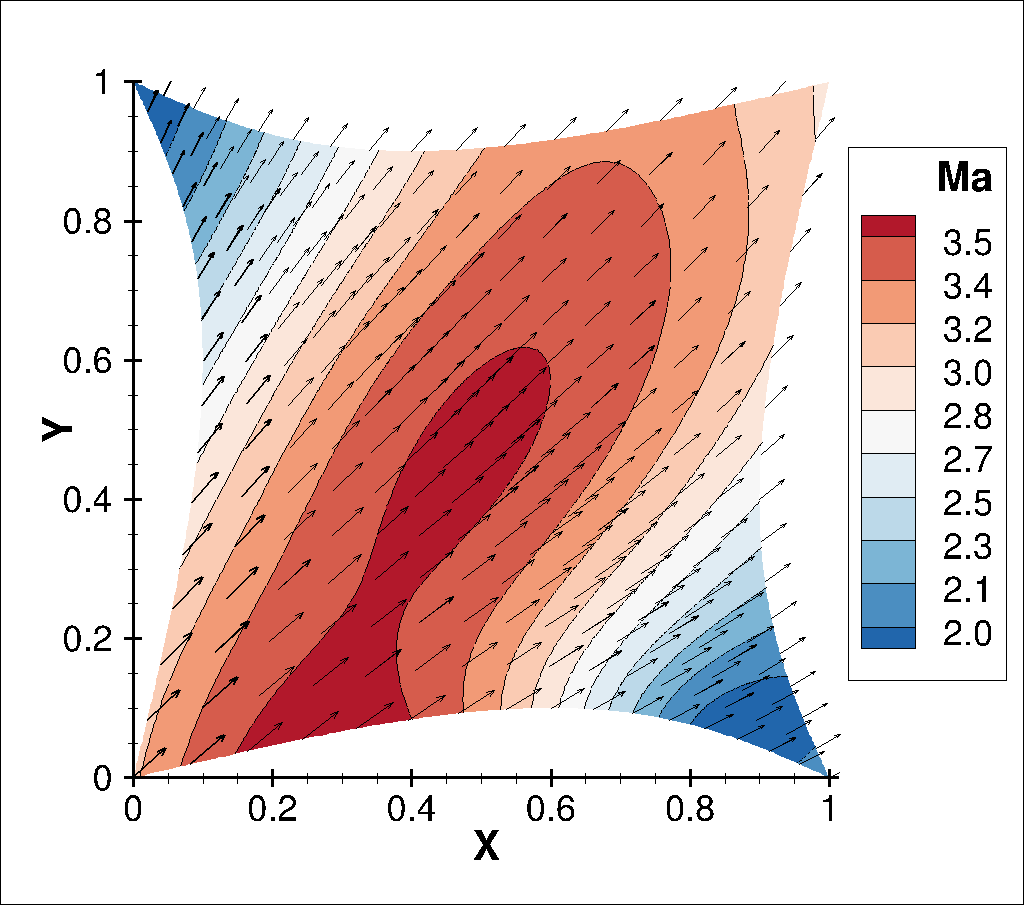}}
\caption{Manufactured solution MS-2}
\label{fig:MS-2}
\end{figure}

\begin{figure}[!hbt]
\centering
\subfloat[$\rho$]{
\includegraphics[trim = 5mm 2mm 18mm 13mm, clip,width=0.32\linewidth]
{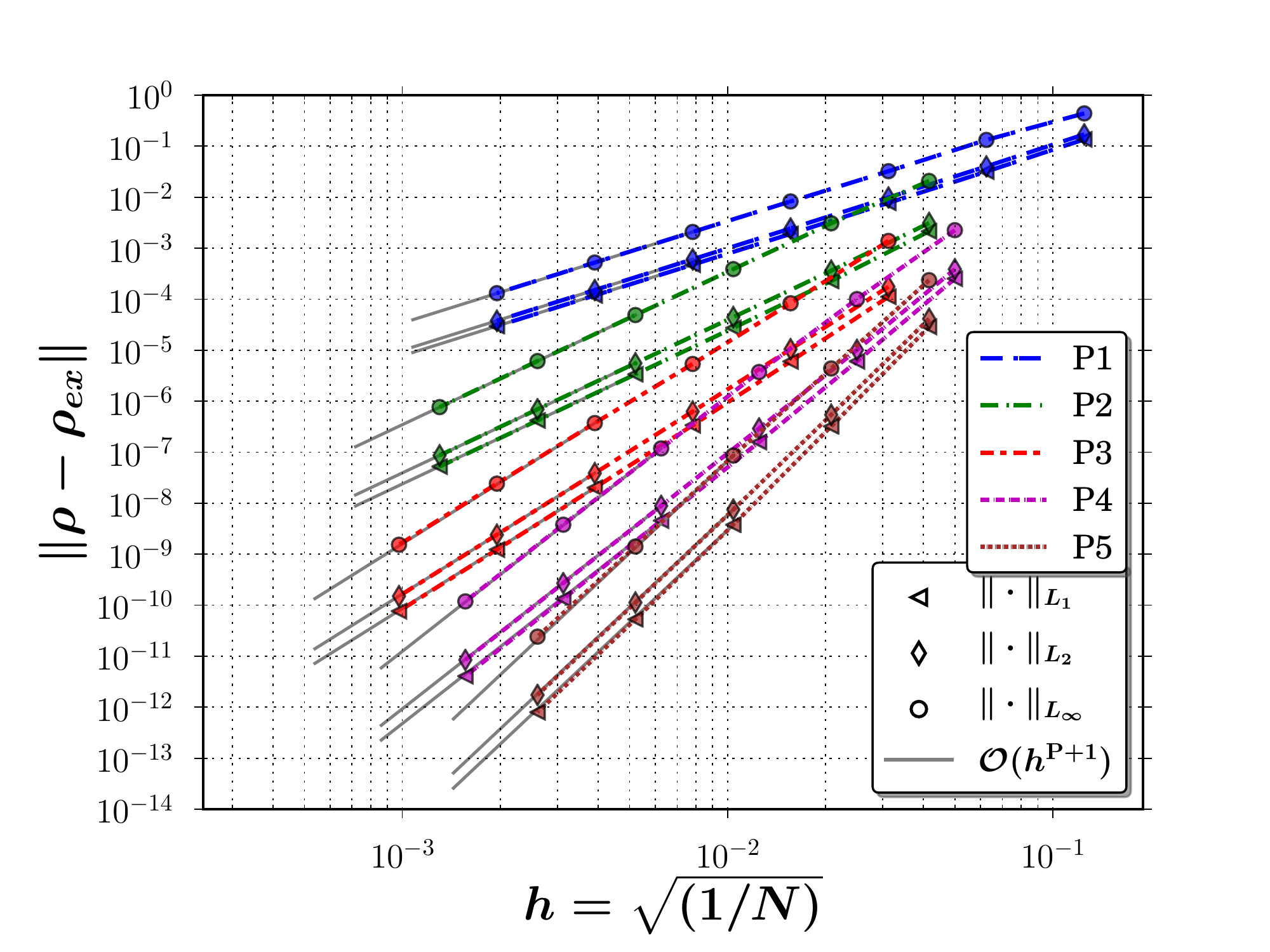}}
~~~
\subfloat[$\rho u$]{
\includegraphics[trim = 5mm 2mm 18mm 13mm, clip,width=0.32\linewidth]{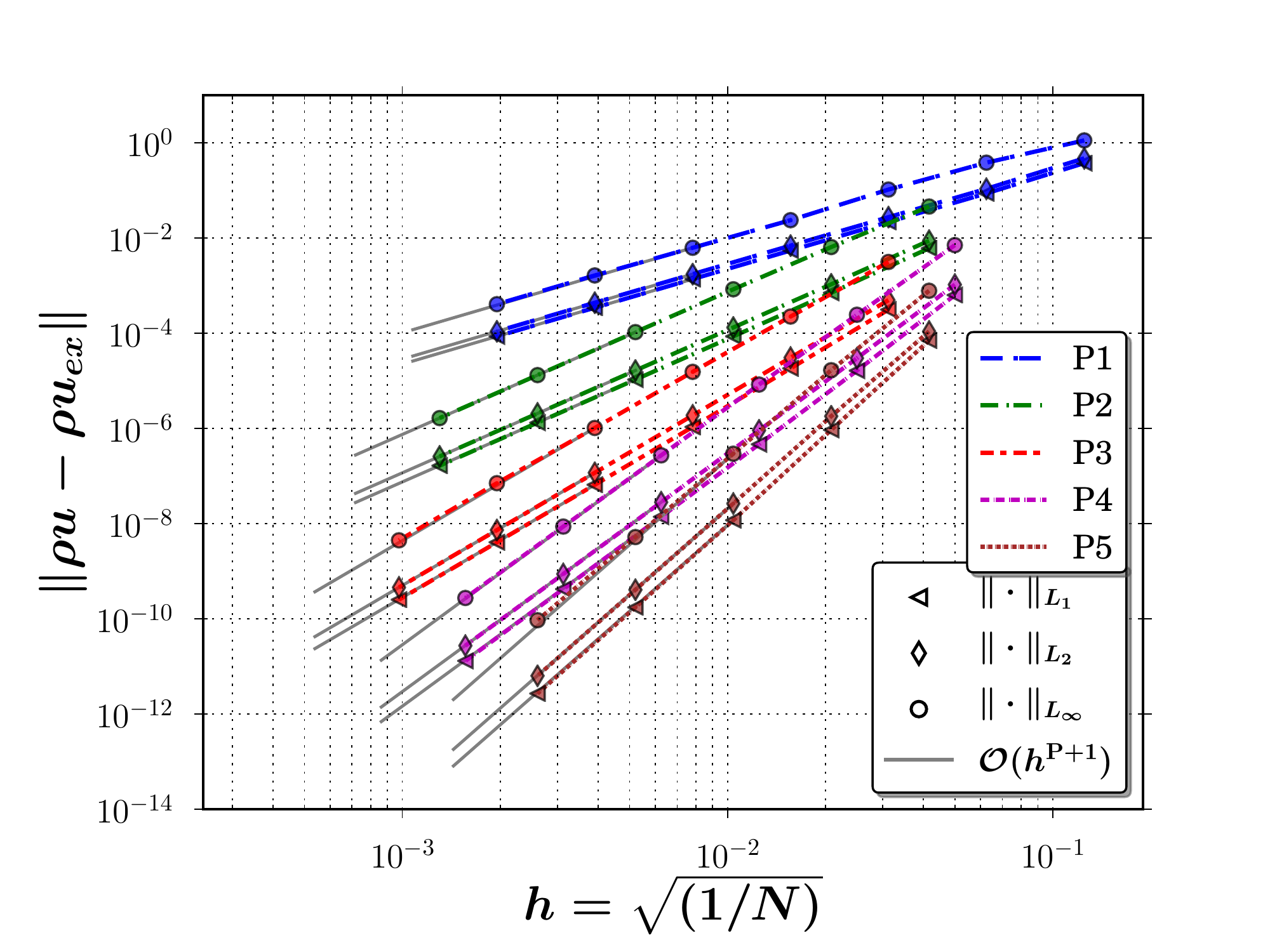}}
\vfill
\subfloat[$\rho v$]{
\includegraphics[trim = 5mm 2mm 18mm 13mm, clip,width=0.32\linewidth]
{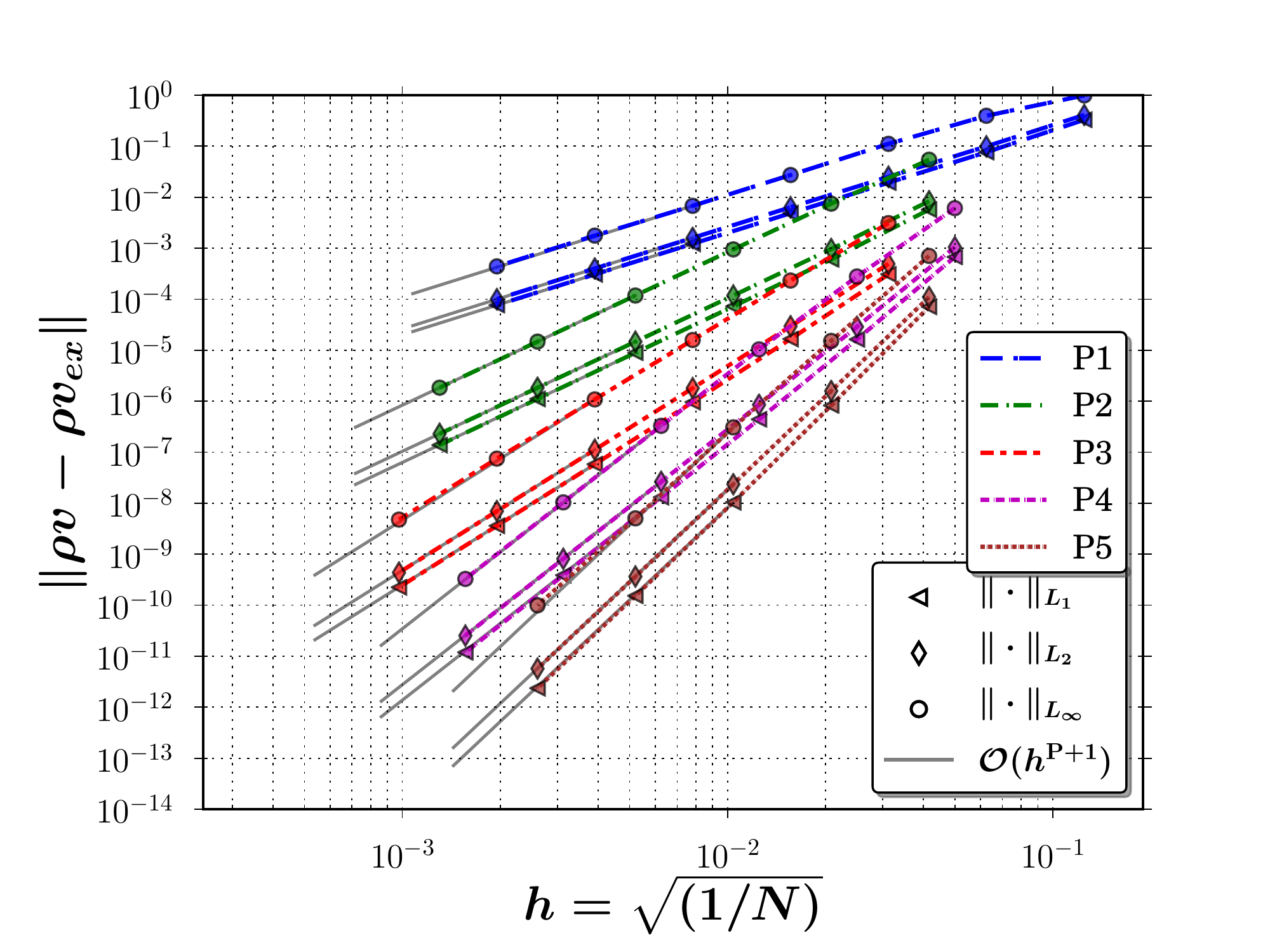}}
~~~
\subfloat[$\rho E$]{
\includegraphics[trim = 5mm 2mm 18mm 13mm, clip,width=0.32\linewidth]{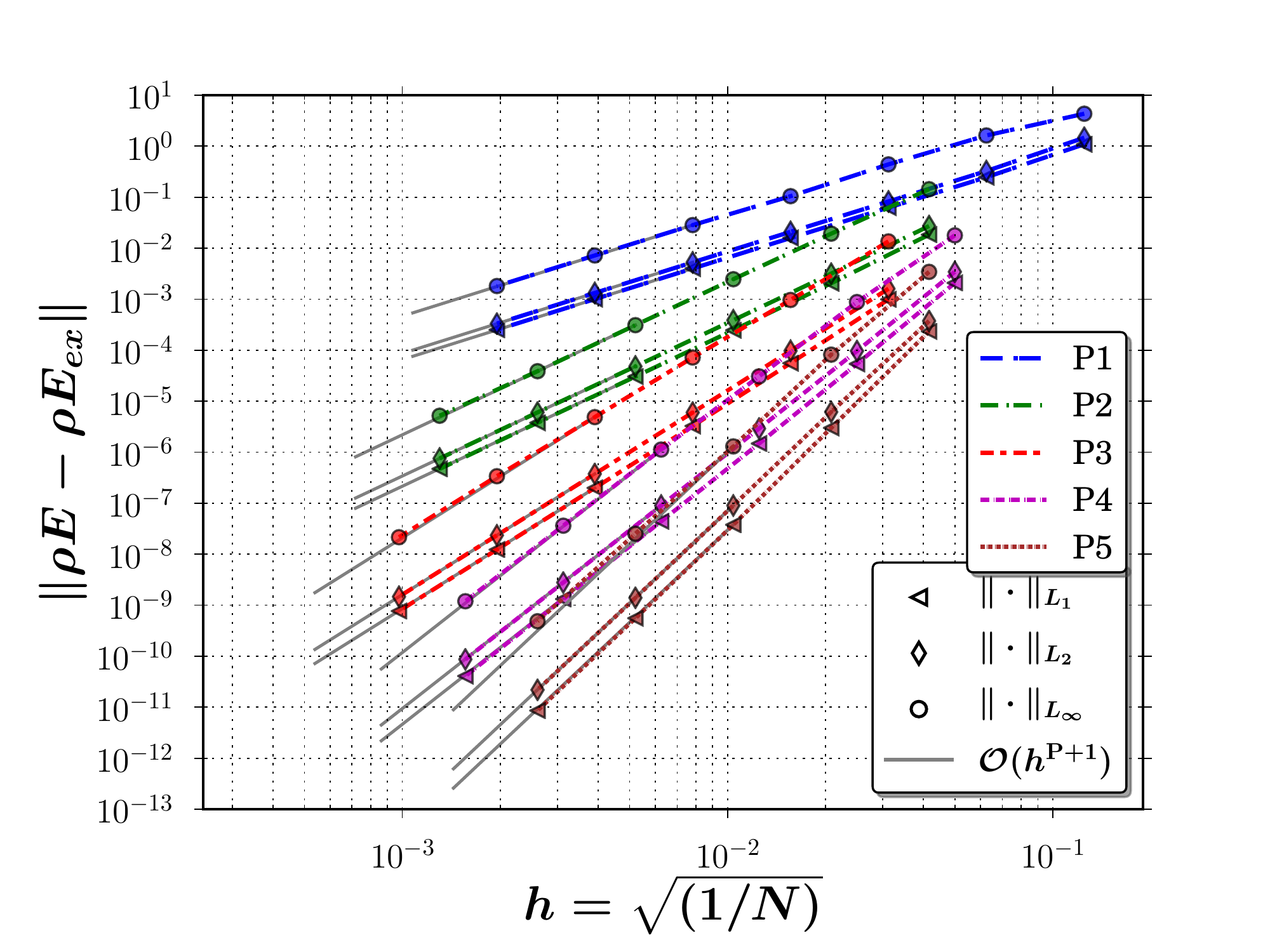}}
\caption{Evolution of the discretization error in $L_1$, $L_2$ and $L_\infty$ norms versus mesh refinement for MS-2 and $\mathrm{P}1$--$\mathrm{P}5$}
\label{fig:Err_allE_allP_MS-2}
\end{figure}

\begin{figure}[!hbt]
\centering
\subfloat[$\rho$]{ 
\includegraphics[trim = 16mm 3mm 18mm 13mm, clip,width=0.30\linewidth]
{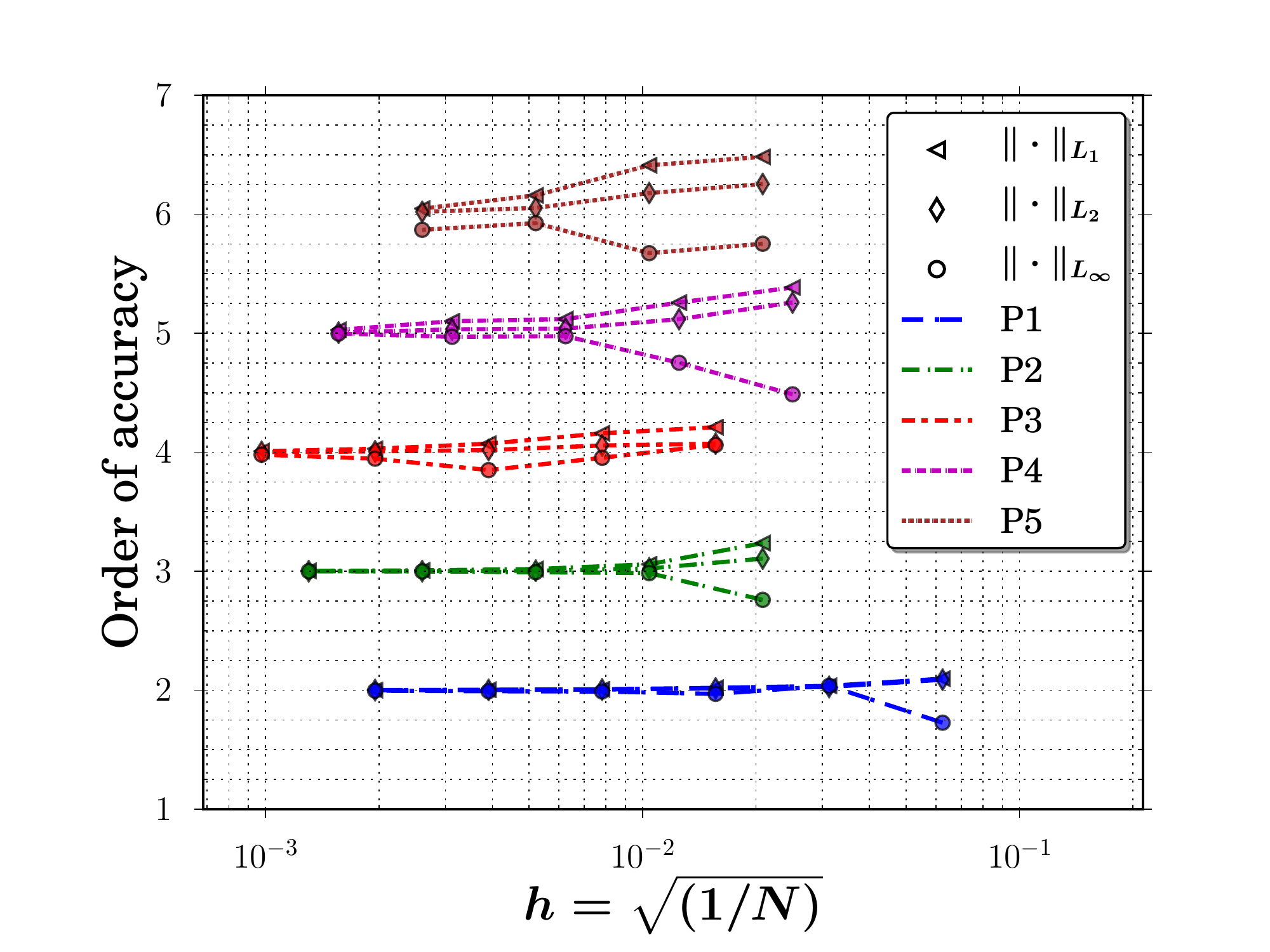}}
~~~
\subfloat[$\rho u$]{
\includegraphics[trim = 16mm 3mm 18mm 13mm, clip,width=0.30\linewidth]{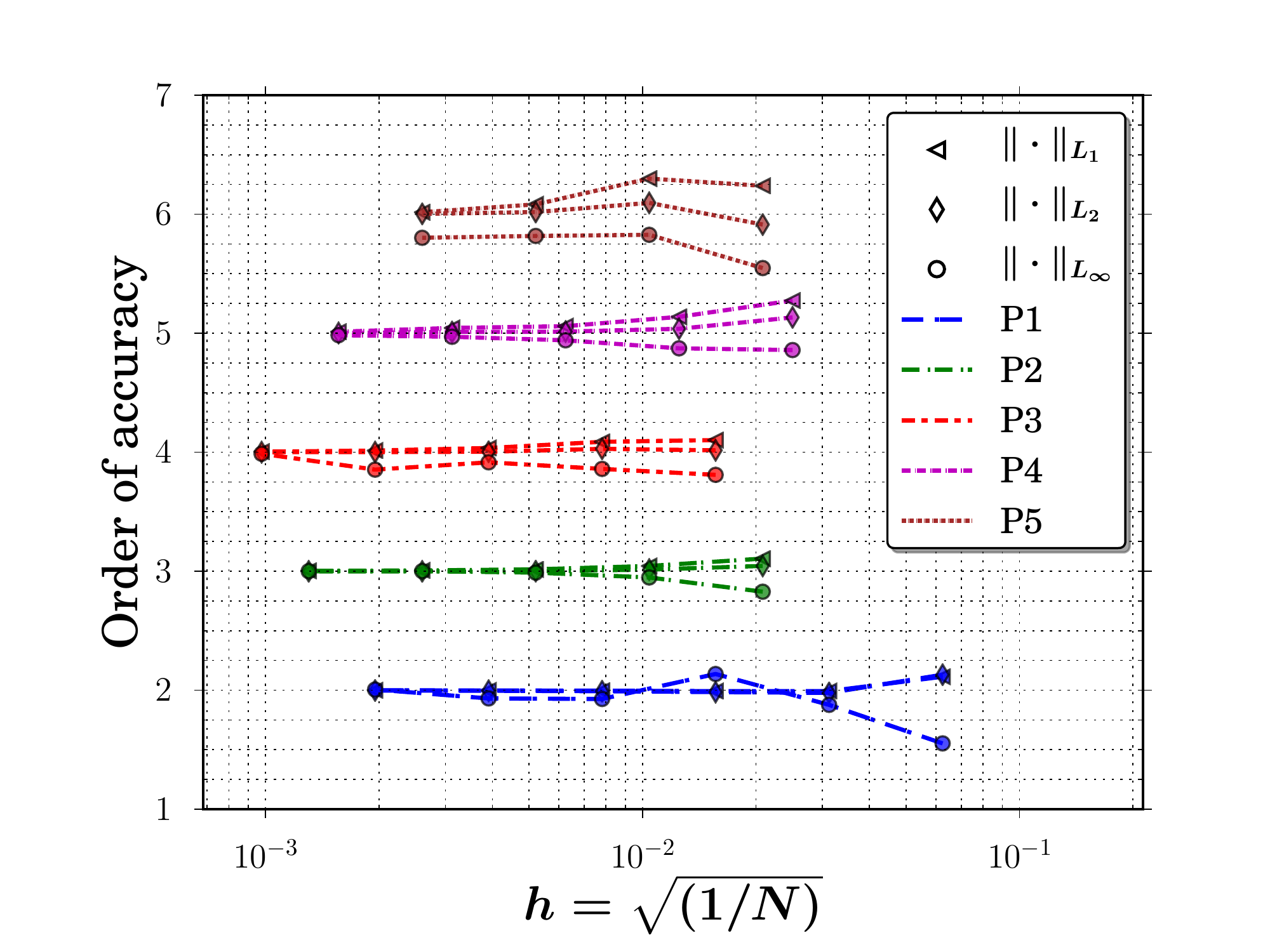}}
\vfill
\subfloat[$\rho v$]{
\includegraphics[trim = 16mm 3mm 18mm 13mm, clip,width=0.30\linewidth]
{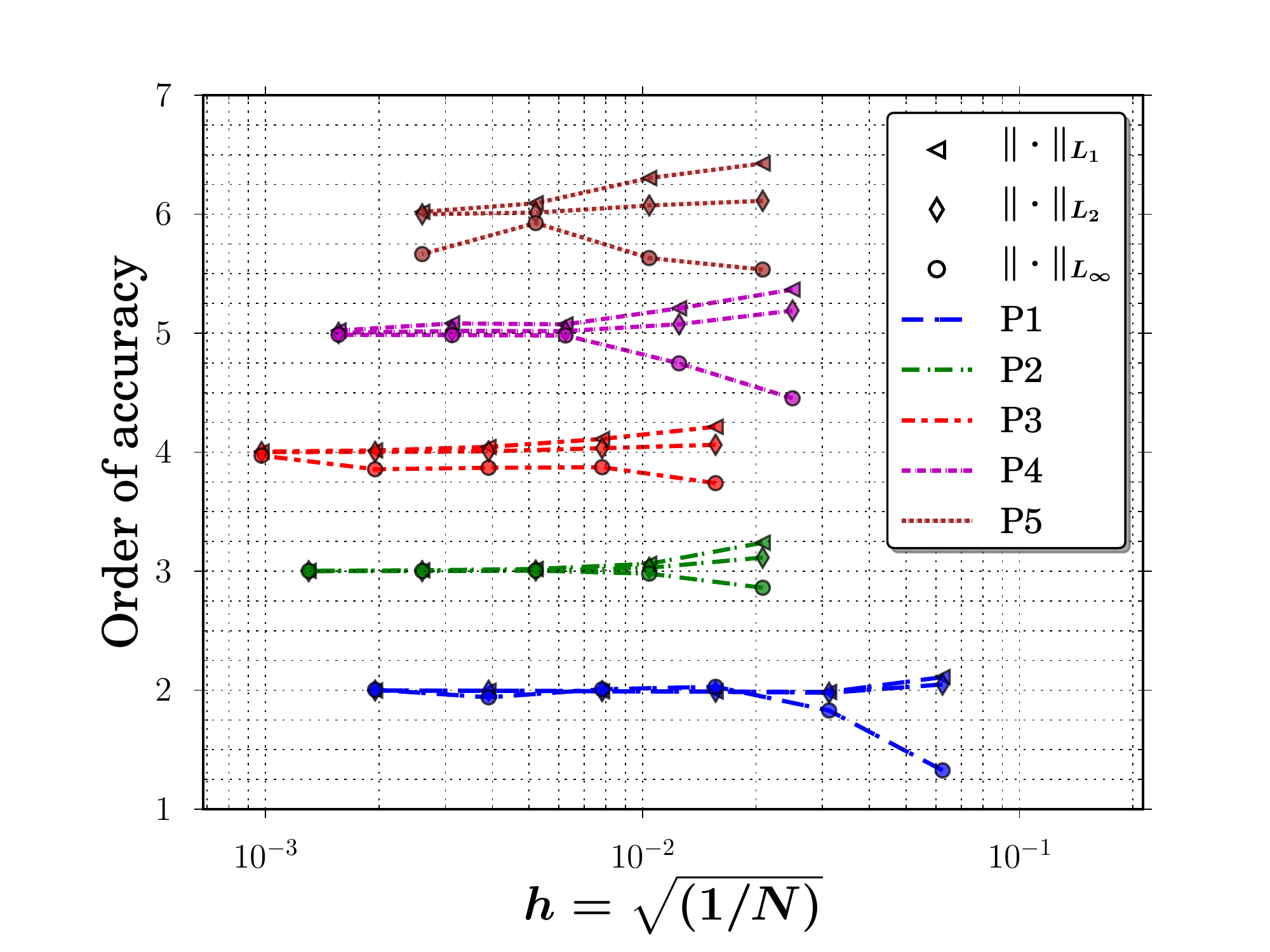}}
~~~
\subfloat[$\rho E$]{
\includegraphics[trim = 16mm 3mm 18mm 13mm, clip,width=0.30\linewidth]{{figs/MMS_022_trigo_Euler_Sup_Userdef_curved_eps/Order_L-norms_w3}.pdf}}
\caption{Evolution of the OOAs in $L_1$, $L_2$ and $L_\infty$ norms versus mesh refinement for MS-2 and $\mathrm{P}1$--$\mathrm{P}5$}
\label{fig:Orders_MS-2}
\end{figure}

\clearpage
\subsection{MS-3}

\begin{figure}[!hbt]
\centering
\subfloat[$\rho^{\mathrm{MS}}$]{
\includegraphics[trim = 0mm 0mm 0mm 0mm, clip,width=0.27\linewidth]
{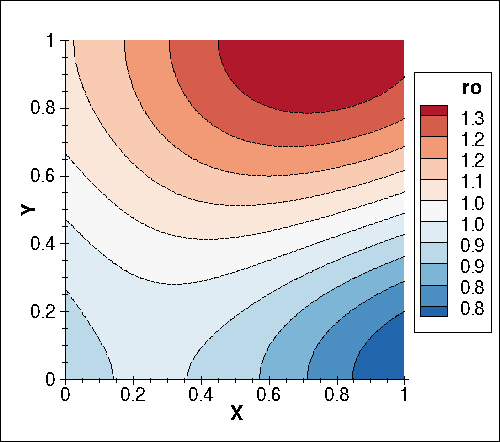}}
~~~
\subfloat[$u^{\mathrm{MS}}$]{
\includegraphics[trim = 0mm 0mm 0mm 0mm, clip,width=0.27\linewidth]
{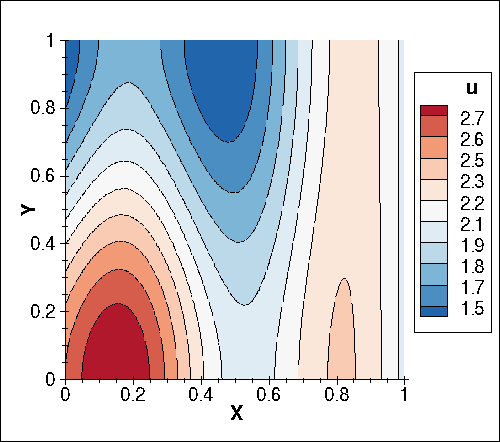}}
\vfill
\subfloat[$v^{\mathrm{MS}}$]{
\includegraphics[trim = 0mm 0mm 0mm 0mm, clip,width=0.27\linewidth]{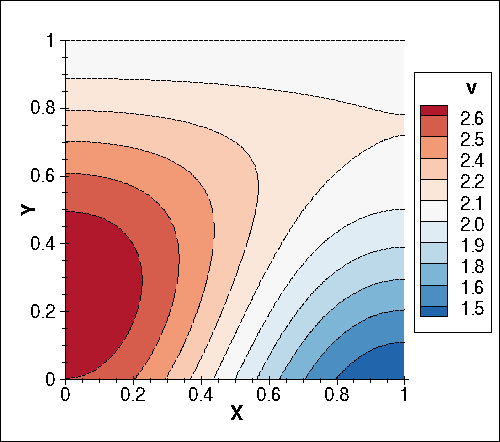}}
~~~
\subfloat[$p^{\mathrm{MS}}$]{
\includegraphics[trim = 0mm 0mm 0mm 0mm, clip,width=0.27\linewidth]{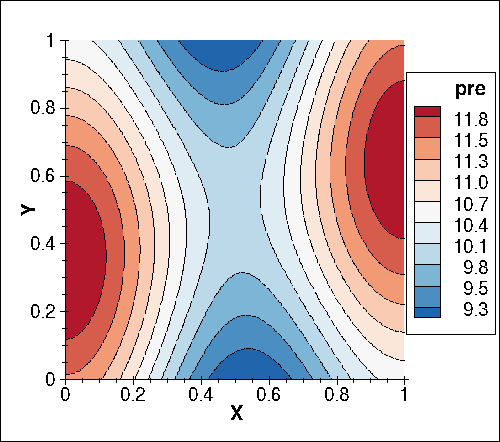}}
\vfill
\subfloat[$\mbox{Ma}^{\mathrm{MS}}$ and ${\bm{u}}^{\mathrm{MS}}$]{
\includegraphics[trim = 0mm 0mm 0mm 0mm, clip,width=0.27\linewidth]{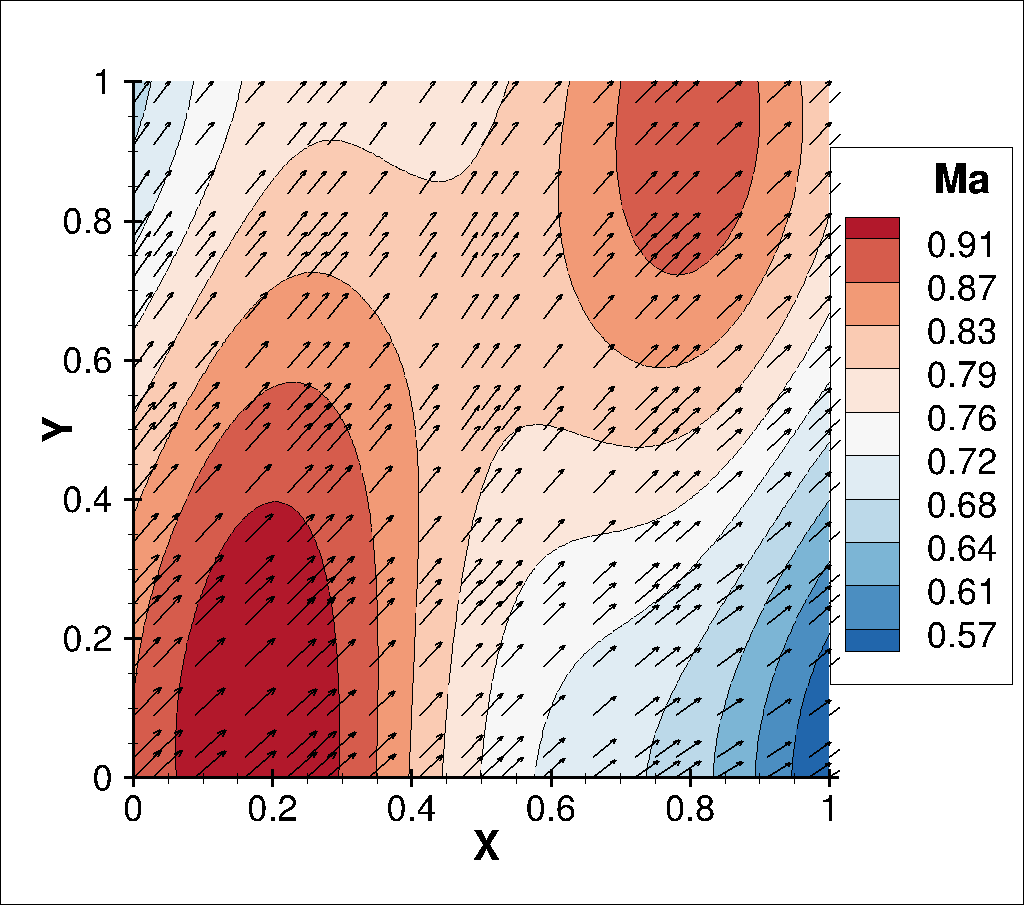}}
\caption{Manufactured solution MS-3}
\label{fig:MS-3}
\end{figure}

\begin{figure}[!hbt]
\centering
\subfloat[$\rho$]{
\includegraphics[trim = 5mm 2mm 18mm 13mm, clip,width=0.32\linewidth]
{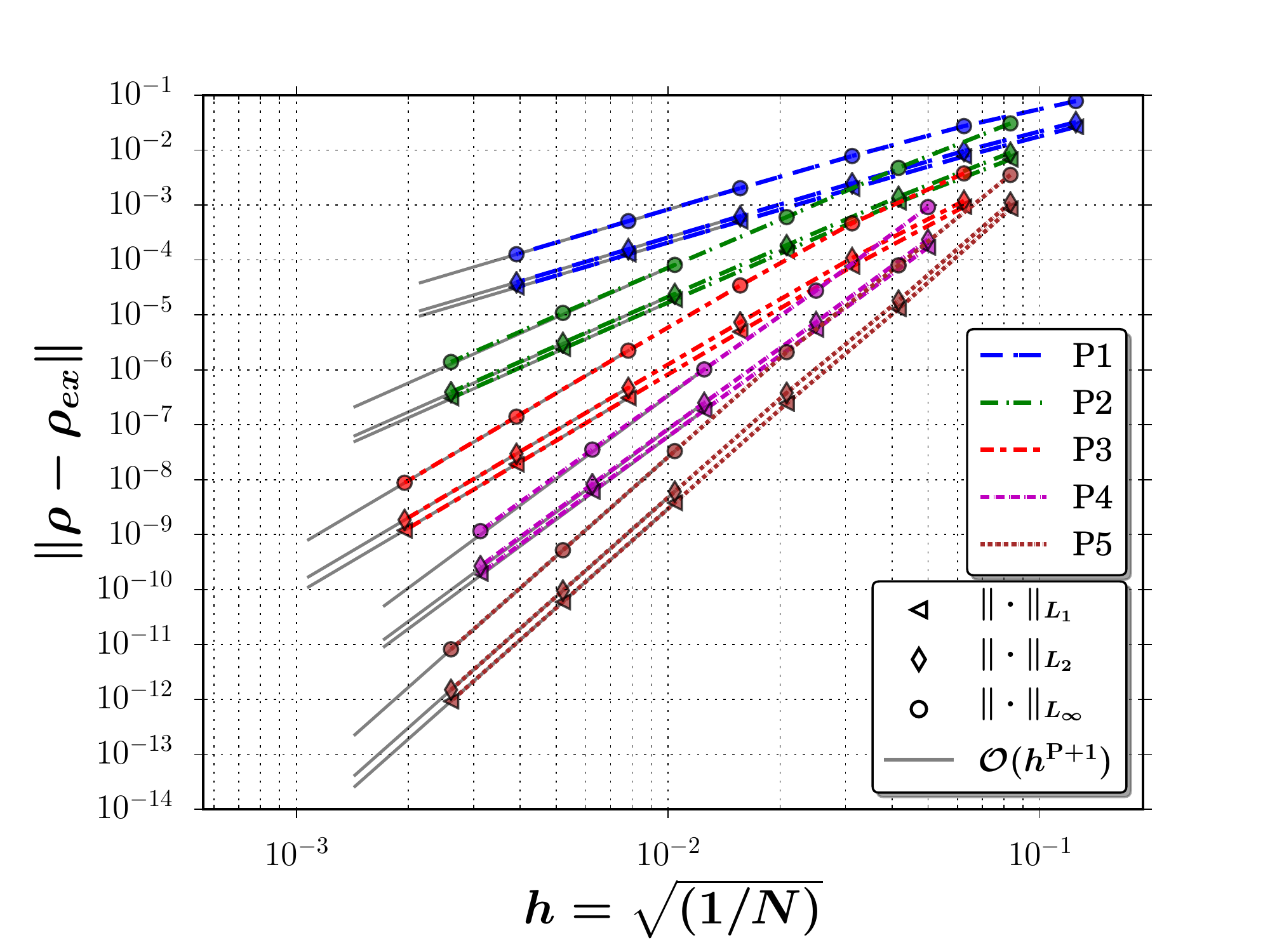}}
~~~
\subfloat[$\rho u$]{
\includegraphics[trim = 5mm 2mm 18mm 13mm, clip,width=0.32\linewidth]{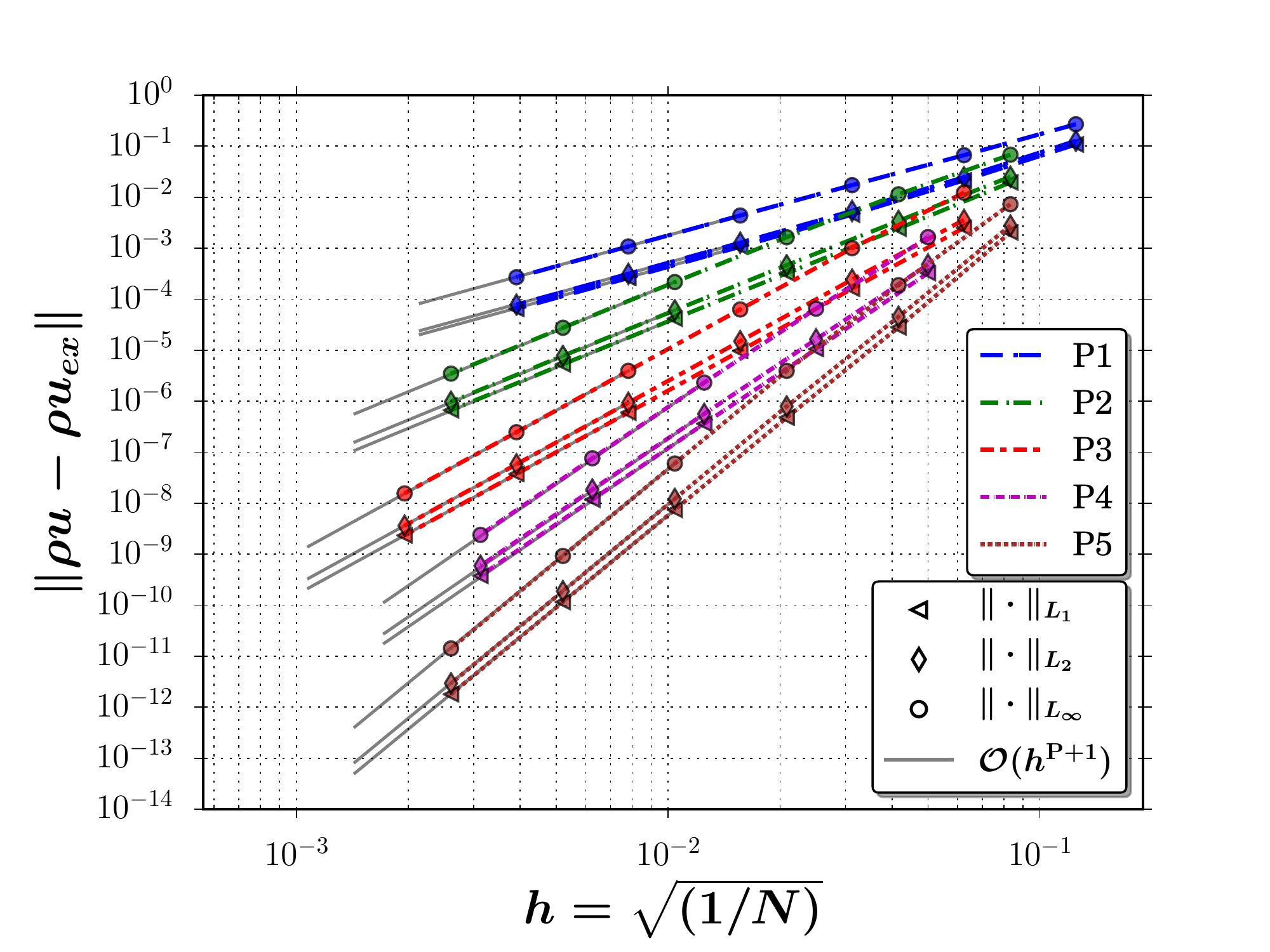}}
\vfill
\subfloat[$\rho v$]{
\includegraphics[trim = 5mm 2mm 18mm 13mm, clip,width=0.32\linewidth]
{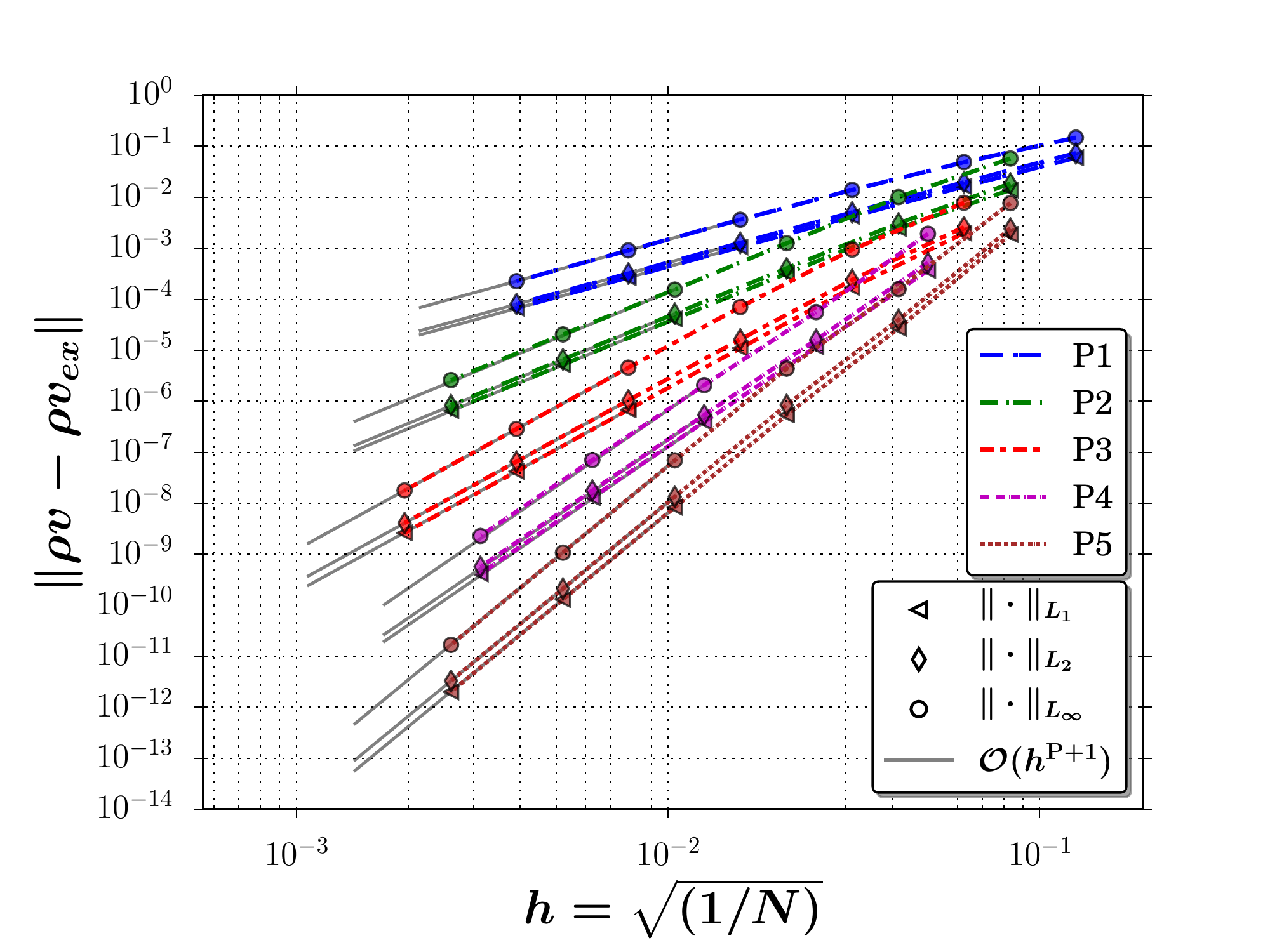}}
~~~
\subfloat[$\rho E$]{
\includegraphics[trim = 5mm 2mm 18mm 13mm, clip,width=0.32\linewidth]{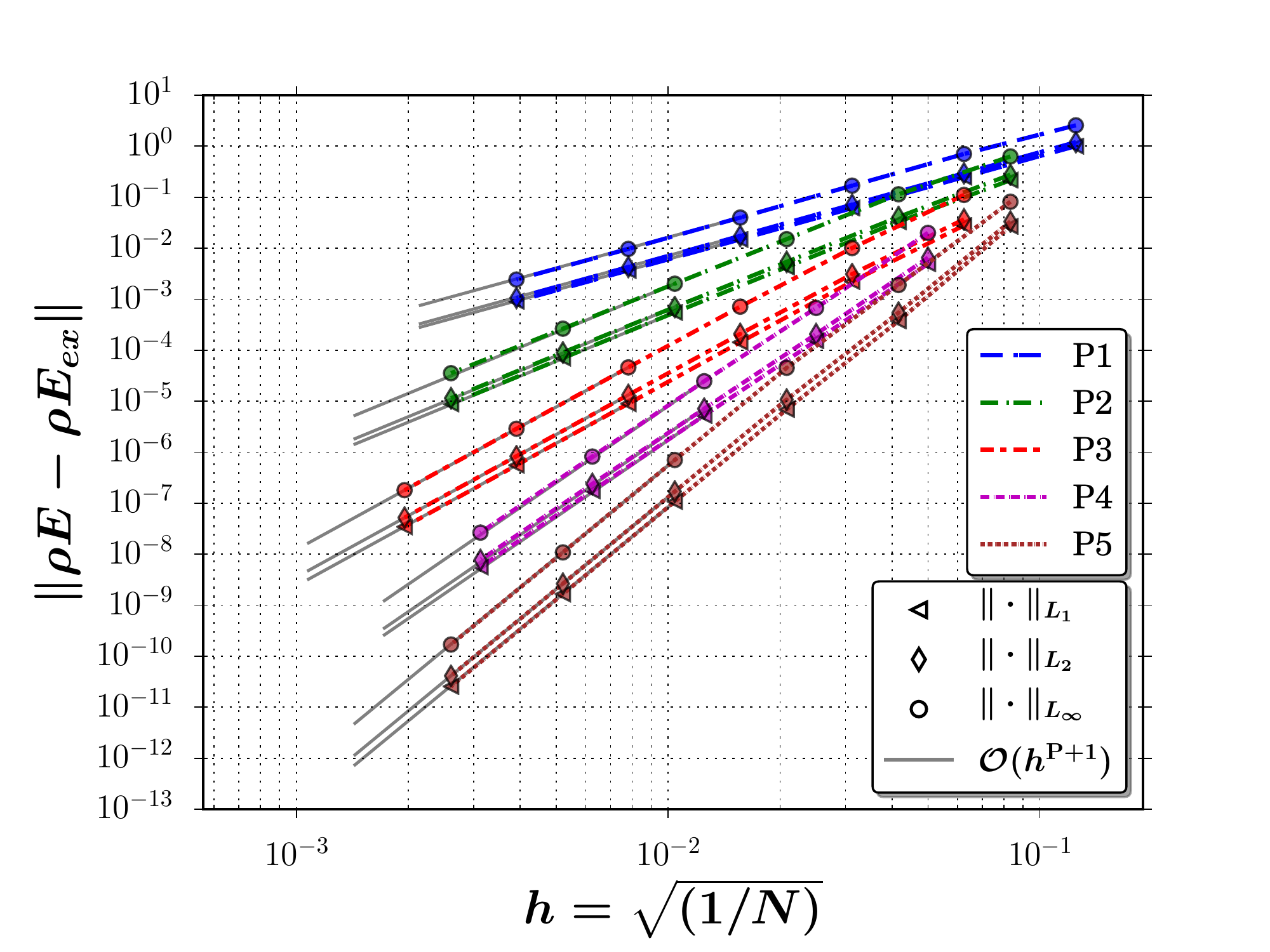}}
\caption{Evolution of the discretization error in $L_1$, $L_2$ and $L_\infty$ norms versus mesh refinement for MS-3, $\mu = 1\times10^{-1}$ and $\mathrm{P}1$--$\mathrm{P}5$}
\label{fig:Err_allE_allP_MS-3}
\end{figure}
\begin{figure}[!hbt]
\centering
\subfloat[$\rho$]{ 
\includegraphics[trim = 16mm 3mm 18mm 13mm, clip,width=0.30\linewidth]
{{figs/MMS_112_trigo_NS_Sub_mu1e-1_eps/Order_L-norms_w0}.pdf}}
~~~
\subfloat[$\rho u$]{
\includegraphics[trim = 16mm 3mm 18mm 13mm, clip,width=0.30\linewidth]{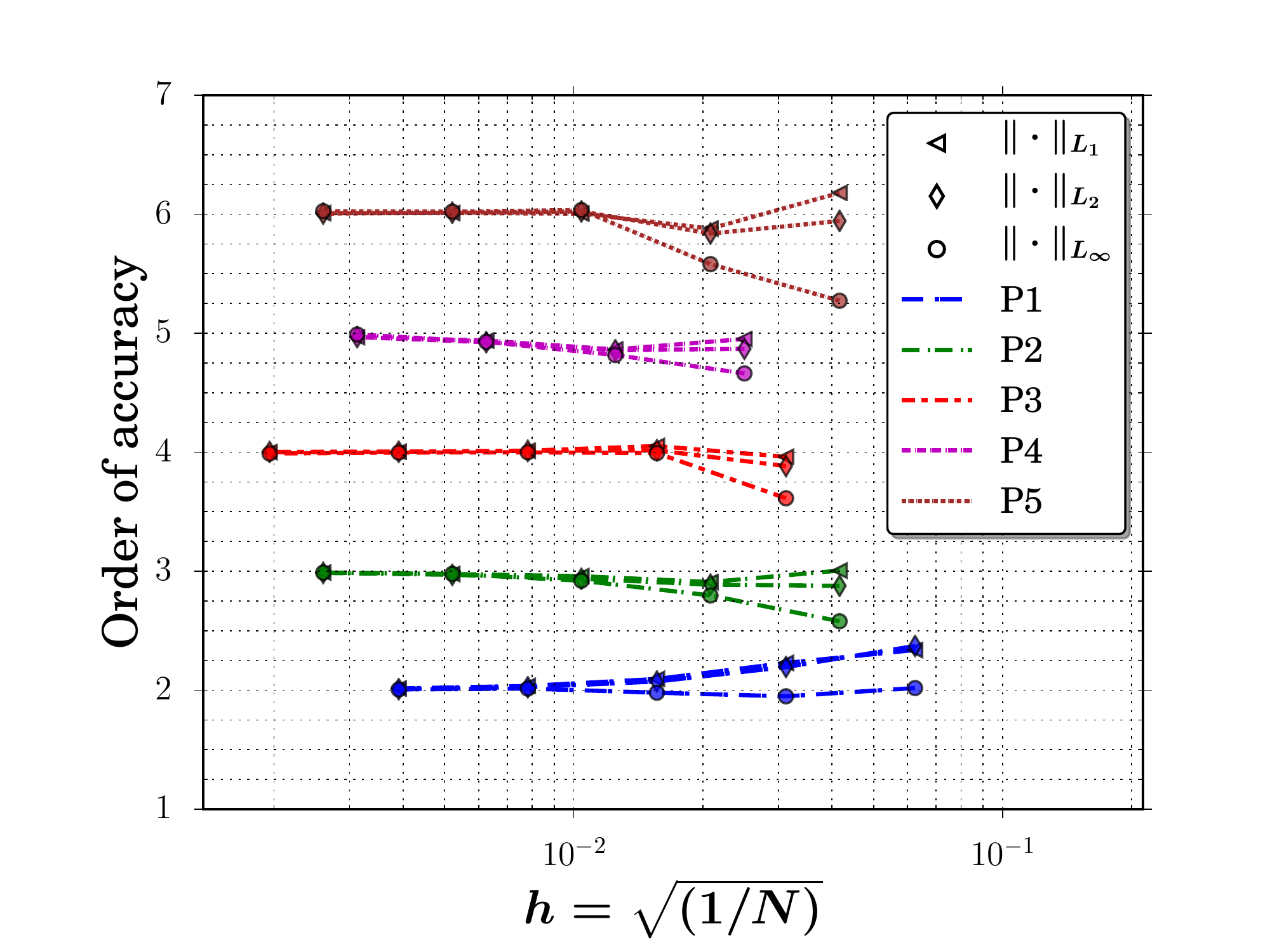}}
\vfill
\subfloat[$\rho v$]{
\includegraphics[trim = 16mm 3mm 18mm 13mm, clip,width=0.30\linewidth]
{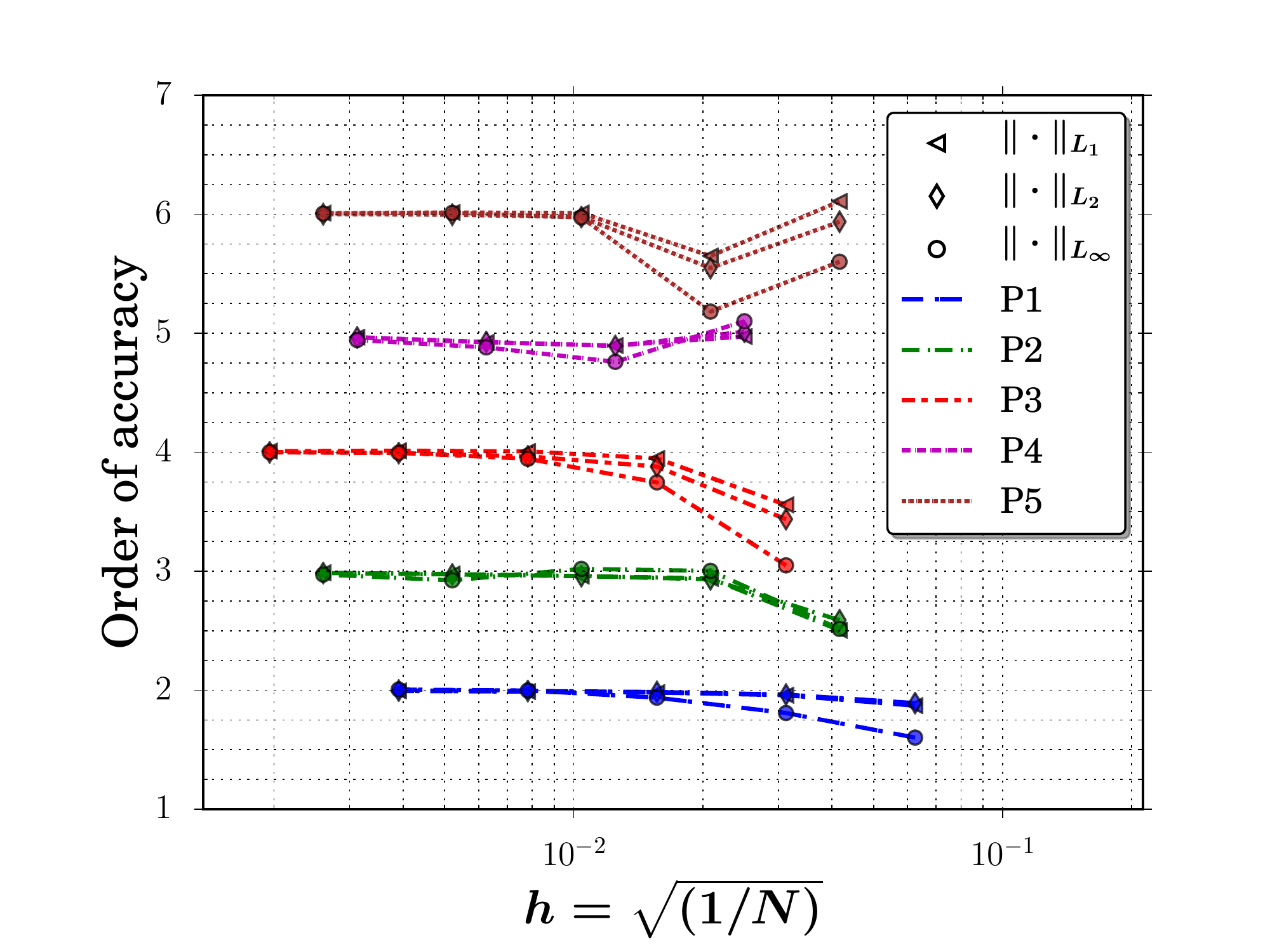}}
~~~
\subfloat[$\rho E$]{
\includegraphics[trim = 16mm 3mm 18mm 13mm, clip,width=0.30\linewidth]{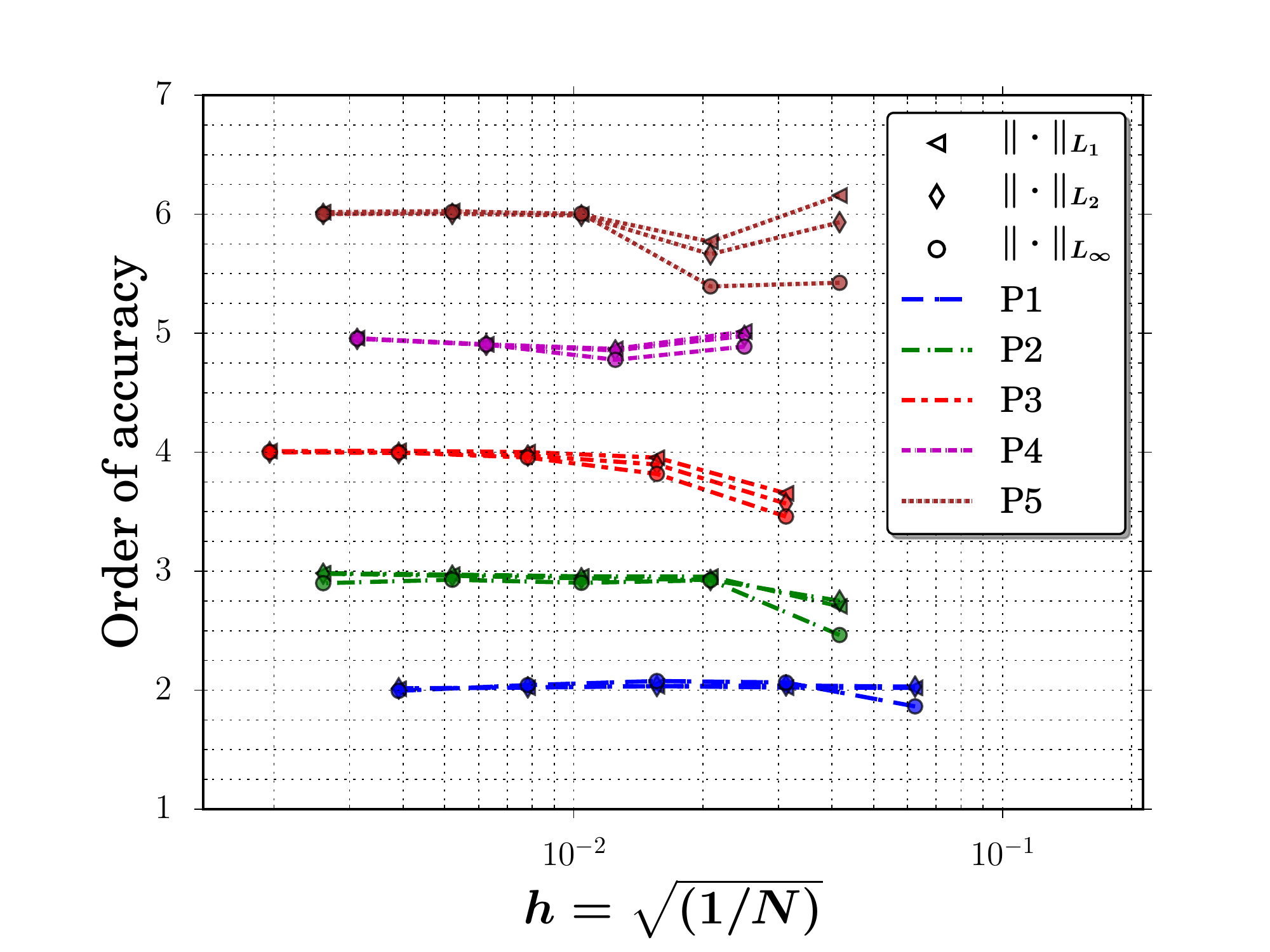}}
\caption{Evolution of the OOAs in $L_1$, $L_2$ and $L_\infty$ norms versus mesh refinement for MS-3, $\mu = 1\times10^{-1}$, and $\mathrm{P}1$--$\mathrm{P}5$}
\label{fig:Orders_MS-3}
\end{figure}

\begin{figure}[!hbt]
\centering
\subfloat[$\rho$]{
\includegraphics[trim = 5mm 2mm 18mm 13mm, clip,width=0.32\linewidth]
{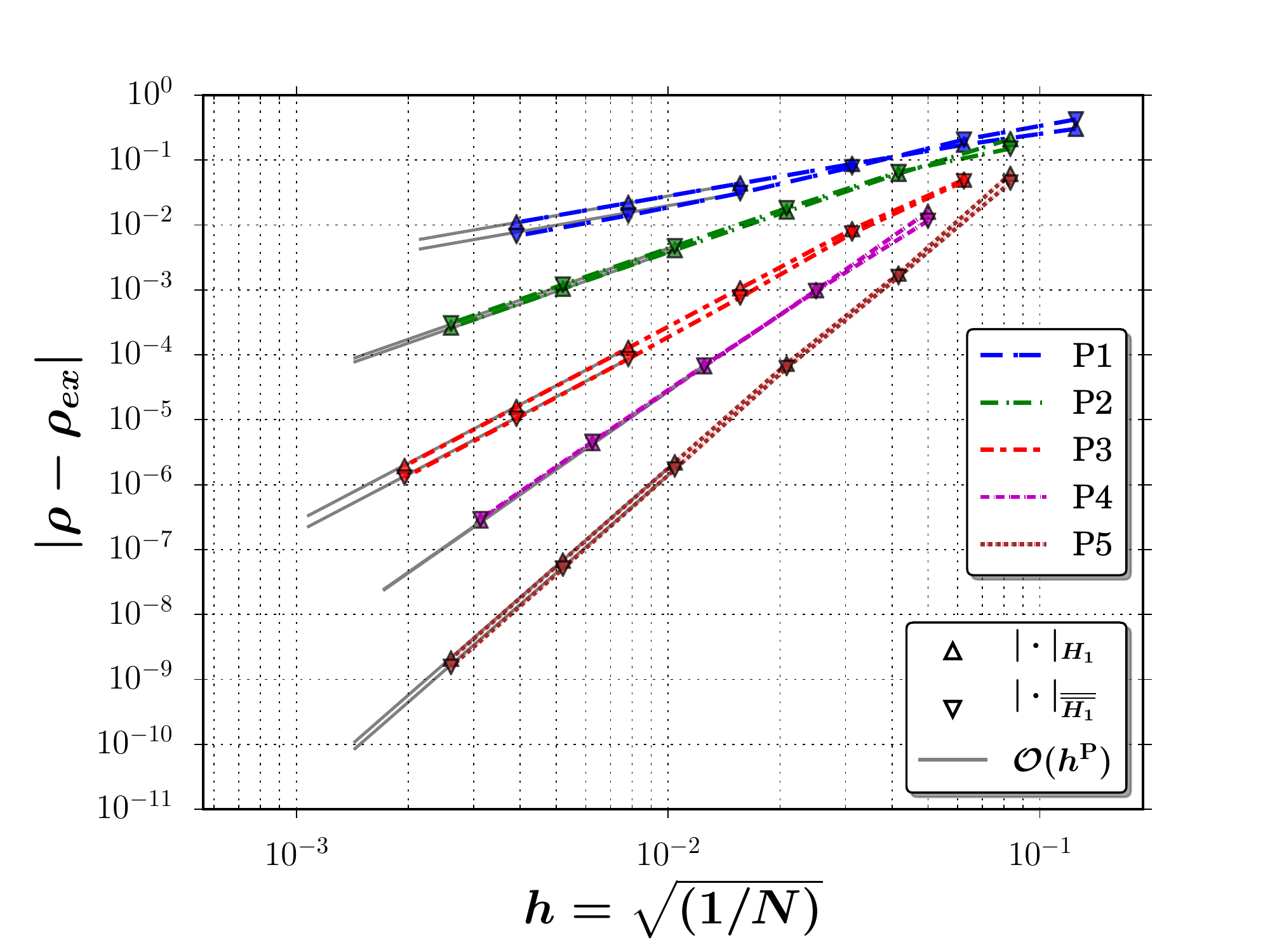}}~~~
\subfloat[$\rho u$]{
\includegraphics[trim = 5mm 2mm 18mm 13mm, clip,width=0.32\linewidth]{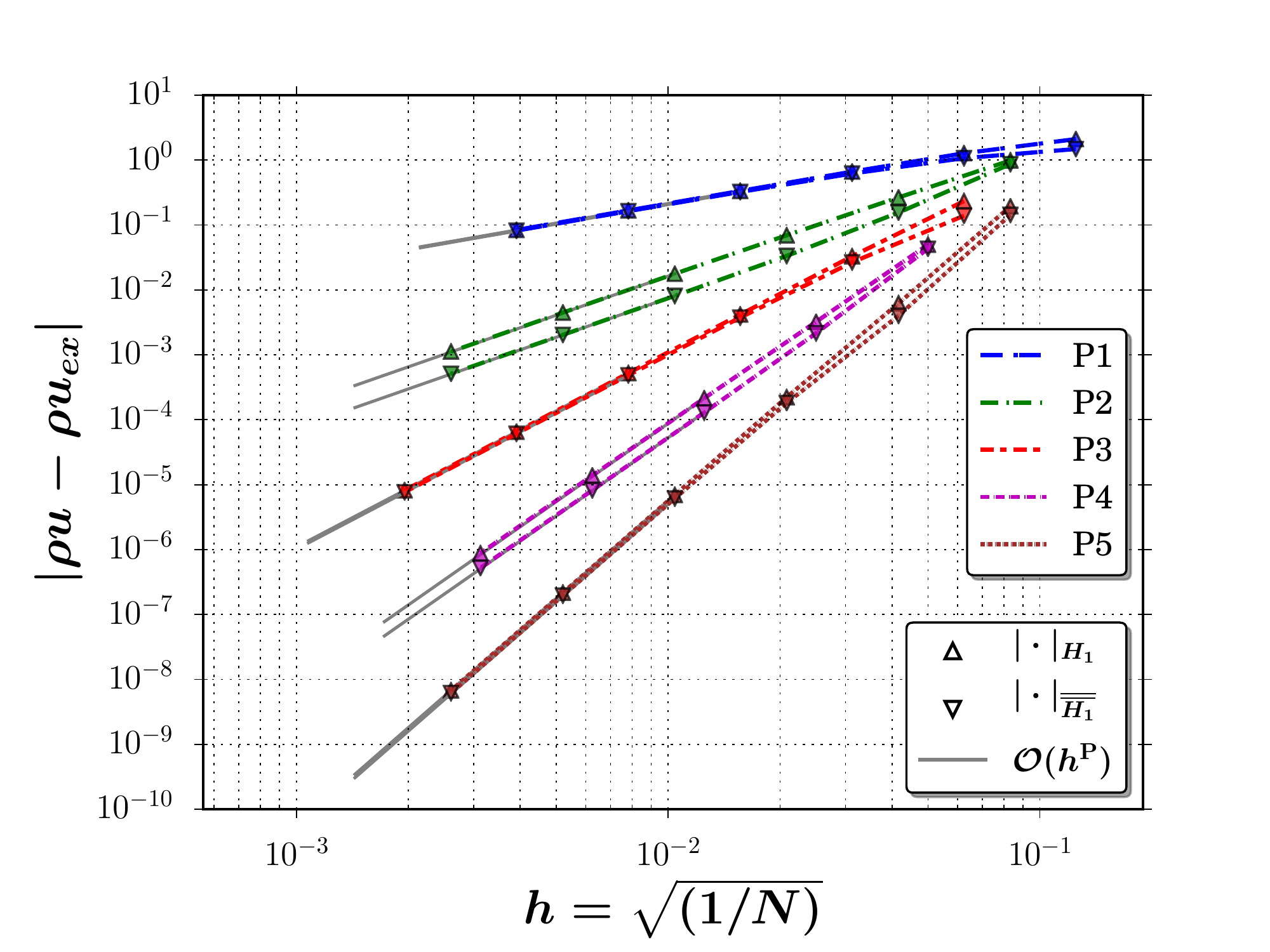}}
\vfill
\subfloat[$\rho v$]{
\includegraphics[trim = 5mm 2mm 18mm 13mm, clip,width=0.32\linewidth]
{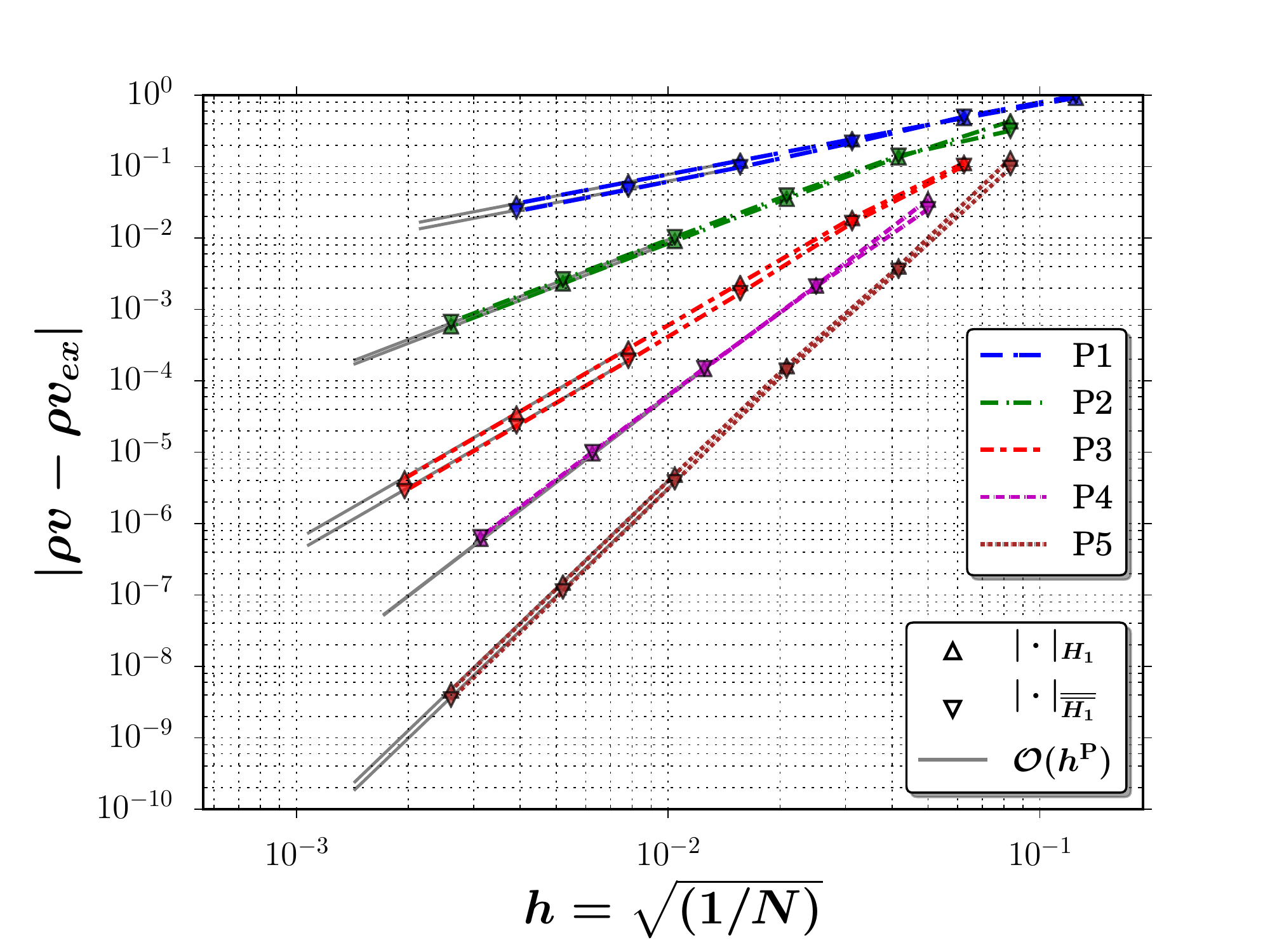}}~~~
\subfloat[$\rho E$]{
\includegraphics[trim = 5mm 2mm 18mm 13mm, clip,width=0.32\linewidth]{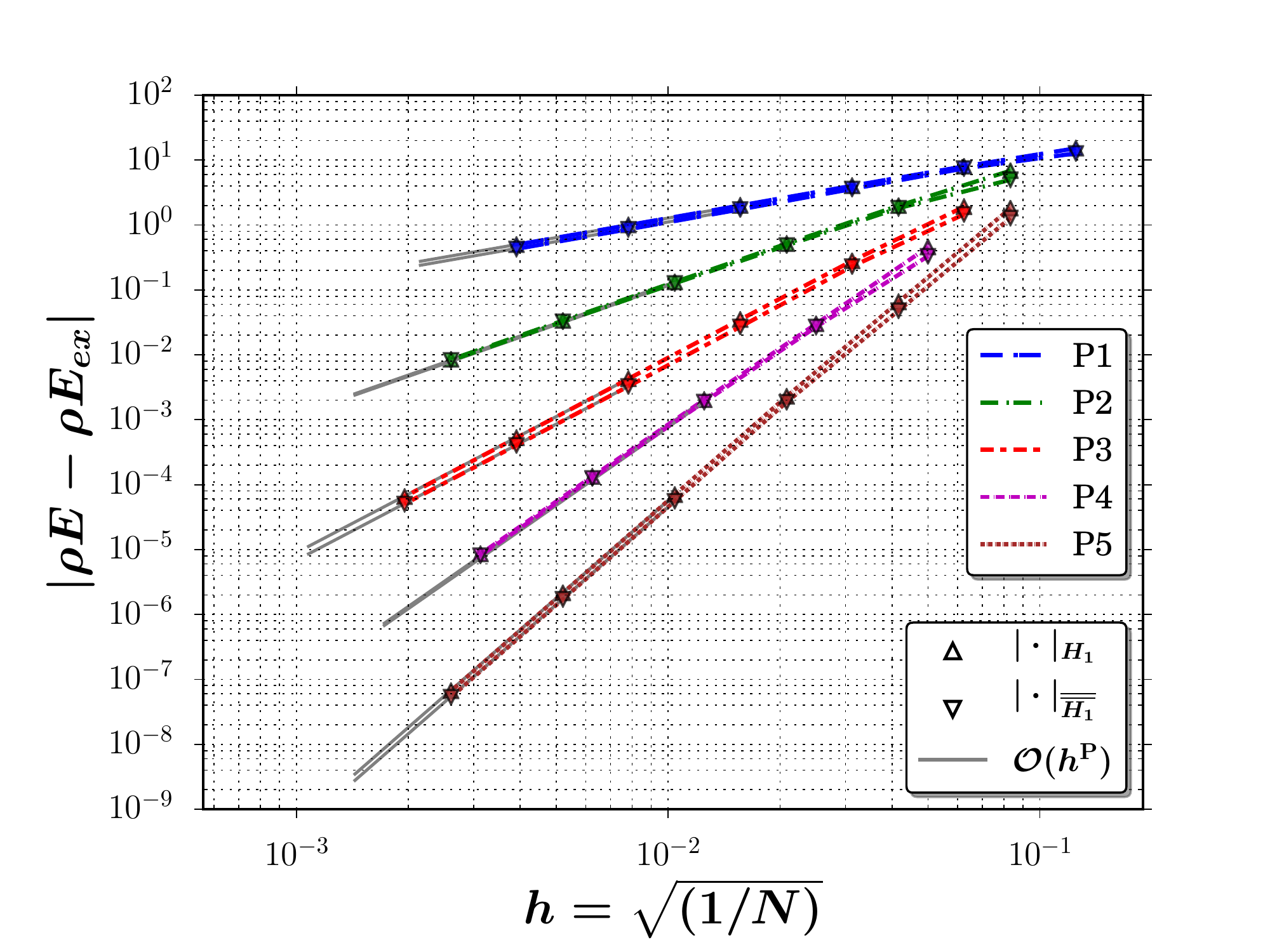}}
\caption{Evolution of the discretization error in $H_1$ semi-norm (for uncorrected and  fully corrected derivatives) versus mesh refinement for MS-3, $\mu = 1\times10^{-1}$ and $\mathrm{P}1$--$\mathrm{P}5$}
\label{fig:Err_allE_allP_H_MS-3}
\end{figure}

\begin{figure}[!hbt]
\centering
\subfloat[$\rho$]{ 
\includegraphics[trim = 16mm 3mm 18mm 13mm, clip,width=0.3\linewidth]
{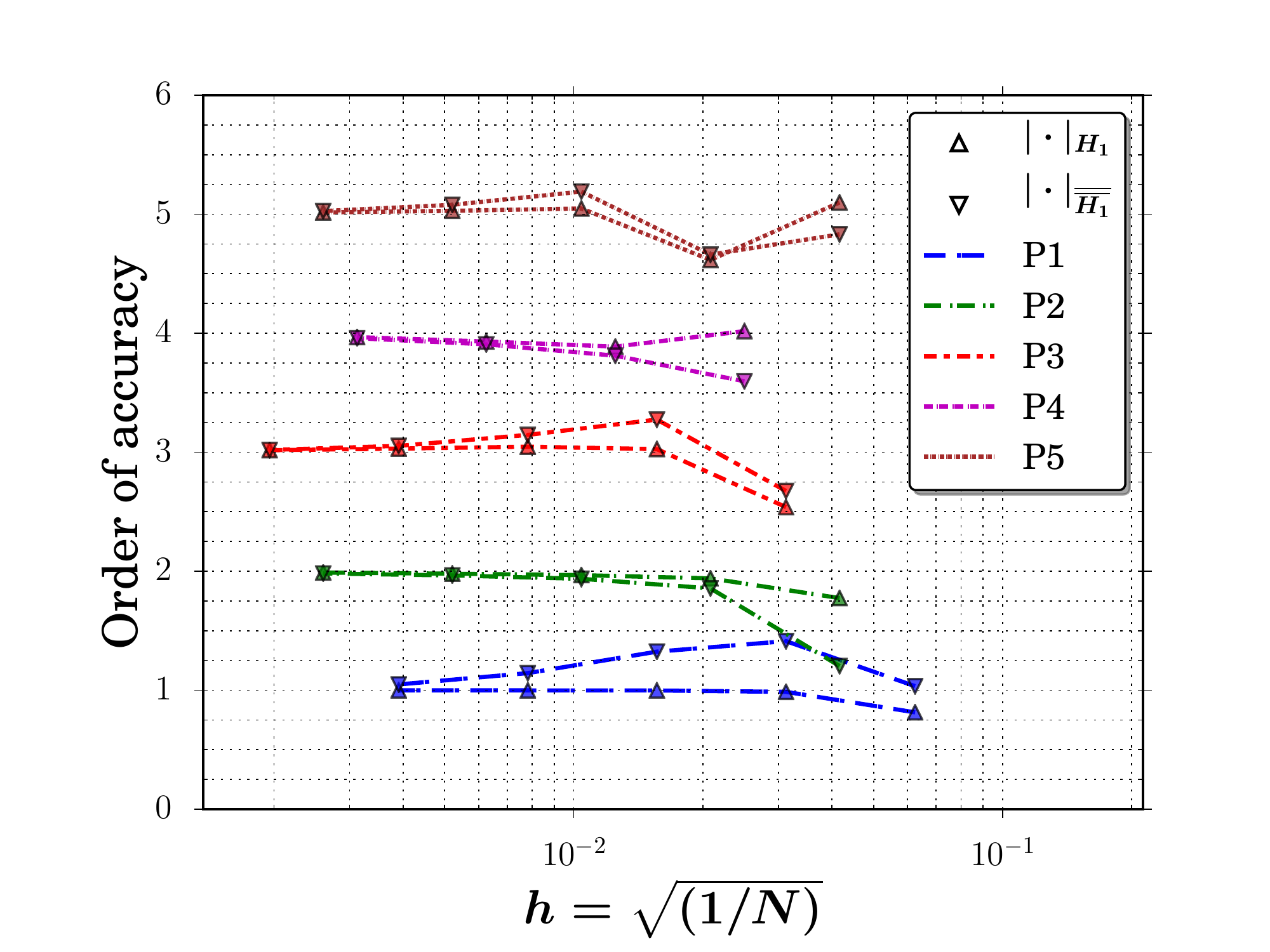}}~~~
\subfloat[$\rho u$]{
\includegraphics[trim = 16mm 3mm 18mm 13mm, clip,width=0.3\linewidth]{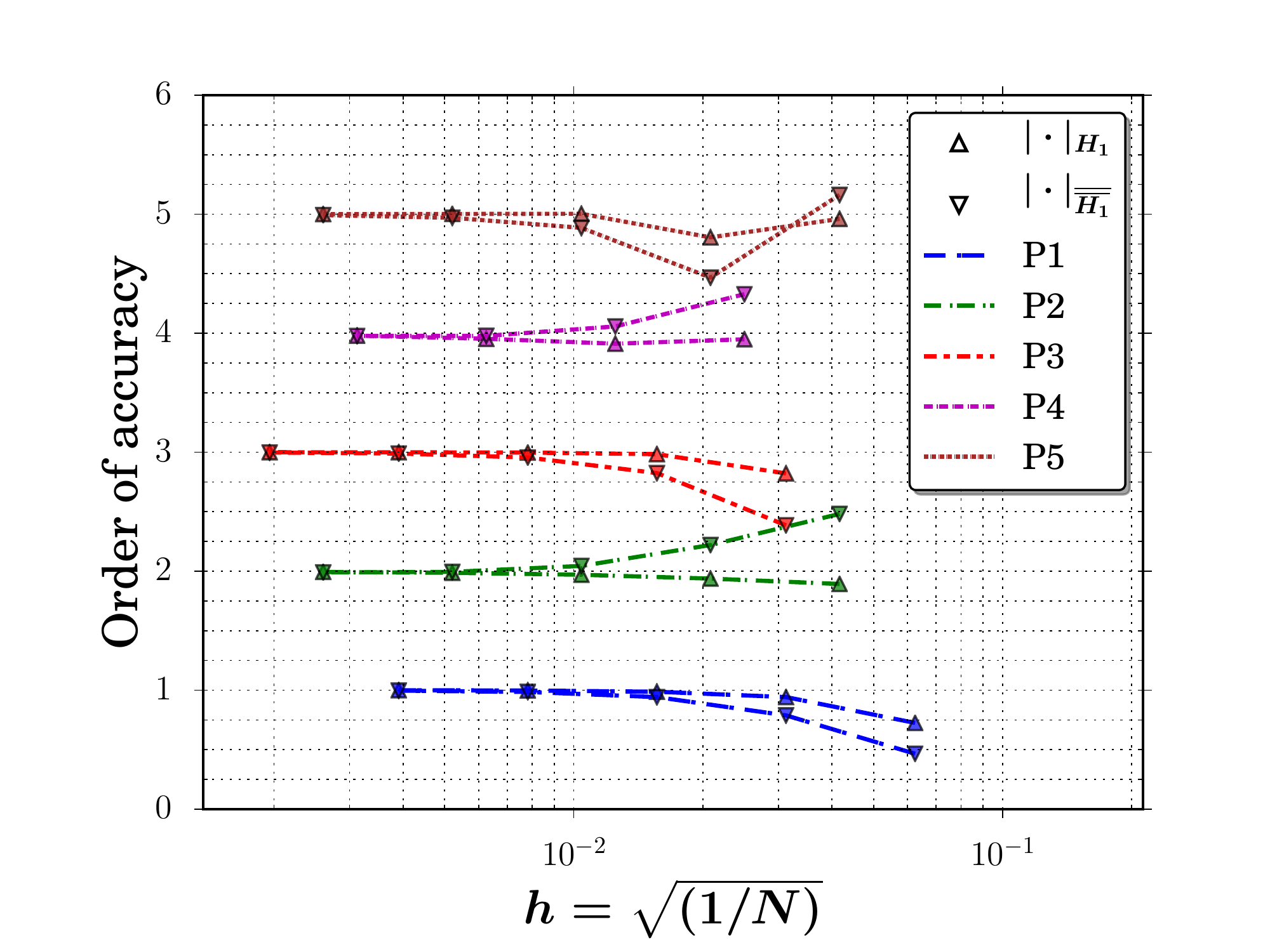}}
\vfill
\subfloat[$\rho v$]{
\includegraphics[trim = 16mm 3mm 18mm 13mm, clip,width=0.3\linewidth]
{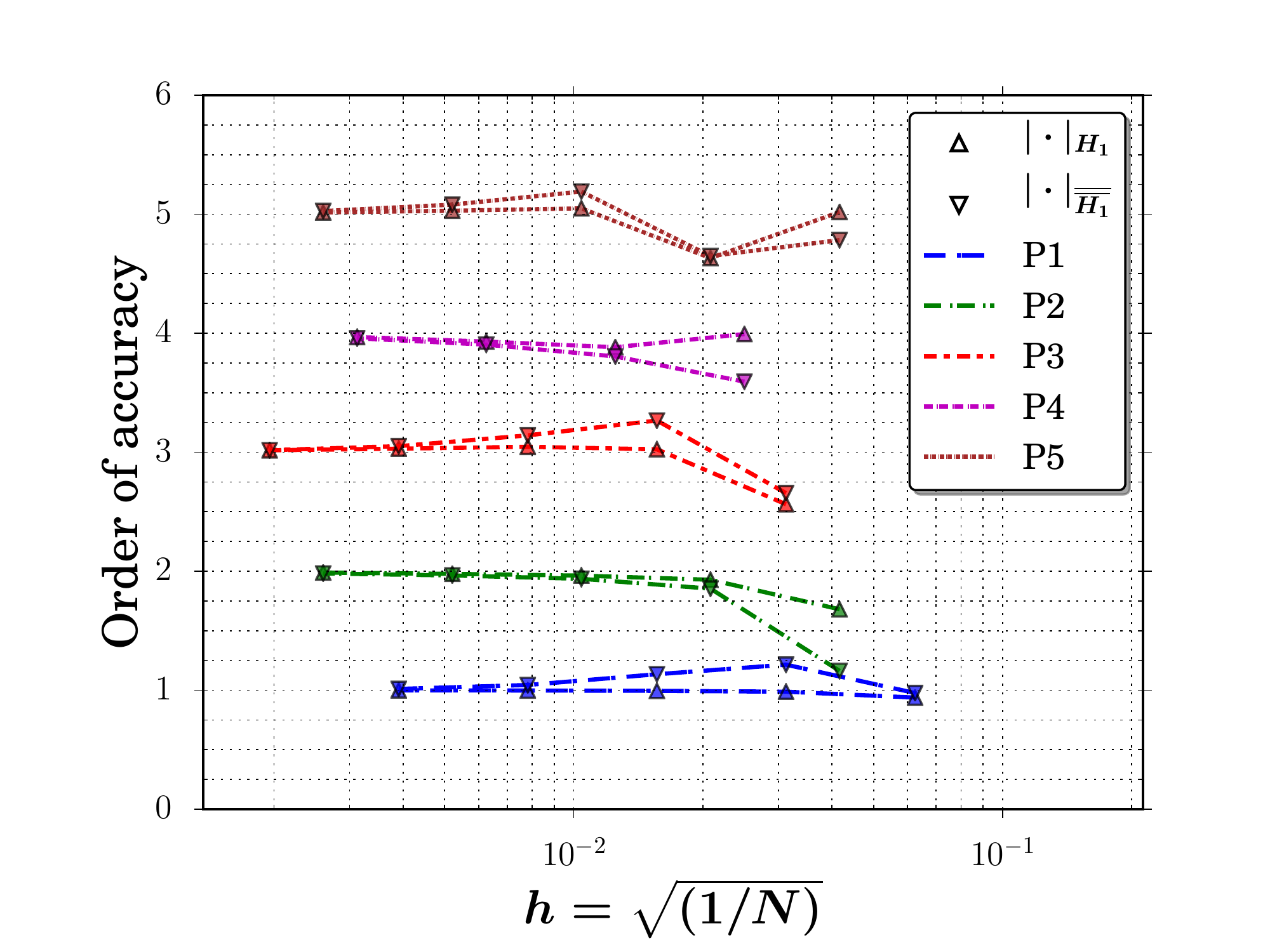}}~~~
\subfloat[$\rho E$]{
\includegraphics[trim = 16mm 3mm 18mm 13mm, clip,width=0.3\linewidth]{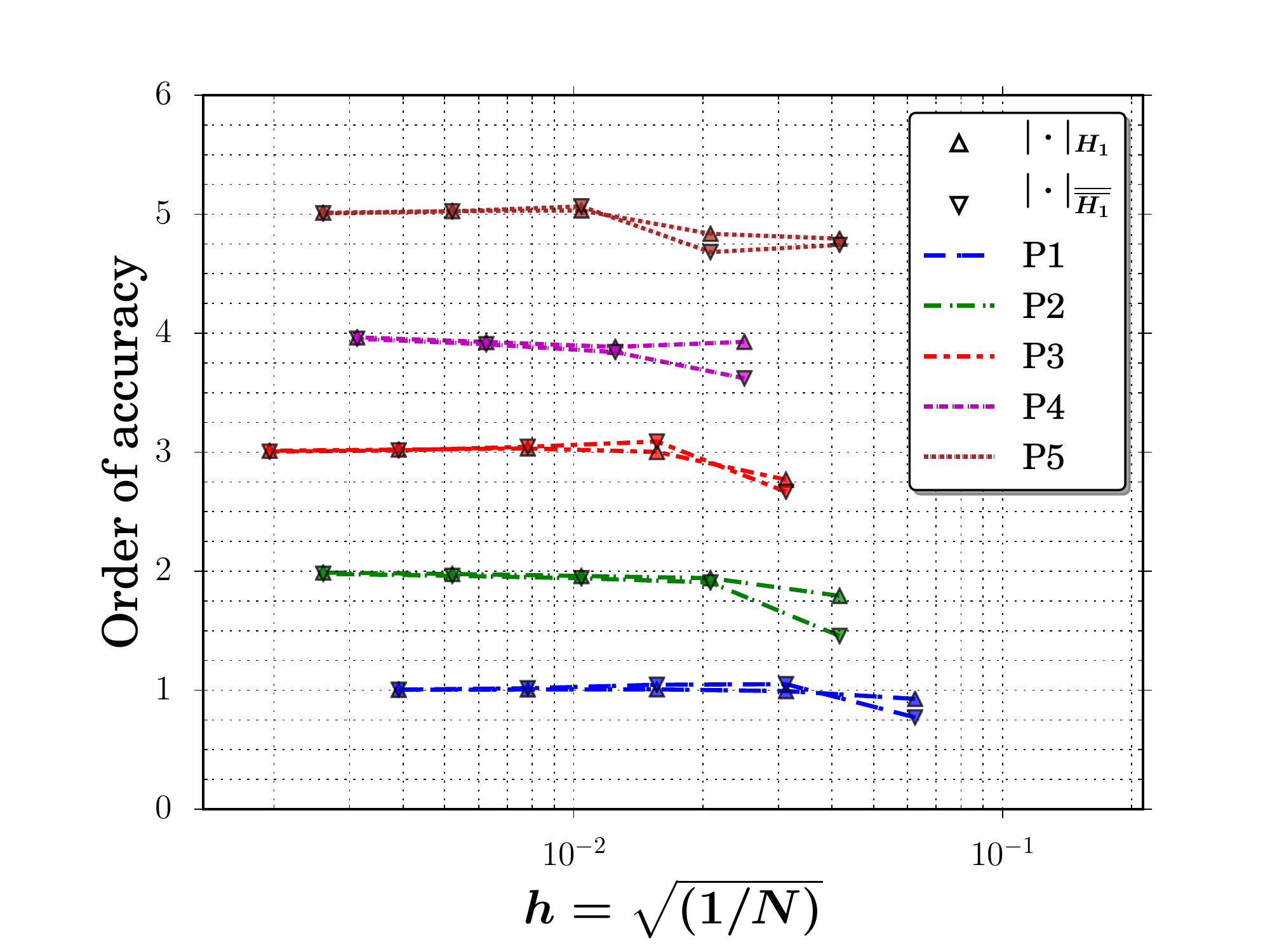}}
\caption{Evolution of the OOAs in $H_1$ semi-norm (for uncorrected and  fully corrected derivatives) versus mesh refinement for MS-3, $\mu = 1\times10^{-1}$,  and $\mathrm{P}1$--$\mathrm{P}5$}
\label{fig:Orders_H_MS-3}
\end{figure}

\clearpage
\subsection{MS-4}

\begin{figure}[!hbt]
\centering
{\includegraphics[trim = 0mm 0mm 0mm 0mm clip,width=0.27\linewidth]
{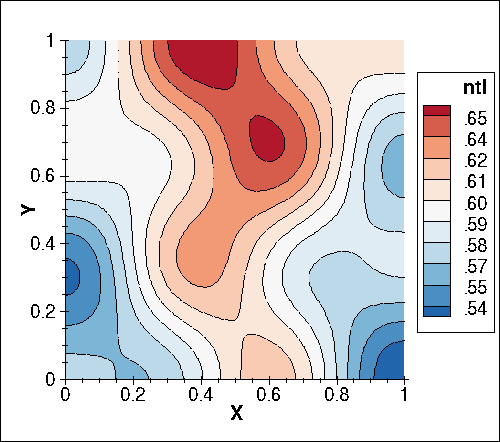}}
\caption{Manufactured solution MS-4 for $\tilde{\nu}$}
\label{fig:MS-4}
\end{figure}

\begin{figure}[!hbt]
\centering
\subfloat[$\rho$]{
\includegraphics[trim = 5mm 2mm 18mm 13mm, clip,width=0.32\linewidth]
{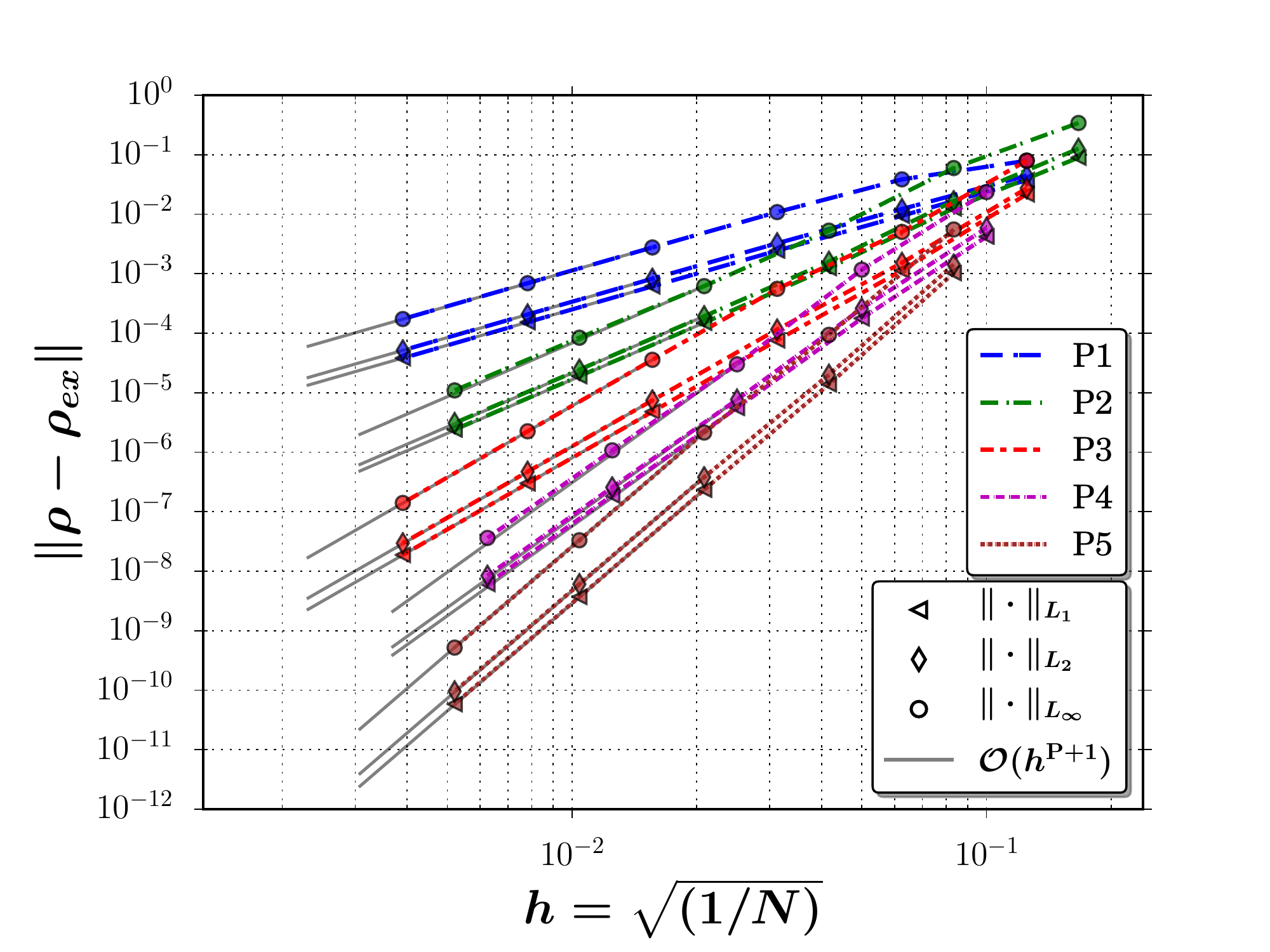}}
~~~
\subfloat[$\rho u$]{
\includegraphics[trim = 5mm 2mm 18mm 13mm, clip,width=0.32\linewidth]{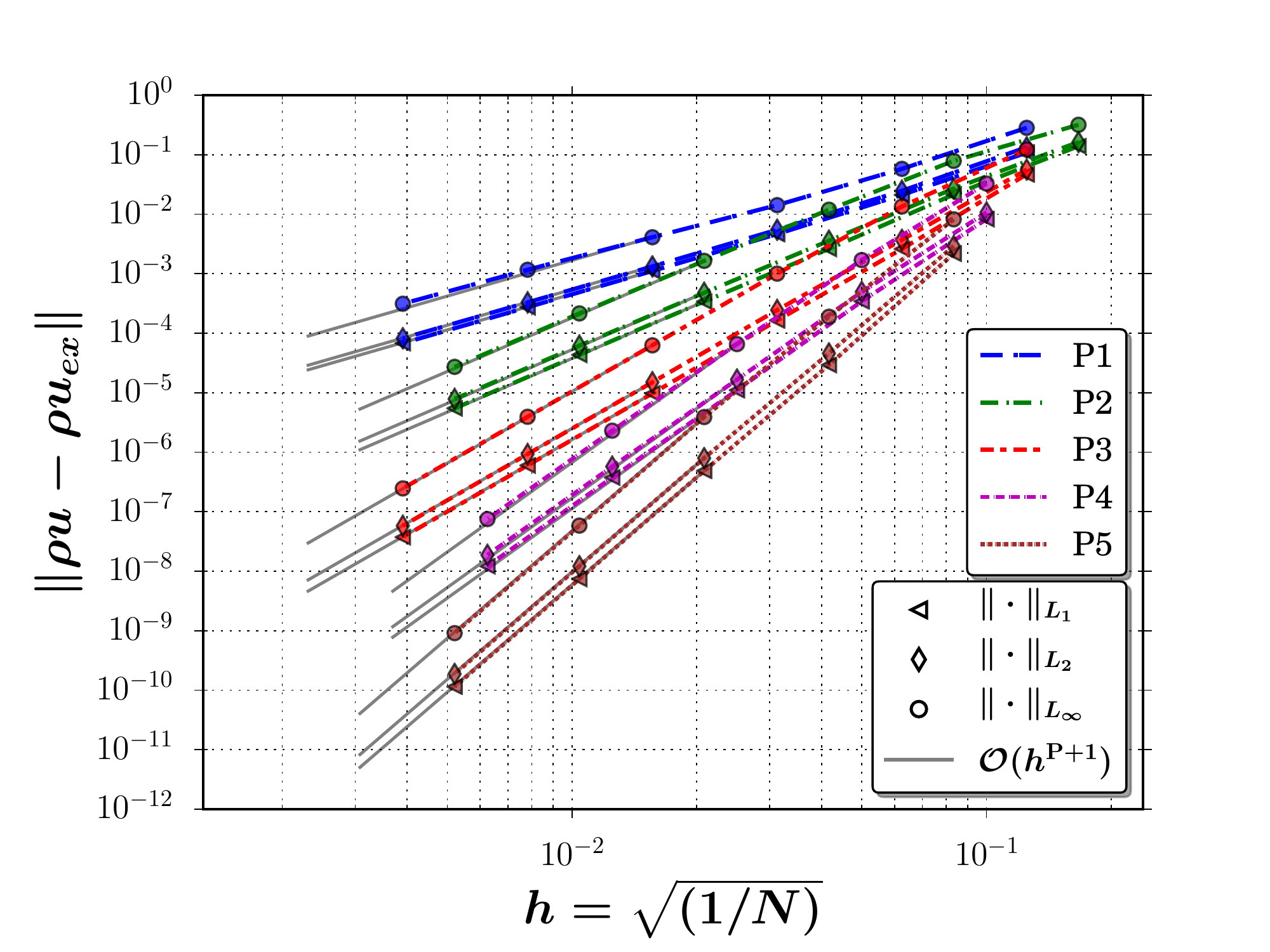}}
\vfill
\subfloat[$\rho v$]{
\includegraphics[trim =5mm 2mm 18mm 13mm, clip,width=0.32\linewidth]
{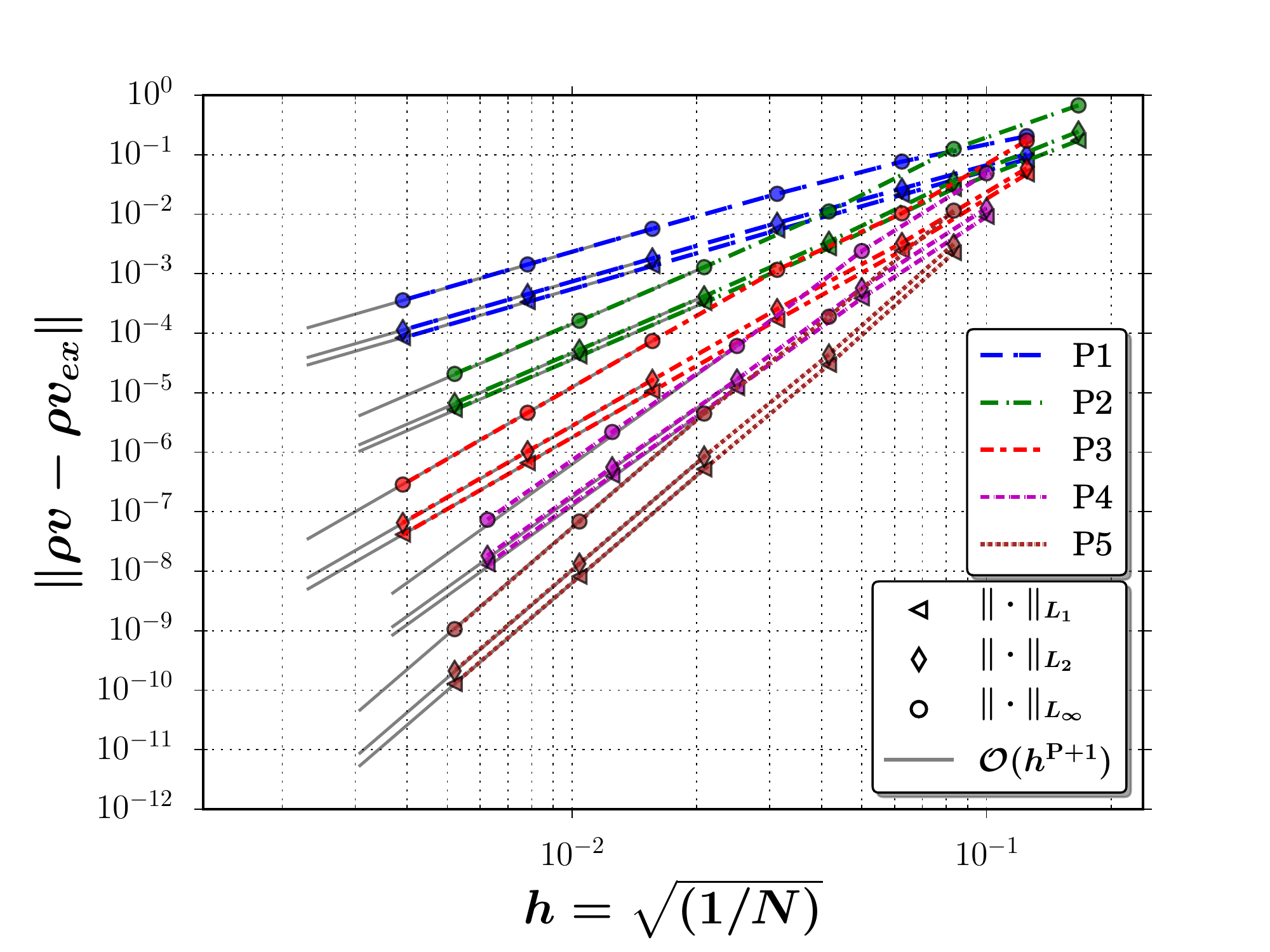}}
~~~
\subfloat[$\rho E$]{
\includegraphics[trim = 5mm 2mm 18mm 13mm, clip,width=0.32\linewidth]{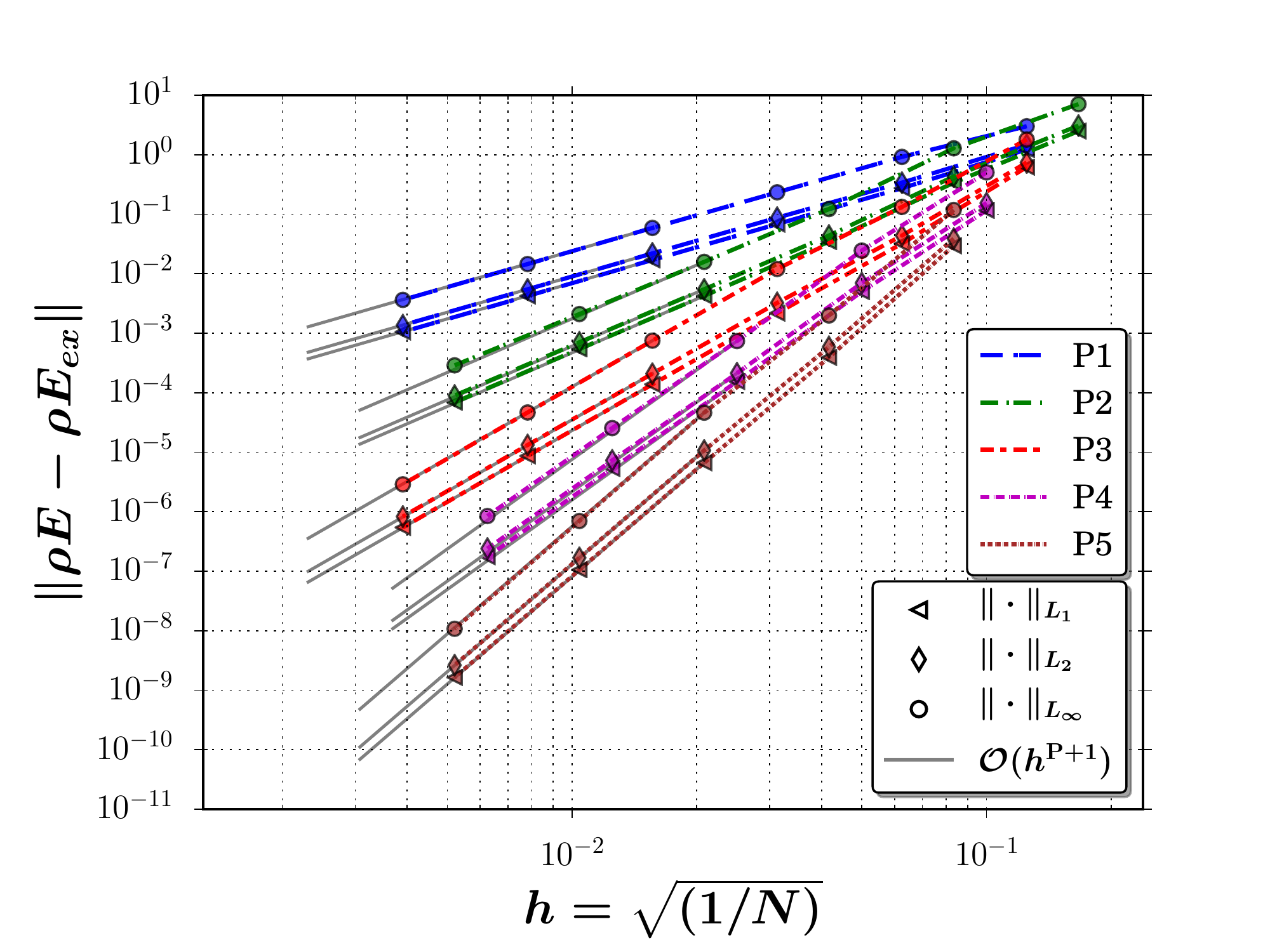}}
\vfill
\subfloat[$\rho \tilde{\nu}$]{
\includegraphics[trim = 5mm 2mm 18mm 13mm, clip,width=0.32\linewidth]
{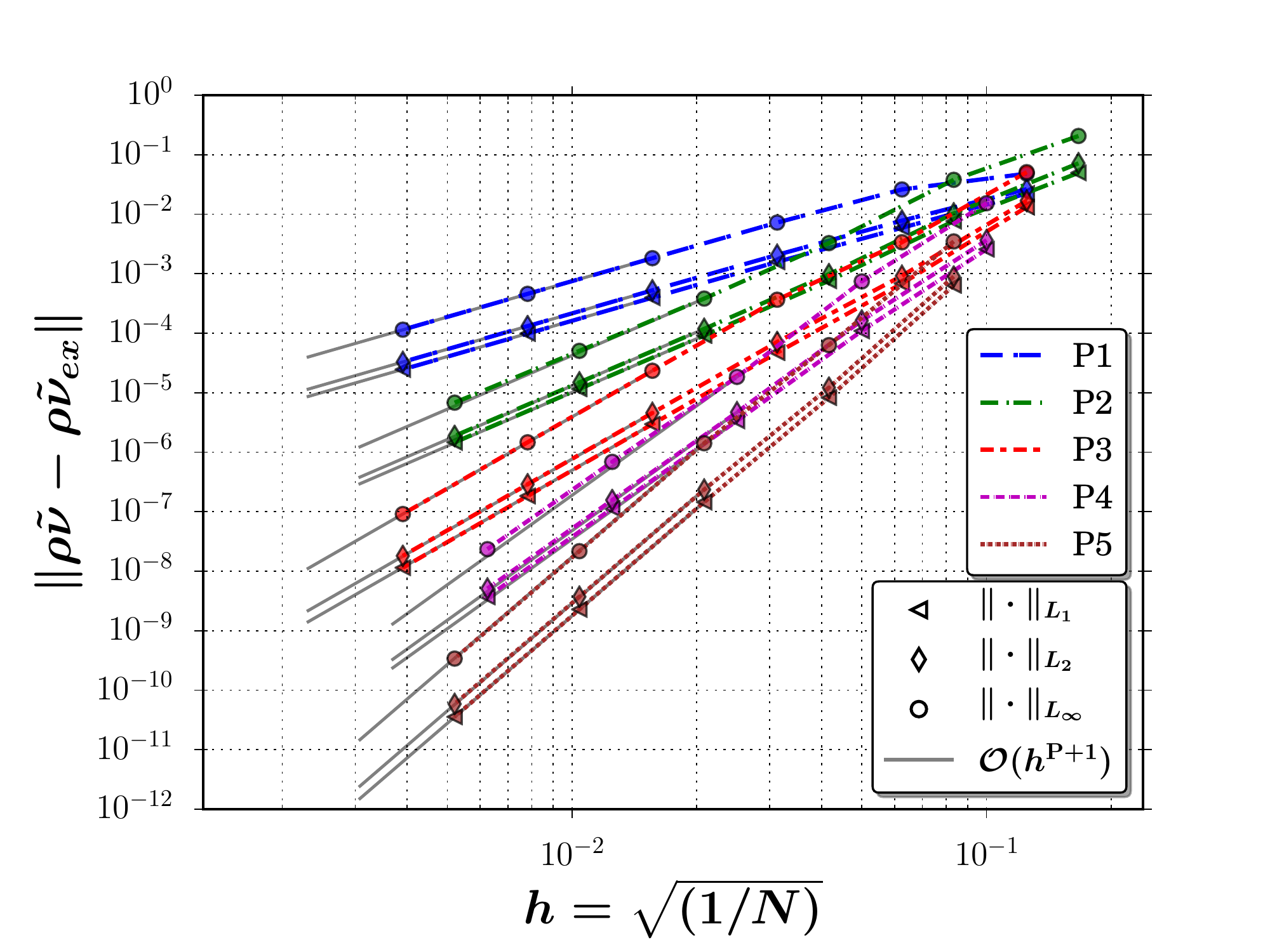}}
\caption{Evolution of the discretization error in $L_1$, $L_2$ and $L_\infty$ norms versus mesh refinement for MS-4 and $\mathrm{P}1$--$\mathrm{P}5$}
\label{fig:Err_allE_allP_MS-4}
\end{figure}

\begin{figure}[!hbt]
\vspace{3.3cm}
\centering
\subfloat[$\rho$]{ 
\includegraphics[trim = 16mm 3mm 18mm 13mm, clip,width=0.32\linewidth]
{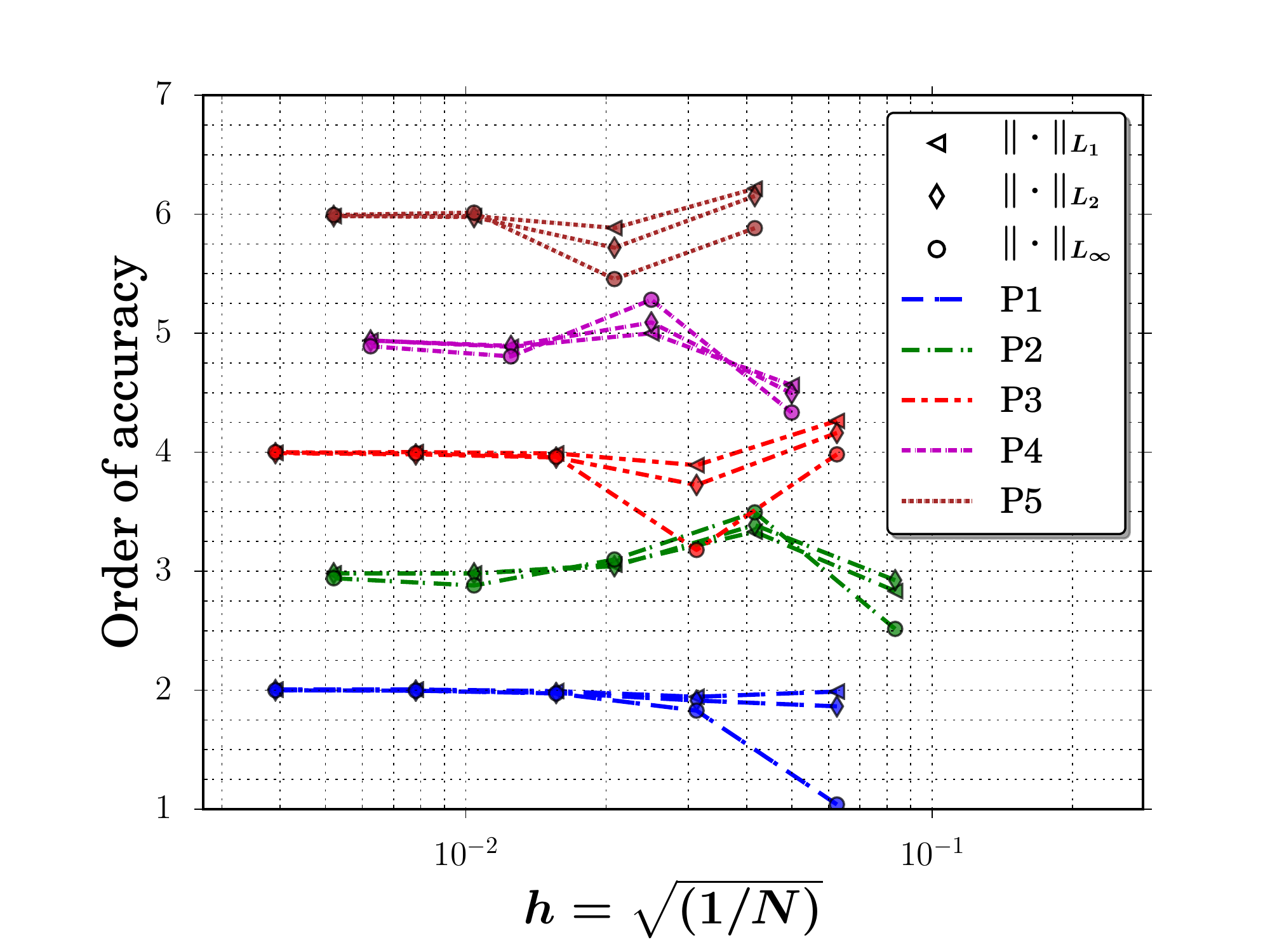}}
~~~
\subfloat[$\rho u$]{
\includegraphics[trim = 16mm 3mm 18mm 13mm, clip,width=0.32\linewidth]{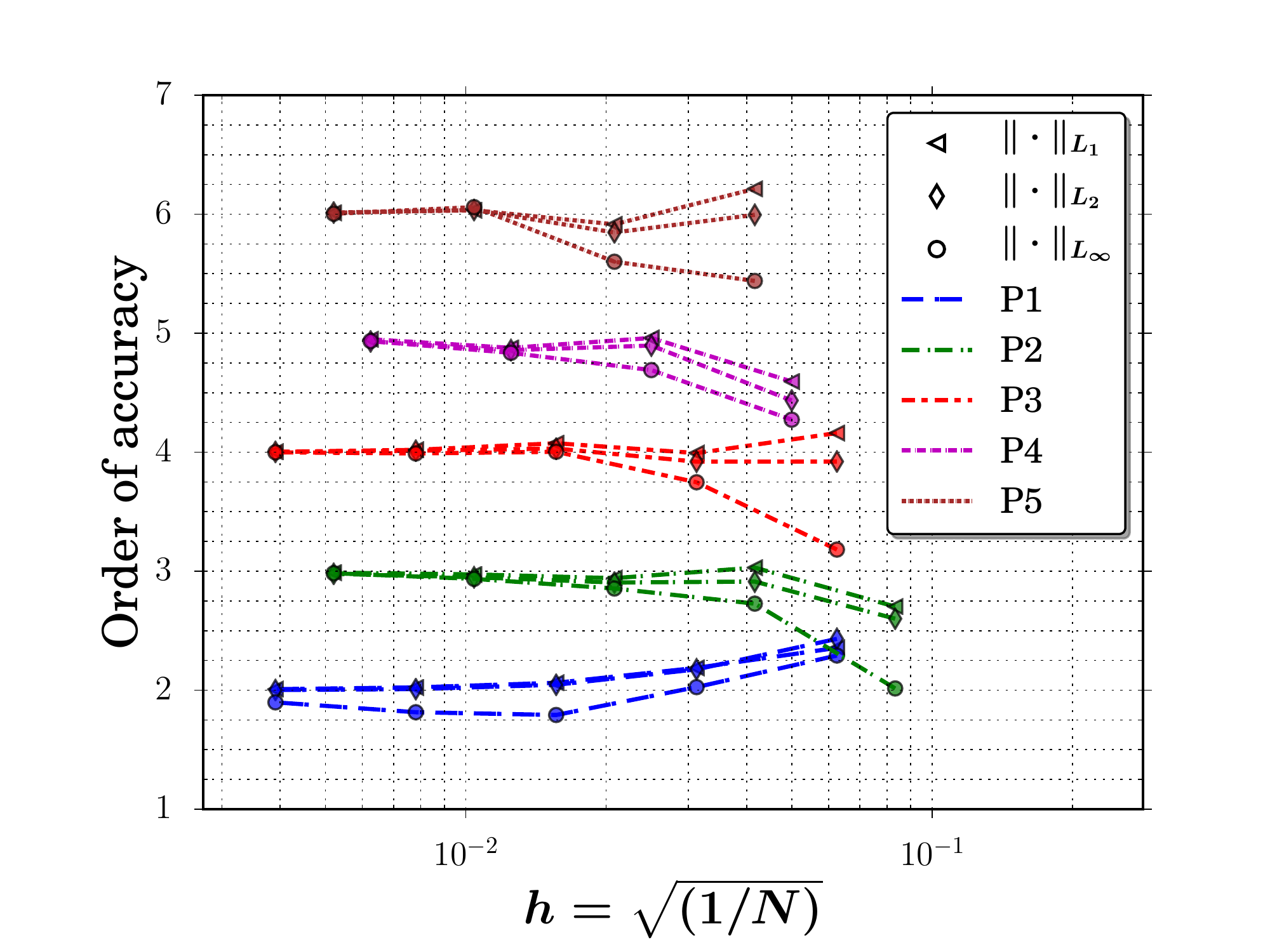}}
\vfill
\subfloat[$\rho v$]{
\includegraphics[trim = 16mm 3mm 18mm 13mm, clip,width=0.32\linewidth]
{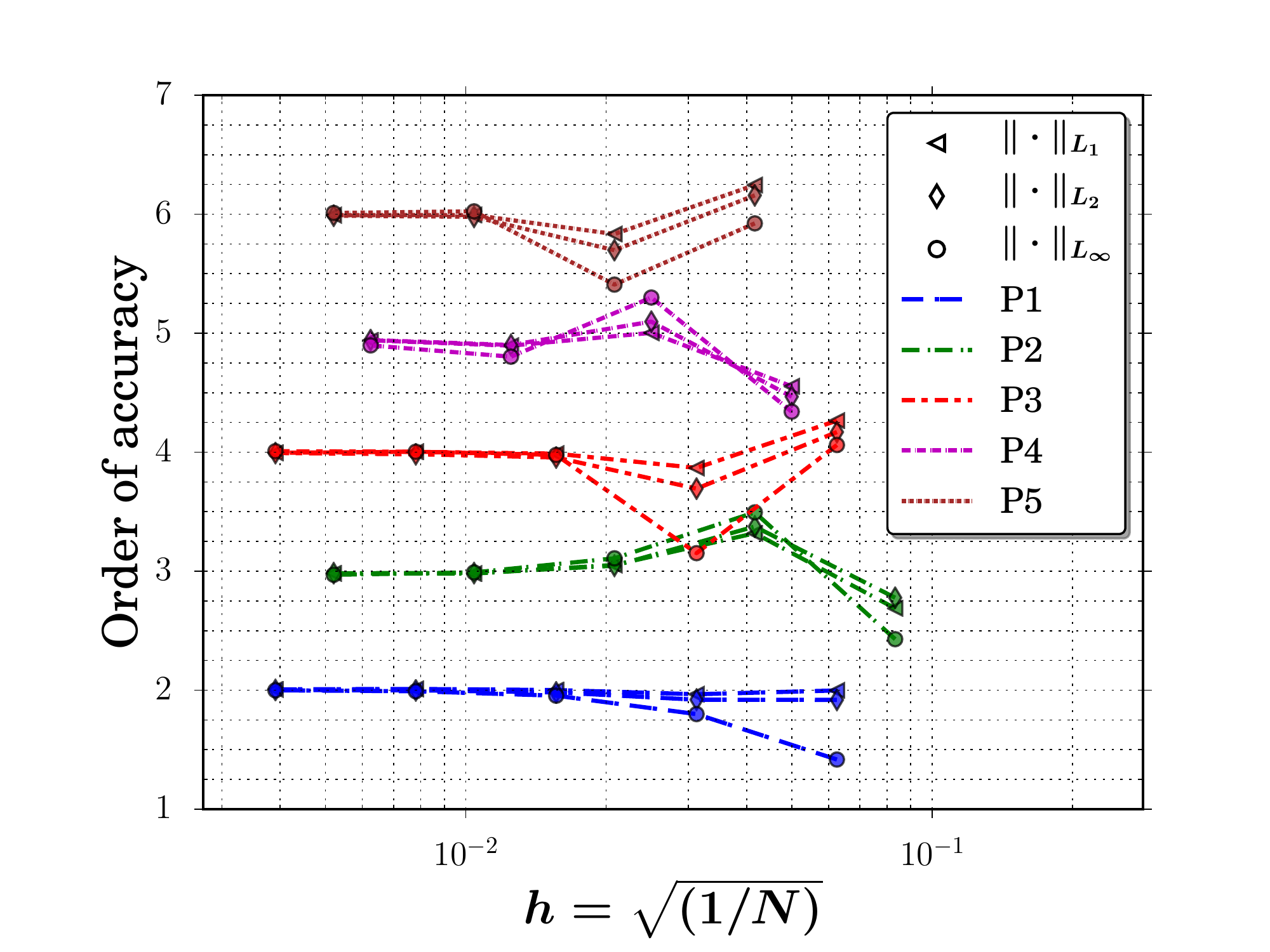}}
~~~
\subfloat[$\rho E$]{
\includegraphics[trim = 16mm 3mm 18mm 13mm, clip,width=0.32\linewidth]{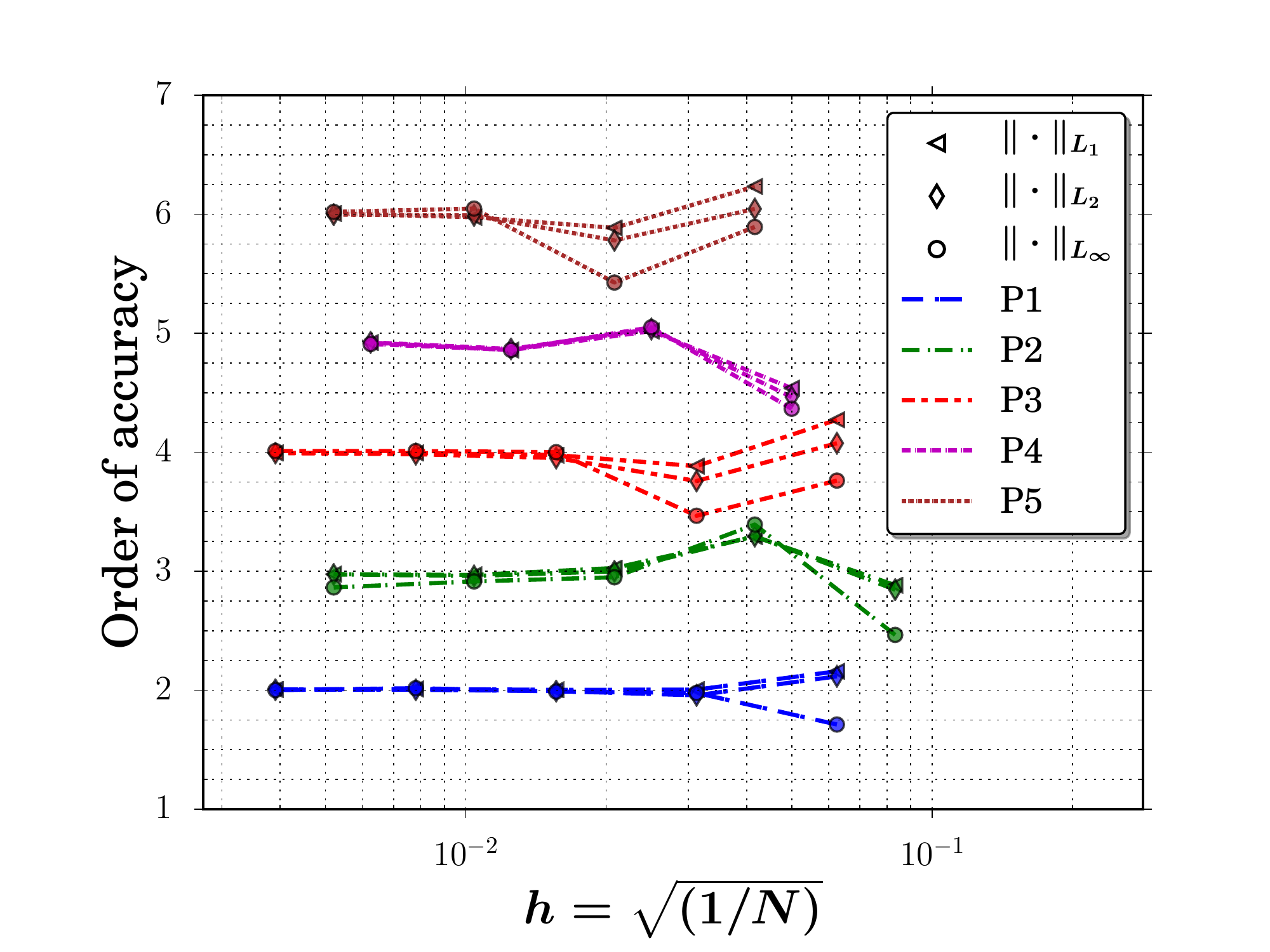}}
\vfill
\subfloat[$\rho \tilde{\nu}$]{
\includegraphics[trim = 16mm 3mm 18mm 13mm, clip,width=0.32\linewidth]{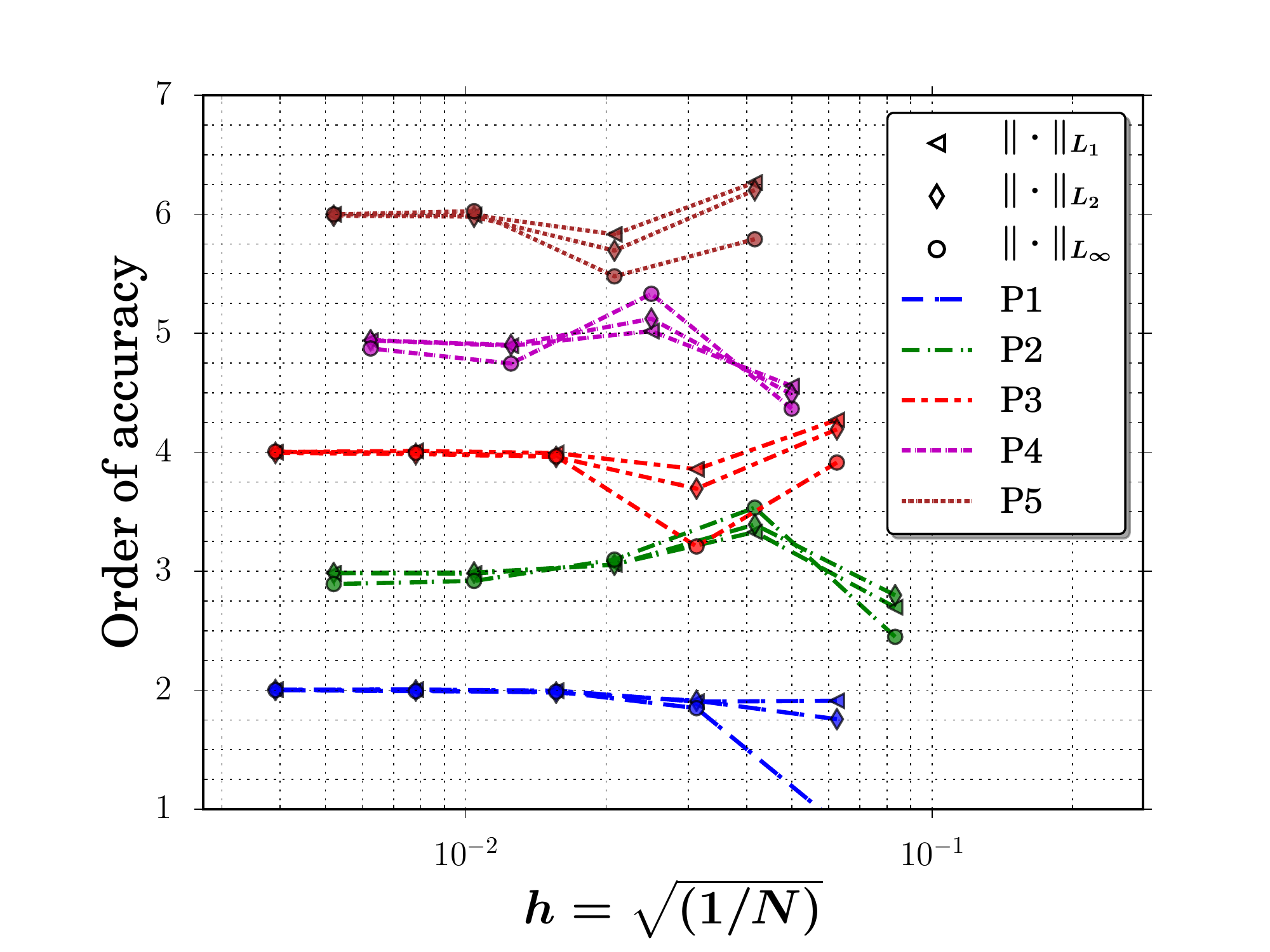}}
\caption{Evolution of the OOAs in $L_1$, $L_2$ and $L_\infty$ norms versus mesh refinement for MS-4 and $\mathrm{P}1$--$\mathrm{P}5$}
\label{fig:Orders_MS-4}
\end{figure}

\begin{figure}[!hbt]
\centering
\subfloat[$\rho$]{
\includegraphics[trim = 5mm 2mm 18mm 13mm, clip,width=0.32\linewidth]
{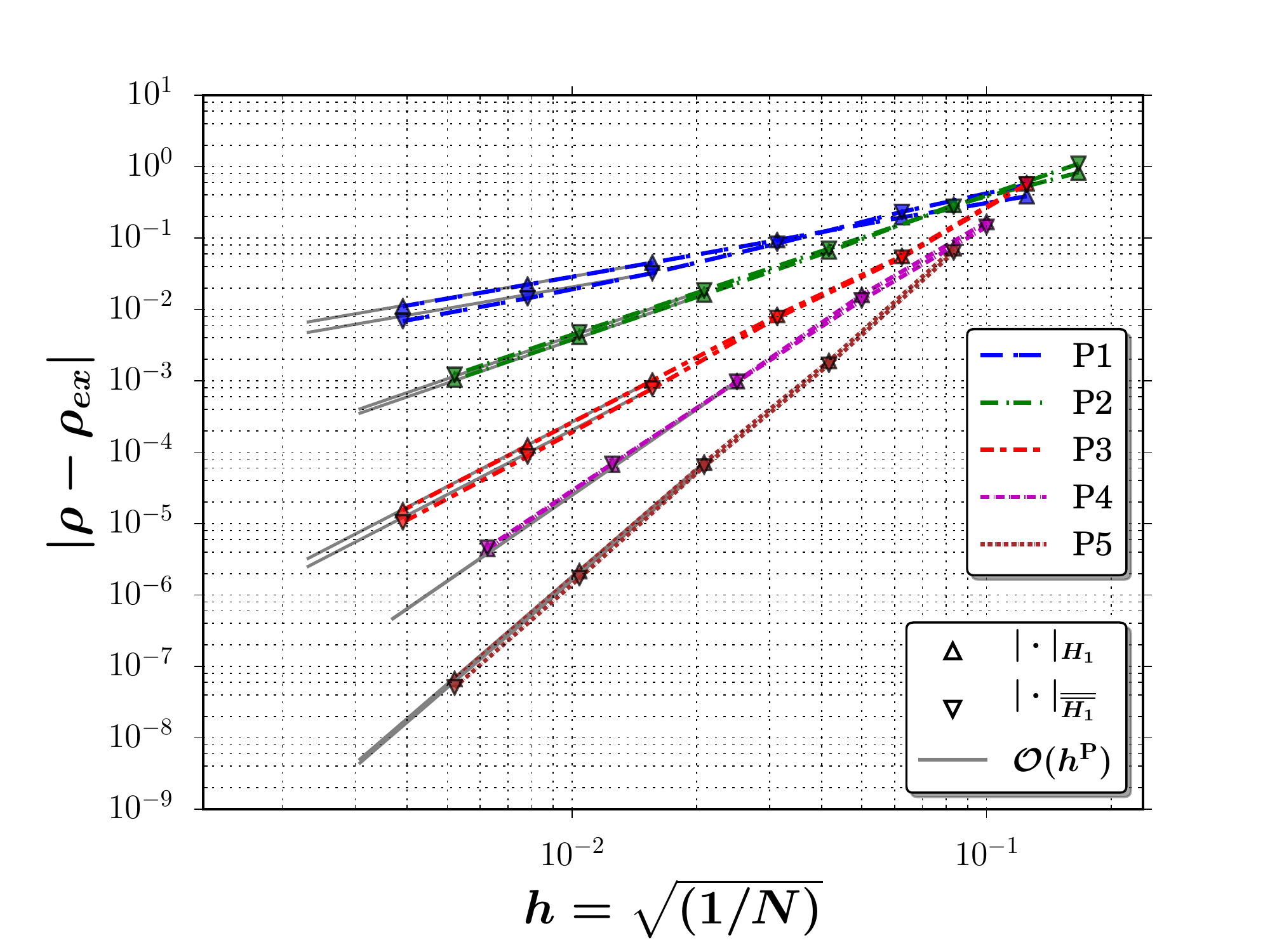}}
~~~
\subfloat[$\rho u$]{
\includegraphics[trim = 5mm 2mm 18mm 13mm, clip,width=0.32\linewidth]{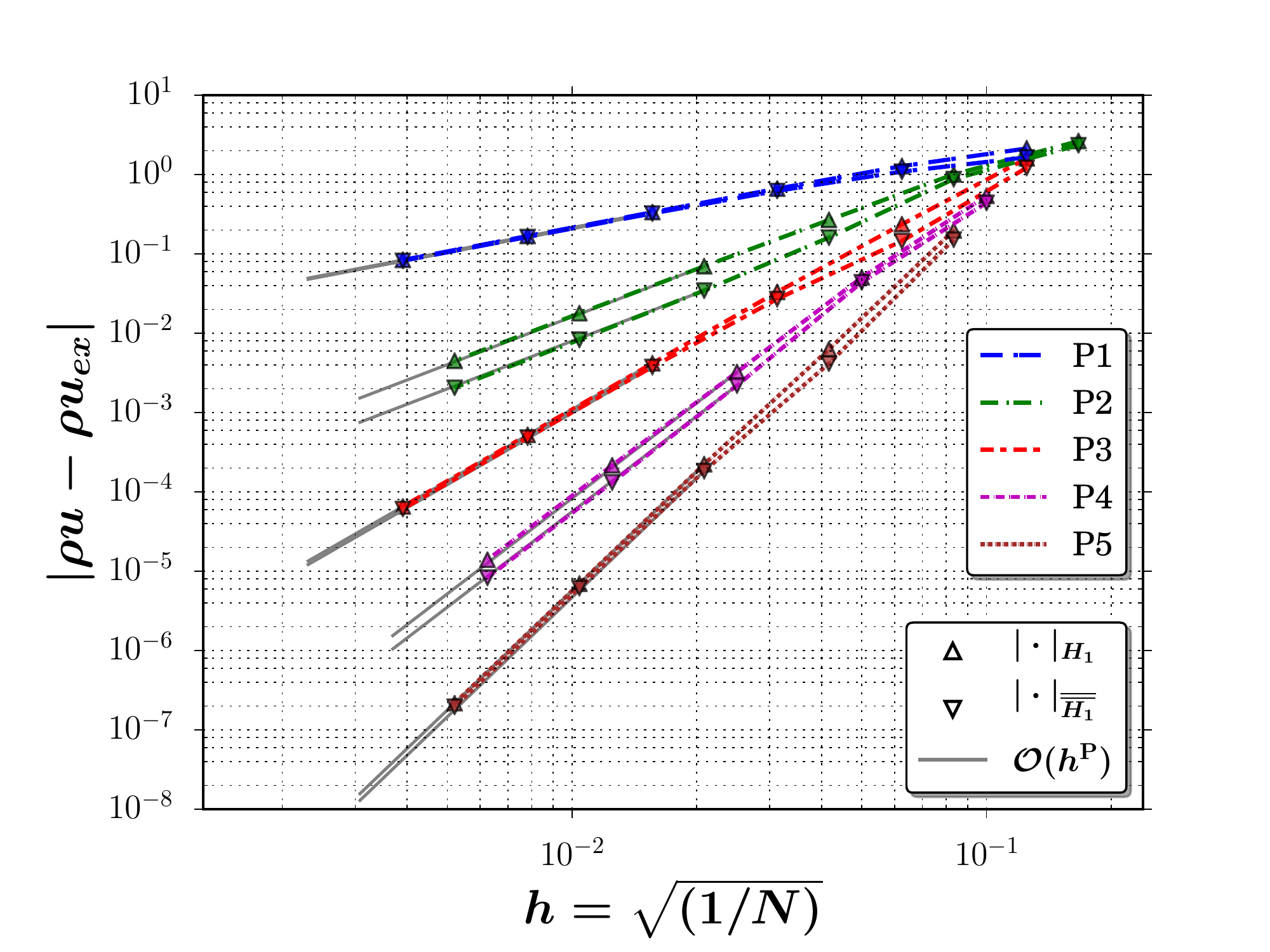}}
\vfill
\subfloat[$\rho v$]{
\includegraphics[trim = 5mm 2mm 18mm 13mm, clip,width=0.32\linewidth]
{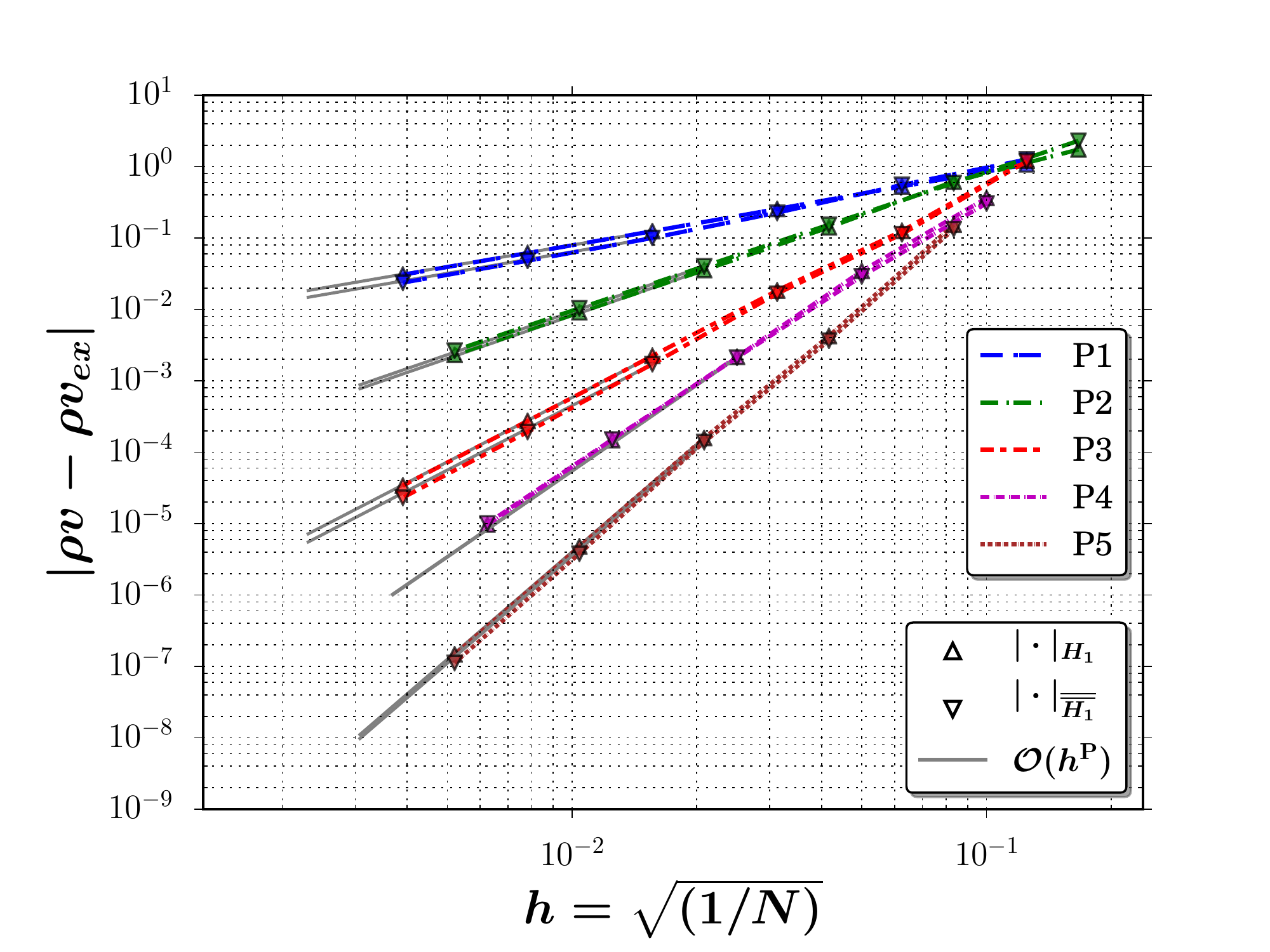}}
~~~
\subfloat[$\rho E$]{
\includegraphics[trim = 5mm 2mm 18mm 13mm, clip,width=0.32\linewidth]{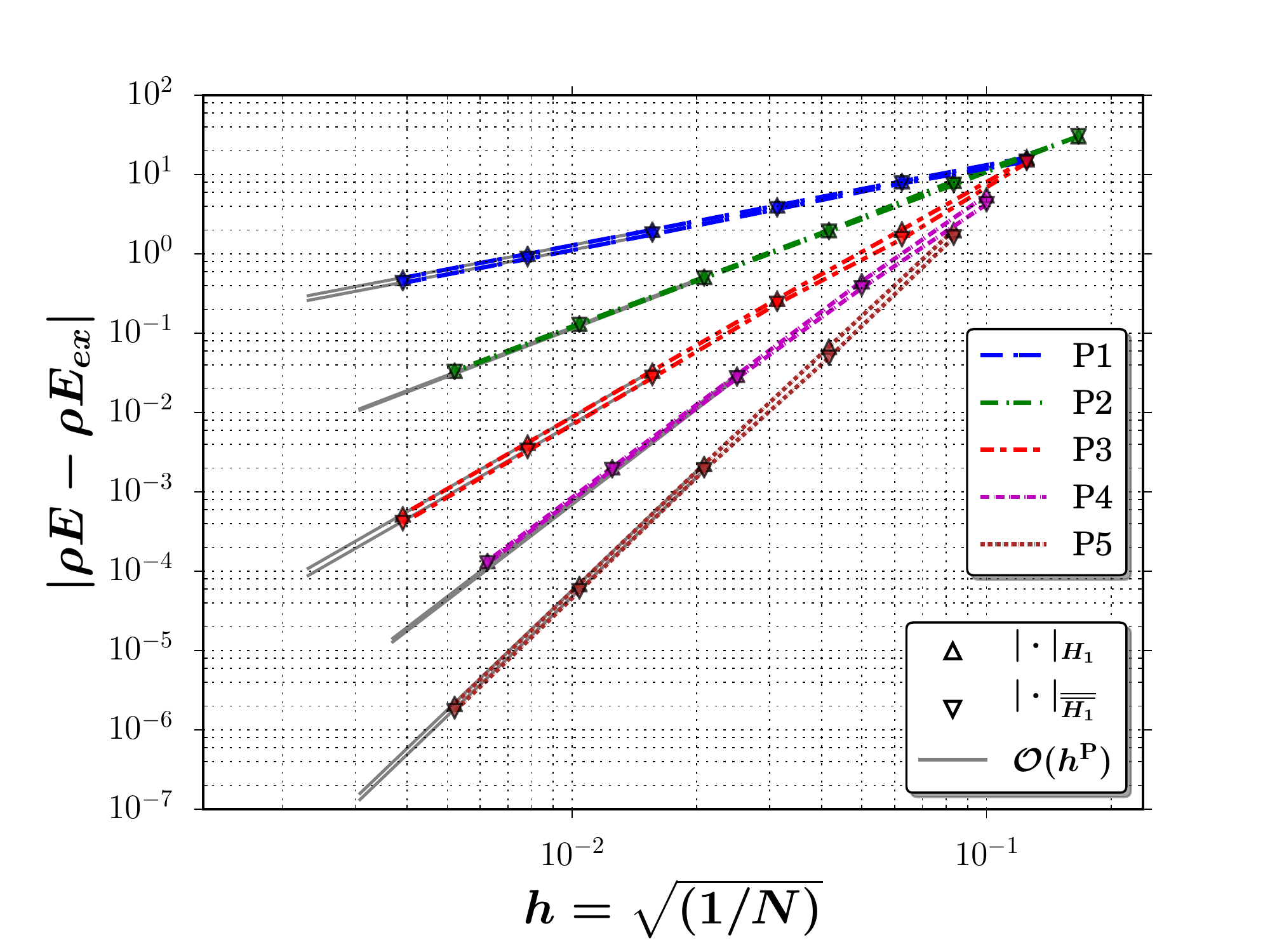}}
\vfill
\subfloat[$\rho \tilde{\nu}$]{
\includegraphics[trim = 5mm 2mm 18mm 13mm, clip,width=0.32\linewidth]{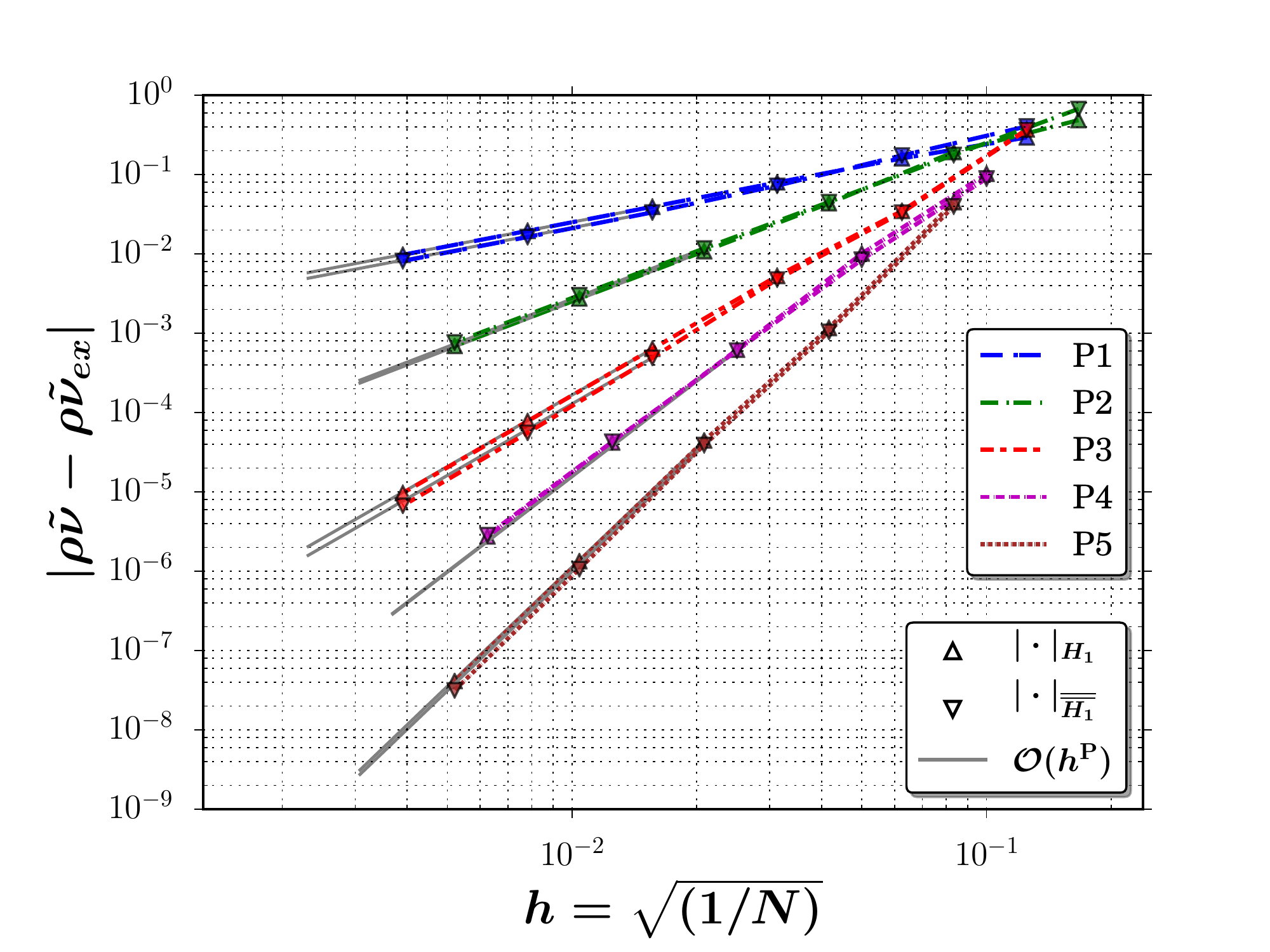}}
\caption{Evolution of the discretization error in $H_1$ semi-norm (for uncorrected and  fully corrected derivatives) versus mesh refinement for MS-4 and $\mathrm{P}1$--$\mathrm{P}5$}
\label{fig:Err_allE_allP_H_MS-4}
\end{figure}

\begin{figure}[!hbt]
\centering
\subfloat[$\rho$]{ 
\includegraphics[trim = 16mm 3mm 18mm 13mm, clip,width=0.32\linewidth]
{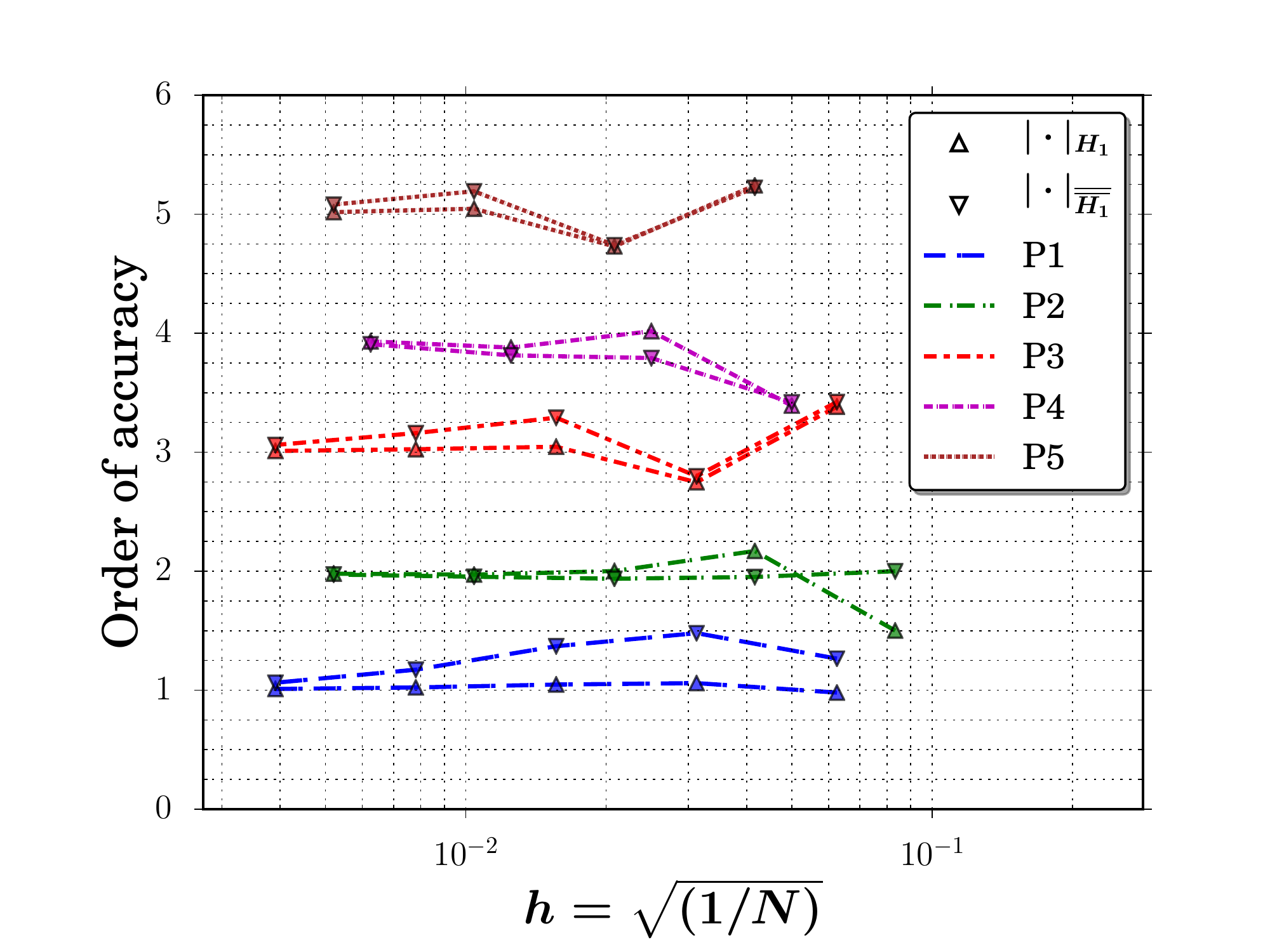}}
~~~
\subfloat[$\rho u$]{
\includegraphics[trim = 16mm 3mm 18mm 13mm, clip,width=0.32\linewidth]{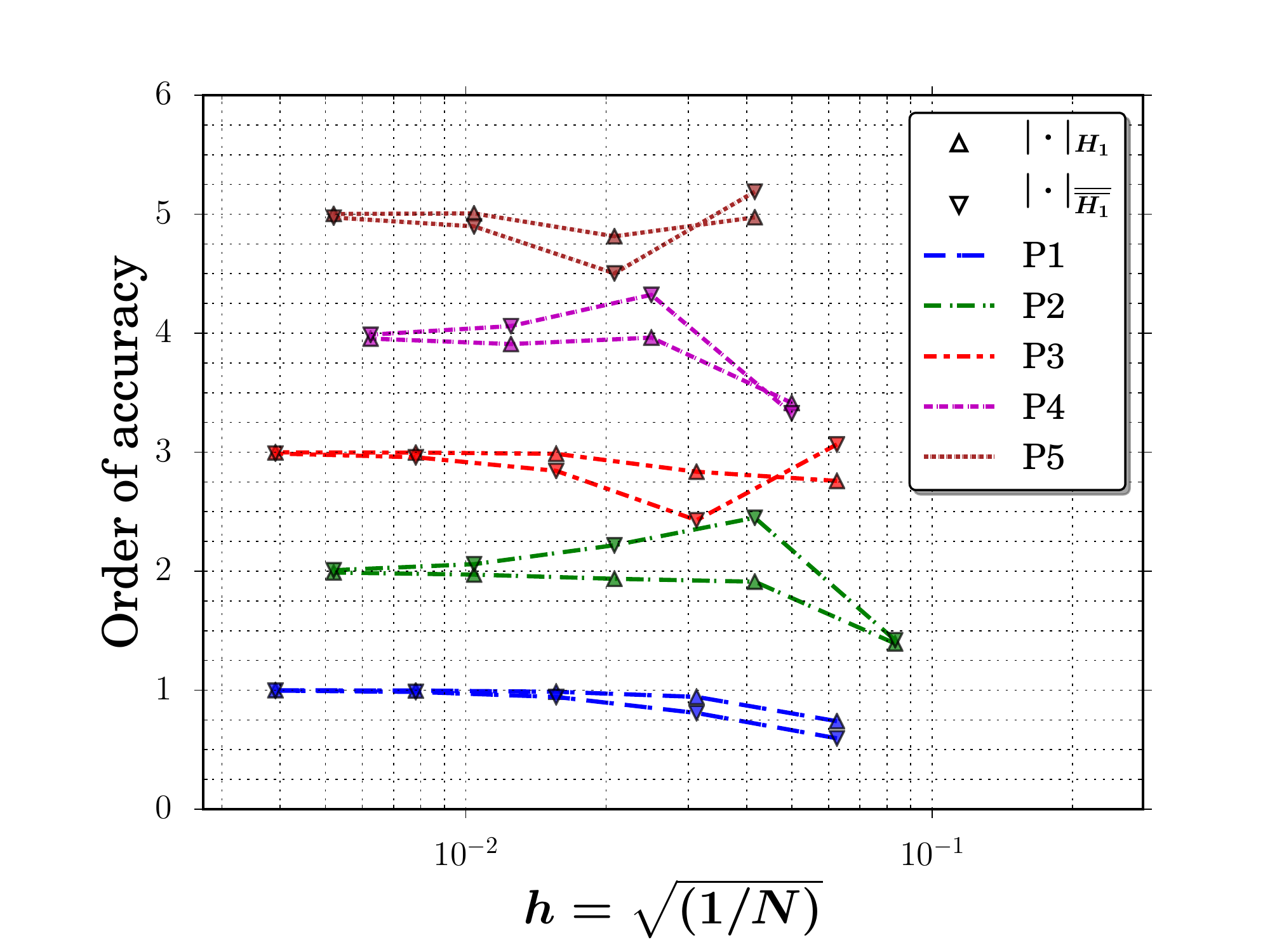}}
\vfill
\subfloat[$\rho v$]{
\includegraphics[trim = 16mm 3mm 18mm 13mm, clip,width=0.32\linewidth]
{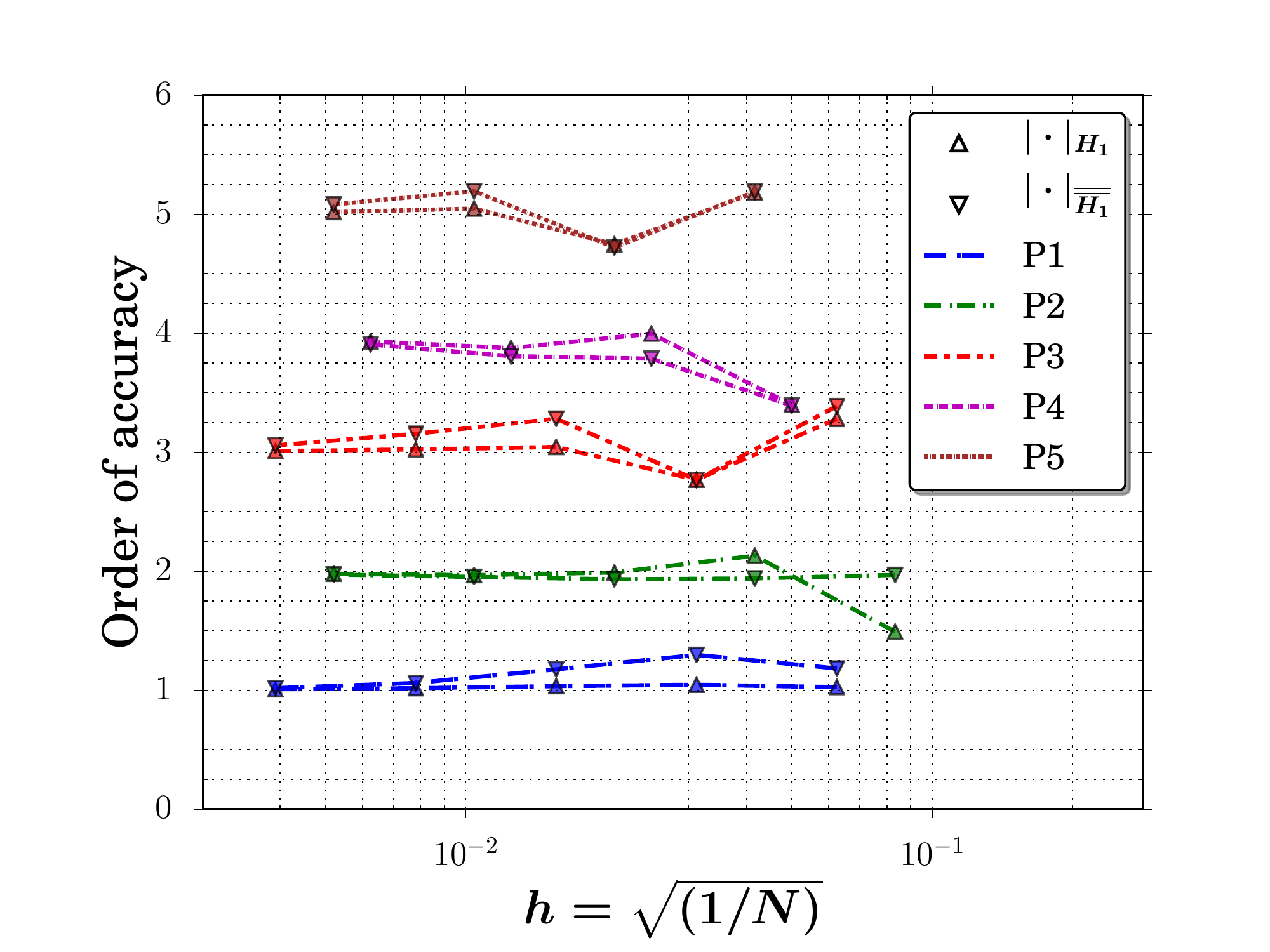}}
~~~
\subfloat[$\rho E$]{
\includegraphics[trim = 16mm 3mm 18mm 13mm, clip,width=0.32\linewidth]{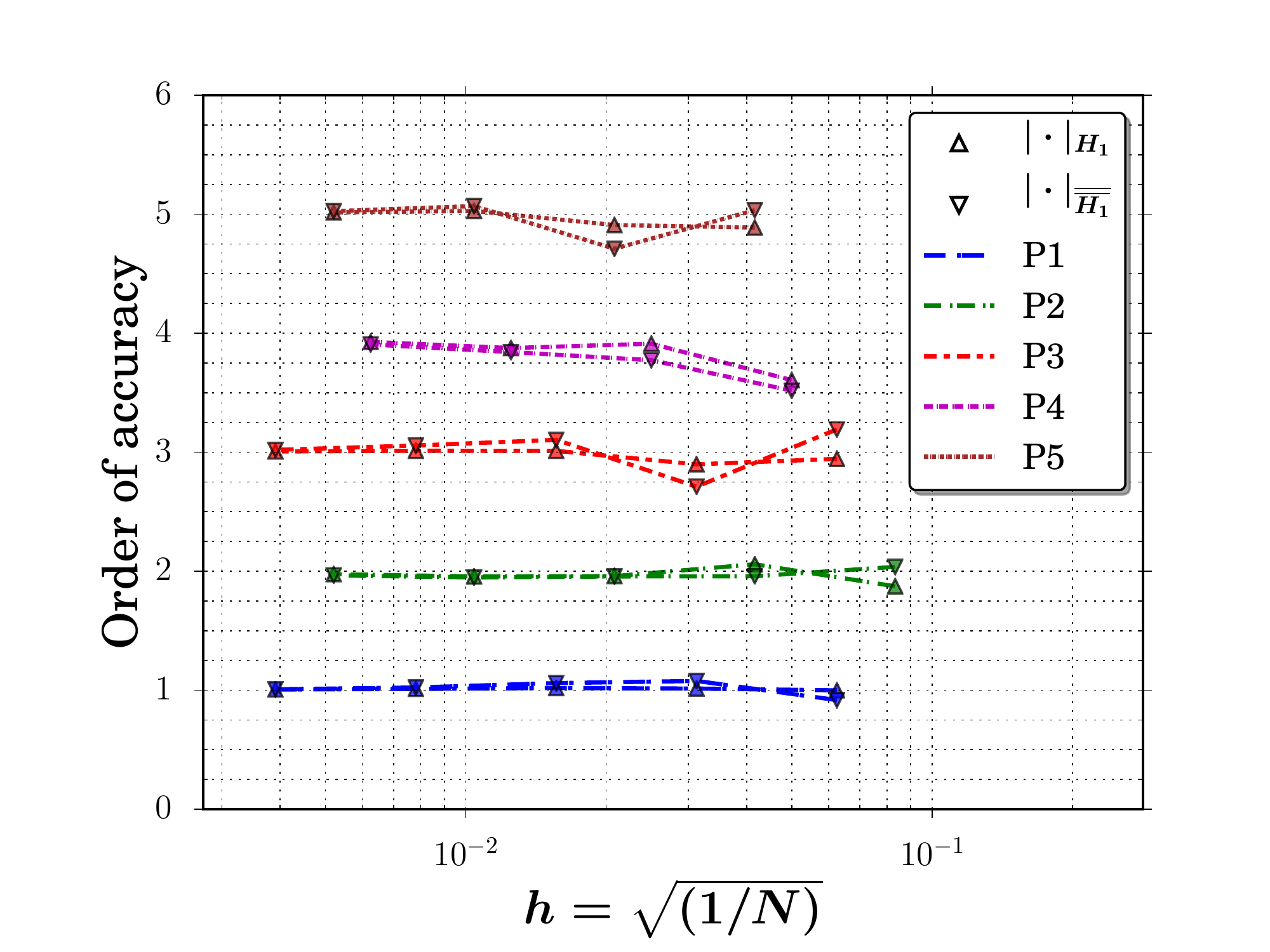}}
\vfill
\subfloat[$\rho \tilde{\nu}$]{
\includegraphics[trim = 16mm 3mm 18mm 13mm, clip,width=0.32\linewidth]{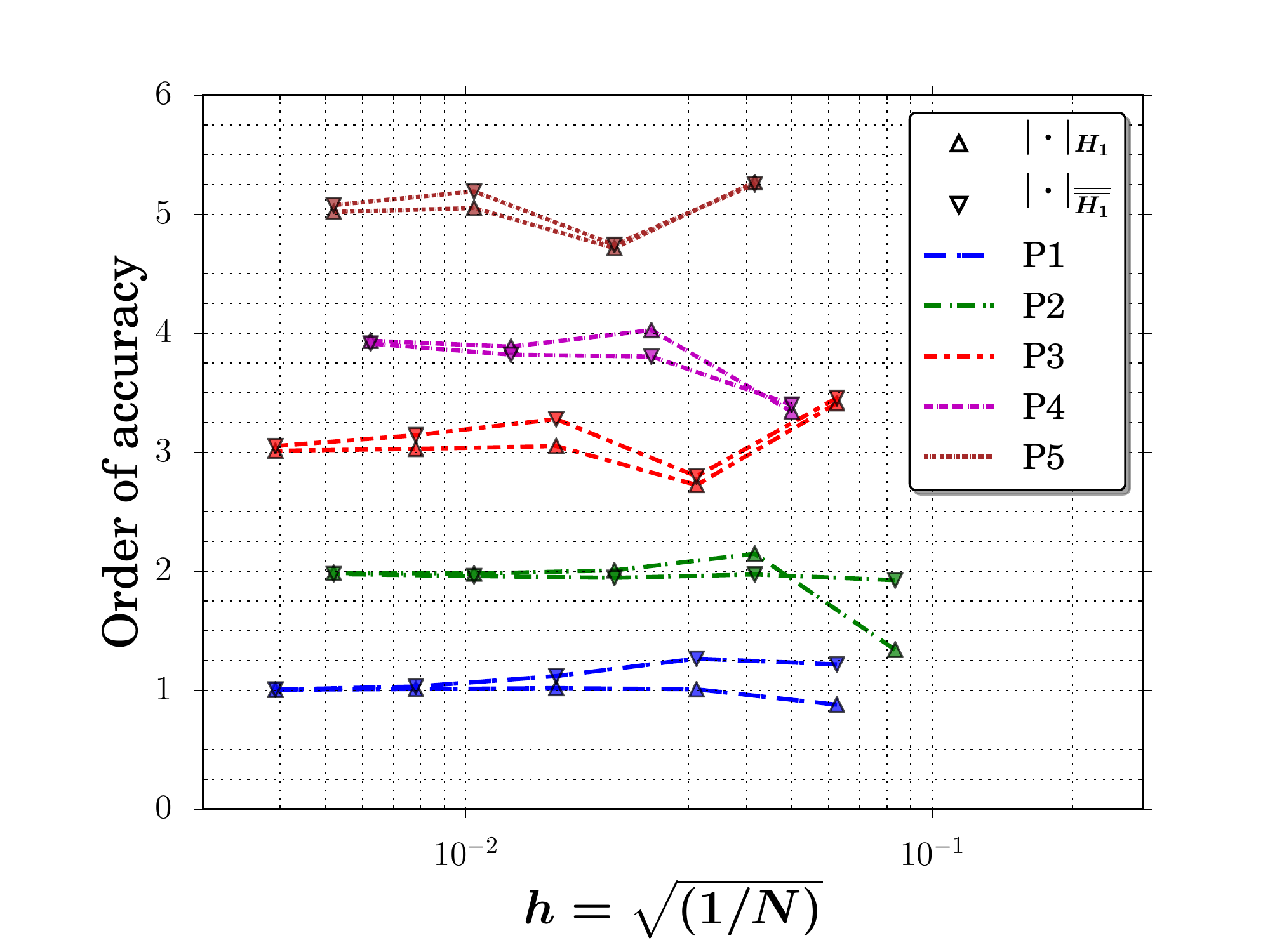}}
\caption{Evolution of the OOAs in $H_1$ semi-norm (for uncorrected and  fully corrected derivatives) versus mesh refinement for MS-4 and $\mathrm{P}1$--$\mathrm{P}5$}
\label{fig:Orders_H_MS-4}
\end{figure}

\clearpage
\subsection{MS-5}

\begin{figure}[!hbt]
\centering
{\includegraphics[trim = 0mm 0mm 0mm 0mm clip,width=0.27\linewidth]
{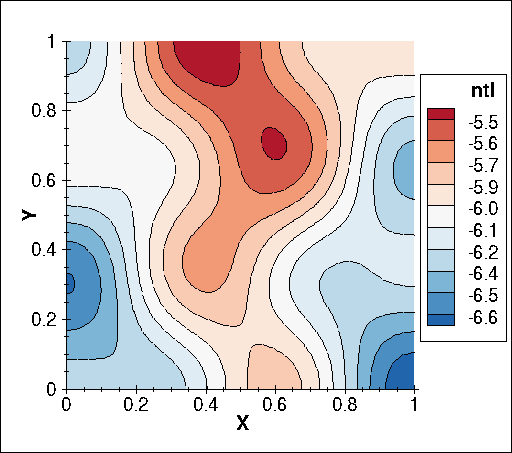}}
\caption{Manufactured solution MS-5 for $\tilde{\nu}$}
\label{fig:MS-5}
\end{figure}

\begin{figure}[!hbt]
\centering
\subfloat[$L$ norms]{
\includegraphics[trim = 5mm 2mm 18mm 13mm, clip,width=0.32\linewidth]{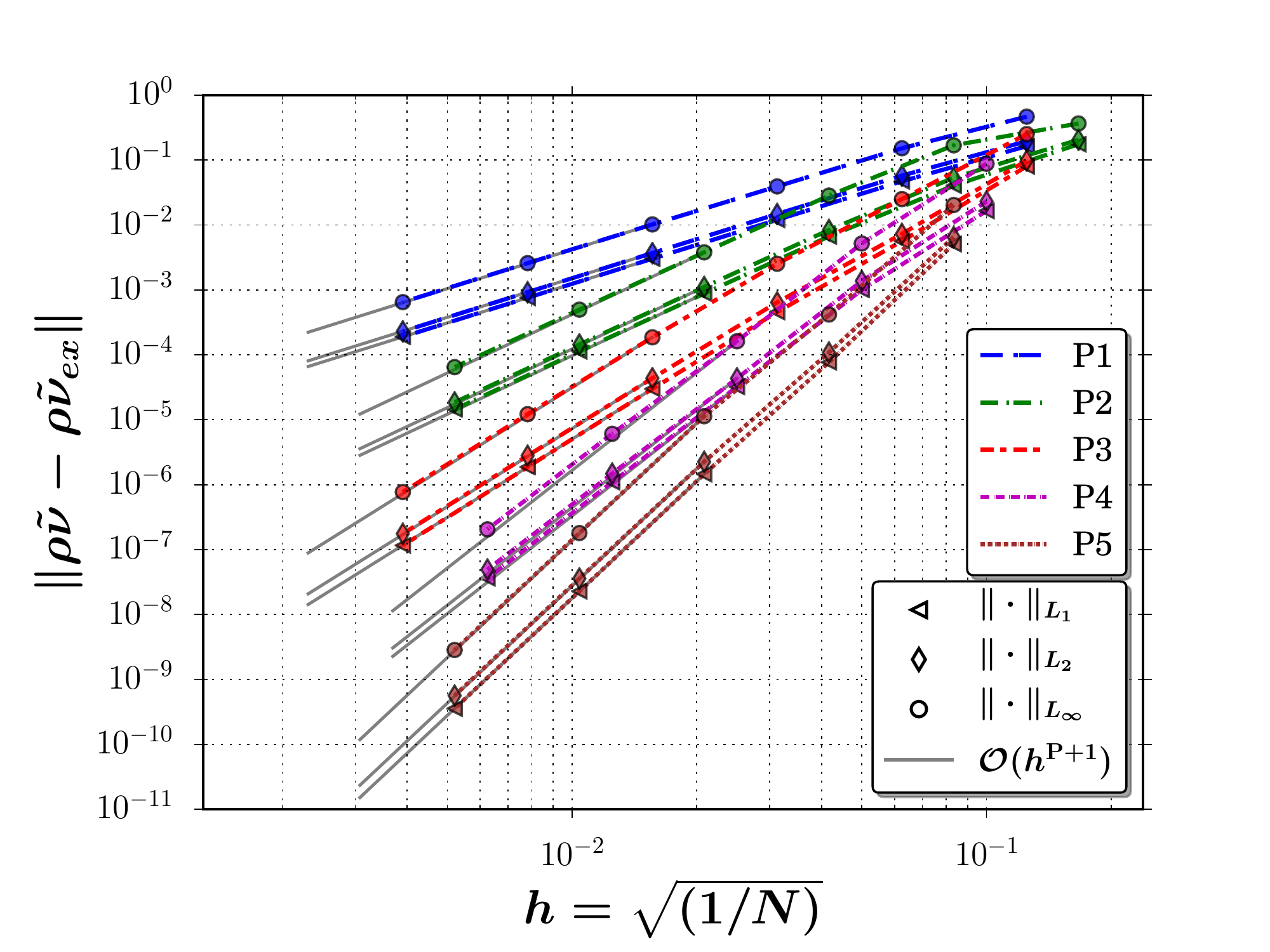}}
~~~~
\subfloat[$H_1$ semi-norm]{
\includegraphics[trim = 5mm 2mm 18mm 13mm, clip,width=0.32\linewidth]{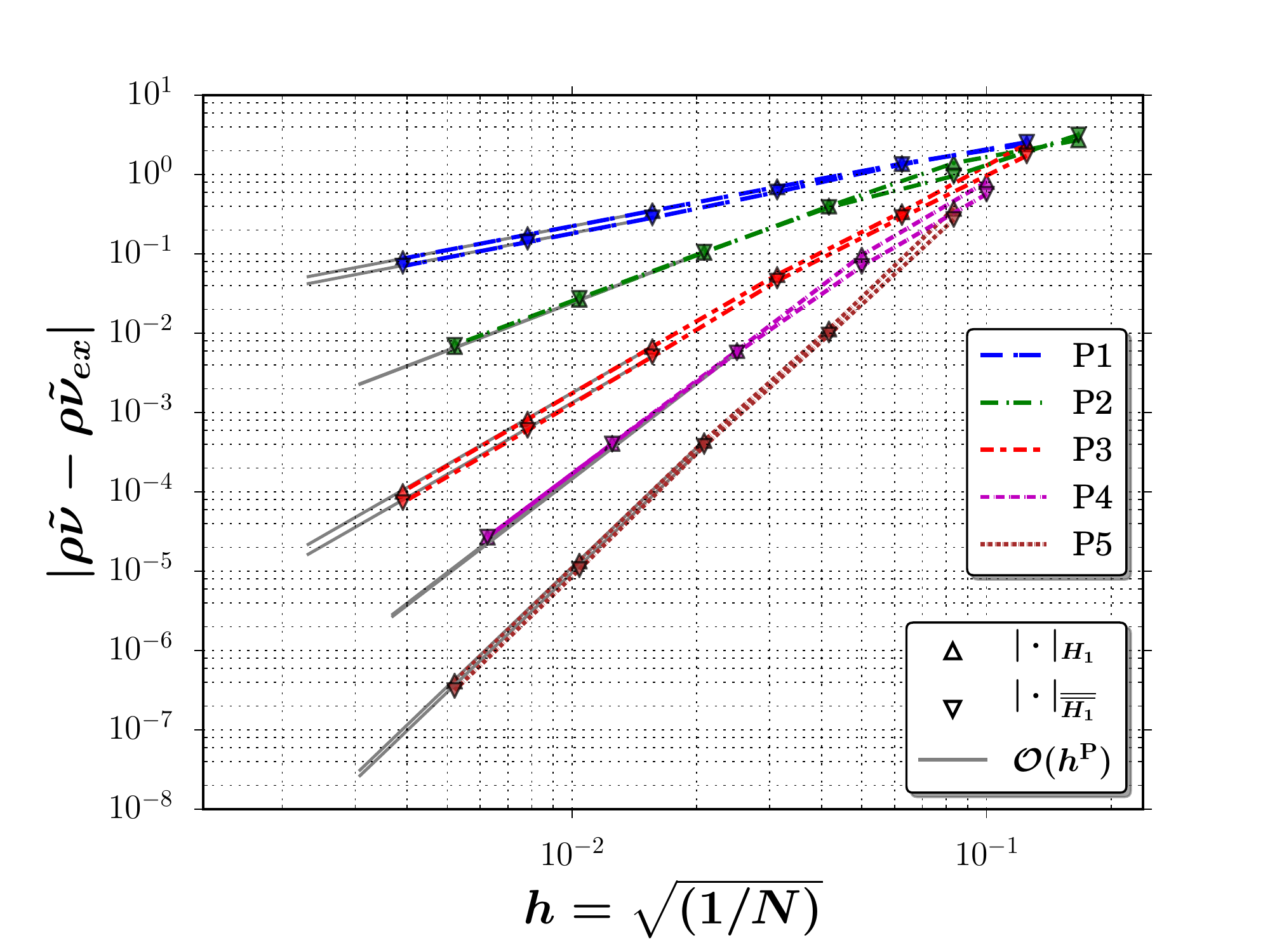}}
\caption{Evolution of the discretization error of $\rho\tilde{\nu}$  versus mesh refinement for MS-5 and $\mathrm{P}1$--$\mathrm{P}5$}
\label{fig:L_MS-5}
\end{figure}

\begin{figure}[!hbt]
\centering
\subfloat[$L$ norms]{
\includegraphics[trim = 16mm 3mm 18mm 13mm, clip,width=0.32\linewidth]{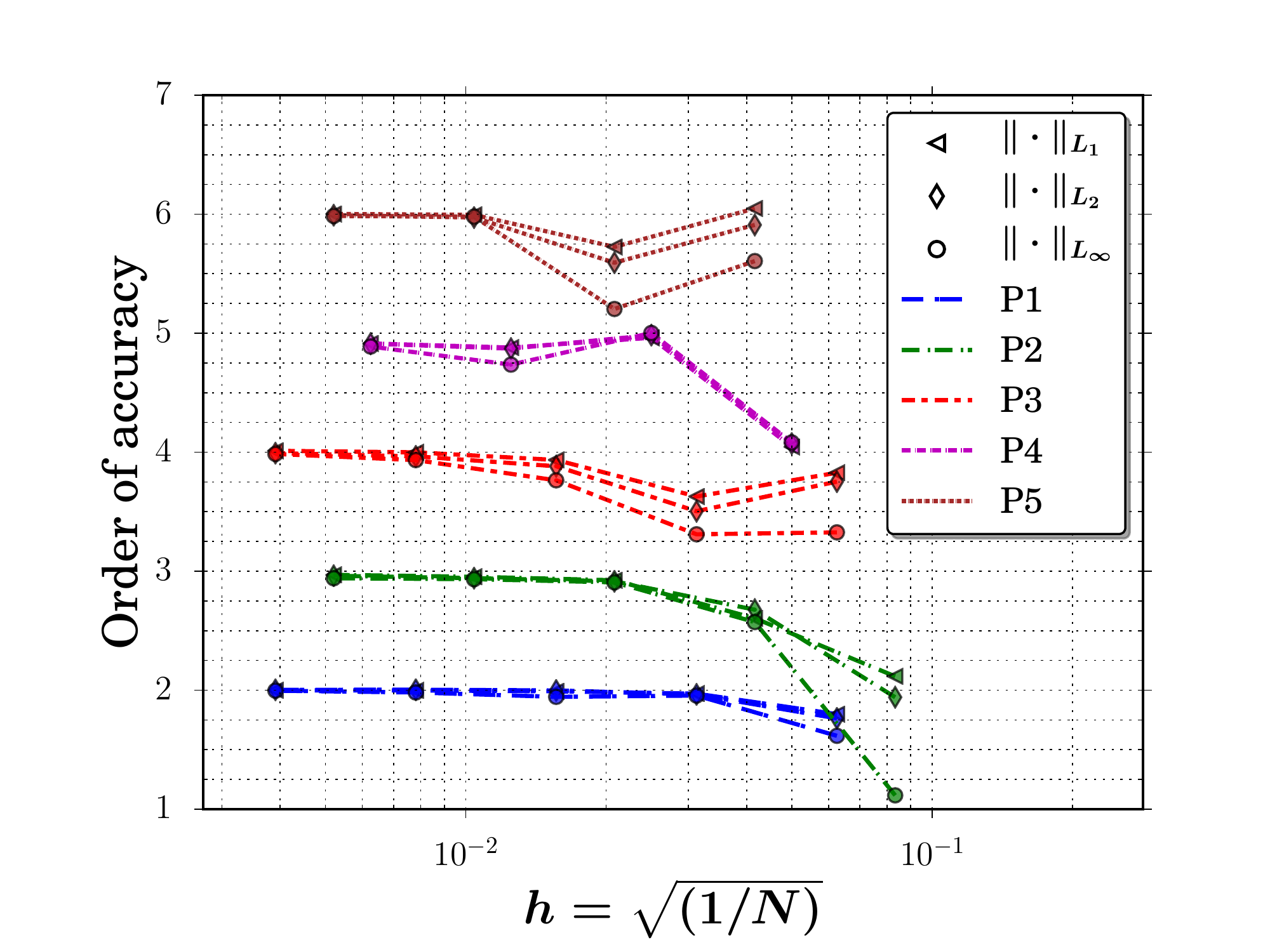}}
~~~
\subfloat[$H_1$ semi-norm]{
\includegraphics[trim = 16mm 3mm 18mm 13mm, clip,width=0.32\linewidth]{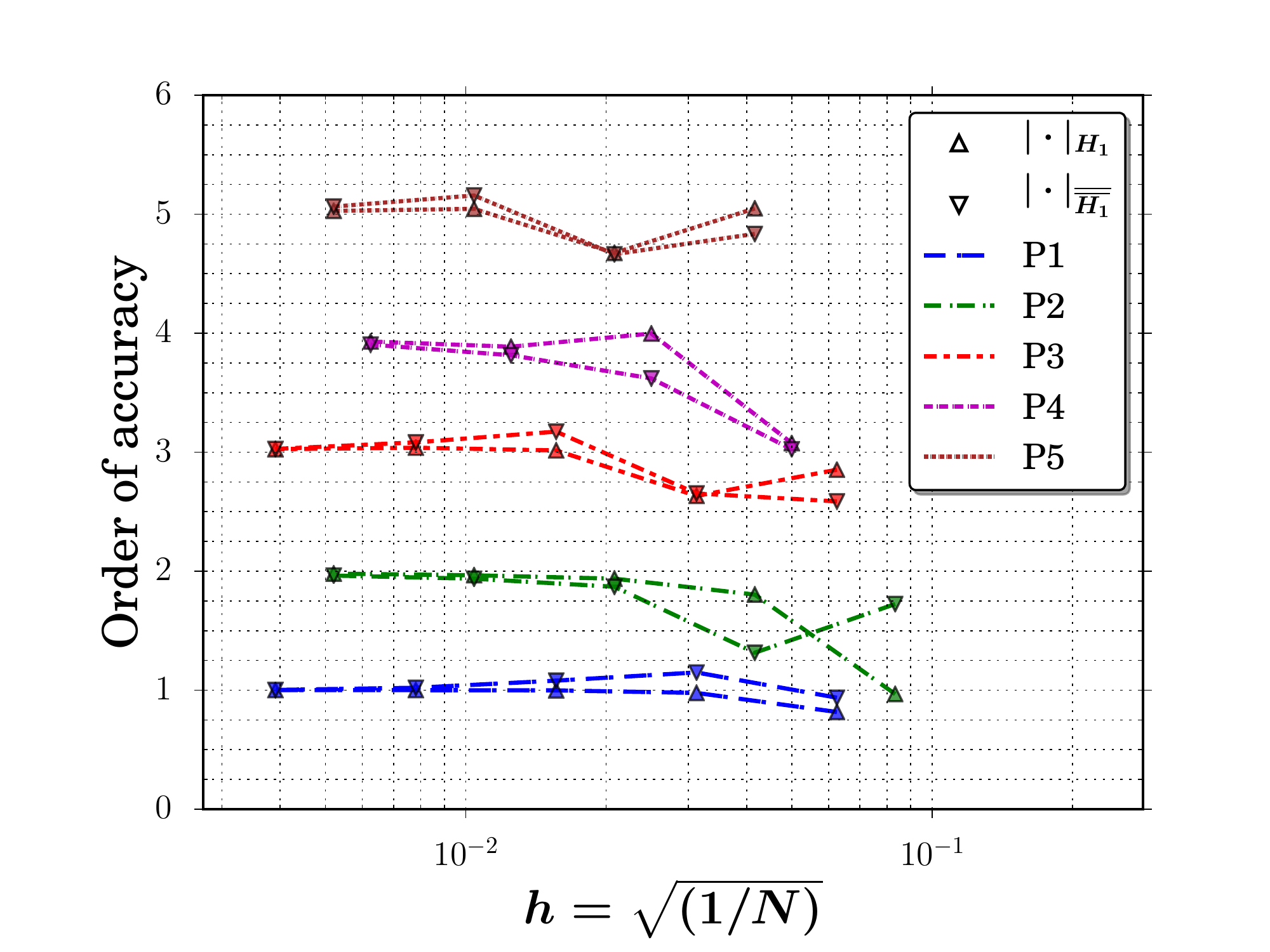}}
\caption{Evolution of the OOAs of $\rho\tilde{\nu}$ versus mesh refinement for MS-5 and $\mathrm{P}1$--$\mathrm{P}5$}
\label{fig:H_MS-5}
\end{figure}

\clearpage

\section{Grid sensitivity results}
\label{sec:gridsens}

\begin{figure}[!hbt]
\vspace{-5mm}
\centering
\subfloat[P1]{
\includegraphics[trim = 5mm 3mm 18mm 13mm, clip,width=0.29\linewidth]{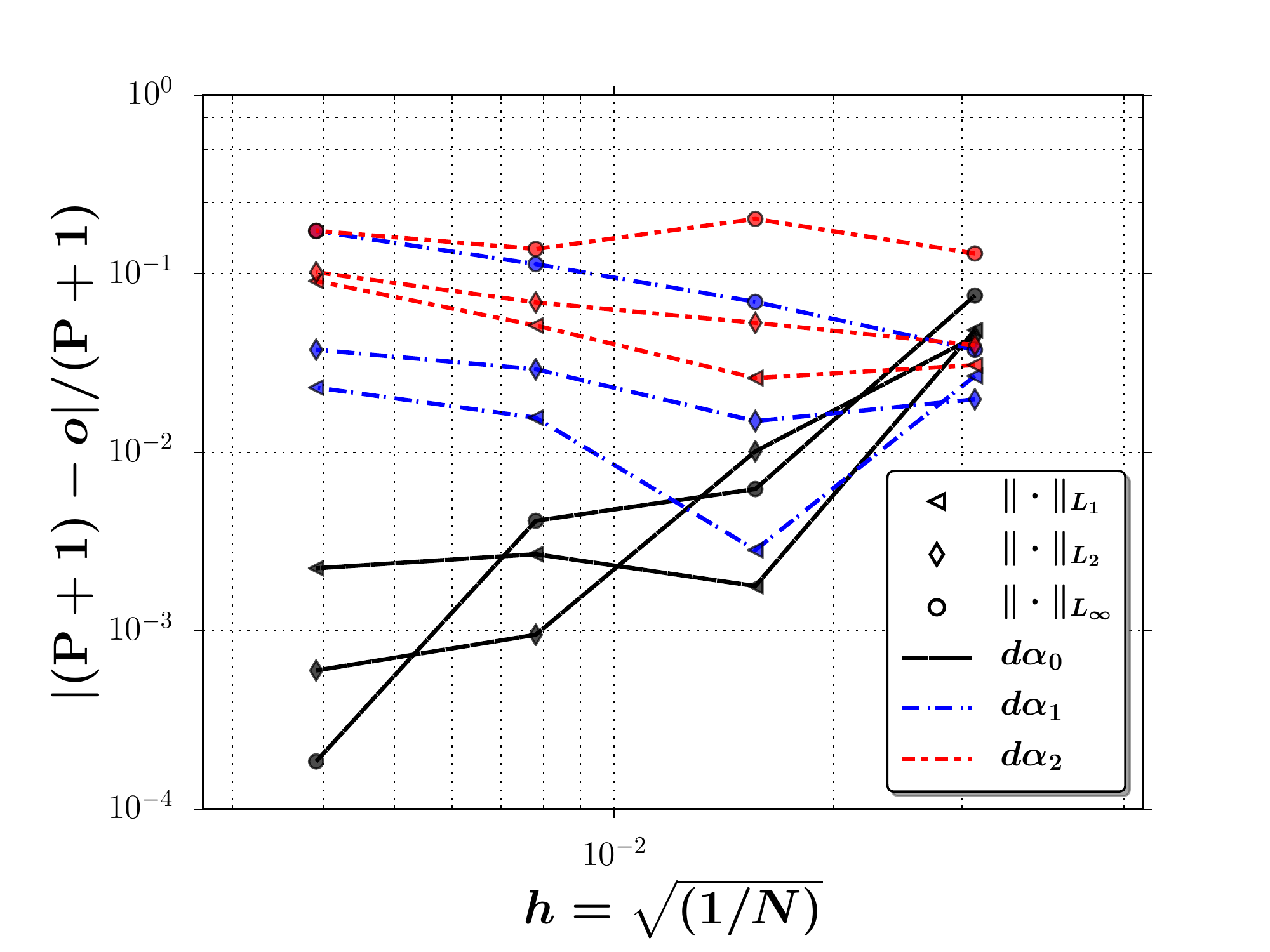}}~~~
\subfloat[P2]{
\includegraphics[trim = 5mm 2mm 18mm 13mm, clip,width=0.29\linewidth]{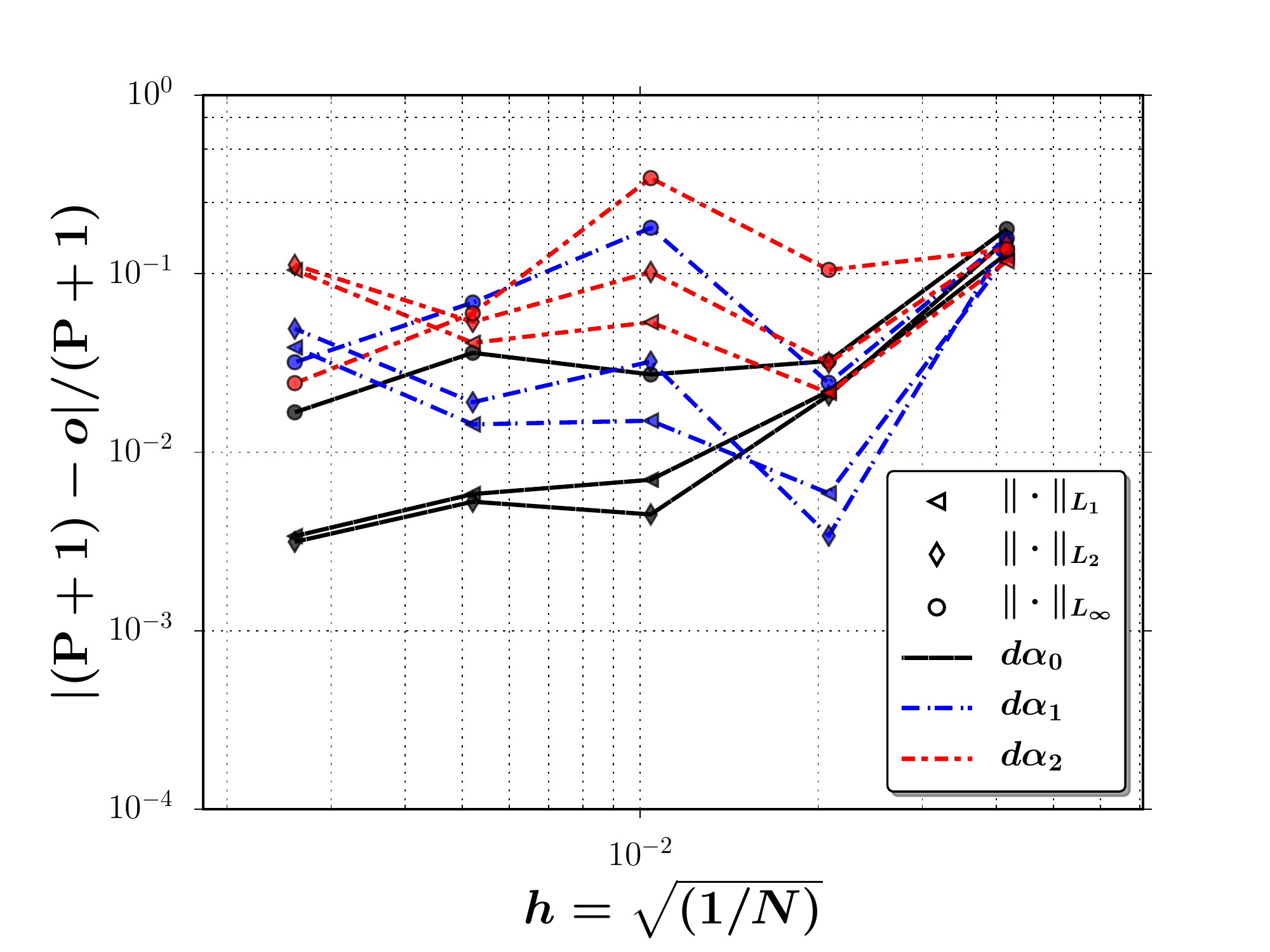}}
\vfill
\subfloat[P3]{
\includegraphics[trim = 5mm 2mm 18mm 13mm, clip,width=0.29\linewidth]{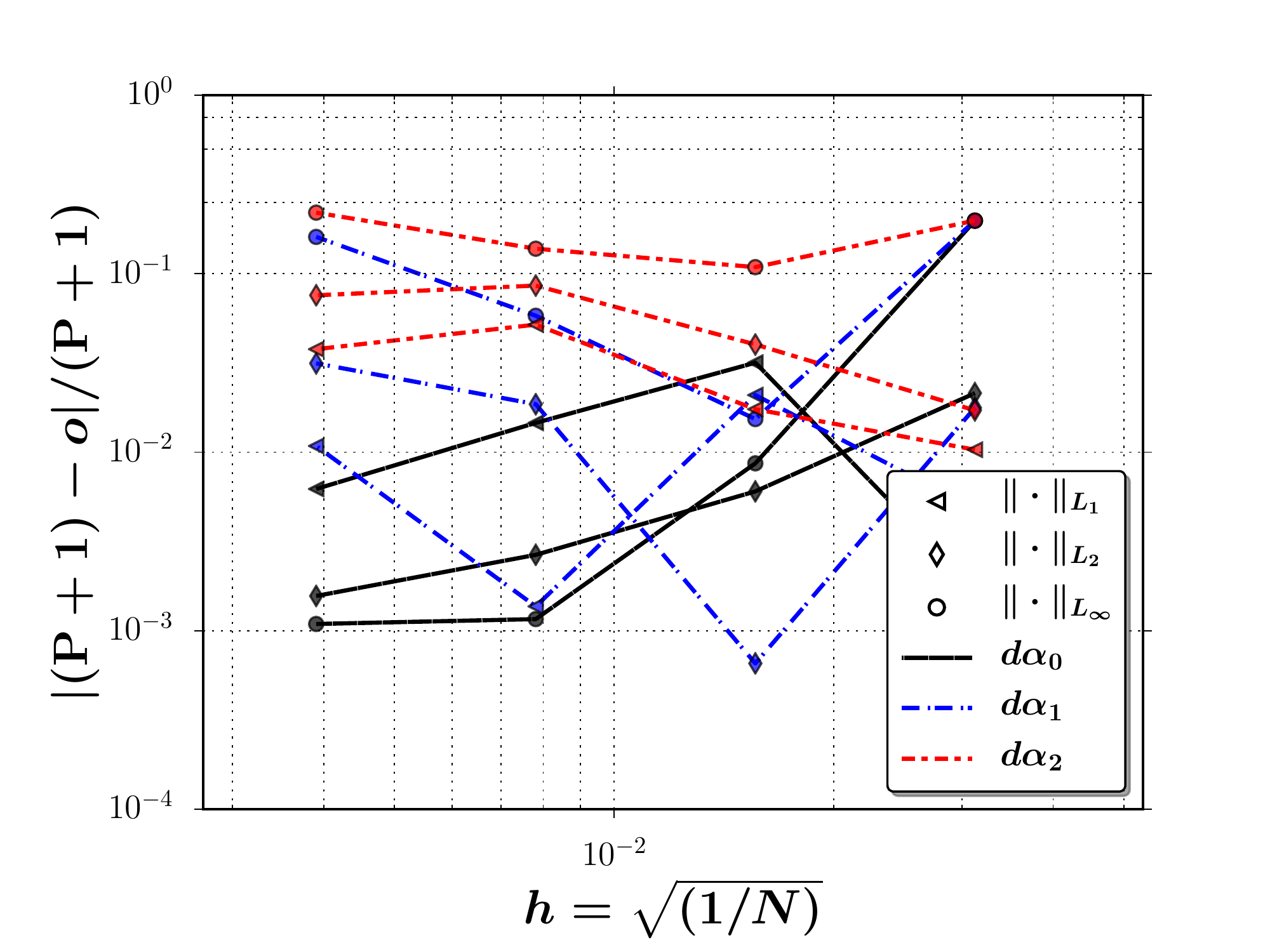}}~~~
\subfloat[P4]{
\includegraphics[trim = 5mm 2mm 18mm 13mm, clip,width=0.29\linewidth]{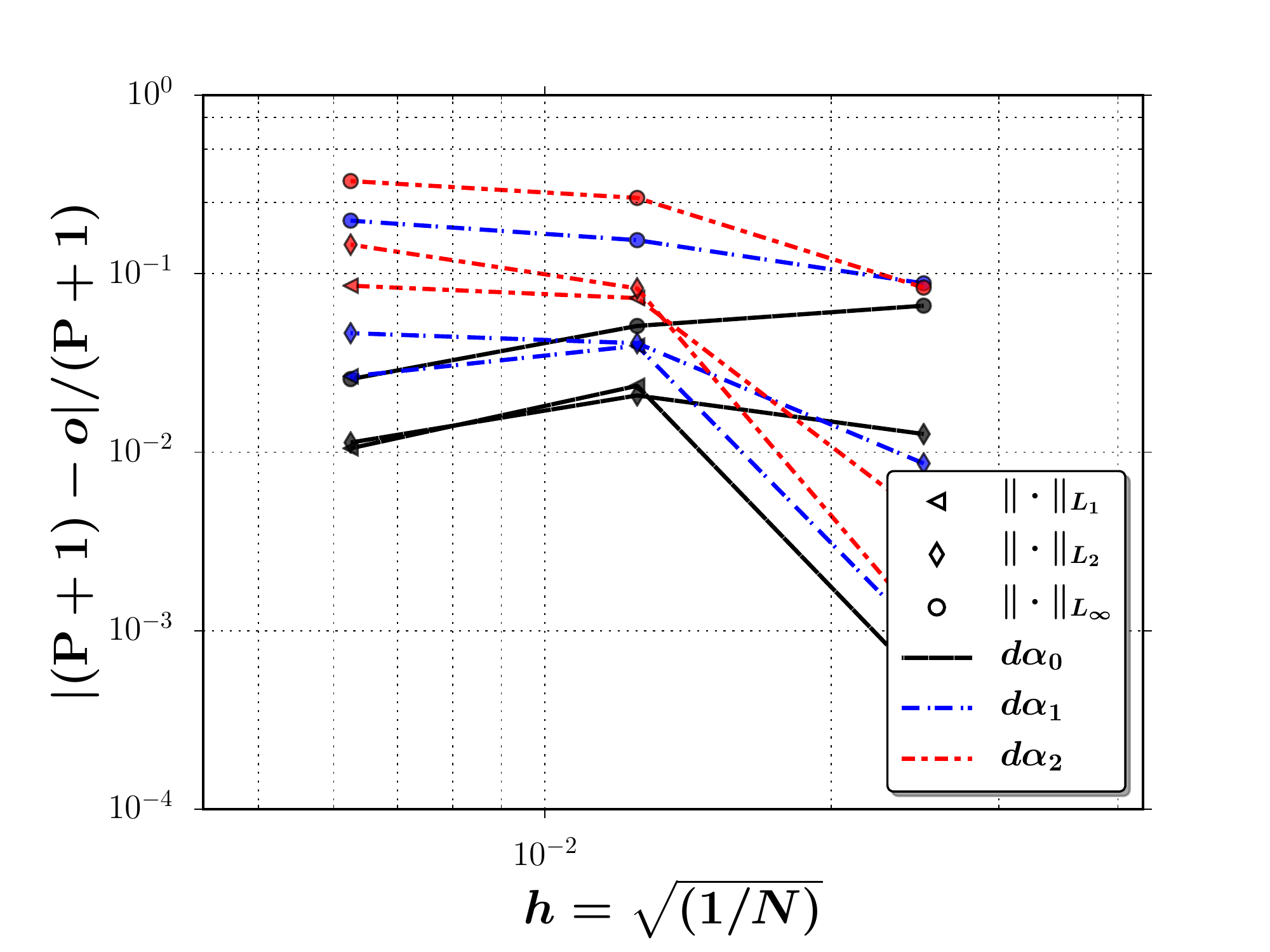}}
\caption{Evolution of the normalized discrepancy between the observed and theoretical OOAs in $L_1$, $L_2$ and $L_\infty$ norms versus mesh refinement for $\rho \tilde{\nu}$ of MS-5 on uniform grid sets}
\label{fig:grid_sens_unif_L}
\end{figure}

\begin{figure}[!hbt]
\vspace{-3mm}
\centering
\subfloat[P1]{
\includegraphics[trim = 5mm 2mm 18mm 13mm, clip,width=0.29\linewidth]{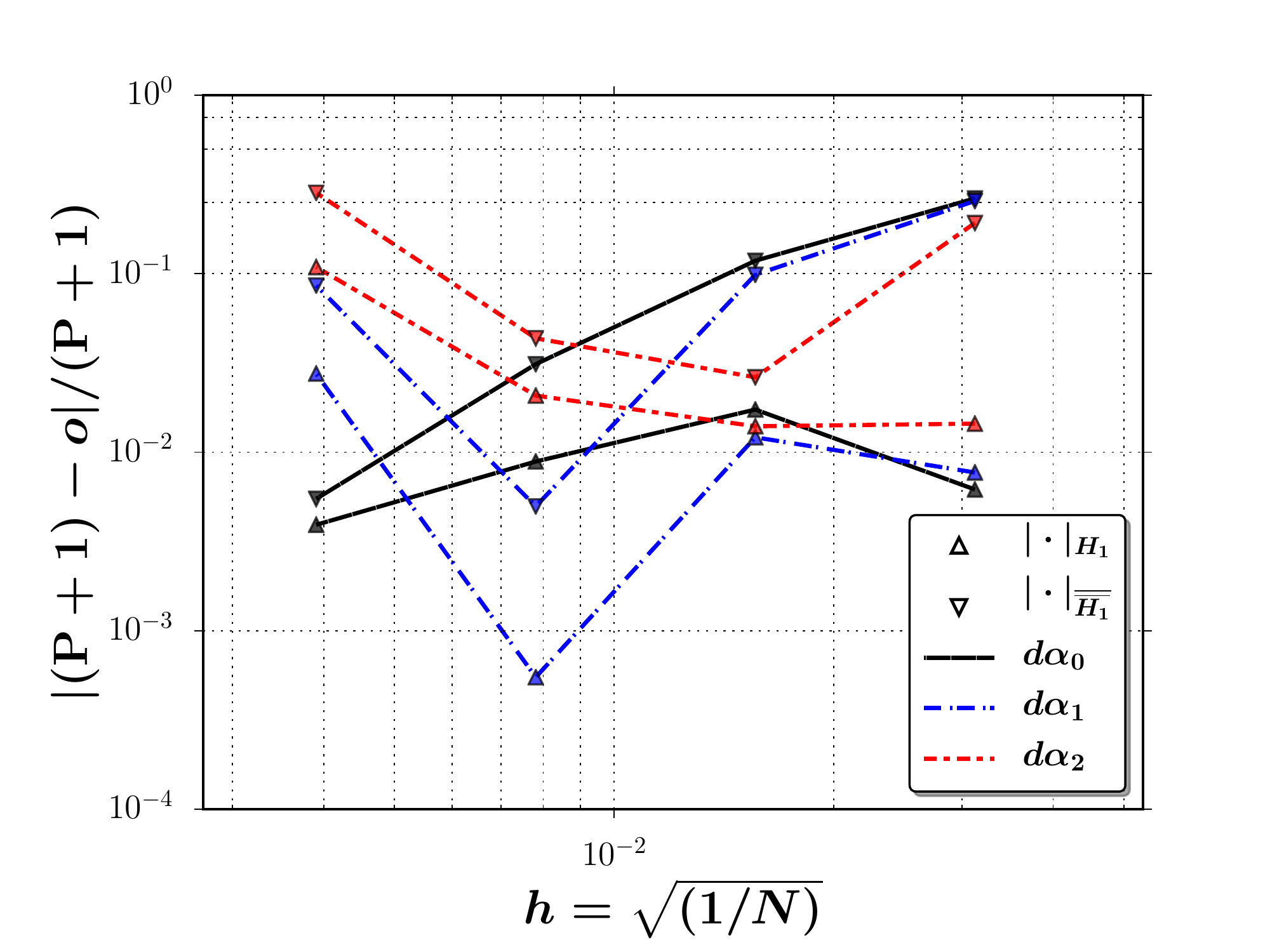}}~~~
\subfloat[P2]{
\includegraphics[trim = 5mm 2mm 18mm 13mm, clip,width=0.29\linewidth]{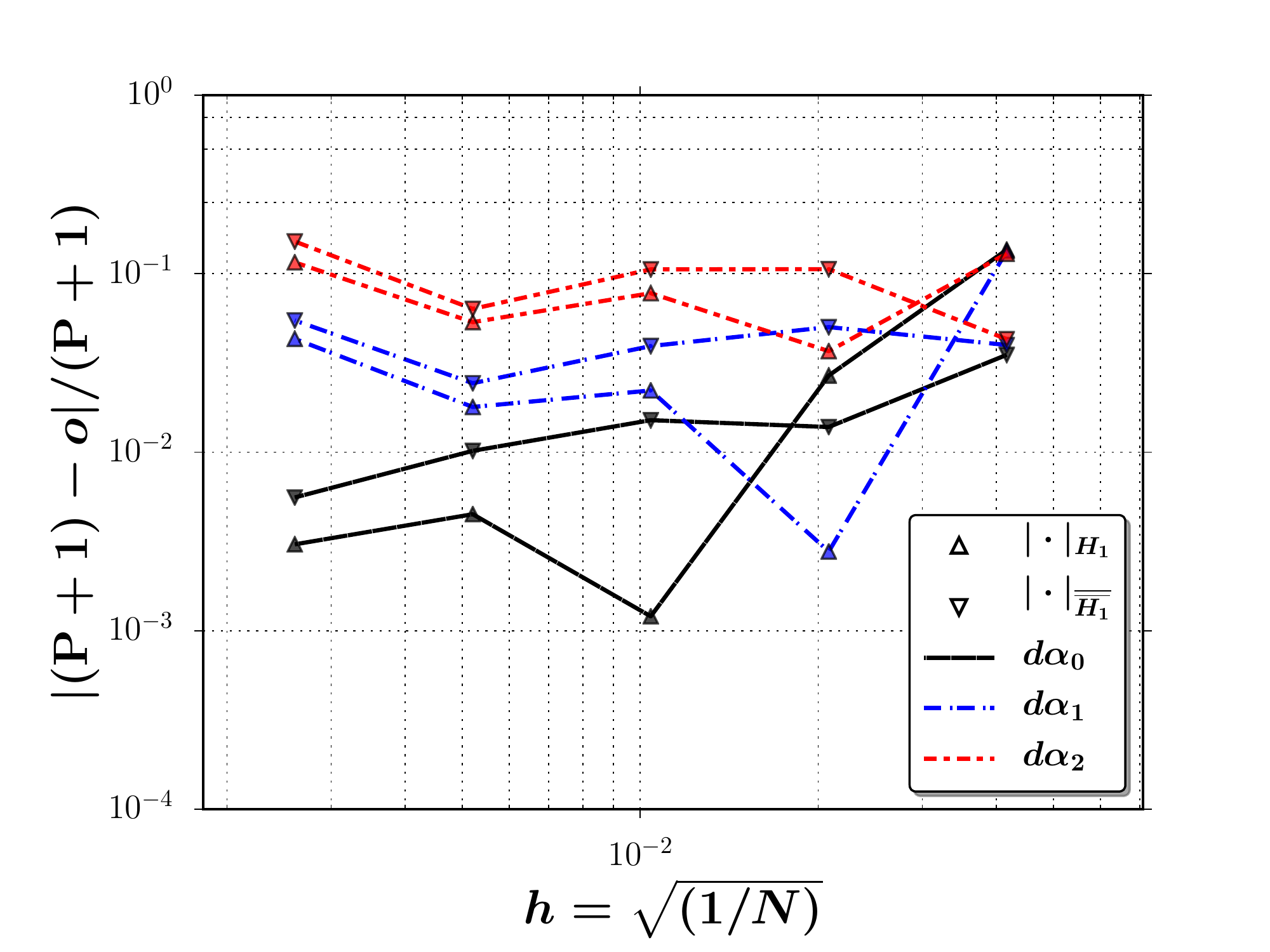}}
\vfill
\subfloat[P3]{
\includegraphics[trim = 5mm 2mm 18mm 13mm, clip,width=0.29\linewidth]{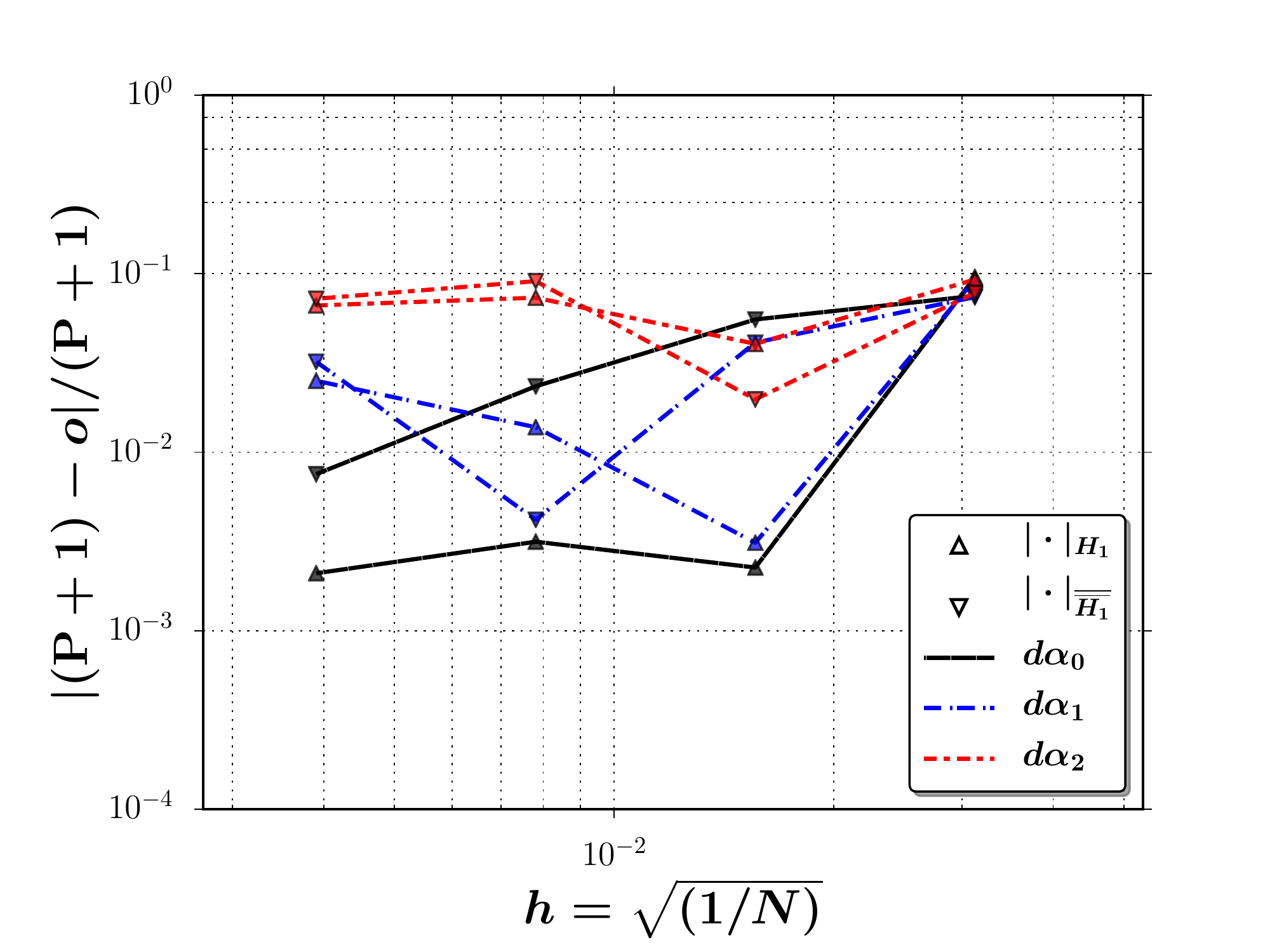}}~~~
\subfloat[P4]{
\includegraphics[trim = 5mm 2mm 18mm 13mm, clip,width=0.29\linewidth]{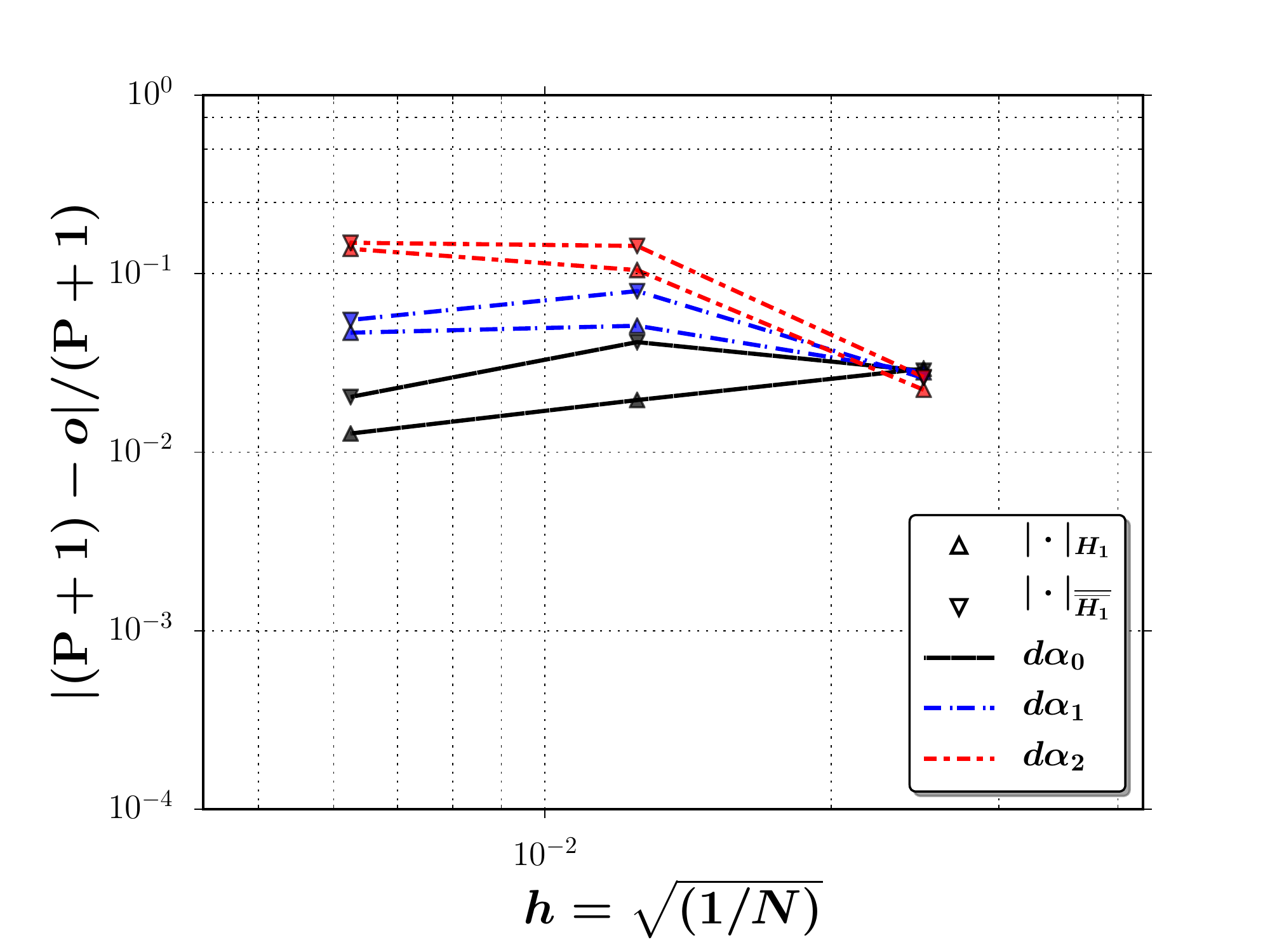}}
\caption{Evolution of the normalized discrepancy between the observed and theoretical OOAs in $H_1$ semi-norm (for uncorrected and  fully corrected derivatives) versus mesh refinement for $\rho \tilde{\nu}$ of MS-5 on uniform grid sets}
\label{fig:grid_sens_unif_H}
\end{figure}

\begin{figure}[!hbt]
\centering
\subfloat[P1]{
\includegraphics[trim = 5mm 2mm 18mm 13mm, clip,width=0.29\linewidth]{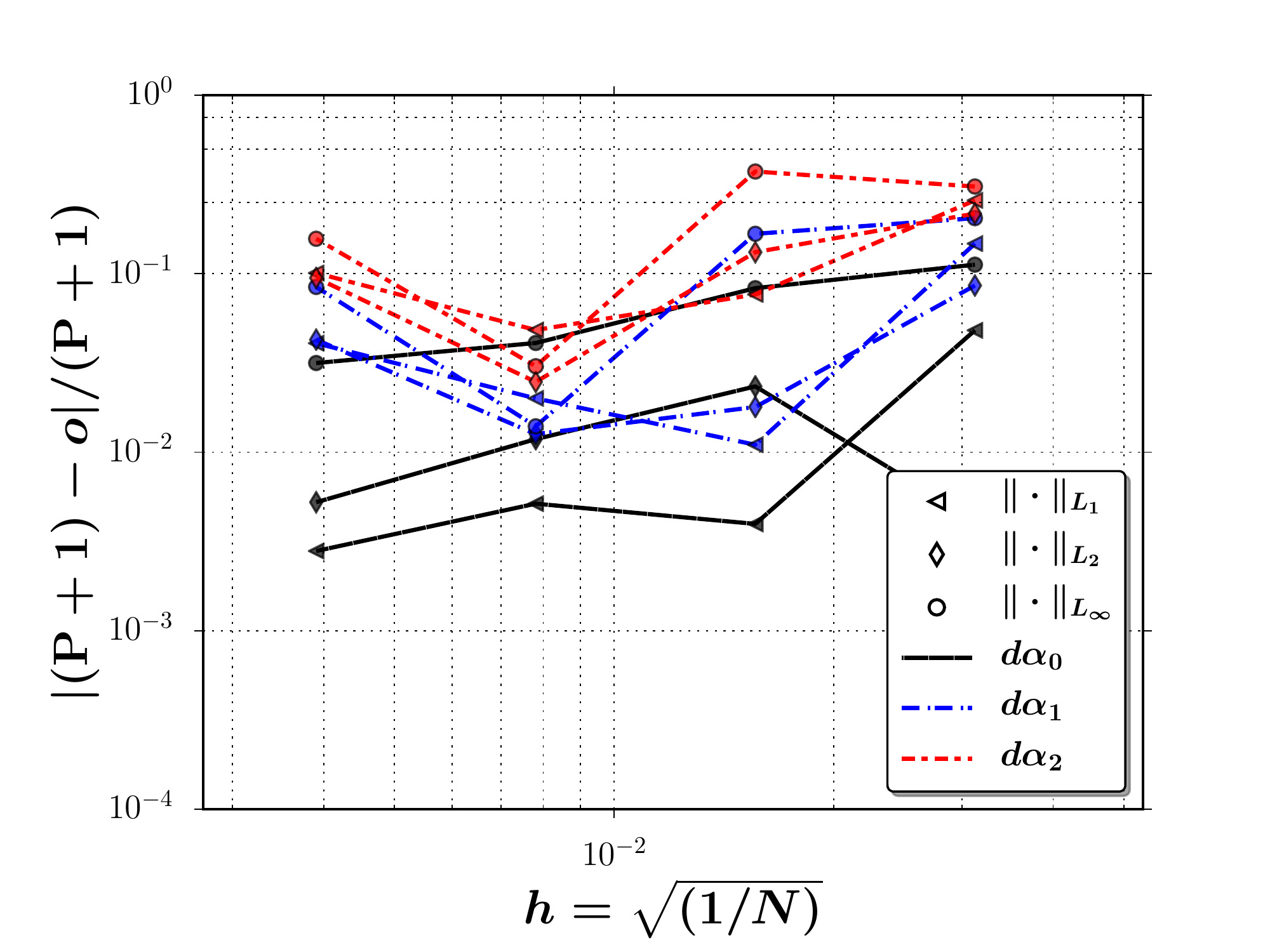}}~~~
\subfloat[P2]{
\includegraphics[trim = 5mm 2mm 18mm 13mm, clip,width=0.29\linewidth]{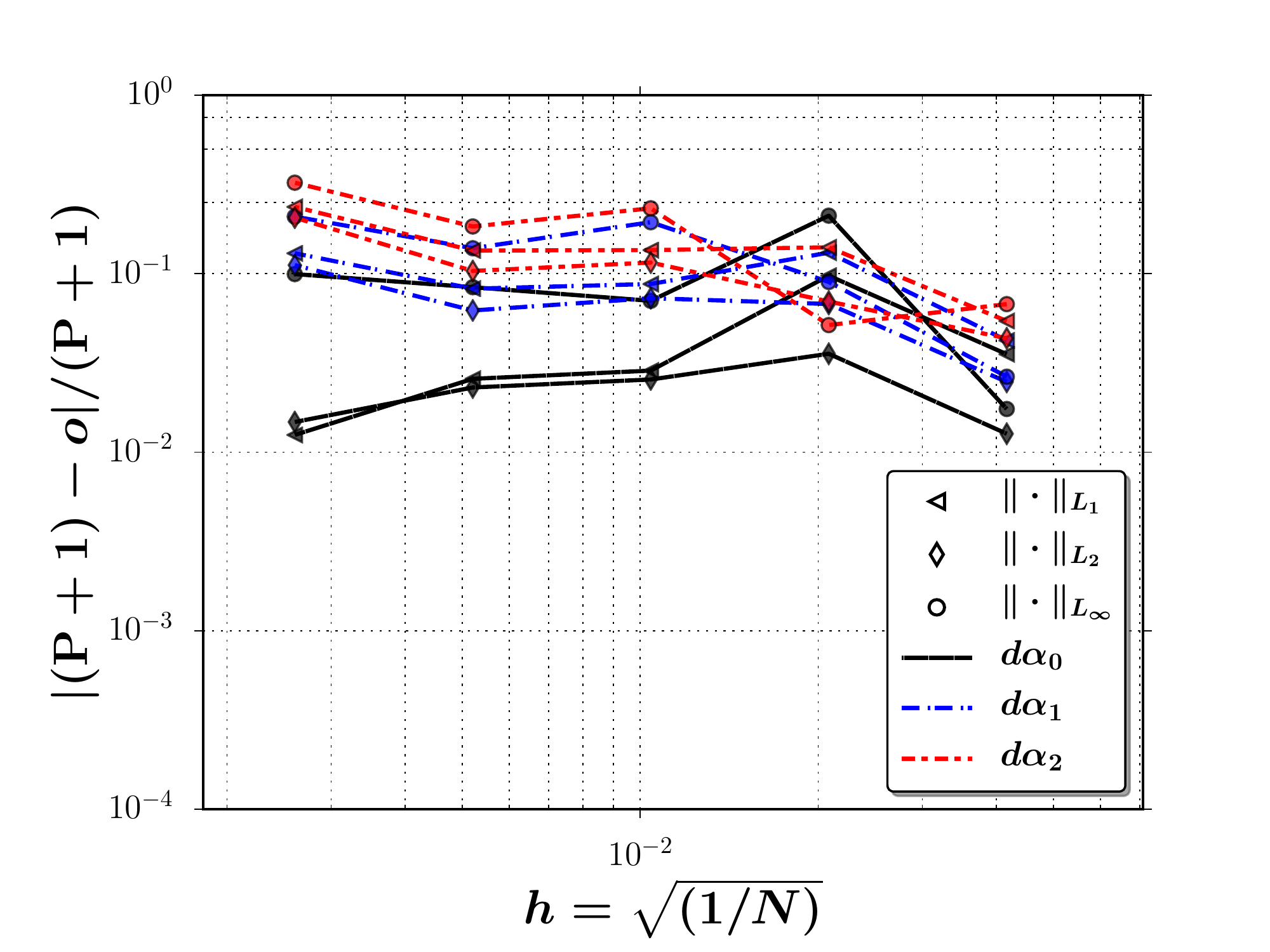}}
\vfill
\subfloat[P3]{
\includegraphics[trim = 5mm 2mm 18mm 13mm, clip,width=0.29\linewidth]{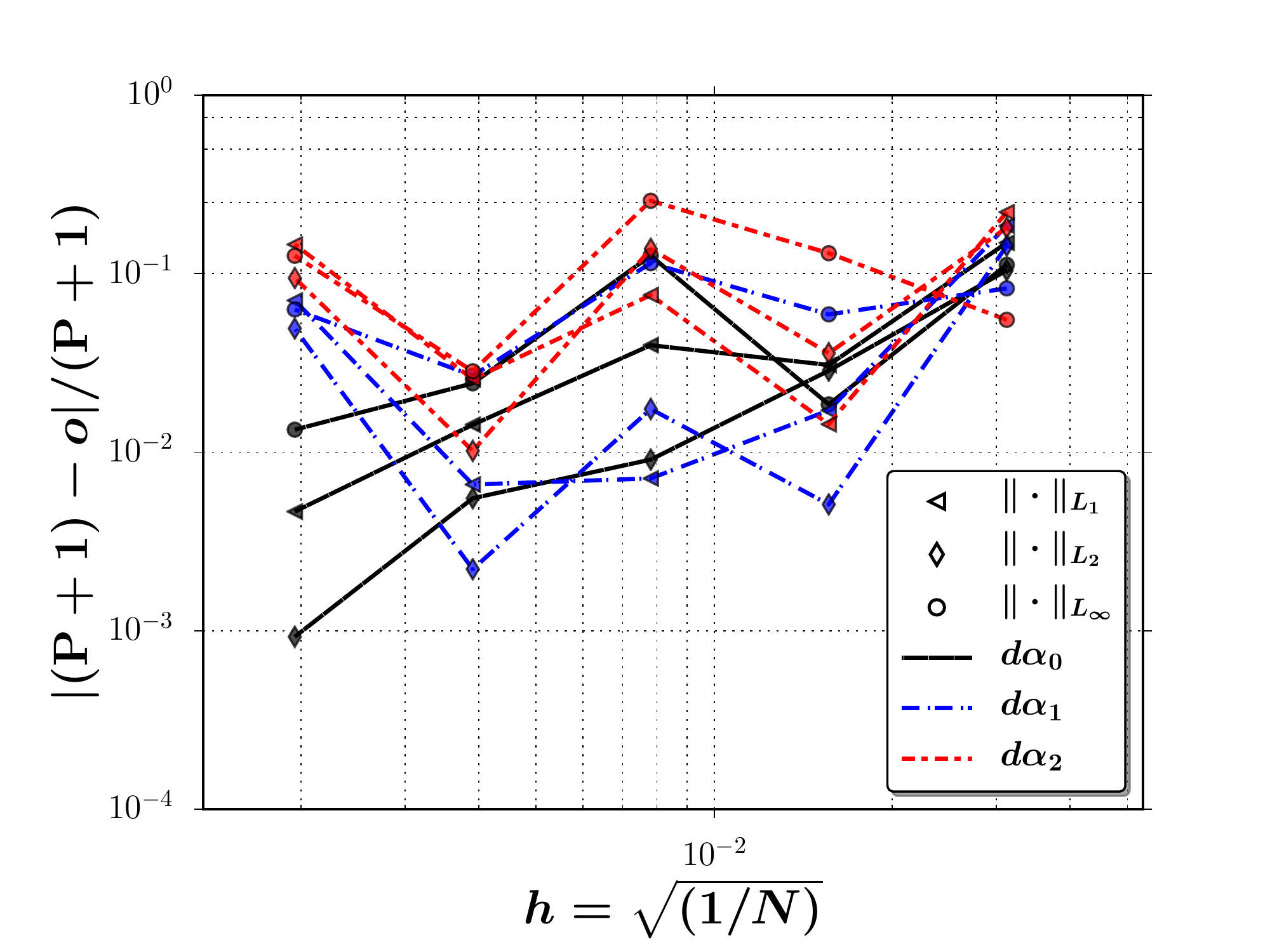}}~~~
\subfloat[P4]{
\includegraphics[trim = 5mm 2mm 18mm 13mm, clip,width=0.29\linewidth]{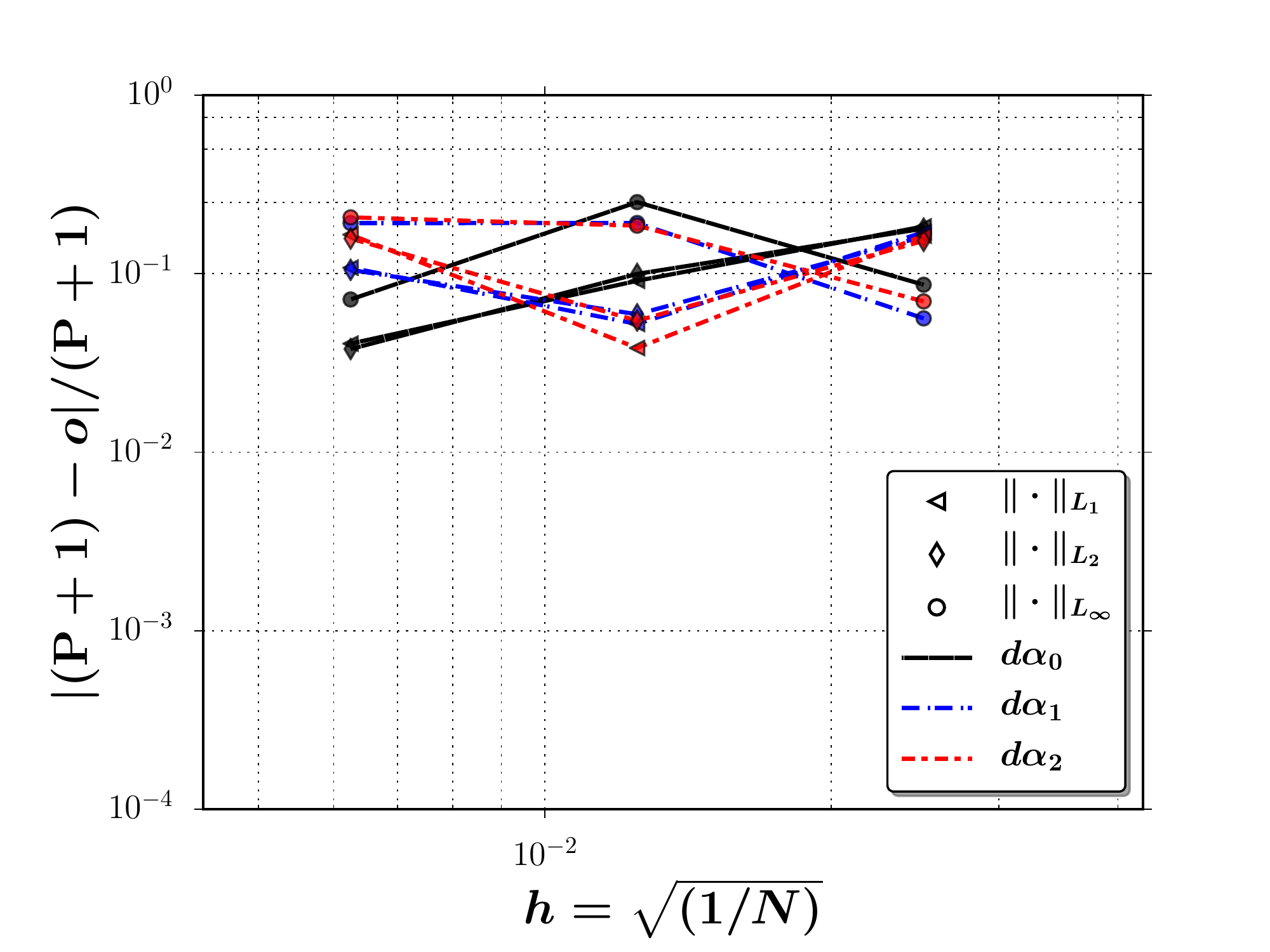}}
\caption{Evolution of the normalized discrepancy between the observed and theoretical OOAs in $L_1$, $L_2$ and $L_\infty$ norms versus mesh refinement for $\rho \tilde{\nu}$ of MS-5 on expanded grid sets}
\label{fig:grid_sens_expnd_L}
\end{figure}

\begin{figure}[!hbt]
\vspace{-6mm}
\centering
\subfloat[P1]{
\includegraphics[trim = 5mm 2mm 18mm 13mm, clip,width=0.29\linewidth]{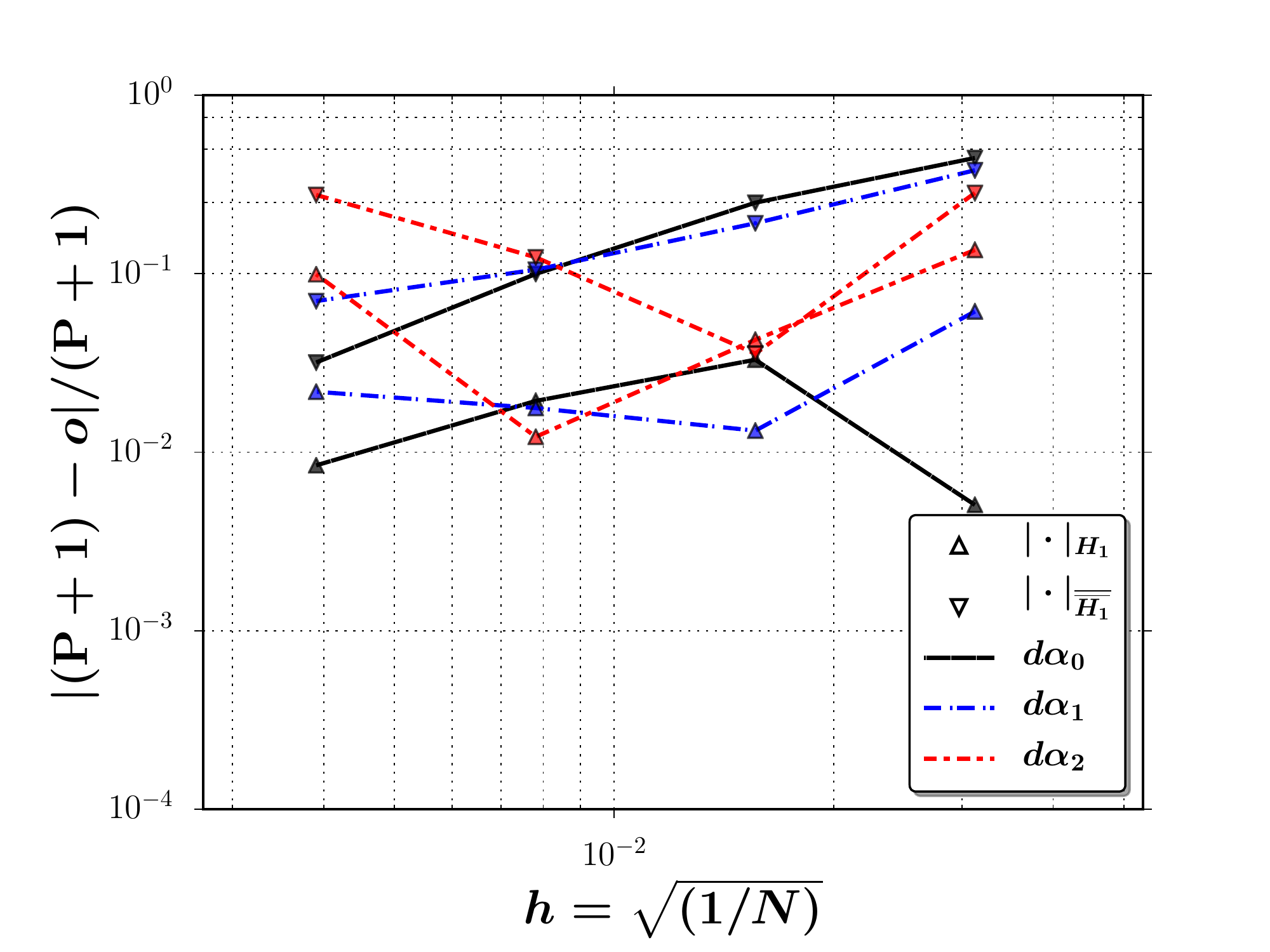}}~~~
\subfloat[P2]{
\includegraphics[trim = 5mm 2mm 18mm 13mm, clip,width=0.29\linewidth]{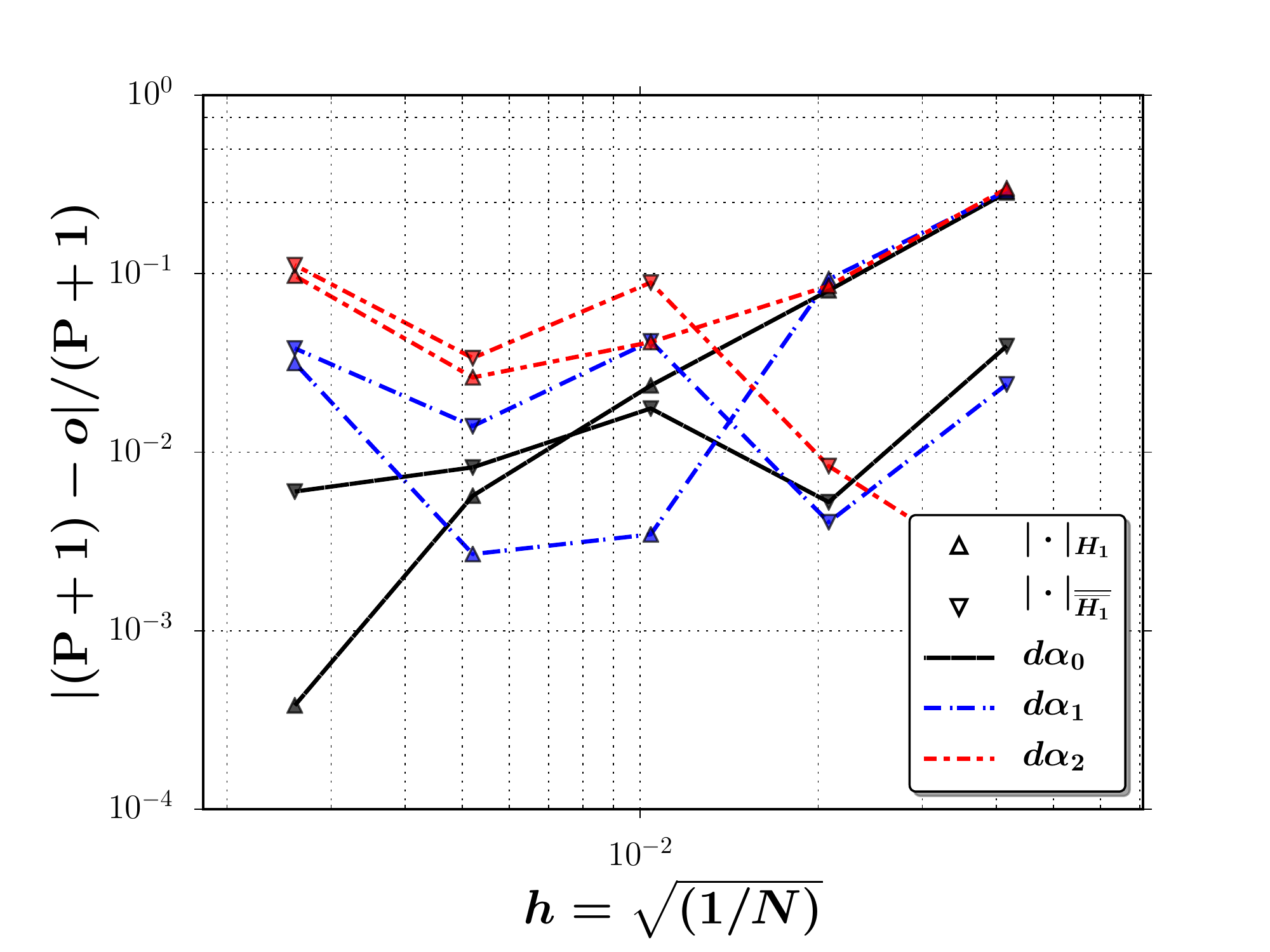}}
\vfill
\subfloat[P3]{
\includegraphics[trim = 5mm 2mm 18mm 13mm, clip,width=0.29\linewidth]{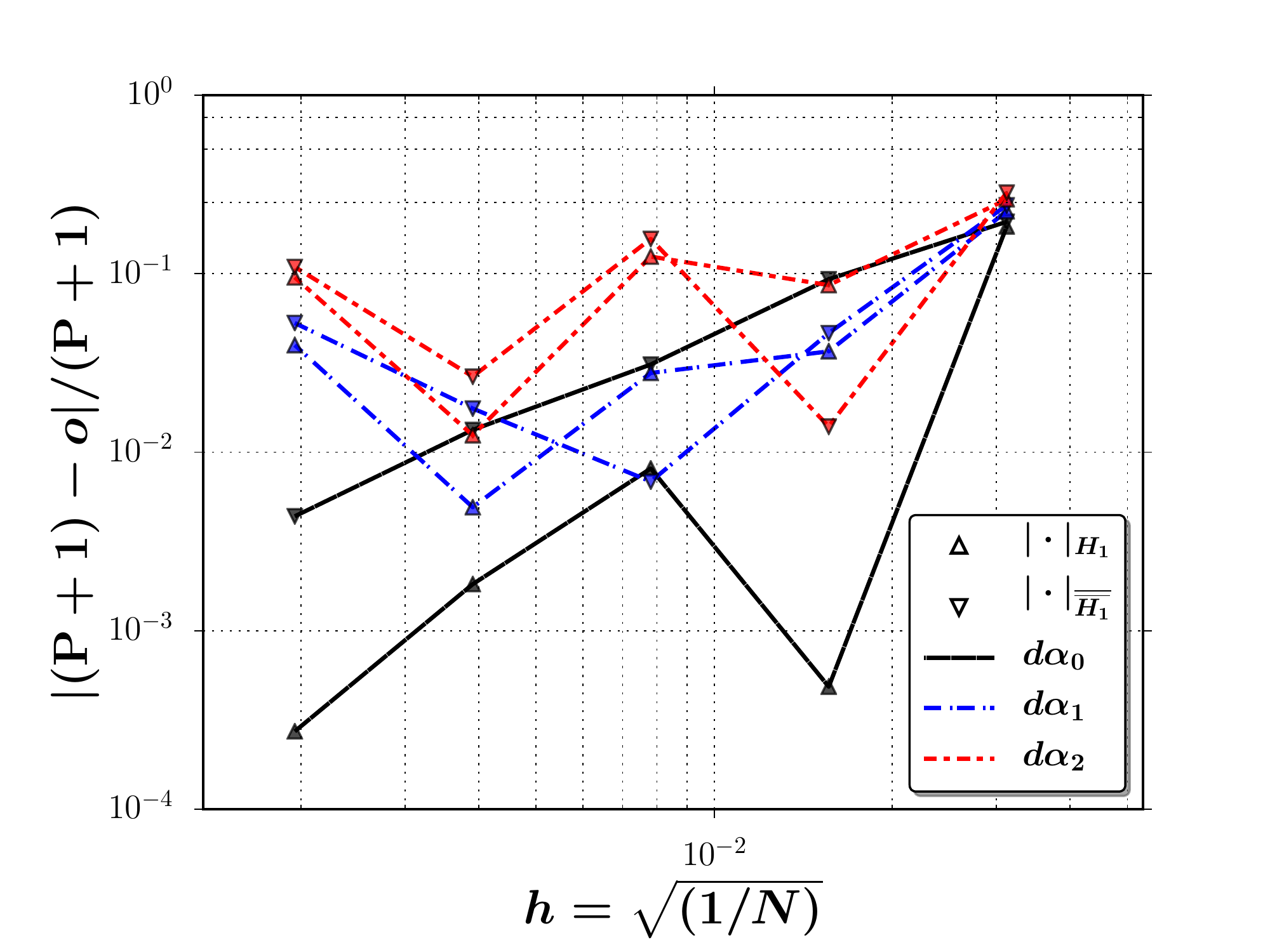}}~~~
\subfloat[P4]{
\includegraphics[trim = 5mm 2mm 18mm 13mm, clip,width=0.29\linewidth]{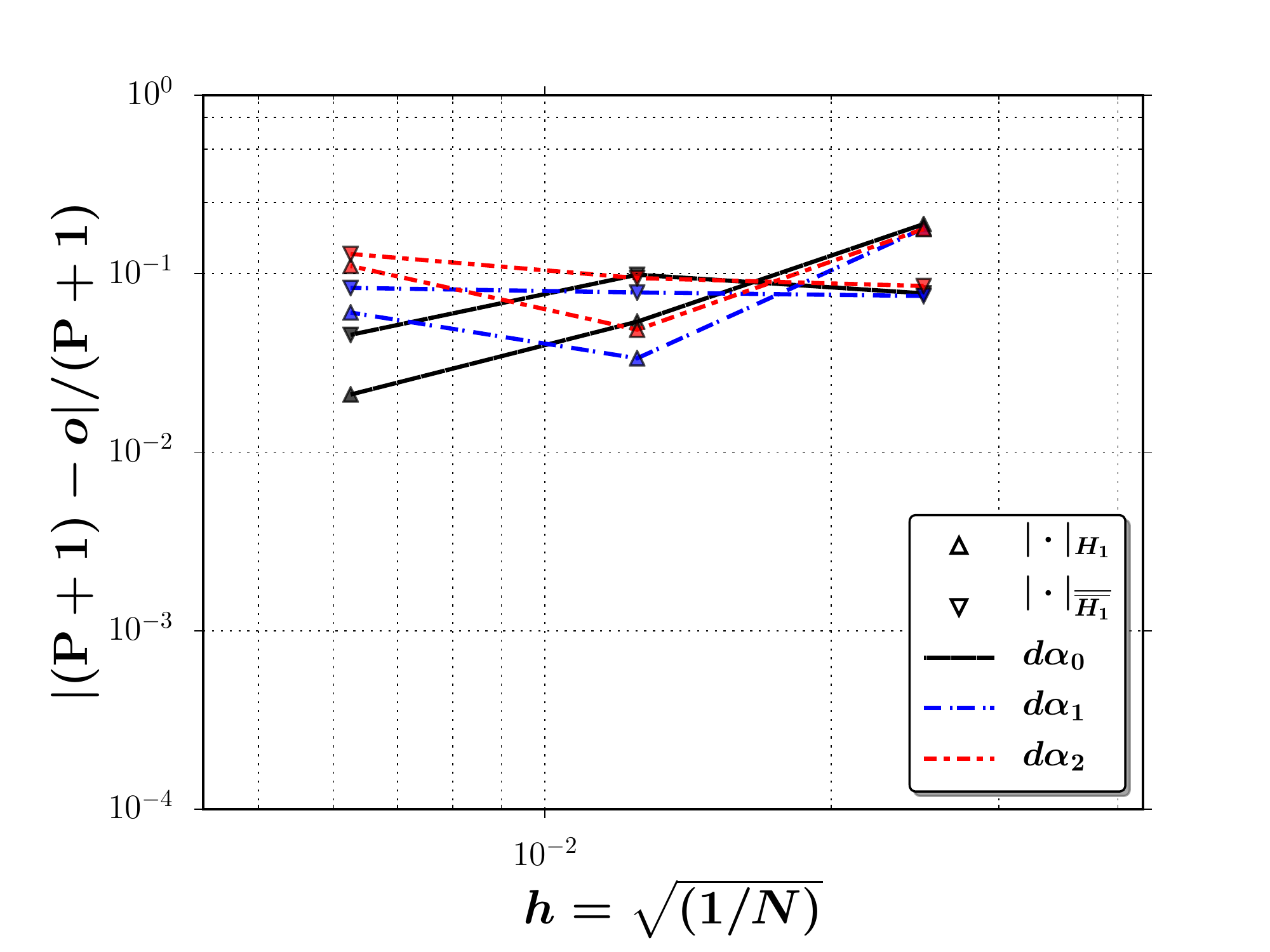}}
\caption{Evolution of the normalized discrepancy between the observed and theoretical OOAs in $H_1$ semi-norm (for uncorrected and  fully corrected derivatives) versus mesh refinement for $\rho \tilde{\nu}$ of MS-5 on expanded grid sets}
\label{fig:grid_sens_expnd_H}
\end{figure}

\clearpage
\bibliography{library}

\begin{thebibliography}{10}
\expandafter\ifx\csname url\endcsname\relax
  \def\url#1{\texttt{#1}}\fi
\expandafter\ifx\csname urlprefix\endcsname\relax\def\urlprefix{URL }\fi
\expandafter\ifx\csname href\endcsname\relax
  \def\href#1#2{#2} \def\path#1{#1}\fi

\bibitem{navah017_github}
F.~Navah, {Tools for the verification of high-order CFD solvers via the method
  of manufactured solutions (MMS)},
  \url{https://github.com/fanav/Verification_MMS_Tools}, [Online; accessed
  26-December-2017] (2017).

\bibitem{PER-GRA:2007}
F.~P{\'{e}}rez, B.~E. Granger, \href{http://ipython.org}{{{IP}ython: a System
  for Interactive Scientific Computing}}, Computing in Science and Engineering
  9~(3) (2007) 21--29.
\newblock \href {http://dx.doi.org/10.1109/MCSE.2007.53}
  {\path{doi:10.1109/MCSE.2007.53}}.
\newline\urlprefix\url{http://ipython.org}

\bibitem{Oberkampf2002}
W.~L. Oberkampf, T.~G. Trucano,
  \href{http://linkinghub.elsevier.com/retrieve/pii/S0376042102000052}{{Verification
  and validation in computational fluid dynamics}}, Progress in Aerospace
  Sciences 38~(3) (2002) 209--272.
\newblock \href {http://dx.doi.org/10.1016/S0376-0421(02)00005-2}
  {\path{doi:10.1016/S0376-0421(02)00005-2}}.
\newline\urlprefix\url{http://linkinghub.elsevier.com/retrieve/pii/S0376042102000052}

\bibitem{vermeire-2014}
B.~C. Vermeire, {Adaptive Implicit-Explicit Time Integration and High-Order
  Unstructured Methods for Implicit Large Eddy Simulation}, Ph.D. thesis,
  McGill (2014).

\bibitem{Arnold-et-al_2002}
D.~N. Arnold, F.~Brezzi, B.~Cockburn, L.~D. Marini, {Unified analysis of
  discontinuous {G}alerkin methods for elliptic problems}, SIAM Journal on
  Numerical Analysis 39~(5) (2002) pp. 1749--1779.

\bibitem{RalfHartmann2008}
{Ralf Hartmann}, \href{http://elib.dlr.de/57074/1/Har08b.pdf}{{Numerical
  Analysis of Higher Order Discontinuous Galerkin Finite Element Methods}}, in:
  VKI LS 2008-08: CFD - ADIGMA course on very high order discretization
  methods, Oct. 13-17, 2008, 2008.
\newline\urlprefix\url{http://elib.dlr.de/57074/1/Har08b.pdf}

\bibitem{roache-1998}
P.~J. Roache, {Verification and Validation in Computational Science and
  Engineering}, Hermosa publishers, Albuquerque, NM, 1998.

\bibitem{oberkampf2010verification}
W.~Oberkampf, C.~Roy,
  \href{https://books.google.ca/books?id=7d26zLEJ1FUC}{Verification and
  Validation in Scientific Computing}, Cambridge University Press, 2010.
\newline\urlprefix\url{https://books.google.ca/books?id=7d26zLEJ1FUC}

\bibitem{Salari}
K.~Salari, P.~Knupp,
  \href{http://prod.sandia.gov/techlib/access-control.cgi/2000/001444.pdf}{{Code
  Verification by the Method of Manufactured Solutions}}, Tech. rep., Sandia
  National Laboratories (2000).
\newline\urlprefix\url{http://prod.sandia.gov/techlib/access-control.cgi/2000/001444.pdf}

\bibitem{Roy2002}
C.~J. Roy, T.~M. Smith, C.~C. Ober,
  \href{http://arc.aiaa.org/doi/abs/10.2514/6.2002-3110}{{Verification of a
  Compressible CFD Code Using the Method of Manufactured Solutions}}, 32$^{nd}$
  AIAA Fluid Dynamics Conference and Exhibit (2002) 1--14\href
  {http://dx.doi.org/10.2514/6.2002-3110} {\path{doi:10.2514/6.2002-3110}}.
\newline\urlprefix\url{http://arc.aiaa.org/doi/abs/10.2514/6.2002-3110}

\bibitem{Roy2005}
C.~J. Roy, {Review of code and solution verification procedures for
  computational simulation}, Journal of Computational Physics 205 (2005)
  131--156.
\newblock \href {http://dx.doi.org/10.1016/j.jcp.2004.10.036}
  {\path{doi:10.1016/j.jcp.2004.10.036}}.

\bibitem{Roy2004}
C.~J. Roy, C.~C. Nelson, T.~M. Smith, C.~C. Ober, {Verification of
  Euler/Navier-Stokes codes using the method of manufactured solutions},
  International Journal for Numerical Methods in Fluids 44~(6) (2004) 599--620.
\newblock \href {http://dx.doi.org/10.1002/fld.660}
  {\path{doi:10.1002/fld.660}}.

\bibitem{Bond2007}
R.~B. Bond, C.~C. Ober, P.~M. Knupp, S.~W. Bova, {Manufactured Solution for
  Computational Fluid Dynamics Boundary Condition Verification}, AIAA Journal
  45~(9) (2007) 2224--2236.
\newblock \href {http://dx.doi.org/10.2514/1.28099}
  {\path{doi:10.2514/1.28099}}.

\bibitem{Shunn2007}
L.~Shunn, F.~Ham, {Method of Manufactured Solutions Applied to variable density
  flow solvers}, Center for Turbulence Research, Annual Research Briefs (2007)
  155--167.

\bibitem{Vedovoto2011}
J.~M. Vedovoto, A.~da~{Silveira Neto}, A.~Mura, L.~F. {Figueira da Silva},
  {Application of the method of manufactured solutions to the verification of a
  pressure-based finite-volume numerical scheme}, Computers and Fluids 51~(1)
  (2011) 85--99.
\newblock \href {http://dx.doi.org/10.1016/j.compfluid.2011.07.014}
  {\path{doi:10.1016/j.compfluid.2011.07.014}}.

\bibitem{Ulerich2012}
R.~Ulerich, K.~C. Estacio-Hiroms, N.~Malaya, R.~D. Moser, {A Transient
  Manufactured Solution for the Compressible Navier-Stokes Equations With a
  Power Lawviscosity}, in: $10^{th}$ World Congress on Computational Mechanics,
  2012, pp. 1--16.
\newblock \href {http://dx.doi.org/10.5151/meceng-wccm2012-16661}
  {\path{doi:10.5151/meceng-wccm2012-16661}}.

\bibitem{Roy2007}
C.~Roy, E.~Tendean, S.~Veluri, R.~Rifki, E.~Luke, S.~Hebert,
  \href{https://arc.aiaa.org/doi/abs/10.2514/6.2007-4203}{{Verification of RANS
  Turbulence Models in Loci-CHEM using the Method of Manufactured Solutions}},
  in: $18^{th}$ AIAA Computational Fluid Dynamics Conference, no. June, Miami,
  FL, 2007, p.~16.
\newblock \href {http://dx.doi.org/10.2514/6.2007-4203}
  {\path{doi:10.2514/6.2007-4203}}.
\newline\urlprefix\url{https://arc.aiaa.org/doi/abs/10.2514/6.2007-4203}

\bibitem{Veluri2012}
S.~P. Veluri, C.~J. Roy, E.~A. Luke,
  \href{http://dx.doi.org/10.1016/j.compfluid.2012.04.028}{{Comprehensive code
  verification techniques for finite volume CFD codes}}, Computers and Fluids
  70 (2012) 59--72.
\newblock \href {http://dx.doi.org/10.1016/j.compfluid.2012.04.028}
  {\path{doi:10.1016/j.compfluid.2012.04.028}}.
\newline\urlprefix\url{http://dx.doi.org/10.1016/j.compfluid.2012.04.028}

\bibitem{Oliver-et-al_2012}
T.~Oliver, K.~Estacio-Hiroms, N.~Malaya, G.~Carey,
  \href{http://arc.aiaa.org/doi/10.2514/6.2012-80}{{Manufactured Solutions for
  the Favre-Averaged Navier-Stokes Equations with Eddy-Viscosity Turbulence
  Models}}, in: 50$^{th}$ {AIAA} Aerospace Sciences Meeting including the New
  Horizons Forum and Aerospace Exposition, American Institute of Aeronautics
  and Astronautics, Nashville, Tennessee, 2012, pp. 1--19.
\newblock \href {http://dx.doi.org/10.2514/6.2012-80}
  {\path{doi:10.2514/6.2012-80}}.
\newline\urlprefix\url{http://arc.aiaa.org/doi/10.2514/6.2012-80}

\bibitem{tremblay2006}
D.~Tremblay, S.~Etienne, D.~Pelletier,
  \href{http://arc.aiaa.org/doi/10.2514/6.2006-3218}{{Code Verification and the
  Method of Manufactured Solutions for Fluid-Structure Interaction Problems}},
  in: $36^{th}$ AIAA Fluid Dynamics Conference and Exhibit, American Institute
  of Aeronautics and Astronautics, 2006, pp. 1--11.
\newblock \href {http://dx.doi.org/10.2514/6.2006-3218}
  {\path{doi:10.2514/6.2006-3218}}.
\newline\urlprefix\url{http://arc.aiaa.org/doi/10.2514/6.2006-3218}

\bibitem{Choudhary2016}
A.~Choudhary, C.~J. Roy, J.~F. Dietiker, M.~Shahnam, R.~Garg, J.~Musser, {Code
  verification for multiphase flows using the method of manufactured
  solutions}, International Journal of Multiphase Flow 80 (2016) 150--163.
\newblock \href {http://dx.doi.org/10.1016/j.ijmultiphaseflow.2015.12.006}
  {\path{doi:10.1016/j.ijmultiphaseflow.2015.12.006}}.

\bibitem{Eca2007a}
L.~E{\c{c}}a, M.~Hoekstra, A.~Hay, D.~Pelletier,
  \href{http://www.tandfonline.com/doi/abs/10.1080/10618560701553436}{{A
  manufactured solution for a two-dimensional steady wall-bounded
  incompressible turbulent flow}}, International Journal of Computational Fluid
  Dynamics 21~(3-4) (2007) 175--188.
\newblock \href {http://dx.doi.org/10.1080/10618560701553436}
  {\path{doi:10.1080/10618560701553436}}.
\newline\urlprefix\url{http://www.tandfonline.com/doi/abs/10.1080/10618560701553436}

\bibitem{Eca2007}
L.~E{\c{c}}a, M.~Hoesktra, a.~Hay, D.~Pelletier, {On the construction of
  manufactured solutions for one and two-equation eddy-viscosity models},
  International Journal for Numerical Methods in Fluids 54~(November 2006)
  (2007) 119--154.
\newblock \href {http://dx.doi.org/10.1002/fld.1387}
  {\path{doi:10.1002/fld.1387}}.

\bibitem{Eca2012}
L.~E{\c{c}}a, M.~Hoekstra, G.~Vaz, {Manufactured solutions for steady-flow
  Reynolds-averaged Navier–Stokes solvers}, International Journal of
  Computational Fluid Dynamics 26~(5) (2012) 313--332.
\newblock \href {http://dx.doi.org/10.1080/10618562.2012.717617}
  {\path{doi:10.1080/10618562.2012.717617}}.

\bibitem{Eca2014}
L.~E{\c{c}}a, M.~Hoekstra,
  \href{http://dx.doi.org/10.1016/j.jcp.2014.01.006}{{A procedure for the
  estimation of the numerical uncertainty of CFD calculations based on grid
  refinement studies}}, Journal of Computational Physics 262 (2014) 104--130.
\newblock \href {http://dx.doi.org/10.1016/j.jcp.2014.01.006}
  {\path{doi:10.1016/j.jcp.2014.01.006}}.
\newline\urlprefix\url{http://dx.doi.org/10.1016/j.jcp.2014.01.006}

\bibitem{Eca2016}
L.~E{\c{c}}a, C.~Klaij, G.~Vaz, M.~Hoekstra, F.~Pereira,
  \href{http://linkinghub.elsevier.com/retrieve/pii/S0021999116000036}{{On code
  verification of RANS solvers}}, Journal of Computational Physics 310 (2016)
  418--439.
\newblock \href {http://dx.doi.org/10.1016/j.jcp.2016.01.002}
  {\path{doi:10.1016/j.jcp.2016.01.002}}.
\newline\urlprefix\url{http://linkinghub.elsevier.com/retrieve/pii/S0021999116000036}

\bibitem{Malaya2013}
N.~Malaya, K.~C. Estacio-Hiroms, R.~H. Stogner, K.~W. Schulz, P.~T. Bauman,
  G.~F. Carey, {MASA: A library for verification using manufactured and
  analytical solutions}, Engineering with Computers 29 (2013) 487--496.
\newblock \href {http://dx.doi.org/10.1007/s00366-012-0267-9}
  {\path{doi:10.1007/s00366-012-0267-9}}.

\bibitem{Fidkowski2013}
K.~J. Fidkowski,
  \href{https://pdfs.semanticscholar.org/9eef/590097b43801cbcbcc3e83fdb70235b4515c.pdf}{{High-Order
  Output-Based Adaptive Methods for Steady and Unsteady Aerodynamics}}, in:
  $37^{th}$ Advanced CFD Lectures series, Von Karman Institute for Fluid
  Dynamics, 2013, pp. 1--108.
\newline\urlprefix\url{https://pdfs.semanticscholar.org/9eef/590097b43801cbcbcc3e83fdb70235b4515c.pdf}

\bibitem{Manuel2014}
R.~L. Manuel, J.~Bull, J.~Crabill, J.~Romero, A.~Sheshadri, J.~E.~W. Ii,
  D.~Williams, F.~Palacios, A.~Jameson, {Verification and Validation of HiFiLES
  : a High-Order LES unstructured solver on multi-GPU platforms}, AIAA
  Aviation~(June) (2014) 1--27.
\newblock \href {http://dx.doi.org/10.2514/6.2014-3168}
  {\path{doi:10.2514/6.2014-3168}}.

\bibitem{Galbraith2015}
M.~C. Galbraith, S.~Allmaras, D.~L. Darmofal,
  \href{https://arc.aiaa.org/doi/abs/10.2514/6.2015-0818}{{A Verification
  Driven Process for Rapid Development of CFD Software}}, in: $53^{rd}$ AIAA
  Aerospace Sciences Meeting, American Institute of Aeronautics and
  Astronautics, Kissimmee, Florida, 2015.
\newblock \href {http://dx.doi.org/10.2514/6.2015-0818}
  {\path{doi:10.2514/6.2015-0818}}.
\newline\urlprefix\url{https://arc.aiaa.org/doi/abs/10.2514/6.2015-0818}

\bibitem{Navah_eccomas_016}
F.~Navah, S.~Nadarajah,
  \href{https://www.eccomas2016.org/proceedings/pdf/10822.pdf}{{On the
  verification of high-order CFD solvers}}, in: European Congress on
  Computational Methods in Applied Sciences and Engineering, Hersonissos,
  Crete, Greece, 2016.
\newline\urlprefix\url{https://www.eccomas2016.org/proceedings/pdf/10822.pdf}

\bibitem{silva2010}
H.~G. Silva, L.~F. Souza, M.~A.~F. Medeiros,
  \href{http://doi.wiley.com/10.1002/fld.2156}{{Verification of a mixed
  high-order accurate DNS code for laminar turbulent transition by the method
  of manufactured solutions}}, International Journal for Numerical Methods in
  Fluids 64~(3) (2010) 336--354.
\newblock \href {http://arxiv.org/abs/fld.2156} {\path{arXiv:fld.2156}}, \href
  {http://dx.doi.org/10.1002/fld.2156} {\path{doi:10.1002/fld.2156}}.
\newline\urlprefix\url{http://doi.wiley.com/10.1002/fld.2156}

\bibitem{Allmaras-et-al_2012}
S.~R. Allmaras, F.~T. Johnson, P.~R. Spalart, {Modifications and Clarifications
  for the Implementation of the Spalart-Allmaras Turbulence Model}, in:
  7$^{th}$ International Conference on Computational Fluid Dynamics, 2012.

\bibitem{Wang-Gao_2009}
Z.~Wang, H.~Gao,
  \href{http://linkinghub.elsevier.com/retrieve/pii/S0021999109004239}{{A
  unifying lifting collocation penalty formulation including the discontinuous
  Galerkin, spectral volume/difference methods for conservation laws on mixed
  grids}}, Journal of Computational Physics 228~(21) (2009) 8161--8186.
\newblock \href {http://dx.doi.org/10.1016/j.jcp.2009.07.036}
  {\path{doi:10.1016/j.jcp.2009.07.036}}.
\newline\urlprefix\url{http://linkinghub.elsevier.com/retrieve/pii/S0021999109004239}

\bibitem{Huynh_2009a}
H.~T. Huynh, \href{https://arc.aiaa.org/doi/abs/10.2514/6.2009-403}{{A
  Reconstruction Approach to High-Order Schemes Including Discontinuous
  Galerkin for Diffusion}}, in: 47$^{th}$ {AIAA} Aerospace Sciences Meeting
  including The New Horizons Forum and Aerospace Exposition, American Institute
  of Aeronautics and Astronautics, Orlando, Florida, 2009, pp. 1--42.
\newblock \href {http://dx.doi.org/10.2514/6.2009-403}
  {\path{doi:10.2514/6.2009-403}}.
\newline\urlprefix\url{https://arc.aiaa.org/doi/abs/10.2514/6.2009-403}

\bibitem{Gao-Wang_2013}
H.~Gao, Z.~Wang,
  \href{http://linkinghub.elsevier.com/retrieve/pii/S0021999112004767}{{A
  conservative correction procedure via reconstruction formulation with the
  Chain-Rule divergence evaluation}}, Journal of Computational Physics 232~(1)
  (2013) 7--13.
\newblock \href {http://dx.doi.org/10.1016/j.jcp.2012.08.030}
  {\path{doi:10.1016/j.jcp.2012.08.030}}.
\newline\urlprefix\url{http://linkinghub.elsevier.com/retrieve/pii/S0021999112004767}

\bibitem{Roe_1981}
P.~Roe,
  \href{http://linkinghub.elsevier.com/retrieve/pii/0021999181901285}{{Approximate
  Riemann solvers, parameter vectors, and difference schemes}}, Journal of
  Computational Physics 43~(2) (1981) 357--372.
\newblock \href {http://dx.doi.org/10.1016/0021-9991(81)90128-5}
  {\path{doi:10.1016/0021-9991(81)90128-5}}.
\newline\urlprefix\url{http://linkinghub.elsevier.com/retrieve/pii/0021999181901285}

\bibitem{Burgess2011}
N.~K. Burgess, {An Adaptive Discontinuous Galerkin Solver for Aerodynamic
  Flows}, Ph.D. thesis, University of Wyoming (2011).

\bibitem{Hesthaven-Warburton_2008}
J.~S. Hesthaven, T.~Warburton, {Nodal Discontinuous Galerkin Methods}, Vol.~54
  of Texts in Applied Mathematics, Springer, 2008.

\bibitem{Bassi-Rebay_2000a}
F.~Bassi, S.~Rebay, {A high order discontinuous {G}alerkin method for
  compressible turbulent flows}, in: Discontinuous Galerkin Methods, Springer,
  2000, pp. 77--88.

\bibitem{Wang-et-al_2011a}
Z.~J. Wang, H.~Gao, T.~Haga,
  \href{http://www.worldscientific.com/worldscibooks/10.1142/7792}{{A unifying
  discontinuous formulation for hybrid meshes}}, in: Adaptive High-Order
  Methods in Computational Fluid Dynamics, World Scientific Pub., 2011, Ch.~15,
  pp. 423--454.
\newline\urlprefix\url{http://www.worldscientific.com/worldscibooks/10.1142/7792}

\bibitem{Mengaldo2014}
G.~Mengaldo, D.~{De Grazia}, J.~Peiro, A.~Farrington, F.~Witherden, S.~J.
  Sherwin, {A Guide to the Implementation of Boundary Conditions in Compact
  High-Order Methods for Compressible Aerodynamics}, in: 7$^{th}$ AIAA
  Theoretical Fluid Mechanics Conference AIAA, Atlanta, GA, 2014.
\newblock \href {http://dx.doi.org/10.2514/6.2014-2923}
  {\path{doi:10.2514/6.2014-2923}}.

\bibitem{Martins-2000}
J.~Martins, I.~Kroo, J.~Alonso,
  \href{http://arc.aiaa.org/doi/10.2514/6.2000-689}{{An automated method for
  sensitivity analysis using complex variables}}, in: $38^{th}$ Aerospace
  Sciences Meeting and Exhibit, American Institute of Aeronautics and
  Astronautics, 2000.
\newblock \href {http://dx.doi.org/10.2514/6.2000-689}
  {\path{doi:10.2514/6.2000-689}}.
\newline\urlprefix\url{http://arc.aiaa.org/doi/10.2514/6.2000-689}

\bibitem{petsc-efficient}
S.~Balay, W.~D. Gropp, L.~C. McInnes, B.~F. Smith, {Efficient Management of
  Parallelism in Object Oriented Numerical Software Libraries}, in: E.~Arge,
  A.~M. Bruaset, H.~P. Langtangen (Eds.), Modern Software Tools in Scientific
  Computing, Birkh{{\"{a}}}user Press, 1997, pp. 163--202.

\bibitem{salas2007}
M.~D. Salas,
  \href{http://linkinghub.elsevier.com/retrieve/pii/S0045793006000053}{{Some
  observations on grid convergence}}, Computers and Fluids 35~(7) (2006)
  688--692.
\newblock \href {http://arxiv.org/abs/fld.1} {\path{arXiv:fld.1}}, \href
  {http://dx.doi.org/10.1016/j.compfluid.2006.01.003}
  {\path{doi:10.1016/j.compfluid.2006.01.003}}.
\newline\urlprefix\url{http://linkinghub.elsevier.com/retrieve/pii/S0045793006000053}

\bibitem{eca-et-al_2013}
L.~E{\c{c}}a, M.~Hoekstra, J.~F. {Beja Pedro}, J.~A.~C. {Falc{\~{a}}o de
  Campos}, \href{http://doi.wiley.com/10.1002/fld.3737}{{On the
  characterization of grid density in grid refinement studies for
  discretization error estimation}}, International Journal for Numerical
  Methods in Fluids 72~(1) (2013) 119--134.
\newblock \href {http://arxiv.org/abs/fld.1} {\path{arXiv:fld.1}}, \href
  {http://dx.doi.org/10.1002/fld.3737} {\path{doi:10.1002/fld.3737}}.
\newline\urlprefix\url{http://doi.wiley.com/10.1002/fld.3737}

\end{thebibliography}

%

\end{document}